%% file: traffic_aware_topology.tex
\documentclass[acmsmall]{acmart}
\usepackage{graphicx}
\usepackage[english]{babel}
\usepackage[utf8]{inputenc}
\usepackage{blindtext}
\usepackage{afterpage}
\usepackage{amssymb} 
\usepackage{pgfplots, pgfplotstable}
\pgfplotsset{width=7cm,compat=newest}

\usepgfplotslibrary{statistics, fillbetween}
\setcopyright{none}
\renewcommand\footnotetextcopyrightpermission[1]{} 
\usepackage[outdir=figures/]{epstopdf}
\usepackage{enumitem}
\acmDOI{}

\acmISBN{}

\usepackage{url}

\usepackage{breakurl}

\acmPrice{}
\usepackage{appendix}
\usepackage[linesnumbered, boxed]{algorithm2e}
\usepackage{multirow}
\usepackage{subcaption}
\usepackage{tikz}
\usepackage{mathtools}
\usetikzlibrary{arrows, calc, pgfplots.groupplots, patterns,backgrounds, spy, shadings}

\theoremstyle{definition}

\pgfplotsset{every axis/.append style={
	        ylabel near ticks,
	        xlabel near ticks,
	        ylabel style={font=\scriptsize},
	        xlabel style={font=\scriptsize},
                    xticklabel style={font=\tiny},
                    yticklabel style={font=\tiny},
            }}

\begin{document}
\newcommand{\metteor}{{COUDER}}

\title[]{\LARGE COUDER: Robust Topology Engineering for Optical Circuit Switched Data Center Networks}


\author{Min Yee Teh}
\affiliation{%
  \institution{Columbia University}}

\author{Shizhen Zhao}
\affiliation{%
  \institution{Shanghai Jiao Tong University}}

\author{Peirui Cao}
\affiliation{%
  \institution{Shanghai Jiao Tong University}}

\author{Keren Bergman}
\affiliation{%
  \institution{Columbia University}}


\begin{abstract}
Many optical circuit switched data center networks (DCN) have been proposed in the past to attain higher capacity and topology reconfigurability, though commercial adoption of these architectures have been minimal. One major challenge these architectures face is the difficulty of handling uncertain traffic demands using commercial optical circuit switches (OCS) with high switching latency. Prior works have generally focused on developing fast-switching OCS prototypes to quickly react to traffic changes through frequent reconfigurations. This approach, however, adds tremendous complexity to the control plane, and raises the barrier for commercial adoption of optical circuit switched data center networks.

We propose {\metteor}, a robust topology and routing optimization framework for reconfigurable optical circuit switched data centers. {\metteor} optimizes topology and routing based on a convex set of traffic matrices, and offers strict throughput guarantees for any future traffic matrices bounded by the convex set. For the bursty traffic demands that are unbounded by the convex set,  we employ a desensitization technique to reduce performance hit. This enables {\metteor} to generate topology and routing solutions capable of handling unexpected traffic changes without relying on frequent topology reconfigurations. Our extensive evaluations based on Facebook's production DCN traces show that, even with daily reconfiguration, {\metteor} achieves about 20\% higher throughput, and about 32\% lower average hop count compared to cost-equivalent static topologies. Our work shows that adoption of reconfigurable topologies in commercial DCNs is feasible even without fast OCSs.

\end{abstract}
\maketitle
\newcommand{\floor}[1]{\left\lfloor #1 \right\rfloor}
\newcommand{\ceil}[1]{\left\lceil #1 \right\rceil}

\section{Introduction}\label{section_introduction}
Given the explosive growth in data center traffic, building networks that meet the requisite bandwidth has also become more challenging. Modern data center networks (DCN) typically employ multi-rooted tree topologies~\cite{leiserson1985fat}, which have a regular structure and redundant paths to support high availability. However, uniform multi-rooted trees are inherently suboptimal for carrying highly skewed traffic that is common in DCNs~\cite{kandula2009flyways, roy2015facebook}. This has motivated several works on using optical circuit switches (OCS) to enable more performant data center architectures~\cite{farrington2011helios, cthrough_wang_2011}. Compared to conventional electrical packet switches, OCSs offer much higher bandwidth with lower power consumption. Most importantly, OCSs introduces the possibility of Topology Engineering (ToE), giving reconfigurability to the DCN topology for dynamic link-allocation between ``hotspots’’ to alleviate congestion.

Despite having shown immense promisein prior works, optical circuit-switched data centers have not been widely deployed even after a decade’s worth of research efforts. The main challenge lies in performing ToE under bursty traffic demands. Early works on ToE proposed reconfiguring topology preemptively using a single estimated traffic matrix (TM)~\cite{farrington2011helios, cthrough_wang_2011}. However, the bursty nature of DCN traffic makes forecasting TMs accurately very difficult~\cite{benson2010network, kandula2009nature}. An inaccurate prediction may lead to further congestion. Even if predictions were accurate, the forecast could still turn stale if reconfiguration takes tens of milliseconds. Subsequent works have thus focused on designing OCSs capable of microsecond-level switching to enable faster reaction to traffic variations~\cite{ProjectSirius, ghobadi2016projector, porter2013integrating, rotornet_mellette2017}. However, these architectures rely on frequent topology and routing updates, which adds significant complexity to the control plane, and disincentivizes adoption by large vendors.

We tackle bursty DCN traffic from a different perspective, using a robust optimization-based ToE framework called {\metteor} (\textbf{C}onvex hull \textbf{O}ptimized with \textbf{U}ncertainty \textbf{D}esensitization for \textbf{E}nhanced \textbf{R}obustness). While prior works optimize topology for a \emph{single} estimated TM~\cite{cthrough_wang_2011, halperin2011augmenting}, our approach optimizes topology based on a predicted \emph{convex set} of TMs. For any traffic matrix that is bounded by this convex set, {\metteor} offers strict throughput guarantees; for outlier TMs not bounded by the convex set, {\metteor} employs a desensitization technique to reduce the performance degradation caused by unexpected traffic bursts. Owing to its robustness to traffic uncertainty,  our approach is less reliant on frequent OCS reconfiguration to handle traffic changes. To our knowledge, {\metteor} is the first framework that tackles ToE from a robust optimization perspective.


Contrary to prior ToE solutions based on non-commercial OCS prototypes that require sophisticated controls~\cite{ghobadi2016projector, hamedazimi2014firefly}, the source of {\metteor}’s complexity is its algorithm’s design. We first discuss the assumed network architecture and system-level considerations for OCS reconfiguration in \S\ref{section_challenges_overall_approach}. Next, using an initial naive formulation of {\metteor}, we analyze the fundamental but subtle shortcomings of this approach in \S\ref{InsightsLearned}. The insights and analyses done here were instrumental to the refinement of {\metteor}’s final implementation, to the point where it is currently able to provide strong throughput guarantees for bounded TMs, while maintaining solution robustness for out-of-bound TMs. Still, the entire topology-routing optimization problem is an NP-complete integer programming problem. To make the problem tractable, we relax the complexity by computing a fractional topology as an intermediate solution first, and then round the intermediate solution to an integer one using a lagrangian method. These steps are detailed in \S\ref{section_overall_methodology}.

In \S\ref{section_performance_evaluation}, we evaluate {\metteor}'s performance using both production DCN traces from Facebook~\cite{roy2015facebook} and synthetically-generated traffic matrices. Performance is measured using two metrics: maximum link utilization (MLU) and average hop count (AHC). Although there is a clear gap in performance between {\metteor} and an instantaneously-reconfigurable ideal ToE solution with \emph{a-priori} traffic knowledge, {\metteor}’s performance is attained under daily reconfiguration, which can be readily achieved using current commercial OCSs,  with only a minor increase management overhead to the SDN controller. Our evaluation results also show that daily reconfiguration is sufficient for {\metteor} to outperform other static DCN topologies\footnote{We posit that more infrequent reconfigurations might be feasible, though we have sufficient data to validate this conjecture, as Facebook’s traces contains only one day's worth of traffic. }. Compared to a static uniform mesh topology, {\metteor} reduces the MLU by about $20\%$, and the AHC by about $32\%$. Compared to a fat tree of comparable cost, {\metteor} reduces the MLU by about $50\%$, and the AHC by about $60\%$. Finally, we use packet level simulations to study how operator-centric metrics like MLU and AHC may affect user-centric application-level performance, such as flow completion time, in \S\ref{subsection:fine_grained_netbench_simulation}. 

In short, the contributions of our work are as follows:
\begin{itemize}
\item We present a topology engineering framework that is robust to traffic uncertainties called {\metteor}. {\metteor} optimizes topology and routing based on a convex traffic set to deliver strong throughput guarantees for traffic bounded by the convex set, and uses a desensitization technique to handle out-of-bounds traffic.
\item We run thorough analysis on production inter-pod traffic, and validate the feasibility of predicting future TMs with a convex set. Specifically, we found that about 92\% of traffic matrices can be bounded by under 30 minutes’ worth of historical traffic.
\item Our approach is robust to traffic changes, and is able to achieve good performance without relying on frequent topology-reconfigurations. This greatly reduces the implementation complexity for commercial adoption of ToE.
\end{itemize}

\begin{figure}[!tp]
\centering
\includegraphics[draft=false, width=2.7in, height=1.4in, trim={0.5cm 0.5cm 0.5cm 0.5cm}]{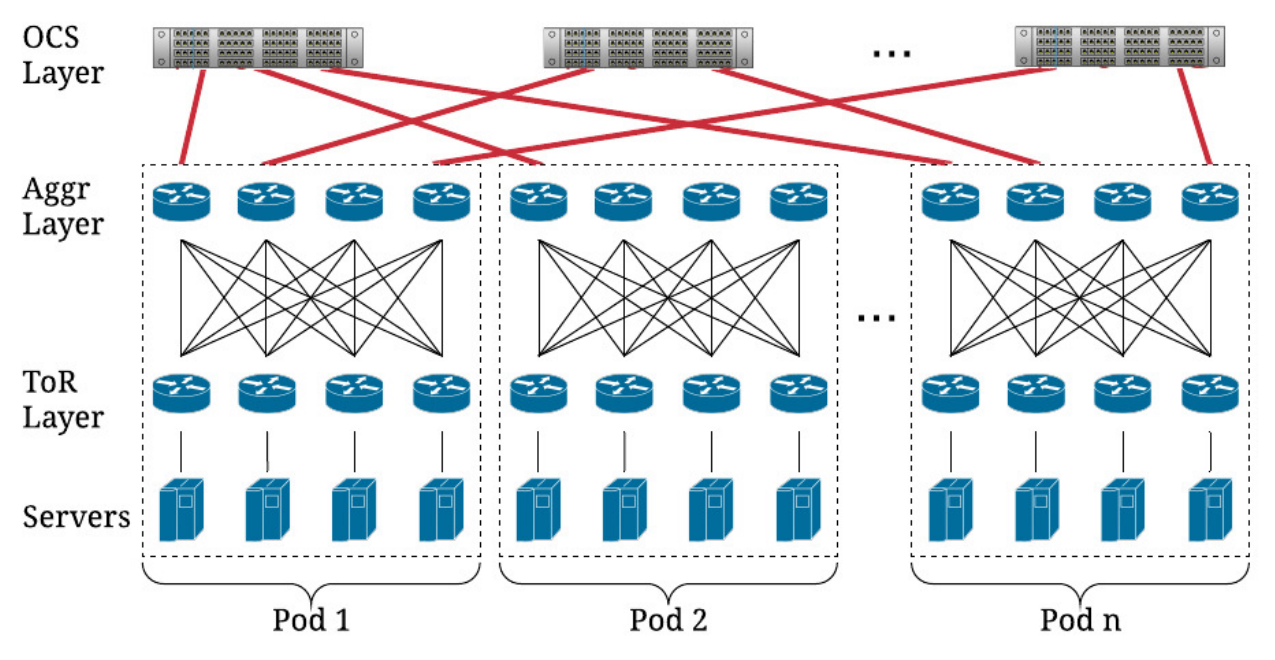}
\caption{\small An illustration of {\metteor}’s DCN network physical topology. A pod is the basic unit of network deployment. Pods are fully-interconnected physically via a layer of OCSs at the core layer. Reconfiguring the OCS switch states realizes a different logical pod-level topology.}
\label{helios_topology}
\vspace{-8pt}
\end{figure}

\section{Related Work}\label{section_related_works}
\subsection{Traffic-Agnostic DCN Topology}
DCN topologies have been traditionally designed to be static and traffic-agnostic, focusing on bisection bandwidth, scalability, failure resiliency, etc. They can be divided into either Clos-like and mesh-like topologies. Clos topology (e.g., Fat-Tree~\cite{al2008scalable, liu2013f10}) is more widely-adopted in large-scale data centers (e.g.,  Google~\cite{singh2015jupiter}, Facebook~\cite{farrington2013facebook}, Cisco~\cite{cisco2016}, and Microsoft~\cite{greenberg2009vl2}), as its regular hierarchical structure simplifies routing and congestion control. Mesh-like expander topology also shows great promise due to its flat hierarchy that, unlike Clos, eliminates the need for the spine layer. This saves cost and offers rich path diversity~\cite{singla2012jellyfish, valadarsky2015xpander, yu2016space}. 


However, DCN traffic is inherently skewed. A study from Microsoft~\cite{kandula2009flyways} showed that only a few top-of-rack (ToR) switches are ``hot'' in a small (1500-server) production data center. Facebook~\cite{roy2015facebook} reported that the inter-pod traffic in one of their data centers varies over more than seven orders of magnitude. As a result, static and traffic-agnostic network topologies can be inherently suboptimal when subject to skewed DCN traffic.

\subsection{Traffic-Aware DCN Topology}\label{section_reconfig_topo}
To handle fast-changing, high-skewed traffic patterns, some researchers have argued for reconfigurable DCN topologies based on optical circuit switches (OCS)~\cite{vahdat2011emerging, liu2010scaling, fields2010transceivers, zhou2017datacenter}. The pioneering work,  Helios~\cite{farrington2011helios}, proposed reconfiguring pod-to-pod topology using OCSs based on a \emph{single} estimated traffic matrix. However, reconfiguring Helios incurs a significant delay (about 30ms), a problem that most commercial OCSs today still face~\cite{calient}. Given that 50\% of DCN flows lasting below 10ms~\cite{kandula2009nature}, a 30ms reconfiguration latency could mean that the topology optimized for pre-switching traffic may no longer be a good fit for post-switching demands.

The need to cope with rapid traffic changes motivated subsequent works aimed at decreasing reconfiguration latency for OCSs. Some of these have focused on providing ToR-level reconfigurability~\cite{ reactor_liu2014, cthrough_wang_2011, singla2010proteus}, potentially reducing latency to microseconds level using sophisticated hardware.  However, these approaches might not scale to data centers with thousands of ToRs, due to the low radix of ToRs and the finite size of OCSs. Others have proposed scaling up reconfigurable networks with steerable wireless transceivers~\cite{ghobadi2016projector, hamedazimi2014firefly, zhou2012mirror}, but these architectures face serious deployment challenges related to environmental conditions in real DCNs, and to the need for sophisticated steering mechanisms.
The Opera architecture~\cite{mellette2019expanding}, built using rotor switches from~\cite{rotornet_mellette2017}, forms a mesh-like expander topology by multiplexing a set of preconfigured matchings in the time-domain. Unfortunately, frequently changing OCS connections may overload the SDN controller, and undermines data center availability as a result.

Some prior works have also looked into circuit-scheduling algorithms in the presence of reconfiguration delays~\cite{bojja2016costly, liu2015scheduling, wang2018neural}. Their problem setup fundamentally differs from that of ours, as we are interested in designing a \emph{single} topology optimized for \emph{many} different traffic demands, while circuit-scheduling algorithms are typically concerned with sequencing circuit configurations to better serve a \emph{single} traffic demand. 


\subsection{Traffic Engineering} \label{section_traffic_engineering_related_works}
Careful traffic engineering (TE) is required to fully realize the potential of reconfigurable topologies. TE generally consists of 2 components, namely path-selection and load-balancing. Path-selection phase selects a set of candidate paths for carrying traffic. and the load-balancing phase calculates the splitting-weights for the candidate paths. 

Path-selection in data centers typically rely on the K-shortest-path algorithm~\cite{yen_ksp, singla2012jellyfish, valadarsky2015xpander}. As for load balancing, nearly all prior works on optical circuit-switched data centers compute routing weights by solving a multi-commodity flow (MCF) problem on a single predicted traffic matrix~\cite{hamedazimi2014firefly, farrington2011helios, cthrough_wang_2011}. However, predicting a TM accurately can be difficult, and an inaccurate traffic prediction may incur unexpected congestion. Rotornet~\cite{rotornet_mellette2017} load-balances traffic using Valiant load-balancing (VLB)~\cite{zhang2008designing}. VLB has several desirable properties, such as being a lightweight traffic-agnostic routing algorithm, robust under demand uncertainties by splitting traffic via indirect paths, and having a guaranteed worst-case throughput-reduction of 2$\times$. However, DCN operators have access to a wealth of historical traffic data to know the range within which traffic will likely occur. This makes VLB overly conservative. Some literature use robust optimization to strike a balance between routing performance and robustness to traffic uncertainty~\cite{wang2006cope, zhang2005optimal, chang2017robust}. Although these solutions are mainly designed for wide area networks (WAN), the core ideas are equally applicable to DCNs.

\section{Weak Traffic Stability}\label{section_proof_of_concept}

\begin{figure}[!tp]
\begin{subfigure}[c]{0.42\linewidth}
\input{plot_data/facebook_data/boundability_traffic_snapshots.pgf}
\vspace{-5pt}
\caption{\small Traffic boundability of Facebook’s traces. Dotted vertical line marks 30-minute lookback.}
\label{fig:boundability}
\end{subfigure}
~
\begin{subfigure}[c]{0.42\linewidth}
\includegraphics[draft=false, height=1.5in, scale=0.78, trim={0cm 0cm 0cm 0.cm}]{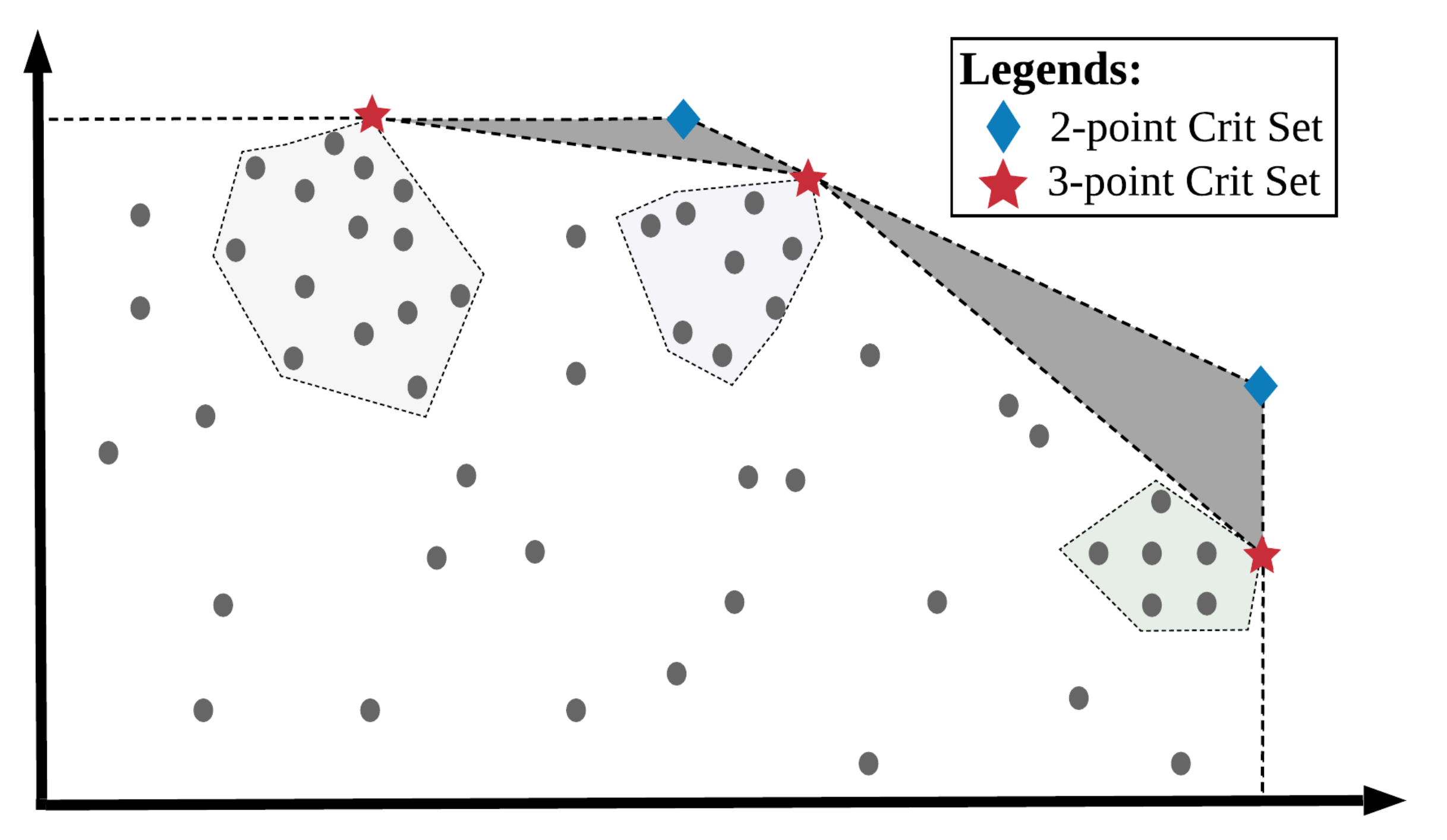}
\vspace{-10pt}
\caption{\small Topology engineering based on convex set of critical traffic matrices.}
\label{fig:convex_set_prediction}
\end{subfigure}
\vspace{-10pt}
\caption{a) Length of lookback window needed to bound inter-pod TMs derived from Facebook DCN clusters . b) A convex set that bounds all other TMs. In general, a smaller convex set forms a larger formed.}
\vspace{-9pt}
\label{fig:facebook_traffic_analysis}
\end{figure}

The belief in a lack of temporal stability in DCN traffic has driven much work on designing faster OCSs and control planes in the past. However, DCN traffic is not entirely random, especially at the pod level. Although pod-level traffic matrices (TM) certainly do not generally exhibit strong temporal stability, they do exhibit a weak form of temporal stability. This makes it possible to find a range (or bound) that would contain most TMs, even though accurate per-traffic matrix prediction may be extremely difficult.

We demonstrate \emph{weak stability} with a simple case study on Facebook’s published data center traces, which contains a worth of one-day's of TM snapshots obtained from a Facebook data center~\cite{roy2015facebook}. This Facebook's data center consists of three clusters of pods (a database cluster, a web search cluster and a hadoop cluster), with intra-cluster communication dominating the entire network. The packet traces were collected with a 1:30000 sampling rate. The packet traces were aggregated into a sequence of 1-second-averaged inter-pod TMs. Then, for each TM in the sequence, we gradually increase the lookback window into the past until the current snapshot can be bounded by traffic snapshots in the lookback window.

Our approach in computing the convex set is motivated by that in~\cite{zhang2005finding}. Using $k$-means clustering, we first group all the considered TMs into $K$ clusters of TMs, and for every cluster compute a component-wise max TM. It is obvious that these $K$ component-wise max TMs form a convex set, which we refer to interchangeably as \emph{critical TMs}. Note that finding the smallest convex set (i.e. the convex hull) is significantly more challenging, especially for high-dimensional TMs.

Let $\{T_1,T_2,...,T_K\}$ be the set of critical TMs for the current interval of TMs. For every TM $T$ in the next interval, we say $T$ is bounded by the critical TMs if there exist $\lambda_1,\lambda_2,...,\lambda_K\geq 0, \sum_{k=1}^K\lambda_k \leq 1$, such that
$T=\lambda_1 T_1+\lambda_2 T_2+\cdots+\lambda_K T_K.$

Fig. \ref{fig:boundability} shows the CDF of TM snapshots bounded by the critical TMs in the preceding interval, as a function of the lookback duration. We generated three curves for the three DCN clusters that run different application mixes respectively, and also one curve by combining the individual clusters. Clearly, more TMs become boundable as the longer the lookback window. Specifically, slightly over 92\% of TMs are boundable given just a 30-minute lookback window. If we extend the lookback window to 3 hours, then nearly all TMs are boundable.

This finding has motivated us to design a robust topology engineering based on the convex set generated by critical TMs. Exploiting the weak temporal stability would in theory allow us to optimize the topology to be sufficiently good for the general case traffic. As long as the new TM is bounded by this convex set, topology reconfiguration is not necessary. However, since there are always TMs that are outside of the convex set, employing proper techniques to ensure solution robustness for out-of-bound TMs is both a major challenge and a key contribution of this work.

\section{System Level Overview of COUDER}\label{section_challenges_overall_approach}

\begin{figure*}[!tp]
\centering
\includegraphics[trim={0.2cm 0.7cm 0.2cm 0.7cm}, width=0.95\linewidth, height=1.05in] {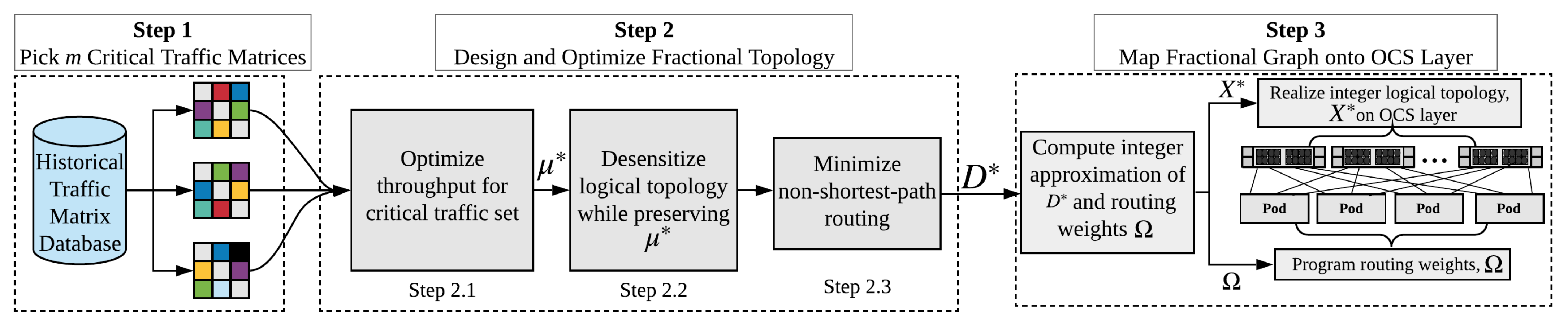}
\caption{\small Illustrating the complete software workflow of {\metteor}.}
\vspace{-19pt} 
\label{topology_engineering_overall_workflow}
\end{figure*}

\subsection{Network Architecture}\label{section_network_architecture}

The assumed DCN architecture is shown in Fig. \ref{helios_topology}, with a layer of OCSs interconnecting a number of pods. A pod is a typical deployment unit for data centers, whose network fabric can be built from either monolithic switches like CE12800~\cite{huawei}, or small-radix switches organized in a non-blocking, Clos-like structure~\cite{singh2015jupiter, farrington2013facebook}. An OCS is a fully optical component that sends incoming optical signals directly to a reconfigurable egress port without packet decoding. Although OCS reconfiguration could take tens of milliseconds, upon completion a new inter-pod topology is realized. Since OCSs do not buffer packets, all paths through the established OCS circuits are transparent to in-flight packets. In contrast to many flexible network architectures with inter-ToR reconfigurability~\cite{ghobadi2016projector, hamedazimi2014firefly, porter2013integrating, cthrough_wang_2011, zhou2012mirror}, however, we focus on inter-pod reconfigurability mainly for the following reasons: 
\begin{itemize}[leftmargin=3pt]
\item \emph{Scalability: } Using pods with hundreds of uplinks to the OCSs, our architecture could support up to about 100 pods. Since each pod could contain $\Theta(1000)$ servers, our architecture can scale up to over 100k servers.  

\item \emph{Traffic stability: } Inter-pod traffic shows more noticeable locality~\cite{roy2015facebook}, and is more stable than inter-ToR traffic~\cite{delimitrou2012echo, kandula2009nature} due to averaging effects from the aggregation switches. This stability makes it much more feasible to not rely on on-demand circuit switching.

\item \emph{High fan-out\footnote{Ability to form direct links with many destinations,.}: } Pods have much higher fan-out than ToRs. Combined with multi-hop routing, every pod is reachable within one or two hops, making it possible for \emph{one} logical topology to serve several, possibly dense\footnote{While inter-ToR traffic matrices are generally quite sparse~\cite{ghobadi2016projector}, inter-pod traffic matrices tend to be more dense, with mostly non-zero entries.}, traffic matrices.

\item \emph{Compatibility with current technology: } Interconnecting pod uplinks with spines using single-mode fibers and optical transceivers is already commonplace in current fat tree-based DCNs~\cite{singlemodefiber2019}. Given that single-mode transceivers with link margin of $>5$dB, which is much higher than the insertion loss of $\leq 3$dB common to commercial OCSs, our proposed network architecture (see Fig. \ref{helios_topology} can be easily realized by simply replacing the spine switches with commercial OCSs. This makes for a smoother transition into optically-connected data centers. 

\end{itemize}

In this paper, we refer to the (fixed) physical connections between the pod and OCSs as the \emph{physical topology}. Topology engineering reconfigures the OCSs to realize a specific \emph{logical topology} as an overlay on the physical topology. 

\subsection{Computing Logical Topology}\label{section_computing_logical_topology}
Prior works have designed reconfigurable topology based on a \emph{single} estimated traffic matrix, obtained either from switch measurements (e.g. Hedera~\cite{al2010hedera}) or from end-host buffer occupancy~\cite{cthrough_wang_2011}. However, due to the bursty nature of DCN traffic \cite{kandula2009nature}, even inter-pod traffic can be difficult to predict accurately, which fundamentally limits the robustness of such an approach.

{\metteor} takes as an input a set of multiple TMs that form a convex set, and outputs a logical topology and routing configuration optimized for the input TMs. The first step is to obtain multiple critical TMs whose convex hull covers historical traffic snapshots (see Step 1 of Fig. \ref{topology_engineering_overall_workflow}). \S\ref{section_proof_of_concept} presents an approach to compute a set of critical TMs, and Fig. \ref{fig:boundability} shows that a majority of future traffic snapshots can be bounded by these critical TMs. 


The next step is to optimize topology for the convex TM set, which presents the main algorithmic challenge of this paper. We begin the discussion using a straightforward but naive formulation in \S\ref{section_model}. We flesh out the shortcomings of our initial formulation in \S\ref{InsightsLearned}, a process which has also taught us valuable insights that were used to refine {\metteor}’s final implementation. However, our topology optimization problem remains an NP-Complete integer programming problem. To reduce algorithmic complexity, we develop a number of relaxation techniques detailed in \S\ref{section_overall_methodology} to make the problem polynomial-time solvable. Our relaxation techniques result in much lower optimality loss than those used in prior works, which is validated numerically in Appendix \ref{appendix_reconfiguration_algo_analysis}.

\subsection{Reconfiguring Logical Topology Safely}

Existing topology engineering solutions generally rely on frequent topology-reconfigurations to react to traffic changes. However, reconfiguring topology frequently, even when done properly, could place a tremendous amount of workload on the SDN controller. When done improperly, a misstep in reconfiguration could compromise DCN availability~\cite{govindan2016evolve, mogul2017thinking}. For instance, a poor-choice of switching configuration, or even a bug in the network controller, carries the risk of failing entire DCN blocks; an admittedly rare risk, but one that increases with the rate of reconfiguration. 

To our knowledge, {\metteor} is the first topology engineering approach that does not rely on frequent reconfigurations to react to traffic changes. Our evaluation results in \S\ref{subsection:reconfig_frequency} shows that {\metteor} can provide sufficiently good performance with just daily reconfiguration, and reconfiguring topology more frequently only brings marginal improvements. This opens up the possibility of prioritizing ``reconfiguring safely’’ over ``reconfiguring quickly’’.


There are two major safety considerations when reconfiguring topology. First, topology reconfiguration must be carefully sequenced to avoid routing packets into ``black holes''.  The SDN controller must first ``drain’’ links by informing packet switches not to route traffic through the optical links that are about to be switched. Only upon verifying that no traffic flows through these links can physical switching take place. After switching completes, the SDN controller can then ``undrain’’ links and start sending traffic through them again. 

Second, topology reconfiguration needs to be staged to maintain sufficient network capacity, especially when traffic demands are high. For instance, if 60\% of links need to be reconfigured when network utilization is at 80\%, the entire reconfiguration process should take at least 3 stages (switching 20\% of links in each stage) to avoid congestion. Note that it is possible to lower the number of reconfiguration stages by applying the minimal rewiring optimization developed in~\cite{zhao2018minimal}, though this is beyond the scope of this work.

\section{Preliminary Mathematics}\label{section_model}
In this section, we introduce the recurring mathematical notations and definitions used throughout this paper. All notations are tabulated in Table \ref{table:notations}.

\subsection{Logical Topology}
\noindent Let $\mathcal{S}=\{s_1,..,s_N\}$ be the set of pods, $\mathcal{O}=\{o_1,..,o_M\}$ be the set of OCSs, and $x_{ij}^m$ be the number of directed links from pod $s_i$ to pod $s_j$ through OCS $o_m$. A logical topology is represented by $X=[x_{ij}], i,j=1,...,n$, where $x_{ij} = \sum_{m=1}^M  x_{ij}^m $ denotes the number of links between pods $s_i$ and $s_j$. Logical topology $X$ can be feasibly realized by a physical topology, \emph{i.f.f.} it satisfies the following:\\
\textbf{OCS-level (Hard) Physical Constraints}: 
\begin{equation}\label{constraint:ocslevel}
\begin{aligned}
&\text{1)} \;\; \sum_{j=1}^N x_{ji}^m \leq h_{\text{ig}}^m(i), \sum_{j=1}^N x_{ij}^m \leq h_{\text{eg}}^m(i),\; \forall  i=1,..,N, \; m = 1,..,M;&\\
&\text{2)} \;\; x_{ij} = \sum_{m=1}^M  x_{ij}^m, \hspace{2mm}x_{ij}\text{ and }x_{ij}^m \text{ are all integers;}
\end{aligned}
\end{equation}
where $h_{\text{ig}}^m(i), h_{\text{eg}}^m(i)$ are the number of ingress/egress links of pod $s_i$ through OCS $o_m$. 

\subsection{Path Selection for Routing}\label{subsection:path_selection}
\begin{figure}[!tp]
\begin{subfigure}[c]{0.51\textwidth}
\centering
\includegraphics[draft=false, width=1.1\textwidth, height=1.4in, trim={0.cm 0.cm 0.cm 0.5cm}]{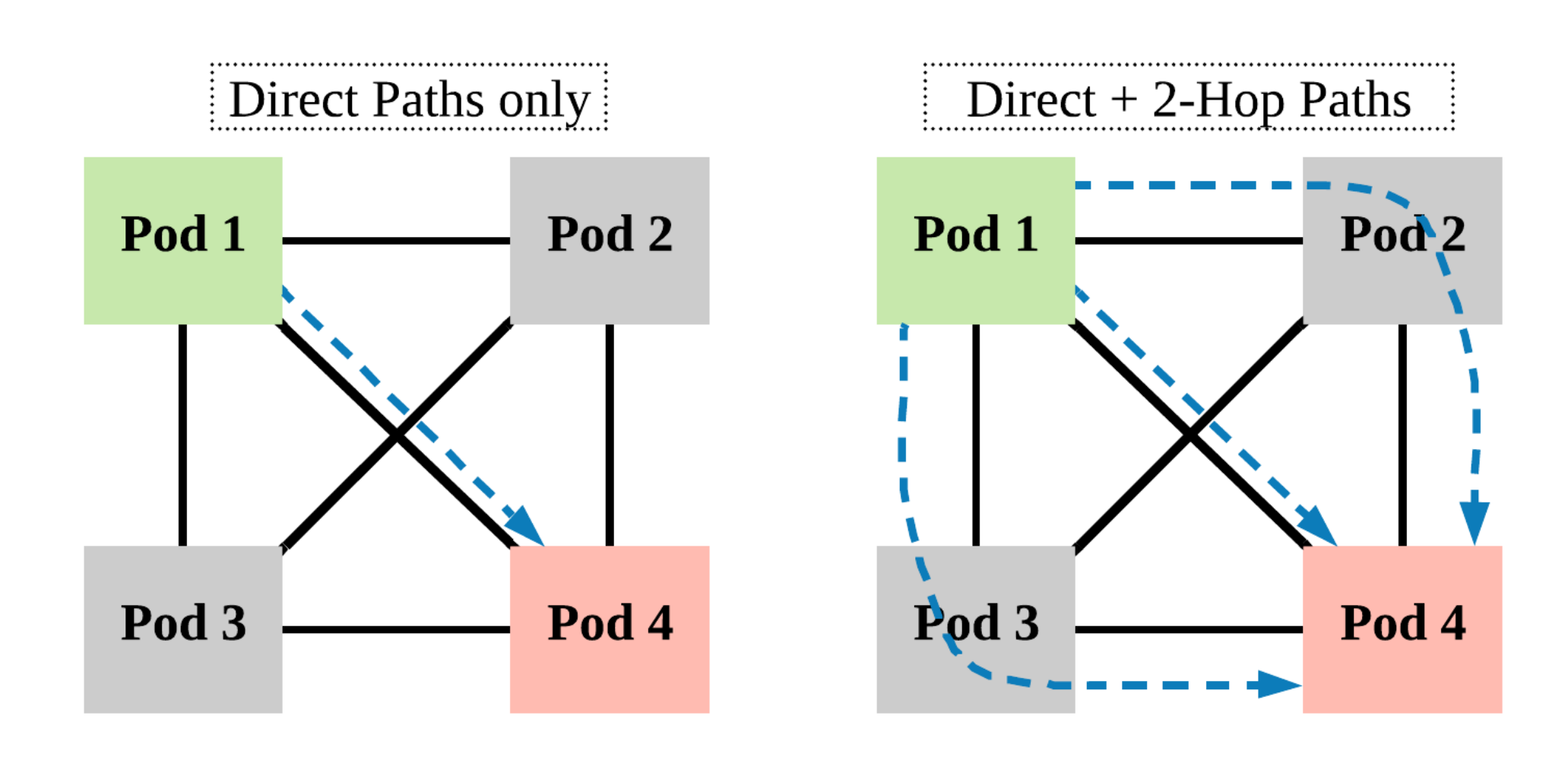}
\vspace{-14pt}
\caption{\small 1 and 2 hop paths.}
\end{subfigure}
~
\begin{subfigure}[c]{0.4\textwidth}
\centering
\vspace{4pt}
\input{plot_data/facebook_data/path_len_total_capacity.pgf}
\vspace{-16pt}
\caption{\small Average inter-pod path capacity}
\end{subfigure}
\vspace{-11pt}
\caption{\small a) {\metteor} routes traffic along all inter-pod paths with a maximum of 2 hops. b) 1 and 2 inter-pod hop paths can sufficiently utilize the most of the entire DCN capacity without incurring significant complexity.}
\vspace{-13pt}
\label{fig:two_hop_routing}
\end{figure}

Just as effective routing strategies are important to efficiently utilize the provisioned DCN capacity, path-selection is an integral part of any routing strategy. {\metteor} considers all node-disjoint paths with at most two-hops between any source-destination pods (see Fig.~\ref{fig:two_hop_routing}a). The 1-hop paths are referred to as direct paths, while the 2-hop paths from source to destination via an intermediate pod are referred to as indirect paths. The rationales for choosing two-hop paths are:
\begin{enumerate}
\item Compared to only the direct paths, considering both direct and 2-hop paths greatly increases the total path capacity between all pod pairs.
\item Considering paths of lengths greater than 2 hops drastically increases both computational and implementation complexity, but does not bring much additional path capacity.
\end{enumerate}
We validate our arguments with a simple experiment. A typical production data center has tens of pods, and each pod generally has hundreds of uplinks~\cite{farrington2013facebook, andreyev2014introducing}. We randomly generate 200 pod-level topology instances in our experiment, with each instance having 20 pods with 128 uplinks of unit bandwidth. For each of the generated instances, we compute the average path capacity between all pod-pairs by considering all node-disjoint paths with hop counts up to $x$, where $x$ is a variable between 1 and 4. As seen in Fig. \ref{fig:two_hop_routing}, considering 2-hop paths greatly increases the available path capacity over just 1-hop direct paths, although further considering 3- and 4-hop paths only yield minor improvements. 


\subsection{Defining Throughput for One Traffic Matrix}
Let $T= [t_{ij}], i,j=1,...,n$ be a TM, where $t_{ij}$ is the traffic rate (in Gbps) from pod $s_i$ to pod $s_j$. Given a logical topology $X$, then, the throughput of $T$ over a logical topology, $X$, is the maximum $\mu$ such that the $\mu T$ can be feasibly routed over $X$. Routing feasibility is defined as follows. 

Let  $\mathcal{P}_{ij}=\{[s_i, s_j], [s_i, s_1, s_j], …, [s_i, s_n, s_j]\}$ be the set of all paths between $(s_i, s_j)$, and $\mathcal{P}= \cup_{(i,j)} \mathcal{P}_{ij}$. (Note that OCSs are transparent to in-fly packets, so passing through an OCS does not constitute an extra hop.) The feasibility of routing $T$ over $X$ can be verified using:
\begin{eqnarray}\label{feasible_constraints}
&&\hspace{-6mm} \text{There exist } \Omega=\{\omega_p\}, p\in \mathcal{P} \text{ such that }\\
&&\hspace{-6mm} 1) \sum_{p \in \mathcal{P}_{ij}} \omega_p = 1, \;  \forall \; i, j=1,...,N \nonumber\\
&&\hspace{-6mm} 2) \sum_{p \in \mathcal{P}, \; (s_i, s_j) \in p} \omega_p t_{\text{src}_p\text{dst}_p} \leq x_{ij} b_{ij}, \; \forall \; i, j=1,...,N\nonumber
\end{eqnarray}
where $b_{ij}$ is the link capacity between $s_i$ and $s_j$, and $\omega_p$ is the fraction of the traffic $t_{\text{src}_p\text{dst}_p}$ routed (in Gbps) via path $p$. The first constraint splits the traffic amongst all the candidate paths, while the second constraint ensures that no link can carry traffic at a rate that exceeds its bandwidth.

When computing throughput, we scale $T$ until the maximum link utilization (MLU) hits 1, where link utilization is the ratio of a link’s traffic flow rate to its capacity. This problem is related to minimizing max link utilization (MLU) for routing an unscaled $T$ over $X$, since throughput is related to MLU as a reciprocal. 

\subsection{Defining Max-min Throughput for Multiple Traffic Matrices}
Given $K$ critical traffic matrices, $\{T_1, .. , T_K\}$, let $\{\mu_1, .. , \mu_K\}$ be the throughputs of routing $\{T_1, .. , T_K\}$ over $X$. We aim to design $X$ that maximizes $\min (\mu_1, .. , \mu_K)$ (max-min) throughput: 
\vspace{-2pt}
\begin{eqnarray}\label{overall_formulation}
&& \max_{X}\mu=\min\{\mu_1, .. ,\mu_K\} \text{, s. t}\\
&& 1) \text{ } X\text{ is an integer matrix that satisfies (\ref{constraint:ocslevel})} \nonumber\\
&& 2) \text{ } (X, \mu_k T_k)\text{ satisfies (\ref{feasible_constraints})}, \; \forall \; k \in \{1, .., K\} \nonumber
\end{eqnarray}

Alternatively, we could maximize the average throughput of all critical TMs, but we avoid this as it gives the optimization solver freedom to selectively maximize the throughputs of the “easier” TMs. Instead, solving (\ref{overall_formulation}) ensures that the optimal (max-min) logical topology solution maximizes all TM throughputs as evenly as possible.


\begin{table}[!t]
\small
\begin{tabular}{|p{3cm}|p{7.06cm}|}
\hline
\textbf{Notation} & \textbf{Description} \\
\hline
$\mathcal{S} = \{s_1, .., s_N\}$ & Set of all $N$ pods \\
\hline
$\mathcal{O} = \{o_1, .., o_M\}$& Set of all $M$ circuit switches\\
\hline
$x_{ij}^m$ & Integer number of pod $i$'s egress links connected to pod $j$'s ingress links through $o_m$\\ 
\hline
$X = [x_{ij}] \in \mathbb{N}^{n\times n}$ & Inter-pod topology; $x_{ij}$ denotes the number (integer) of $s_i$ egress links connected to ingress links of $s_j$\\ 
\hline
$T = [t_{ij}] \in \mathbb{R}^{N \times N}$ & Traffic matrix, where $t_{ij}$ denotes the traffic rate (Gbps) sent from $s_i$ to $s_j$\\ 
\hline
$D = [d_{ij}] \in \mathbb{R}^{N \times N}$ & Fractional topology; $d_{ij}$ denotes the number (fractional) of $s_i$ egress links connected to ingress links of $s_j$\\ 
\hline
$h_{eg}^m(s_i), h_{ig}^m(s_i)$ & Number of physical egress and ingress links, respectively, connecting $s_i$ to $o_m$\\
\hline
$r_{\text{eg}}^i, r_{\text{ig}}^i$ & Number of egress and ingress links, respectively, of $s_i$\\
\hline
$b_{ij}$ & Link capacity (Gbps) between $s_i$ and $s_j$\\
\hline
$\mathcal{P}_{ij}$ & Set of candidate paths from $s_i$ to $s_j$ \\
\hline
$\text{src}_{p}, \text{dst}_{p}$ & Source and destination pod index of path $p$ \\
\hline
$\omega_{p}$ & Fraction of traffic $t_{\text{src}_{p} \text{dst}_{p}}$ allocated to path $p$ \\
\hline
$\mu$ & Traffic scale-up factor (a. k. a. throughput)\\
\hline
\end{tabular}
\caption{\small Table of recurring mathematical notations used in this paper.} 
\label{table:notations}
\vspace{-15pt}
\end{table}

\section{Potential Pitfalls \& Lessons Learned in Topology Engineering}\label{InsightsLearned}
Our initial formulation of {\metteor} takes the form of (\ref{overall_formulation}), which we will use as a starting point for discussion. Over the course of {\metteor}'s development, this initial formulation led us into several pitfalls, from which we have learned key insights that allowed us to gradually refine {\metteor}. Here, we share a few notable ones based on anecdotal experience. 

\subsection{Lesson 1: Use a single routing weight for optimization}\label{lesson1}
(\ref{overall_formulation}) offers the following performance guarantee:

\begin{lemma}\label{guarantee1}
Let $X^{*}$ be the optimal solution of (\ref{overall_formulation}), and $\mu^{*}$ be the max-min throughput of the critical TMs $\{T_1,...,T_K\}$. For any traffic matrix $T$ bounded by $\{T_1,...,T_K\}$, routing $\mu^{*}T$ over $X^{*}$ is feasible.
\end{lemma} 

\begin{proof}
Given $T=\lambda_1 T_1+\lambda_2 T_2+\cdots+\lambda_K T_K,$
where $0 \leq \lambda_1,...,\lambda_K\leq 1$, and let $\Omega^{k}$ be a feasible routing allocation for $\mu^{*}T_{k}, \; k=1,2,...,K,$ over the logical topology $X^{*}$. A routing allocation $\Omega$ for $\mu^{*}T$ can be constructed as follows:
\begin{equation}\label{construct_routing}
\omega_p=\frac{\sum_{k=1}^K \lambda_{k} \omega_p^k t_{\text{src}_p\text{dst}_p}^k}{t_{\text{src}_p\text{dst}_p}}.
\end{equation}
It is easy to verify that  $\omega_p, \mu^{*}T, X^{*}$ satisfy the constraint (\ref{feasible_constraints}), which completes the proof.
\end{proof}

The biggest problem with (\ref{overall_formulation}) is that it gives the solver free reign to use different routing solutions for different TMs. This makes the optimal routing weights a function of each TM in the convex set. As a result, we would require \emph{a-priori} knowledge of future TMs and the ability to update the routing weights optimized for each future TM for Lemma \ref{guarantee1} to hold, even if the future TMs are bounded by the convex set. These requirements are unrealistic in practice, and thus greatly diminishes the value of the formulation (\ref{overall_formulation}). By introducing a small change to (\ref{overall_formulation}), we arrive at the following:
\begin{eqnarray}\label{formulation2}
&& \max_{X,\Omega=\{\omega_p\}}\mu=\min\{\mu_1, .. ,\mu_K\} \text{, s. t}\\
&& 1) \text{ } X\text{ is an integer matrix that satisfies (\ref{constraint:ocslevel})} \nonumber\\
&& 2) \text{ } (X, \mu_k T_k, \Omega)\text{ satisfy (\ref{feasible_constraints})}, \; \forall \; k \in \{1, .., K\} \nonumber
\end{eqnarray}

(\ref{formulation2}) differs from (\ref{overall_formulation}) in a subtle but important way: its optimization uses a \emph{single} set of routing weights, $\Omega$. By solving for the max-min throughput with a \emph{single} set of routing weights, the routing weight solution (\ref{construct_routing}) becomes a constant function of TMs, and thus Lemma \ref{guarantee2} holds.

\begin{lemma}\label{guarantee2}
Let $X^{**}, \Omega^{**}$ be the optimal solution of (\ref{formulation2}), and $\mu^{**}$ be the max-min throughput value for the critical TMs $\{T_1,...,T_K\}$. Then, for any traffic matrix $T$ bounded by these critical TMs, routing $\mu^{**}T$ over $X^{**}$ is feasible under the routing weight $\Omega^{**}$.
\end{lemma} 

Lemma (\ref{guarantee2}) guarantees that for the logical topology $X^{**}$ and the routing weight solution $\Omega^{**}$, any traffic matrix bounded by the convex set will achieve a throughput no less than $\mu^{**}$, without needing \emph{a-priori} knowledge of the exact traffic matrices, or the ability to update routing weights for every TM. Since we restrict all critical TMs using the same routing weight solution, $\mu^{**}$ would be no larger than $\mu^*$. However, as we will see in the following experiment, $\mu^{**}$ is much easier to achieve, and thus Lemma (\ref{guarantee2}) can be much more useful in practice.

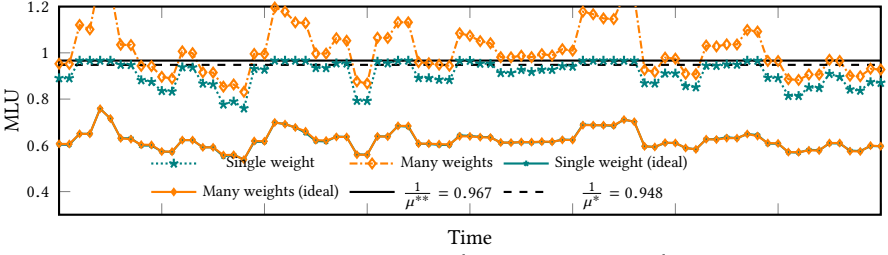
\begin{figure}[!tp]
\pgfplotstableread{plot_data/facebook_data/single_vs_multiple_routingweights_experiment.txt} \mluperf
\centering
\begin{tikzpicture}
\begin{axis}[height=1.7in, width=0.9\linewidth, ylabel near ticks, xlabel near ticks, ylabel={\footnotesize MLU}, ylabel shift = -4pt, xlabel={\footnotesize Time}, xlabel shift = -5pt, xmin=0, xmax=80, ymin=0.3, ymax=1.2, legend columns=3, legend style={at={(0.1,0.34)},anchor=north west, cells={align=left}, draw=none, fill=none}, xticklabels=\empty, ytick pos=left, ylabel style={align=center}, thick]
\addplot+[densely dotted, teal,  mark=star, mark options={solid}, mark repeat={1}, mark size=1.8pt] table[x=x_range, y=single_mlu_unsorted] from \mluperf;
\addlegendentry{\tiny Single weight\;};
\addplot+[densely dashdotted, orange,  mark=diamond, mark options={solid}, mark repeat={1}, mark size=1.8pt] table[x=x_range, y=many_mlu_point_unsorted] from \mluperf;
\addlegendentry{\tiny Many weights\;};
\addplot+[teal,  mark=star, mark options={solid}, mark repeat={1}, mark size=1.1pt] table[x=x_range, y=single_mlu_ideal_unsorted] from \mluperf;
\addlegendentry{\tiny Single weight (ideal)\;};
\addplot+[orange,  mark=diamond, mark options={solid}, mark repeat={1}, mark size=1.1pt] table[x=x_range, y=many_mlu_ideal_unsorted] from \mluperf;
\addlegendentry{\tiny Many weights (ideal)\;};
\addplot+[mark=none, black, domain = 0:80, thick] {0.9665}; 
\addlegendentry{\tiny $\frac{1}{\mu^{**}} = 0.967$\;};
\addplot+[mark=none, black, domain = 0:80, dashed, thick] {0.9480}; 
\addlegendentry{\tiny $\frac{1}{\mu^*} = 0.948$\;};
\end{axis}
\end{tikzpicture}
\vspace{-13pt}
\caption{\small MLU time series comparing optimization techniques using a single \emph{v.s.} many routing weights.}
\label{fig:routing_weights_optimization_single_vs_many}
\vspace{-12pt}
\end{figure}

We conduct an experiment based on the 1-second inter-pod traffic matrix snapshots obtained from~\cite{roy2015facebook}. We select at random a contiguous sequence of TMs, extract $K$ critical TMs (here $K=5$), and then compute two topologies $X^*$ and $X^{**}$ based on (\ref{overall_formulation}) and (\ref{formulation2}), respectively. The MLU performances of the two topologies are evaluated using the same sequence of TMs so that the evaluated TMs are bounded by the convex set. These results are shown in Fig. \ref{fig:routing_weights_optimization_single_vs_many}.

If we had a-priori knowledge of the evaluation TMs, it would be possible to compute the offline-optimal routing weights that minimizes MLU for every TM. In that case, both topologies have almost identical MLU performance. However, assuming \emph{a-priori} knowledge of future TMs is unrealistic. For the topology $X^*$, we could do the next best thing for each TM by optimizing its routing weights based on its previous TM. Fig. \ref{fig:routing_weights_optimization_single_vs_many} shows that by and large, $X^*$ performs poorly in terms of MLU, and even violates the upper bound $\frac{1}{\mu^*}$ offered by Lemma \ref{guarantee1}. For the topology $X^{**}$, even without \emph{a-priori} knowledge, we could use $\Omega^{**}$ to route all the TMs ($\Omega^{**}$ is obtained alongside with $X^{**}$ when solving (\ref{formulation2})). Fig. \ref{fig:routing_weights_optimization_single_vs_many} shows that the MLUs of $X^{**}$ are strictly no greater than $\frac{1}{\mu^{**}}$, indicating that Lemma \ref{guarantee2} holds for the bounded TMs.

\subsection{Lesson 2: Desensitization is necessary in Topology Engineering}\label{lesson2}
Lemma \ref{guarantee2} is useful as it provides an achievable theoretical guarantee for throughput when the evaluated TMs are bounded by the convex TM set. It does not, however, offer any guarantee for outlier TMs that are outside of the convex set. In practice, outlier TMs are inevitable due to unpredictable traffic bursts. If not handled properly, these unbounded TMs could cause severe network congestion, and undermine network availability.

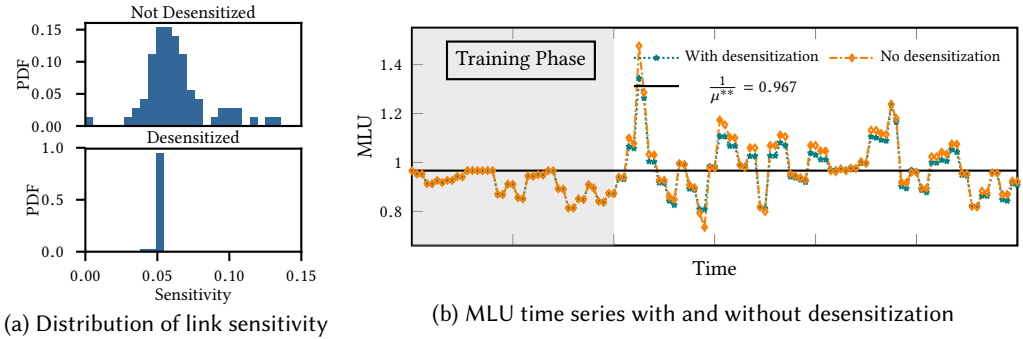
\begin{figure}[!tp]
\begin{subfigure}[c]{0.31\linewidth} 
\centering
        \input{plot_data/facebook_data/sensitivity_pdf.pgf}
\vspace{-5pt}
\caption{\small Distribution of link sensitivity}
\end{subfigure}
~
\begin{subfigure}[c]{0.66\linewidth} 
\vspace{7pt}
\pgfplotstableread{plot_data/facebook_data/sensitivity_with_vs_without_experiment.txt} \mluperf
\begin{tikzpicture}
\begin{axis}[height=1.75in, width=1.05\columnwidth, ylabel near ticks, xlabel near ticks, ylabel={\footnotesize MLU}, ylabel shift = -4pt, xlabel={\footnotesize Time}, xlabel shift = -5pt, xmin=40, xmax=160, legend columns=2, legend style={at={(0.35,0.95)},anchor=north west, cells={align=left}, draw=none, fill=none}, xticklabels=\empty, ytick pos=left, ylabel style={align=center}, thick]
      \fill[lightgray, opacity=0.3]
        (rel axis cs:0,0)--(rel axis cs:0.333,0)--
        (rel axis cs:0.333,1)--(rel axis cs:0,1)--cycle;
\node[draw,align=left] (textbox) at (rel axis cs:0.18,0.85)  {\footnotesize Training Phase};

\addplot+[densely dotted, teal,  mark=star, mark options={solid}, mark repeat={1}, mark size=1.2pt] table[x=x_range, y=mlu_optimized_unsorted] from \mluperf;
\addlegendentry{\tiny With desensitization\;};
\addplot+[densely dashdotted, orange,  mark=diamond, mark options={solid}, mark repeat={1}, mark size=1.2pt] table[x=x_range, y=mlu_unoptimized_unsorted] from \mluperf;
\addlegendentry{\tiny No desensitization\;};
\addplot+[mark=none, black, domain = 0:160, thick] {0.9665}; 
\addlegendentry{\tiny $\frac{1}{\mu^{**}} = 0.967$\;};
\end{axis}
\end{tikzpicture}
\caption{\small MLU time series with and without desensitization}
\end{subfigure}
\vspace{-10pt}
\caption{\small MLU time series highlighting the effects of traffic uncertainty on performance. Compares the relative performance degradations when sensitivity is optimized to when it is not.}
\label{fig:sensitivity_with_vs_without_optimization}
\end{figure}

We handle outlier TMs using a desensitization step, for which we introduce a sensitivity metric for every link $(s_i, s_j)$. For any path $p$ that traverses the link $(s_i, s_j)$, a demand surge $\Delta$ in $t_{\text{src}_p\text{dst}_p}$ would increase the link $(s_i, s_j)$'s utilization by $\Delta\omega_p/(x_{ij}b_{ij})$. We define sensitivity for link $(s_i, s_j)$ as $$\text{SEN}_{ij}=\max_{p:(s_i, s_j) \in p}\frac{\omega_p}{x_{ij}b_{ij}}.$$
{\metteor}’s desensitization step then aims to minimize the maximum sensitivity of all links (see step 2 of \S\ref{section_fractional_topology} for details), which prevents the routing solution from allocating too much weight on any single link, thereby reducing the increase in link utilization when demand surges. While performing desensitization offers no theoretical guarantees on performance, our preliminary investigations in Fig. \ref{fig:sensitivity_with_vs_without_optimization} and extensive simulations later in \S\ref{section_performance_evaluation} show that the desensitization step offers a marked improvement in solution robustness and overall performance. 

Fig. \ref{fig:sensitivity_with_vs_without_optimization}(a) shows that the link sensitivity distributions without desensitization exhibits a noticeably longer tail and a larger variance than that with desentization. As a result, we can see in Fig. \ref{fig:sensitivity_with_vs_without_optimization}(b) that the MLU performance in the evaluation (online) stage is poorer when desensitization is not performed. In contrast, the MLU time series with desensitization shows consistently better MLU in the evaluation phase. Since the TMs in the evaluation phase may be unbounded by the critical TMs of the training phase TMs, the MLU may exceed $1/\mu^{**}$.

\section{Overall Methodology}\label{section_overall_methodology}

Based on the insights discussed in \S\ref{InsightsLearned}, we now present {\metteor}'s final implementation. The final implementation uses  a single set of routing weights for all critical TMs to attain performance guarantees for bounded-TMs, and incorporates a desensitization step to improve solution robustness under traffic uncertainties. 

Still, the entire topology design problem is a hard integer programming problem (ILP) whose runtime scales exponentially with the number of pods and OCSs. We reduce the complexity by designing a \emph{fractional} logical topology (Step 2 in Fig.  \ref{topology_engineering_overall_workflow}) that optimizes throughput for all TMs first. Computing a fractional topology without the integer requirement is easy using linear programming (LP). Next, we configure the OCSs such that the \emph{integer} logical topology best approximates the fractional topology (Step 3 of Fig. \ref{topology_engineering_overall_workflow}). These steps are detailed in \S\ref{section_fractional_topology} and \S\ref{section_circuit_switch_mapping}, respectively.

\subsection{Computing Fractional Topology}\label{section_fractional_topology}
Before proceeding, a formal definition of \emph{fractional topology} is needed.
\begin{definition}
Given a set of pods $\mathcal{S}=\{s_1, …, s_N\}$ and the number of ingress \& egress links $r_{\text{ig}}^i, r_{\text{eg}}^i$, $D = [d_{ij}] \in \mathbb{R}^{N \times N}$ is a fractional topology \emph{iff} it satisfies:
\begin{equation}
\sum\limits_{j=1}^N d_{ij} \leq r_{\text{eg}}^i, \sum\limits_{i=1}^N d_{ij} \leq r_{\text{ig}}^{j} \; \quad\forall \; i,j = 1, ..., n
\label{constraint:egressandingress}
\end{equation}
\label{defn:fractional_graph}
\end{definition}
Simply put, a fractional topology, $D$, describes the inter-pod (fractional) link count in a way that satisfies each pod’s in/out-degree constraints. This definition ignores the OCSs; as we will consider the OCS layer later when rounding the fractional logical topology into an integer one, accounting for them here unnecessarily increases the number of variables needed for representation\footnote{As the number of ports of an OCS is comparable to a pod’s uplink, the number of OCSs, $M$, must be in the same order as the number of pods $N$. Factoring in the OCS layer will raise the number of variables from $O(N^2)$ to $O(N^2 M)$, which unnecessarily increases the problem’s space and runtime complexity.}. 

{\metteor} computes a fractional topology in three steps. The first step is to find the optimal max-min throughput of $\mu^*$ for all the critical TMs using the following formulation:
\begin{equation}\label{step1}
\boxed {
\begin{aligned}
\textbf{Max-min Throughput:}   \; \;    & \max_{D,\Omega=\{\omega_p\}}\mu=\min\{\mu_1, .. ,\mu_K\} \text{, s. t.} & \\
     & 1) \text{ } D \text{ satisfies (\ref{constraint:egressandingress})} & \\
     & 2) \text{ } (D, \mu_k T_k, \Omega)\text{ satisfies (\ref{feasible_constraints})}, \; \forall \; k \in \{1, .., K\} & \\
\end{aligned}
}
\end{equation}

We have assumed the same routing weights for different critical TMs. However, note that (\ref{step1}) is non-linear, as the second constraint of (\ref{step1}) contains a product term $\mu \omega_p$ of two optimization variables. In Appendix \ref{linearize_step1}, we show how to linearize (\ref{step1}) into a simple LP problem that can be efficiently solved with commercial optimization solvers like Gurobi~\cite{gurobi}. Having computed $\mu^*$, we fix $\mu^*$ and compute $D,\Omega$ that minimizes sensitivity in the second step:
\begin{equation}\label{step2}
\boxed{
\begin{aligned}
\textbf{Desensitization:}   \; \;   & \min_{D,\Omega=\{\omega_p\}}\max_p\frac{\omega_p}{\min_{(s_i,s_j) \in p}d_{ij}} \text{, s. t.} & \\
     & 1) \text{ } D\text{ satisfies (\ref{constraint:egressandingress})} & \\
     & 2) \text{ } (D, \mu^* T_k, \Omega)\text{ satisfy (\ref{feasible_constraints})}, \; \forall \; k \in \{1, .., K\} & \\
\end{aligned}
}
\end{equation}

Similar to (\ref{step1}), (\ref{step2}) is also non-linear as its objective function takes the form of reciprocals of optimization variables. Lacking the means of solving (\ref{step2}) directly with commercial solvers, we instead use an iterative approach. In each iteration, $\beta$ is fixed so that (\ref{step2}) transforms into an LP feasibility problem, whereby we check if there exists $D$ and $\Omega$ such that the maximum sensitivity is no greater than the current $\beta$. If (\ref{step2}) is feasible for the current iteration’s $\beta$ value, then the $\beta$ value for the next iteration is reduced; if (\ref{step2}) is infeasible, then $\beta$ is increased in the next iteration. Using a binary-search scheme, we can quickly converge to the optimal $\beta$ value over several iterations. 

In the third step, we optimize the average hop count by maximizing traffic traversing the direct paths. Specifically, given the direct paths for all pod pairs, $p_{ij}^{|1|}$ for all $i,j=1,...,N$, the max-min throughput $\mu^*$, and the optimal sensitivity $\beta$, we compute $D^*$ and $\Omega^*$ as follows:
\begin{equation}\label{step3}
\boxed{
\begin{aligned}
\textbf{Minimize Avg. Hop Count:}   \; \; & \max\limits_{} \min\limits_{k = 1,..,K } \sum_{i=1}^N\sum_{j=1}^N \omega_{p_{ij}^{|1|}} t_{ij}^k \text{, s. t.} & \\
& 1) \text{ } D\text{ satisfies (\ref{constraint:egressandingress})} & \\
& 2) \text{ } (D, \mu^* T_k, \Omega)\text{ satisfy (\ref{feasible_constraints})}, \; \forall \; k \in \{1, .., K\} & \\
& 3) \text{ } \omega_p\leq \beta d_{ij}, \; \forall \; i\neq j, p\in\mathcal{P}_{ij} & \\
\end{aligned}
}
\end{equation}

\noindent\textbf{Remark for routing: }
Note that the routing weight set solution $\Omega^*$ is paired with the fractional topology $D^*$. Once we convert $D^*$ to an integer topology, $X^*$, the routing weight set based on $X^*$ needs to be recomputed using (\ref{step1})-(\ref{step3}).

\subsection{Realizing $D^*$ on the OCS Layer}\label{section_circuit_switch_mapping}
We now need to realize the integer logical topology, $X$ on the OCS layer such that $X$ best approximates $D^*$. The problem here is to decide the total number of links $x_{ij}^m$ connecting pod $s_i$ to pod $s_j$ through OCS $o_m$, for every $i,j=1,2,...,N$ and $m=1,2,...,M$. Since there are $M$ OCSs, each $d_{ij}^*$ entry in $D^*$ can be split into $M$ integers, $x_{ij}^m, m=1,...,M$, such that $\sum_{m = 1}^{M} x_{ij}^m \approx d_{ij}$, where:
\begin{equation}\label{con_constraint}
\text{\hspace{-19mm} \textbf{Soft / Matching Constraints:\hspace{12mm}}}\floor{d_{ij}^*} \leq \sum_{m=1}^M x_{ij}^m \leq \ceil{d_{ij}^*}, \quad \forall \; i,j=1,...,n
\end{equation}
Then, a logical topology can be found by solving
\begin{equation}\label{map_to_ocs}
\text{Find }\{x_{ij}^m\}\text{ satisfying (\ref{constraint:ocslevel}) and (\ref{con_constraint}).}
\end{equation} 

However, (\ref{map_to_ocs}) is NP-Complete, as the proven NP-Complete 3-Dimensional Contingency Table problem~\cite{irving1994three} can be reduced to our problem. Fortunately, unlike the physical OCS constraints (\ref{constraint:ocslevel}), constraints (\ref{con_constraint}) are ``soft’’, which can be relaxed to reduce algorithmic complexity.

Initially, we tried two natural ideas for relaxing (\ref{map_to_ocs})’s complexity. The first is a naive approach that solves it directly with ILP, but this approach faces two main issues: 1) it has an exponential runtime complexity, and 2) it cannot gracefully relax the soft constraints when satisfying (\ref{con_constraint}) is infeasible. We also tried a greedy maximum matching approach next as done in Helios~\cite{farrington2011helios}, which maps each OCS to a max-weight matching subproblem based on $D^*$, and then greedily solving them. However, this greedy approach would violate many soft constraints such that the resulting logical topology $X$ is no longer a good approximation of $D^*$, causing poor network performance.

We wanted a reliable approach that (a) has low complexity, (b) is mathematically sound, and (c) can \emph{gracefully} relax soft constraints when necessary. The ILP approach only achieves (b), and the greedy algorithm only achieves (a). Inspired by convex optimization theories,  we developed an algorithm that achieves all three criteria.

\subsubsection{Lagrangian Dual Method (LDM)}\label{section_lagrangian_dual}\hfill\\
Our Lagrangian Dual method is motivated by the dual ascent method in~\cite{Boyd2011ADMM}. By transforming the primal problem (\ref{map_to_ocs}) into its Lagrangian dual problem, we can gracefully relax the soft constraints (\ref{con_constraint}) and decouple (\ref{map_to_ocs}) into several easier subproblems. Next, we give a brief overview of LDM. 

LDM first introduces a strictly-convex objective function $U(\mathbf{x})$ for the primal problem: 
\begin{equation}
\begin{aligned}
U(\mathbf{x}) = \sum_{i=1}^N\sum_{j=1}^N \sum_{m=1}^M -\big( x_{ij}^m - h_{ij}^m\big)^2
\end{aligned},
\label{eqn:lagrangian_primal_objective}
\end{equation}
where $h_{ij}^m = \min\big(h_{eg}^m(s_i), h_{ig}^m(s_j)\big)$. We take advantage of the fact that $x_{ij}^m \leq h_{ij}^m \; \forall i, j =1,2,...,N, \; m =1,2,...,M$ to ``coerce'' the logical topology to form more logical links through higher $x_{ij}^m$ values. The monotonically increasing objective function forces the integer logical topology to fully-utilize available physical links. In addition, a quadratic objective function possesses a strong convexity that stabilizes solution convergence.

LDM then introduces a set of dual variables $\mathbf{p}^\pm=[p_{ij}^-, p_{ij}^+, i,j=1,2,...,N]$ for the soft constraint (\ref{con_constraint}). The dual problem is:
\begin{equation}
\begin{aligned}
\min\limits_{\mathbf{p}^\pm}\Bigg\{\max\limits_{\mathbf{x}}U(\mathbf{x}) & + \sum_{i=1}^N\sum_{j=1}^N \bigg[p_{ij}^- \bigg( \sum_{m=1}^M x_{ij}^m - c_{ij}^-\bigg) -p_{ij}^+ \bigg( \sum_{m=1}^M x_{ij}^m - c_{ij}^+ \bigg) \bigg] \Bigg\} \\
\text{s.t } & \mathbf{p^\pm}  \succcurlyeq 0 \text{ and }\mathbf{x}\text{ satisfies (\ref{constraint:ocslevel})}
\end{aligned}
\label{eqn:langrangian_duality}
\end{equation}
The dual variables $\mathbf{p}^\pm$ essentially ``scores'' the cost of forming logical links between pods. For instance, when a particular pod pair $(s_i, s_j)$ is under-provisioned (i.e $\sum_{m=1}^M x_{ij}^m < c_{ij}^-$), $p_{ij}^-$ will increase while $p_{ij}^+$ will decrease such that future OCSs will more likely form links between $(s_i, s_j)$. The opposite is also true when $(s_i, s_j)$ is over-provisioned. Thus, iteratively updating $\mathbf{p}^\pm$ exerts a negative-feedback until the soft constraint is met. Rearranging the objective of (\ref{eqn:langrangian_duality}) gives us:
\begin{equation}
\begin{aligned}
\min_{\mathbf{p}^\pm} \Bigg\{- \sum_{m=1}^M \Bigg[\min_{\mathbf{x}^m}  \sum_{i=1}^N\sum_{j=1}^N \Big(-( x_{ij}^m - h_{ij}^m)^2+(p_{ij}^--p_{ij}^+)x_{ij}^m\Big)
   \Bigg] + \sum_{i=1}^N\sum_{j=1}^N p_{ij}^+ c_{ij}^+ - \sum_{i=1}^N\sum_{j=1}^N p_{ij}^- c_{ij}^- \Bigg\}
\end{aligned}
\label{eqn:lagrangian_duality_mcf}
\end{equation}
(\ref{eqn:lagrangian_duality_mcf}) indicates that for any given $\mathbf{p}^\pm$, finding the optimal $\mathbf{x}$ can be decoupled into $M$ subproblems, each corresponding to one OCS:
\begin{eqnarray}\label{eqn:decomposed_problem}
&\max_{\mathbf{x}^m}&\sum_{i=1}^N\sum_{j=1}^N \bigg(-( x_{ij}^m - h_{ij}^m)^2+(p_{ij}^- - p_{ij}^+)x_{ij}^m\bigg)\\
& \text{s.t: }& \sum_{i=1}^N x_{ij}^m \leq h_{\text{ig}}^m(j), \sum_{j=1}^N x_{ij}^m \leq h_{\text{eg}}^m(i)\quad \forall \; i,j = 1,..,N \nonumber
\end{eqnarray}
These subproblems are still integer programming problems. Fortunately, if we apply first order linear approximation to the quadratic terms in (\ref{eqn:decomposed_problem}), we can convert (\ref{eqn:decomposed_problem}) to a Min-cost Flow (MCF) problem, which can be solved in polynomial time using Goldberg-Tarjan \cite{bunnagel1998efficient} algorithm. 


In order to solve the overall minimization problem w.r.t $\mathbf{p}^{\pm}$, we employ a subgradient method with a harmonic step function. (The dual objective function is not differentiable, thus the typical gradient descent algorithm cannot be used here.) The detailed algorithm for LDM is stated in Appendix \ref{appendix:lagrangian_dual_detailed_walkthrough}, followed by its optimality analysis in Appendix \ref{appendix_reconfiguration_algo_analysis}.

\section{Performance Evaluation}\label{section_performance_evaluation}

We assume a fluid traffic model to help scale our evaluation, while still capturing the essential macroscopic properties. 

\noindent\textbf{Datasets: } Our evaluations are driven by both production traces and synthetically-generated traffic matrices. Processing the production trace from~\cite{roy2015facebook} gives us slightly over 86000 1-second TM snapshots over the course of 1 day. For the sake of brevity, we only show the evaluation results of the combined cluster traces; the remaining evaluation results based on the three individual clusters are available in Appendix \ref{appendix:additional_results}. We placed the aggregated traffic matrices and the software implementation of {\metteor} in an open source repository~\cite{anonrepo} to aid reproducibility of this work.


Facebook's production trace shows a strong clustering effect, with as much as 99\% of the fabric’s traffic being internal to clusters. Though these traffic patterns may be common in large-scale production DCNs, they are not universally-representative, especially for DCNs with disaggregated storage. In DCNs with disaggregated storage generally have more dominant traffic between clusters, particularly between the compute and storage clusters. To simulate these traffic patterns, we synthesize a sequence of traffic matrices by first splitting all the network pods into two evenly-sized clusters of pods: one for storage, the other for compute. Then, for each synthesized TM, we generate a random write/read demand for every compute pod and distribute such demand uniformly among all the storage pods. Storage pods do not communicate amongst themselves. 

\noindent \textbf{Metrics: } 
The main metrics used for evaluation are: 
\begin{itemize}[leftmargin=3pt]
\item \emph{Max Link Utilization (MLU)}: Link utilization is defined as the ratio of traffic demand traversing said link to its capacity. The maximum link utilization (MLU) is a good indicator of the congestion level at the most bottlenecked link; so a lower MLU is preferred. Although MLU cannot exceed 1 in practice because packets can be dropped, we allow MLU to be greater than 1 in our evaluation, as it could reflect how severe the congestion is.

\item \emph{Average Hop Count} (AHC) is the average inter-pod link traversed when sending traffic to its destination. {\metteor} routes traffic along both direct 1-hop and indirect 2-hop paths to attain larger path capacity between pods, which is helpful for absorbing unexpected traffic bursts. That said, a lower AHC is still preferred when possible, because lowering AHC reduces packet latency, DCN bandwidth tax~\cite{mellette2019expanding}, and flow completion time (see \S\ref{subsection:fine_grained_netbench_simulation}).

\end{itemize}


\subsection{Versus Other Topology \& Routing Solutions}\label{subsection:topology_comparison}
\begin{figure*}[!t]
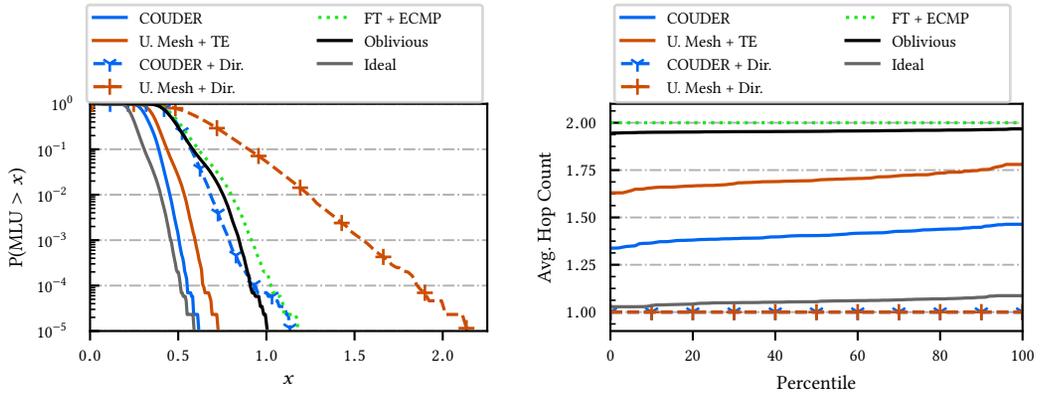

\begin{subfigure}[c]{0.49\textwidth}
\centering
 \input{plot_data/facebook_data/fb_cluster_combined/mlu_decayrate_fb_cluster_combined.pgf}
\end{subfigure}
~
\begin{subfigure}[c]{0.48\textwidth}
\centering
\input{plot_data/facebook_data/fb_cluster_combined/ahc_percentile_fb_cluster_combined.pgf}
\end{subfigure}
\vspace{-8pt}
\caption{\small Performance based on Facebook’s inter-pod DCN traces, aggregated into 1-second traffic matrices.} 
\vspace{-9pt}
\label{fig:metteor_vs_others_facebook_combined}
\end{figure*}

\begin{figure*}[!t]
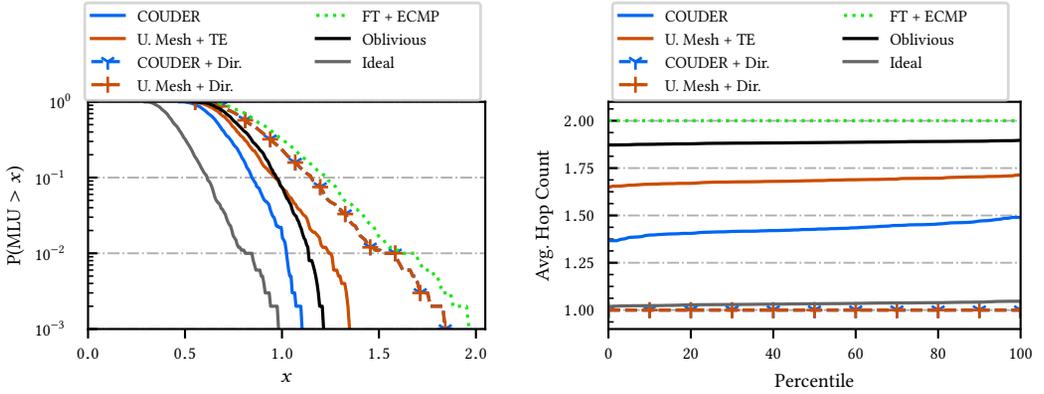

\begin{subfigure}[c]{0.49\textwidth}
\centering
\input{plot_data/synthetic_traffic/bipartite/mlu_decayrate_bipartite.pgf}
\end{subfigure}
~
\begin{subfigure}[c]{0.48\textwidth}
\centering
\input{plot_data/synthetic_traffic/bipartite/ahc_percentile_bipartite.pgf}
\end{subfigure}
\vspace{-8pt}
\caption{\small Performance using synthetically-generated traffic matrices for data centers with disaggregated storage.} 
\vspace{-9pt}
\label{fig:metteor_vs_others_synthetic_bipartite}
\end{figure*}

We first compare {\metteor} against several other representative DCN topologies and routing solutions in terms of MLU and AHC. The results are shown in Figs. \ref{fig:metteor_vs_others_facebook_combined} and \ref{fig:metteor_vs_others_synthetic_bipartite}.


\noindent \textbf{{\metteor }'s setting: }
For TMs derived from Facebook’s DCN traces, {\metteor} computes an inter-pod topology and a single set of routing weights based on 5 critical TMs extracted from the first hour’s traffic matrices. The process is similar for the synthetically-generated desegregated storage TMs, but we extract the critical TMs from the first one-tenth of traffic snapshots instead. The computed topology and routing solution are fixed as we evaluated the performance for remaining traffic snapshots in the sequence.

\noindent \textbf{Versus the optimal performance (Ideal): }
The optimal performance represents the (unrealistic) performance upper bound of ToE, which assumes that an offline-optimal topology and routing for each TM can be computed.  
Given a TM snapshot, The optimal topology and routing solution can be computed using (\ref{overall_formulation}), and the optimal MLU is simply the reciprocal of the optimal throughput. As is evident in Figs. \ref{fig:metteor_vs_others_facebook_combined} and \ref{fig:metteor_vs_others_synthetic_bipartite}, the optimal MLU is the lowest. That said, {\metteor} is still the closest to optimal in terms of MLU compared to other approaches. Note that the AHC of the optimal solution is slightly greater than 1, as we lose some optimality when rounding a fractional topology into an integer one. Thus, a small fraction of traffic must traverse 2-hop paths to ``regain’’ the best MLU. 



\noindent \textbf{Versus fat tree (FT + ECMP): }
Fat tree is the de facto standard for DCN topologies and a non-oversubscribed fat tree is rearrangeably non-blocking. Most operators would oversubscribe to either the aggregation or core layers in practice to save cost~\cite{greenberg2009vl2, benson2010network, farrington2013facebook}. In this work, we compare against a 2:1 oversubscribed fat tree, which has a more comparable deployment cost to a {\metteor} solution than a fully non-blocking fat tree. The fat tree is routed with ECMP.

As shown in Figs. \ref{fig:metteor_vs_others_facebook_combined} and \ref{fig:metteor_vs_others_synthetic_bipartite}, {\metteor} reduces MLU by about $50\%$, and AHC by about $60\%$ over a 2:1 oversubscribed fat tree. Since the fat tree has an additional spine layer of packet switches, any inter-pod traffic must traverse at least 2 hops (one between the source pod and the spine, and the other between the spine and the destination pod), so the AHC is invariably 2.

\noindent \textbf{Versus uniform mesh (U. Mesh + TE \& Oblivious): }
Next, we compare {\metteor} against a uniform mesh that directly connects pods without an OCS layer. Uniform mesh is a class of expander networks. When coupled with multi-path traffic engineering, a uniform mesh can easily attain bandwidth comparable to that of a fat tree at a lower capital cost~\cite{kassing2017beyond}. This, along with their incremental expandability~\cite{zhang2019understanding}, makes mesh expanders an attractive option for building future DCNs. Since {\metteor}’s logical topology is also mesh-like with non-uniform inter-pod connectivity, a uniform mesh is a natural baseline for comparison. 


We first evaluate a uniform mesh's performance when coupled with {\metteor}'s routing strategy (denoted as \textbf{Unif. Mesh + TE}). On average, {\metteor} reduces AHC by about $32\%$ and lowers MLU by about $20\%$ over a uniform mesh + TE. This comparison shows the benefits of having a reconfigurable OCS layer, allowing {\metteor} to reconfigure topology to better fit the traffic patterns by strategically placing links more between hotspots. 

 
Next, we compare {\metteor} against a uniform mesh with VLB routing (denoted as \textbf{oblivious}), which routes a packet to its destination through a randomly chosen intermediate pod. VLB is a highly robust routing scheme as it is traffic-oblivious. However, although inter-pod DCN traffic can be bursty at times, most snapshots can be captured reasonably well by a convex set of historical TMs (see Fig. \ref{fig:boundability}), which makes VLB overly conservative compared to other traffic-aware traffic engineering techniques. As shown in Figs. \ref{fig:metteor_vs_others_facebook_combined} and \ref{fig:metteor_vs_others_synthetic_bipartite}, the uniform mesh$+$VLB (oblivious) option performs significantly worse than {\metteor} in terms of both MLU and AHC.



\noindent \textbf{Versus direct-path routing (U. Mesh + Dir \& {\metteor} + Dir): }
Finally, we evaluate direct-path routing (i.e. AHC is invariably 1) when coupled with {\metteor} and uniform mesh topologies. Recall from Fig. \ref{fig:two_hop_routing}(b) in \S\ref{subsection:path_selection} that routing with direct paths exclusively offers much less path capacity between pod pairs compared to multi-path routing with 2-hop and direct paths. Thus, the MLU tends to be higher due to there being insufficient path capacity to ``absorb’’ traffic surges. Using {\metteor} alleviates this issue somewhat by allocating more link capacity to the hotspots, although the MLU is still much worse than that of {\metteor} with multi-path routing.


\subsection{Impact of Different {\metteor} Parameters on Performance}\label{subsection:parameter_settings}
Next, we study how different parameters may affect {\metteor}’s performance. Specifically, we look at how different numbers of critical TMs (\S\ref{subsection:different_k}) and how different reconfiguration frequencies (\S\ref{subsection:reconfig_frequency}) change the network performance.

\subsubsection{Effects of Critical Traffic Matrix Set Size}\label{subsection:different_k}
\begin{figure}[!tp]
\begin{subfigure}[c]{0.48\linewidth}
    \begin{center}
        \input{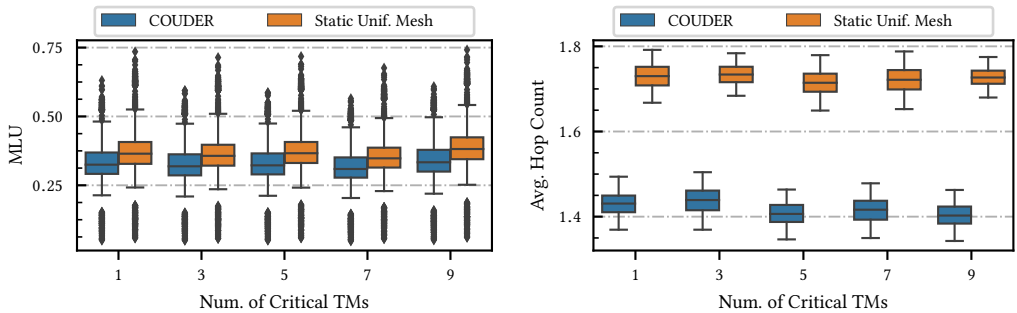}
    \end{center}
\end{subfigure}
~
\begin{subfigure}[c]{0.48\linewidth}
    \begin{center}
        \input{plot_data/facebook_data/fb_cluster_combined/combined_fabric_diff_k_ahc.pgf}
    \end{center}
\end{subfigure}
\vspace{-10pt}
    \caption{\small The effects of choosing different numbers of critical TMs for {\metteor} on MLU and AHC performance. }
\vspace{-16pt}
\label{fig:different_k}
\end{figure}

\hfill\\
To evaluate the impact of different numbers of critical TMs, we design an experiment with the same setup as in \S\ref{subsection:topology_comparison} but with different numbers of critical TMs, and summarize the results in Fig. \ref{fig:different_k}. 

Recall from \S\ref{section_proof_of_concept} that the critical TMs chosen by our algorithm form an outer bound of the historical TMs, and this outer bound becomes tighter as we increase the number of critical TMs (see Fig. \ref{fig:convex_set_prediction}). Choosing a larger bound could cover more grounds to handle traffic bursts, but weakens the performance guarantee for the bounded TMs. For instance, when there is only one critical TM (i.e. the critical TM would be the entry-wise historical max), the resulting outer bound would be the largest. In this case, the MLU performance with $K=1$ turns out to be the worst, as shown in Fig. \ref{fig:different_k}. 
Meanwhile, picking $K=7$ critical TMs offers the best MLU performance, while $K=5$ offers the best AHC performance. At any rate, {\metteor}'s performance is not sensitive to variations of $K$. In this case, choosing any number of critical TMs between 5 and 7 should be fine.



\subsubsection{Impact of Reconfiguration Frequency and Latency}\label{subsection:reconfig_frequency}\hfill\\
The evaluation in \S\ref{subsection:topology_comparison} with Facebook's one-day DCN trace assumes that topology and routing are only reconfigured once. This option has a low implementation and management complexity. However, many research efforts have been dedicated to develop faster optical circuit switches and control loops in the past. A natural question to ask, then, is whether more frequent topology reconfigurations can be more beneficial to performance? 

To answer this question, we apply {\metteor} with various reconfiguration frequencies, ranging from once every 30 seconds, 5 minutes, 1 hour, and 1 day. The initial convex set is computed based on the first hour’s worth of traffic matrix snapshots. Each reconfiguration event will update the current convex TM set by considering the traffic snapshots during the previous reconfiguration window. As a result of this, the convex TM set used for topology-routing optimization is monotonically increasing over the sequence of reconfiguration epochs. 

\begin{figure*}[!t]
\hspace{-6pt}
\begin{subfigure}{0.22\textwidth}
\centering
\vspace{-10pt}
\input{plot_data/facebook_data/reconfiguration_leftover_dcn_capacity.pgf}
\vspace{-17pt}
\caption{\small Leftover capacity}
\end{subfigure}
~
\begin{subfigure}{0.24\textwidth}
\centering
\input{plot_data/facebook_data/fb_cluster_combined/p999_nosense_reconfig_frequency_latency_combined.pgf}
\vspace{-17pt}
\caption{\small Tail MLU (w/o desensitization)}
\end{subfigure}
~ 
\begin{subfigure}{0.24\textwidth}
\centering
\input{plot_data/facebook_data/fb_cluster_combined/p999_reconfig_frequency_latency_combined.pgf}
\vspace{-17pt}
\caption{\small Tail MLU (w/ desensitization)}
\end{subfigure}
~ 
\begin{subfigure}{0.24\textwidth}
\centering
\vspace{-10pt}
\input{plot_data/facebook_data/fb_cluster_combined/ahc_nosense_reconfig_frequency_latency_combined.pgf}
\vspace{-17pt}
\caption{\small Average hop count}
\end{subfigure}
\vspace{-10pt}
\caption{\small Impact of reconfiguration frequency and latency on performance.  } 
\vspace{-12pt}
\label{fig:reconfig_frequency_and_latency}
\end{figure*}
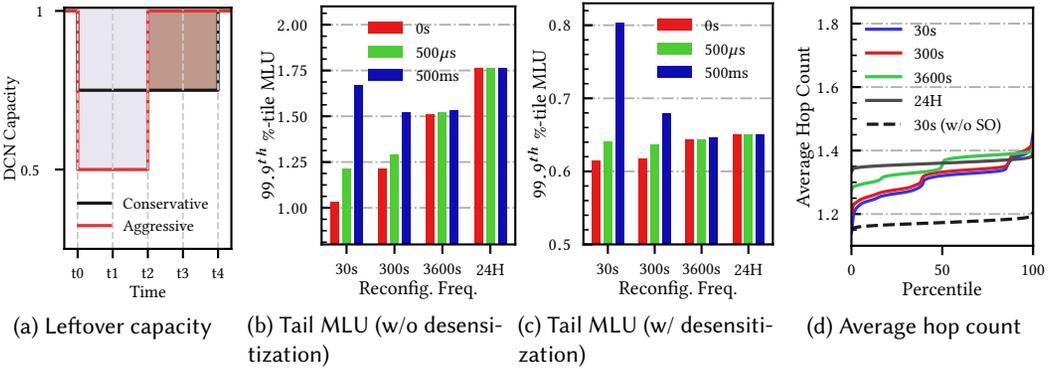



During each topology reconfiguration, we use a policy that no more than $1 - \alpha_{pred}$ fraction of links can be switched in a stage, where $\alpha_{pred}$ is the estimated MLU of the critical traffic matrix set. If $p$ fraction of links need to be reconfigured, $\ceil{\frac{p}{1 - \alpha_{pred}}}$ number of stages are required. In practice, however, each reconfiguration stage incurs a non-negligible latency, during which a DCN will experience reduced capacity (see Fig. \ref{fig:reconfig_frequency_and_latency}a). The capacity reduction during reconfiguration is accounted for in our evaluations. 

Network operators may choose different reconfiguration policies based on their SLOs. Pursuing an aggressive reconfiguration policy (i.e. reconfiguring more number of links per stage) speeds up the topology-switching process, but may risk tearing down a more substantial portion of the network capacity during switching. Conversely, one could pursue a more gradual reconfiguration process using more incremental stages to ensure sufficient capacity is preserved, though this may prolong the total reconfiguration latency. 


Fig. \ref{fig:reconfig_frequency_and_latency} shows the tail MLU statistics\footnote{The non-tail performances for different reconfiguration frequencies are similar, and hence not shown.} and the AHC of {\metteor} at various reconfiguration frequencies and latencies. Note that whether or not performing the desensitization step in {\metteor} makes a big difference. Without desensitization, the $99.9^{\text{th}}$ percentile MLU values are clearly higher, and having a lower reconfiguration frequency makes {\metteor}'s MLU performance even worse. With desensitization, {\metteor} can achieve good MLU even with reconfiguration frequency kept low. Although performing more frequent topology updates could improve {\metteor}'s performance, the improvements are minor, and may not be worth the increased control and management complexity. In fact, when each reconfiguration stage latency is as high as 500ms, frequent topology updates could even lead to poorer tail MLU, as the network would operate longer at reduced-capacity. Therefore, operators pursuing frequent reconfigurations for better performance should examine the cost-benefit, depending on their hardware and software capabilities.

\subsection{Robustness under Unexpected Bursts}\label{subsection:robustness_evaluation}
Finally, we evaluate {\metteor}’s robustness under unexpected bursts in traffic demands. While \S\ref{section_proof_of_concept} showed that DCN traffic does exhibit weak temporal stability, some TMs will inevitably fall outside the bounds of the critical TMs. Therefore, it is crucial for  {\metteor} to be robust under these circumstances. To this end, we design the following experiment.

First, we extract $K=5$ critical TMs, $\{T_1,T_2,...,T_K\}$, from the entire sequence of 1-second inter-pod traffic matrices. The critical TMs are used for topology and routing optimization. To generate traces with different levels of bursts, we first compute a base component-wise max TM $T^b=[t_{ij}^b]$, where $t_{ij}^b$ is the maximum traffic demand among all the $t_{ij}$'s in the TM sequence, and a standard deviation matrix $\Sigma=[\sigma_{ij}]$, where $\sigma_{ij}$ is the standard deviation of the all $t_{ij}$'s in the sequence. We then exhaustively enumerate all the possible burst sets, each of which contains one or two source-destination pod pairs. Then, for each burst set, denoted as $B$, the corresponding burst TM is $\tilde{T}(B)=[\tilde{t}_{ij}(B)]$, where
\begin{equation}\label{synthetic_trace}
\tilde{t}_{ij}(B)=
\left\{\begin{array}{l}
t_{ij}^b + \text{burst\_factor } * \sigma_{ij}, \text{ for }(i,j)\in B,\\
t_{ij}^b, \text{ for }(i,j)\notin B.
\end{array}\right.
\end{equation}
The ``burst\_factor'’ parameter acts as a control knob for the level of burstiness. 

\begin{figure}[!tp]
\vspace{-5pt}
    \begin{center}
        \input{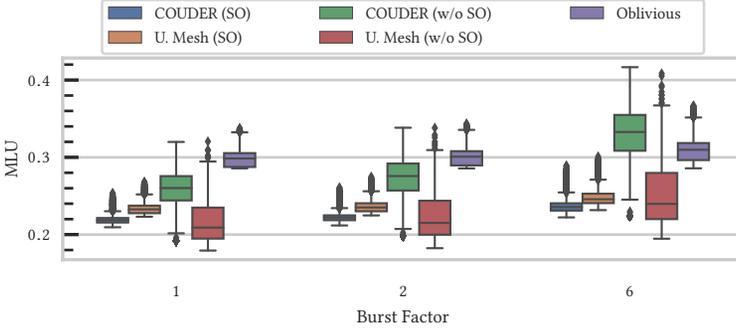}
    \end{center}
\vspace{-14pt}
\caption{\small Statistics MLU performance at various burst factors shown as boxplots. U. Mesh denotes a static uniform mesh, which uses the same routing technique as {\metteor}. All the boxes with \textbf{SO} labels are \textbf{S}ensitivity-\textbf{O}ptimized, meaning that desensitization was involved. Each box captures the $75^{th}$, $25^{th}$ percentiles and median values; the outliers beyond the upper and lower whiskers are dotted.}
\vspace{-12pt}
\label{robustness_analysis}
\end{figure}

Fig. \ref{robustness_analysis} shows the MLU distribution for all the generated burst TMs. The overall MLU distribution increases with an increase in burst factor. Notice that the desensitization step is essential for lowering overall MLU and its variance. For instance, {\metteor}’s overall MLU and its variance without desensitization are the highest. This is because the non-desensitized {\metteor} would allocate fewer links between pods with less predicted traffic. Whenever these pod pairs experience a demand surge that exceeds the bounds set by the critical TMs, the MLU may easily increase. This issue can be mitigated using desensitization on top of {\metteor}, which shows the best overall MLU, consistently beating mesh topology under desensitized TE and valiant load balancing (VLB).

\section{Packet Level Simulations}\label{subsection:fine_grained_netbench_simulation}
The evaluations thus far have focused on link utilization and average hop count (network efficiency), both of which are irrefutably crucial to DCN operators. However, evaluating performance based solely on macroscopic metrics may be insufficient for two important reasons: 1) performance evaluated on aggregated TM snapshots will inevitably smoothen out effects due to small timescale bursts only packet level simulations can capture, and 2) how operator-centric metrics translate into to user-centric, application-level performance (e.g. flow completion time (FCT)~\cite{dukkipati2006flowcompletiontime}) remains unclear. To this end, we aim to bridge this gap by exploring the interplay between the operator-centric metrics and application-centric metrics using a packet simulator called NetBench~\cite{netbench}. This allows us to extend our evaluations by measuring packet latency, packet drops rate, flow completion time, etc., at various levels of MLU and average hop count. 

The simulation uses DCTCP congestion control~\cite{alizadeh2010data}. We assume that inter-pod links have 40Gbps capacity, and a propagation latency of 500ns. We simulate a total of 2 seconds of network events. The simulator is given 0.5 second to warm up and down; only the flows that are initialized within the [0.5, 1.5] second range are considered for analysis.

Next, we chose (at random) a 5-minute time frame to collect an aggregated inter-pod traffic matrix, $T=[t_{ij}]$, using the trace in~\cite{roy2015facebook}. Flows from pod $s_i$ to pod $s_j$ are generated following a Poisson arrival process with rate $\lambda t_{ij}$, where the size of each flow follows uniform distribution. The inter-pod logical topology is computed based on the 5-minute aggregated TM using {\metteor}, while a few routing weights are computed, each with a specific AHC value\footnote{This can be easily done by adding an AHC constraint into \metteor’s formulation (\ref{step1})-(\ref{step3}).}. We vary the MLU for the same TM using the same routing-topology solutions by simply adjusting the flow arrival rates. 

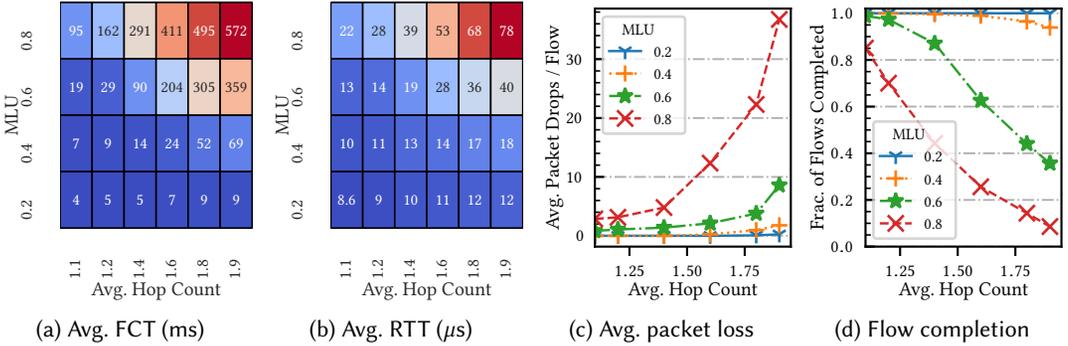
\begin{figure*}[!t]
\hspace{-20pt}
\begin{subfigure}[c]{0.24\textwidth}
\centering
\input{plot_data/packet_level_simulation/packet_level_average_fct.pgf}
\vspace{-14pt}
\caption{\small Avg. FCT (ms)}
\end{subfigure}
~
\begin{subfigure}[c]{0.24\textwidth}
\centering
\input{plot_data/packet_level_simulation/packet_level_average_rtt.pgf}
\vspace{-14pt}
\caption{\small Avg. RTT ($\mu$s)}
\end{subfigure}
~
\begin{subfigure}[c]{0.24\textwidth}
\centering
\input{plot_data/packet_level_simulation/packet_level_average_packet_loss.pgf}
\vspace{-14pt}
\caption{\small Avg. packet loss}
\end{subfigure}
~
\begin{subfigure}[c]{0.24\textwidth}
\centering
\input{plot_data/packet_level_simulation/packet_level_frac_flows_completed.pgf}
\vspace{-14pt}
\caption{\small Flow completion}
\end{subfigure}
\vspace{-10pt}
\caption{\small Packet level simulations showing the effects of MLU and average hop count and fine-grained metrics using a traffic matrix snapshot from Facebook trace.} 
\vspace{-9pt}
\label{fig:packet_level_simulations}
\end{figure*}

The impact of different MLU and AHC combinations to application-centric metrics is summarized  in Fig.~\ref{fig:packet_level_simulations}. As expected, higher MLUs lead to poorer performance (i.e. lower percentage of flows are completed within the simulations timeframes, more packets are dropped, packet round-trip times increase, etc). As higher MLUs indicate more severe link congestion, the average flow throughput will drop as a result. Increasing AHC, like MLU, also leads to packet level performance to deteriorate. The compounding effect of average hop count becomes more evident when congestion is high. 

As shown in Fig.~\ref{fig:packet_level_simulations}c, the packet drop rate increases super-linearly with hop count when MLU is 0.6 and 0.8. This is because when packets traverse longer paths and more packet switches, they leave a larger network footprint that leads to an increase in overall router queue lengths. When congestion is low, a higher AHC may not significantly affect performance, and packets may simply experience only a slight increase in round trip latency. When congestion is high, then queueing delay comes into play, causing a super-linear increase in round trip latency. As buffers become increasingly full, the typically shallow-buffered packet switches will begin dropping packets due to buffer overflow. As a result, TCP will have to throttle its send rate, leading to a drastic increase in flow completion time (see Fig.~\ref{fig:packet_level_simulations}(a)), and fewer flow completions (see Fig.\ref{fig:packet_level_simulations}(d)).

The above results demonstrate that both MLU and AHC play critical roles for application-level performance. Since {\metteor} can achieve both a low MLU and AHC simultaneously (see results in \S\ref{section_performance_evaluation}), it will also lead to better application-level performance.

\section{Conclusion}\label{section_conclusion}

We present {\metteor}, a robust topology engineering approach that does not rely on frequent reconfigurations to react to traffic changes. Unlike previous ToE solutions that attempt to react to traffic variations in real time, {\metteor} designs inter-pod topologies based on multiple critical TMs extracted from historical traffic matrices, and adopts a desensitization technique to further enhance its topologies against unexpected bursts. Compared with existing DCN topologies that do not use OCSs, {\metteor} shows clear performance benefits even with daily reconfiguration. Reconfiguring OCSs at such low frequencies greatly lowers the technological barrier to ToE deployment, thus paving a path towards the incremental adoption of optical circuit switched DCNs.

\bibliographystyle{ACM-Reference-Format}
\clearpage
\bibliography{reference}
\clearpage
\begin{appendices}
\section{Linearize Formulation (\ref{step1})}\label{linearize_step1}
We rewrite (\ref{step1}) as the following equivalent form:
\begin{eqnarray}\label{linearize_formulation1}
&& \max_{D,\Omega=\{\omega_p\}}\mu \text{, s. t}\\
&& 1) \sum\limits_{j=1}^N d_{ij} \leq r_{\text{eg}}^i, \sum\limits_{i=1}^N d_{ij} \leq r_{\text{ig}}^{j} \; \quad\forall \; i,j = 1, ..., N \nonumber\\
&& 2a) \sum_{p \in \mathcal{P}_{ij}} \omega_p = 1, \;  \forall \; i, j=1,...,N\nonumber\\
&& 2b) \sum_{p \in \mathcal{P}, (s_i, s_j) \text{ is a link along } p} \omega_p \mu t_{\text{src}_p\text{dst}_p}^k \leq x_{ij} b_{ij},\nonumber\\
&& \hspace{20mm}\; \forall \; i, j=1,...,N\text{ and }k=1,...,K\nonumber
\end{eqnarray}

The nonlinear terms of (\ref{linearize_formulation1}) are $\omega_p \mu$ in the constraint (2b). We define new variables $$\omega_p^{\prime}=\omega_p \mu.$$
Substituting all the $\omega_p$'s by $\omega_p^{\prime}$'s, we obtain
\begin{eqnarray}\label{linearize_formulation2}
&& \max_{D,\Omega=\{\omega_p\}}\mu \text{, s. t}\\
&& 1) \sum\limits_{j=1}^N d_{ij} \leq r_{\text{eg}}^i, \sum\limits_{i=1}^N d_{ij} \leq r_{\text{ig}}^{j} \; \quad\forall \; i,j = 1, ..., N \nonumber\\
&& 2a) \sum_{p \in \mathcal{P}_{ij}} \omega_p^{\prime} = \mu, \;  \forall \; i, j=1,...,N\nonumber\\
&& 2b) \sum_{p \in \mathcal{P}, (s_i, s_j) \text{ is a link along } p} \omega_p^{\prime} t_{\text{src}_p\text{dst}_p}^k \leq x_{ij} b_{ij},\nonumber\\
&& \hspace{20mm}\; \forall \; i, j=1,...,N\text{ and }k=1,...,K\nonumber
\end{eqnarray}

It is easy to verify that (\ref{linearize_formulation2}) is a linear programming problem, and is equivalent to (\ref{linearize_formulation1}).

\section{Additional Evaluation Results}\label{appendix:additional_results}
\begin{figure*}[!ht]
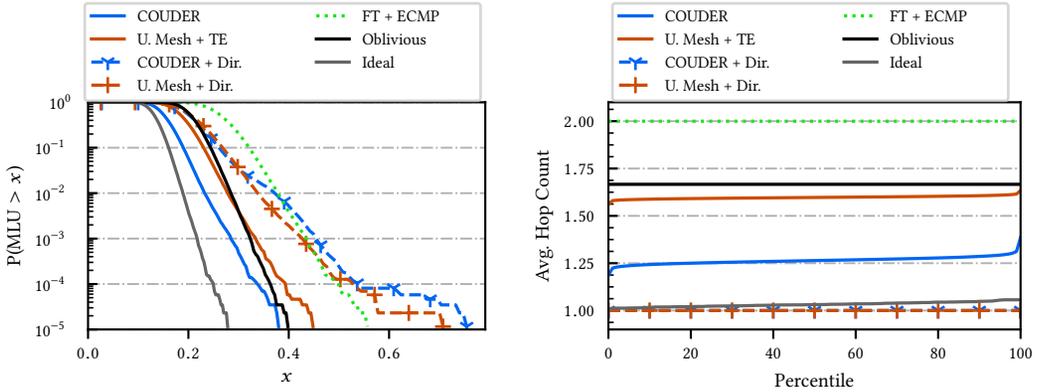

\begin{subfigure}[c]{0.49\textwidth}
\centering
 \input{plot_data/facebook_data/fb_cluster_A/mlu_decayrate_fb_cluster_A.pgf}
\end{subfigure}
~
\begin{subfigure}[c]{0.48\textwidth}
\centering
\input{plot_data/facebook_data/fb_cluster_A/ahc_percentile_fb_cluster_A.pgf}
\end{subfigure}
\vspace{-8pt}
\caption{\small Performance using Facebook’s Cache-follower cluster traces, aggregated into 1-second traffic matrices. } 
\label{fig:cache_follower_cluster_results}
\end{figure*}

\begin{figure*}[!t]
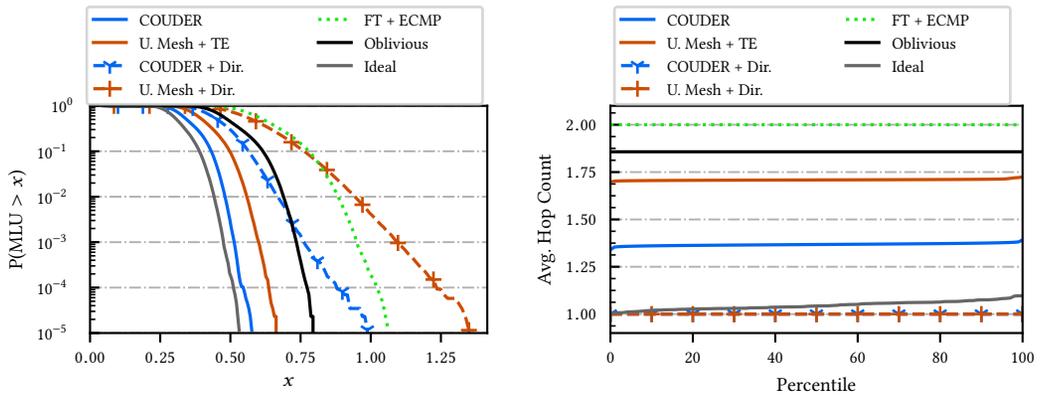

\begin{subfigure}[c]{0.49\textwidth}
\centering
 \input{plot_data/facebook_data/fb_cluster_B/mlu_decayrate_fb_cluster_B.pgf}
\end{subfigure}
~
\begin{subfigure}[c]{0.48\textwidth}
\centering
\input{plot_data/facebook_data/fb_cluster_B/ahc_percentile_fb_cluster_B.pgf}
\end{subfigure}
\vspace{-8pt}
\caption{\small Performance using Facebook’s Websearch cluster traces, aggregated into 1-second traffic matrices. } 
\label{fig:websearch_cluster_results}
\end{figure*}

\begin{figure*}[!t]
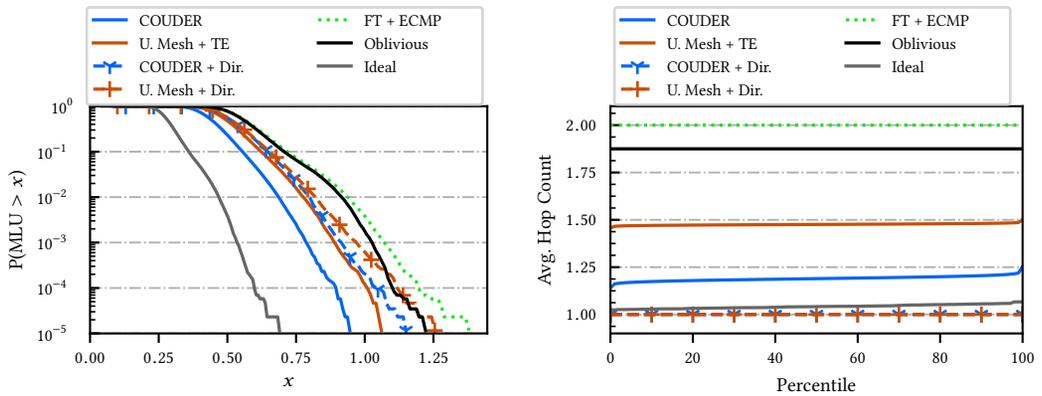

\begin{subfigure}[c]{0.49\textwidth}
\centering
 \input{plot_data/facebook_data/fb_cluster_C/mlu_decayrate_fb_cluster_C.pgf}
\end{subfigure}
~
\begin{subfigure}[c]{0.48\textwidth}
\centering
\input{plot_data/facebook_data/fb_cluster_C/ahc_percentile_fb_cluster_C.pgf}
\end{subfigure}
\vspace{-8pt}
\caption{\small Performance using Facebook’s Hadoop cluster traces, aggregated into 1-second traffic matrices. } 
\label{fig:hadoop_cluster_results}
\end{figure*}

Fig. \ref{fig:cache_follower_cluster_results}-Fig. \ref{fig:hadoop_cluster_results} show the evaluation results for the individual Facebook clusters.

\section{Detailed Walkthrough for LDM}\label{appendix:lagrangian_dual_detailed_walkthrough}
Lagrangian Dual method was motivated by the dual ascent method in~\cite{Boyd2011ADMM}. By introducing dual variables for soft constraints, LDM not only achieves graceful relaxation of soft constraints, but also relaxes the original NP-hard problem to a polynomial-time solvable problem. Nevertheless, LDM differs from the dual ascent method due to integer requirement. In this section, we detail the steps required for LDM to work.

\subsection{Primal Problem}\hfill\\
Our goal is to find an integer solution of $\mathbf{x} = [x_{ij}^m]$ satisfying the soft constraint (\ref{con_constraint}) and the hard constraints in (\ref{constraint:ocslevel}). In theory, there is no need for an objective function of $\mathbf{x}$ in our problem, since the problem itself is more concerned with satisfiability of the soft-constraints. However, this will lead to an algorithm with extremely poor convergence property. To speed up convergence, we introduce a strictly convex objective function for our primal problem, which is written as:
\begin{equation}
\begin{aligned}
    \mathbf{P: } \max_{\mathbf{x}} U(\mathbf{x}) &= \sum_{m=1}^M\sum_{i=1}^n\sum_{j=1}^n U_{ij}^m(x^m_{ij})\\
     \text{s.t : } & \quad (\ref{constraint:ocslevel}), \; (\ref{con_constraint}), \; \text{ are satisfied }\\
\end{aligned}
\label{primal_objective}
\end{equation}

At first, we chose $U_{ij}^m(x^m_{ij})=0$, which is not strictly convex. In this case, the solution does not converge even after running a large number of iterations. We then chose $U_{ij}^m(x^m_{ij})=-(x^m_{ij})^2$, which introduces a sharper objective function landscape that facilitated superior convergence. However, this objective function will result in a solution of $\mathbf{x}$ that connects as fewer links as possible in each OCS, which not only wastes physical resources but also results in an overall decrease in network capacity. Finally, we went with:
\begin{equation}\label{primalU}
U_{ij}^m(x_{ij}^m) = - \Big(x_{ij}^m \; - \; h_{ij}^m \Big)^2
\end{equation}
Where $h_{ij}^m = \min\big(h_{eg}^m(s_i), h_{in}^m(s_j)\big)$, taking advantage of the fact that $h_{ij}^m \geq x_{ij}^m$ to ensure that the optimal solution maximizes the formation of logical links.

\subsection{Dual Problem}\hfill\\
To relax the soft constraint (\ref{con_constraint}), we introduce dual variables $\mathbf{p}^+=[p_{ij}^+] \geq 0, \mathbf{p}^-=[p_{ij}^-] \geq 0$, and the following Lagrangian of the primal problem (\ref{primal_objective}):
$$L(\mathbf{x}, \mathbf{p}^+, \mathbf{p}^-)=\sum_{i=1}^N\sum_{j=1}^N \left[ \sum_{m=1}^M U_{ij}^m(x_{ij}^m) - p_{ij}^+\Bigg( \sum_{m=1}^M x_{ij}^m - \ceil{d_{ij}^*}\Bigg)+ p_{ij}^-\Bigg(\sum_{m=1}^M x_{ij}^m - \floor{d_{ij}^*}\Bigg) \right].$$

Note that for every $\mathbf{x}$ satisfying constraints (\ref{constraint:ocslevel}),  (\ref{con_constraint}), and every
$\mathbf{p}^+ \geq 0$ and $\mathbf{p}^-\geq 0$, the following inequality holds: $$L(\mathbf{x}, \mathbf{p}^+, \mathbf{p}^-)\geq \sum_{i=1}^N\sum_{j=1}^N \sum_{m=1}^M U_{ij}^m(x_{ij}^m).$$ Let
$$g(\mathbf{p}^+, \mathbf{p}^-):=\max_{\mathbf{x}} L(\mathbf{x}, \mathbf{p}^+, \mathbf{p}^-)\quad\text{s.t. }(\ref{constraint:ocslevel}) \text{ is satisfied}$$
We then have
\begin{eqnarray}\label{eqn:duality_gap}
g(\mathbf{p}^+, \mathbf{p}^-)&\geq&\max_{\mathbf{x}} L(\mathbf{x}, \mathbf{p}^+, \mathbf{p}^-)\;\text{satisfying }(\ref{con_constraint}), (\ref{constraint:ocslevel})\\
&\geq&\text{Optimal value of the primal problem (\ref{primal_objective})}\nonumber
\end{eqnarray}

Next, we introduce the dual problem:
\begin{equation}\label{dual_problem}
\mathbf{D: } \min_{\mathbf{p}^+, \mathbf{p}^-} g(\mathbf{p}^+, \mathbf{p}^-)\quad\text{s.t  } \mathbf{p}^+ \geq 0, \; \mathbf{p}^- \geq 0.
\end{equation}
Since the inequality (\ref{eqn:duality_gap}) holds for all $\mathbf{p}^+ \geq 0$ and $\mathbf{p}^-\geq 0$, we must have
$$\min_{\mathbf{p}^+\geq 0, \mathbf{p}^-\geq 0} g(\mathbf{p}^+, \mathbf{p}^-)\geq\text{The maximum value of the primal problem }(\ref{primal_objective}).$$
\emph{Duality gap} is then defined as the difference between the minimum value of the dual problem (\ref{dual_problem}) and the maximum value of the primal problem (\ref{primal_objective}).

If the primal decision variable $\mathbf{x}$ were fractional numbers instead of integers, under mild constraints\footnote{For Slater's Condition: see \S 5.2.3 in~\cite{ConvexOptimization}.}, the duality gap would be $0$. In that case, the optimal primal solution can be obtained by solving the dual problem instead. As we will see shortly, the dual problem (\ref{dual_problem}) is much easier to solve. However, (\ref{primal_objective}) is an integer problem with non-zero duality gap, hence solving the dual problem (\ref{dual_problem}) cannot give us the optimal solution of the primal problem (\ref{primal_objective}). Nevertheless, by optimizing the dual problem, we can still obtain a good sub-optimal solution to (\ref{primal_objective}) that satisfies all the hard constraints and a vast majority of the soft constraints.

\subsection{Subgradient Method}\hfill\\
The key aspect of LDM is the optimization of the dual problem (\ref{dual_problem}). Since the dual objective function is not differentiable, the typical gradient descent algorithm cannot be applied here. Hence, we use the subgradient method~\cite{SubgradientMethod} instead, whose general form is given as follows:

\begin{definition}\label{def:subgradient}
(Subgradient method~\cite{SubgradientForm}): Let $f:\mathbb{R}^n\rightarrow \mathbb{R}$ be a convex function with domain $\mathbb{R}^n$, a classical subgradient method iterates
$$y^{(\tau+1)}=y^{(\tau)} - \alpha_{\tau} \gamma^{(\tau)},$$
where $\gamma^{(\tau)}$ denotes a subgradient of $f$ at $y^{(\tau)}$, where $y^{(\tau)}$ is the $\tau$-th iterate of $y$. If $f$ is differentiable, then the only subgradient is the gradient vector of $f$. It may happen that $\gamma^{(\tau)}$ is not a descent direction for $f$ at $y^{(\tau)}$. We therefore keep a list of $f_{\text{best}}$ to keep track of the lowest objective function value found so far, i.e.,
$$f_{\text{best}}=\min\{f_{\text{best}}, f(y^{(\tau)})\}.$$
\end{definition}

Computing subgradients is the key step of the above subgradient method. The following lemma tells us how to compute a subgradient for the dual objective function $g(\mathbf{p}^+, \mathbf{p}^-)$.

\begin{lemma}\label{lem:subgradient}
For a given $(\hat{\mathbf{p}}^+, \hat{\mathbf{p}}^-)$, let $\hat{\mathbf{x}}$ be an integer solution that maximizes the lagrangian $L(\mathbf{x}, \hat{\mathbf{p}}^+, \hat{\mathbf{p}}^-)$, i.e., $$g(\hat{\mathbf{p}}^+, \hat{\mathbf{p}}^-) = \max_{\mathbf{x}}L(\mathbf{x}, \hat{\mathbf{p}}^+, \hat{\mathbf{p}}^-) = L(\hat{\mathbf{x}}, \hat{\mathbf{p}}^+, \hat{\mathbf{p}}^-).$$
Then, $\left[\ceil{d_{ij}^*} - \sum_{m=1}^M \hat{x}_{ij}^m, \sum_{m=1}^M \hat{x}_{ij}^m - \floor{d_{ij}^*}, i,j = 1,...,N\right]$ is a subgradient of $g(\mathbf{p}^+, \mathbf{p}^-)$ at $(\hat{\mathbf{p}}^+, \hat{\mathbf{p}}^-)$, i.e.,
$$g(\mathbf{p}^+, \mathbf{p}^-)-g(\hat{\mathbf{p}}^+, \hat{\mathbf{p}}^-)\geq \sum_{i=1}^N\sum_{j=1}^N \bigg(\ceil{d_{ij}^*} - \sum_{m=1}^M \hat{x}_{ij}^m\bigg)(p_{ij}^+ - \hat{p}_{ij}^+) + \sum_{i=1}^N\sum_{j=1}^N \bigg(\sum_{m=1}^M \hat{x}_{ij}^m - \floor{d_{ij}^*}\bigg)(p_{ij}^- - \hat{p}_{ij}^-)$$
for any $(\mathbf{p}^+, \mathbf{p}^-)$ in a neighbourhood of $(\hat{\mathbf{p}}^+, \hat{\mathbf{p}}^-)$.
\end{lemma}

\begin{proof}
Consider an arbitrary $(\mathbf{p}^+, \mathbf{p}^-)$. According to the definition of $g(\mathbf{p}^+, \mathbf{p}^-)$, we must have
$$g(\mathbf{p}^+, \mathbf{p}^-) = \max_{\mathbf{x}} L(\mathbf{x}, \mathbf{p}^+, \mathbf{p}^-) \geq L(\hat{\mathbf{x}}, \mathbf{p}^+, \mathbf{p}^-).$$
Then,
\begin{eqnarray}
&&g(\mathbf{p}^+, \mathbf{p}^-)-g(\hat{\mathbf{p}}^+, \hat{\mathbf{p}}^-)\geq L(\hat{\mathbf{x}}, \mathbf{p}^+, \mathbf{p}^-) - L(\hat{\mathbf{x}}, \hat{\mathbf{p}}^+, \hat{\mathbf{p}}^-)\nonumber\\
&=&\sum_{i=1}^N\sum_{j=1}^N \bigg(\ceil{d_{ij}^*} - \sum_{m=1}^M \hat{x}_{ij}^m\bigg)(p_{ij}^+ - \hat{p}_{ij}^+) + \sum_{i=1}^N\sum_{j=1}^N \bigg(\sum_{m=1}^M \hat{x}_{ij}^m - \floor{d_{ij}^*}\bigg)(p_{ij}^- - \hat{p}_{ij}^-),\nonumber
\end{eqnarray}
which completes the proof.
\end{proof}

Note that for each $(\hat{\mathbf{p}}^+, \hat{\mathbf{p}}^-)$, $\hat{\mathbf{x}}$ may not be the only solution that maximizes the Lagrangian $L(\mathbf{x}, \hat{\mathbf{p}}^+, \hat{\mathbf{p}}^-)$, because $L(\mathbf{x}, \hat{\mathbf{p}}^+, \hat{\mathbf{p}}^-)$ has integer variables $\mathbf{x}$. It is thus possible to have multiple subgradients for $g(\mathbf{p}^+, \mathbf{p}^-)$ at $(\hat{\mathbf{p}}^+, \hat{\mathbf{p}}^-)$, in which case $g(\mathbf{p}^+, \mathbf{p}^-)$ is not differentiable at $(\hat{\mathbf{p}}^+, \hat{\mathbf{p}}^-)$. If $g(\mathbf{p}^+, \mathbf{p}^-)$ were differentiable at $(\hat{\mathbf{p}}^+, \hat{\mathbf{p}}^-)$, there would be only one subgradient, which is the gradient of $g(\mathbf{p}^+, \mathbf{p}^-)$.

According to Lemma \ref{lem:subgradient}, the most critical part of calculating subgradient is to find a maximizer for a given Lagrangian. By rearranging the dual objective function $g(\mathbf{p}^+, \mathbf{p}^-)$, we obtain the following:
\begin{eqnarray}
&&g(\mathbf{p}^+, \mathbf{p}^-)\nonumber\\
&=& \max_{\mathbf{x}} L(\mathbf{x}, \mathbf{p}^+, \mathbf{p}^-)\quad\text{s.t. }(\ref{constraint:ocslevel}) \text{ is satisfied }\nonumber\\
&=&\sum_{m=1}^M\max_{\mathbf{x}^m}\left[\sum_{i=1}^N\sum_{j=1}^N \bigg(U_{ij}^m(x_{ij}^m)+(p_{ij}^- - p_{ij}^+)x_{ij}^m\bigg)\right]\nonumber\\
&&+\sum_{i=1}^N\sum_{j=1}^N (p_{ij}^+\ceil{d_{ij}^*} - p_{ij}^- \floor{d_{ij}^*})\quad\text{s.t. }(\ref{constraint:ocslevel}) \text{ is satisfied} \nonumber
\end{eqnarray}
From the above equation, we can see that optimizing the Lagrangian can be decomposed into $M$ subproblems:
\begin{eqnarray}\label{eqn:DecomposedProblem}
&\max_{\mathbf{x}^m}&\sum_{i=1}^N\sum_{j=1}^N \bigg(U_{ij}^m(x_{ij}^m)+(p_{ij}^- - p_{ij}^+)x_{ij}^m\bigg)\\
& \text{s.t: }& \sum_{i=1}^N x_{ij}^m \leq h_{\text{ig}}^m(j), \sum_{j=1}^N x_{ij}^m \leq h_{\text{eg}}^m(i)\quad \forall \; i,j=1,2,...,N\nonumber
\end{eqnarray}
Although these subproblems have significantly fewer decision variables, they are still integer programming problems with quadratic objective function, which can be hard to solve. To further reduce complexity, we apply first-order approximation to the nonlinear terms in (\ref{eqn:DecomposedProblem}), and obtain
\begin{eqnarray}\label{eqn:LinearDecomposedProblem}
&\max_{x^m}&\sum_{i=1}^N\sum_{j=1}^N \bigg(\frac{d U_{ij}^m}{d x_{ij}^m}(\hat{x}_{ij}^m)+(p_{ij}^- - p_{ij}^+)x_{ij}^m\bigg)\\
& \text{s.t: }& \sum_{i=1}^N x_{ij}^m \leq h_{\text{ig}}^m(j), \sum_{j=1}^N x_{ij}^m \leq h_{\text{eg}}^m(i),\nonumber\\
&&\max\{\hat{x}_{ij}^m - 1, 0\}\leq x_{ij}^m\leq \hat{x}_{ij}^m + 1 \quad \forall \; i,j=1,2,...,N\nonumber
\end{eqnarray}
where $\mathbf{\hat{x}}$ is the previous estimate of $\mathbf{x}$. The approximated problem (\ref{eqn:LinearDecomposedProblem}) can be solved in polynomial time using the method in Appendix \ref{appendix:mincost_flow}.

\subsection{Detailed Algorithm}\hfill\\
The detailed algorithm is shown in Algorithm \ref{algorithm:overall_primaldual}. Note that we update dual variables right after computing a configuration for each OCS to hasten solution convergence. Another option is to update dual variables after iterating through all the OCSs for one round. The problem with this option is that OCSs with the same physical striping will be configured exactly the same way in the same iteration, causing the solution to oscillate and slow down convergence.

Note that the harmonic step size function $\delta(\tau)$ is chosen because its sum approaches infinity as we take infinitely many step sizes. This way, we ensure that $\mathbf{p}^+, \mathbf{p}^-$'s growth is not handicapped by the step size if their optimal values are large.

Since LDM cannot guarantee all the soft constraints being satisfied, we also introduce a goodness function for a feasible OCS switch configuration state, $\mathbf{x}$, to keep track of the best solution thus far:
\begin{equation}\label{eqn:goodness_function}
\Psi(\mathbf{x}) = \sum_{i=1}^N\sum_{j=1}^N \psi_{ij},
\end{equation}
where $\psi_{ij}$ is an indicator variable that equals 1 when the $(i, j)$ pod pair’s soft constraints is satisfied, and 0 otherwise.

\begin{algorithm}
 \KwData{
    \begin{itemize}
        \item $D^* = [d_{ij}^*] \in \mathbb{R}^{N \times N}$ - fractional topology
        \item $\tau_{max}$ - number of iterations
    \end{itemize}
 }
 \KwResult{
    $\mathbf{x}^{*}=[x_{ij}^m{}^*] \in \mathbb{Z}^{N^2M}$ - OCS switch states
 }
Initialize: $\hat{\mathbf{x}} := 0, \mathbf{x}^{*} := 0, \mathbf{p}^+ :=0, \mathbf{p}^- :=0$ \;

 \For{$\tau \in \{1, 2, ..., \tau_{max}\}$}{
  Set step size $\delta := \frac{1}{\tau}$  \;
  \For{$m \in \{1,2,...,M\}$}{
      Solve (\ref{eqn:LinearDecomposedProblem}) based on Appendix \ref{appendix:mincost_flow}, and let $x^m$ be the integer solution\;
      Update $\hat{\mathbf{x}}$ in the $m$-th OCS by setting $\hat{x}^m=x^m$\;
      \If{$\Psi(\mathbf{x}^*) < \Psi(\hat{\mathbf{x}})$} {
        $\mathbf{x}^{*} := \hat{\mathbf{x}}$\;
      }
      Update dual variables using $p_{ij}^+:=\max\left\{p_{ij}^+-\delta\Bigg(\ceil{d_{ij}^*}-\sum_{m^{\prime}=1}^M\hat{x}_{ij}^{m^{\prime}}\Bigg), 0\right\}$ and $p_{ij}^-:=\max\left\{p_{ij}^--\delta\Bigg(\sum_{m^{\prime}=1}^M\hat{x}_{ij}^{m^{\prime}}-\floor{d_{ij}^*}\Bigg), 0\right\}$.
   }
 }
 \caption{Lagrangian duality method}\label{algorithm:overall_primaldual}
\end{algorithm}

\section{Mapping (\ref{eqn:LinearDecomposedProblem}) to a Min-Cost Circulation Problem}\label{appendix:mincost_flow}
In this section, we study a general form of (\ref{eqn:LinearDecomposedProblem}) as follows:
\begin{eqnarray}\label{eqn:general_mincostflow}
&\min\limits_{\mathbf{a}=[a_{ij}]}& \sum_{i=1}^I\sum_{j=1}^J C_{ij}a_{ij}\\
& \text{s.t: }& \sum_{i=1}^I a_{ij} \leq P_j, \sum_{j=1}^J a_{ij} \leq Q_i,\nonumber\\
&& L_{ij}\leq a_{ij} \leq U_{ij} \quad \forall \; i=1,...,I,j=1,...,J\nonumber
\end{eqnarray}
where $\mathbf{a}=[a_{ij}]$ is an $I\times J$ integer matrix to be solved, and $\mathbf{C}=[C_{ij}], \mathbf{P}=[P_j], \mathbf{Q}=[Q_i], \mathbf{L}=[L_{ij}], \mathbf{U}=[U_{ij}]$ are predefined constants. We would like to show that (\ref{eqn:general_mincostflow}) can be easily mapped to a min-cost circulation problem, which is polynomial time-solvable with  integer solution guarantees as long as $\mathbf{P}=[P_j], \mathbf{Q}=[Q_i], \mathbf{L}=[L_{ij}], \mathbf{U}=[U_{ij}]$ are all integers.

\subsection{Min-Cost Circulation Problem}
\begin{definition}\label{mincostcirculation}
(Min-Cost Circulation Problem) Given a flow network with
\begin{itemize}
  \item $l(v,w)$, lower bound on flow from node $v$ to node $w$;
  \item $u(v,w)$, upper bound on flow from node $v$ to node $w$;
  \item $c(v,w)$, cost of a unit of flow on $(v,w)$,
\end{itemize}
the goal of the min-cost circulation problem is to find a flow assignment $f(v,w)$ that minimizes
$$\sum_{(v,w)}c(v,w)\cdot f(v,w),$$
while satisfying the following two constraints:
\begin{enumerate}
  \item Throughput constraints: $l(v,w)\leq f(v,w)\leq u(v,w)$;
  \item Flow conservation constraints: $\sum_u f(u,v) = \sum_w f(v,w)$ for any node $v$.
\end{enumerate}
\end{definition}

Note that all the constant parameters $l(v,w),u(v,w)$ are all positive and $c(v,w)$ can be either positive or negative. In addition, min-cost circulation problem has a very nice property that guarantees integer solutions:
\begin{lemma}\label{integralflowtheorem}
(Integral Flow Theorem) Given a feasible circulation problem, if $l(v,w)$'s and $u(v,w)$'s are all integers, then there exists a feasible flow assignment such that all flows are integers.
\end{lemma}

In fact, for feasible circulation problems with integer bounds, most max-flow algorithms, e.g., Edmonds-Karp algorithm \cite{Edmonds1972Theoretical} and Goldberg-Tarjan algorithm \cite{Goldberg1988A}, are guaranteed to generate integer solutions.

\subsection{Detailed Transformation Steps}
\begin{figure}[ht!]
    \centering
    \includegraphics[scale=0.44]{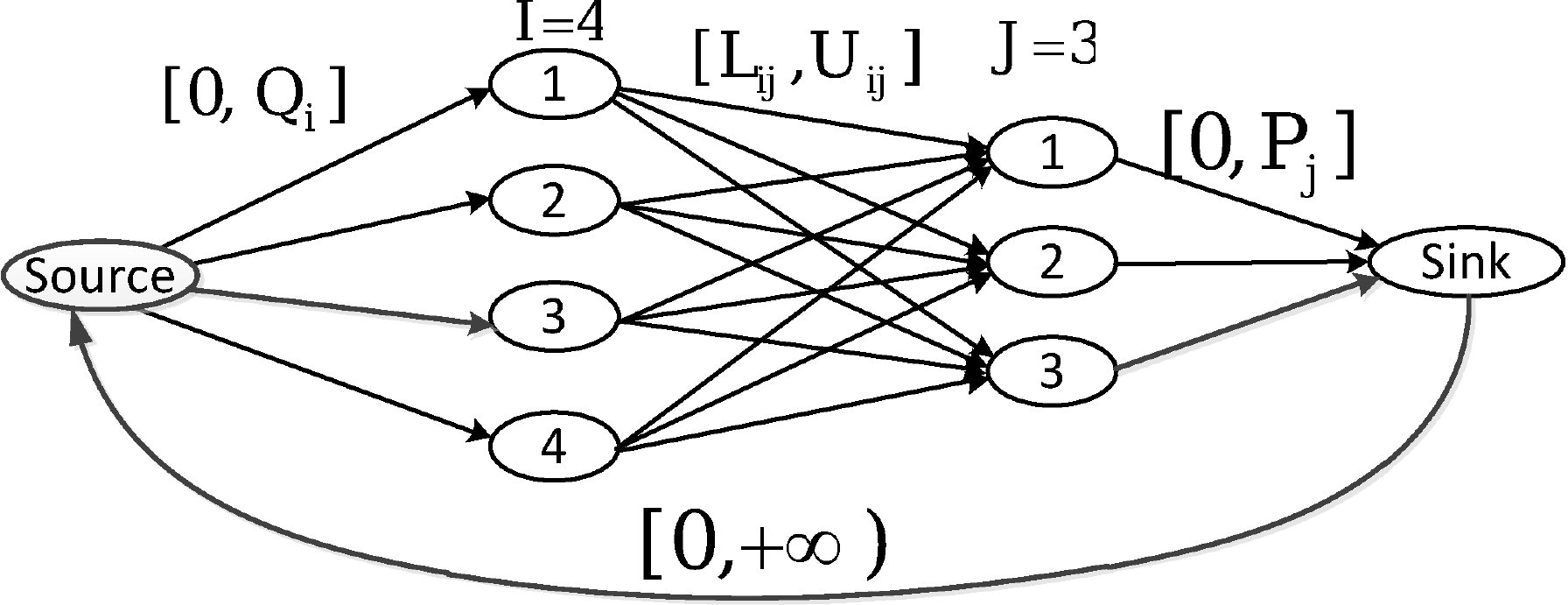}
    \caption{A flow graph example corresponding to equation (\ref{eqn:general_mincostflow}).}
    \label{fig:min_cost_flow}
\end{figure}

We first construct a flow network based on equation (\ref{eqn:general_mincostflow}) as follows
(also see Fig. \ref{fig:min_cost_flow}):
\begin{enumerate}
  \item Create a directed bipartite graph. Note that ${\mathbf{a}}$ is an $I\times J$ matrix. We create $I$ nodes on the left hand side of the bipartite graph, and create $J$ nodes on the right hand side of the bipartite graph. We add a directed link from $i$ to $j$, and set the bounds of this link as $[L_{ij}, U_{ij}]$ and the cost of this link as $C_{ij}$.
  \item Add a source node, and for each of the $I$ left nodes, add a link that connects to this source node. The bounds of the $i$-th link is set as $[0, Q_i]$, and the cost is set to $0$.
  \item Add a sink node and $J$ links from the $J$ right nodes to this sink node. The bounds of the $j$-th link is set as $[0, P_i]$, and the cost is set to $0$.
  \item Add a feedback link from the sink node to the source node. The bounds of this feedback link is set as $[0,\infty)$, and the cost is set as a very small negative value $-\epsilon$, e.g., $-10^{-6}$.
\end{enumerate}

We then assign flows to this flow network.
\begin{enumerate}
  \item For the link from the $i$-th left node to the $j$-th right node, assign $a_{ij}$ amount of flow.
  \item For the link from the source node to the $i$-th left node, assign $\sum_{j=1}^J a_{ij}$ amount of flow.
  \item For the link from the $j$-th right node to the sink node, assign $\sum_{i=1}^I a_{ij}$ amount of flow.
  \item For the feedback link from the sink node to the source node, assign $\sum_{i=1}^I\sum_{j=1}^J a_{ij}$ amount of flow.
\end{enumerate}

It is easy to verify that the above flow assignment satisfies the flow conservation constraints in Definition \ref{mincostcirculation}. Further, by enforcing the throughput constraints in Definition \ref{mincostcirculation}, all the constraints in (\ref{eqn:general_mincostflow}) are also satisfied. Further, the objective function of this min-cost flow problem is
\begin{equation}\label{eqn:mcf_objective}
\sum_{i=1}^I\sum_{j=1}^J C_{ij}a_{ij}+\epsilon \bigg(\sum_{i=1}^I\sum_{j=1}^J a_{ij}\bigg).
\end{equation}
Since $a_{ij}$’s are all integers, $\sum_{i=1}^I\sum_{j=1}^J C_{ij}a_{ij}$ cannot be take on a continuum of values. Then, as long as $\epsilon$ is small enough, minimizing (\ref{eqn:mcf_objective}) will also minimize the objective function in (\ref{eqn:general_mincostflow}). The benefit of having a small negative cost $\epsilon$ is that more flows can be assigned if possible.

\section{OCS mapping - Optimality Analysis}\label{appendix_reconfiguration_algo_analysis}
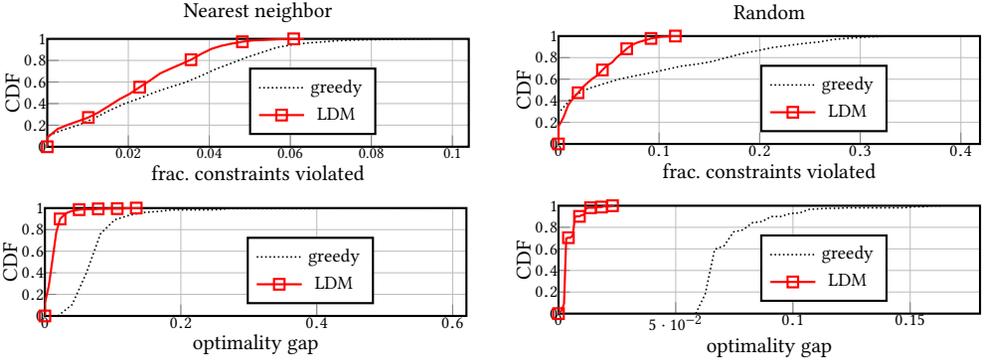
\begin{figure}[ht!]
\pgfplotstableread{plot_data/ocs_reconf_benchmark.txt}
	\datatable
\begin{subfigure}[c]{0.47\textwidth} 
\centering
\begin{tikzpicture}
\begin{axis}[xlabel = \footnotesize frac. constraints violated, xlabel near ticks, cycle list={{color=black, densely dotted, semithick}, {color=red, mark=square}, {color = blue, mark=x}}, width=1.1\linewidth, height=3.0cm, outer sep=-3pt, ymin=0,ymax=1,xmin=0, ylabel= \footnotesize CDF, ylabel near ticks, ylabel shift=-2.pt, title={\footnotesize Nearest neighbor}, xlabel shift=-1.pt, legend style={at={(0.50, 0.65)}, anchor= north west}, xtick pos=left, ytick pos=left, grid, thick, xtick={0.02, 0.04, 0.06, 0.08, 0.1}, xticklabels={0.02, 0.04, 0.06, 0.08, 0.1}]
\addplot+[mark repeat={5}] table[x=violation__knn_greedy_x,y=violation__knn_greedy_y] from \datatable ;
\addplot+[mark repeat={5}] table[x=violation__knn_primaldual_x, y=violation__knn_primaldual_y] from \datatable ;
\legend{\scriptsize greedy, \scriptsize LDM}
\end{axis}
\end{tikzpicture}
\end{subfigure} 
~
\begin{subfigure}[c]{0.47\textwidth}
\centering
\begin{tikzpicture}
\begin{axis}[xlabel = \footnotesize frac. constraints violated, xlabel near ticks, cycle list={{color=black, densely dotted, semithick}, {color=red, mark=square}, {color = blue, mark=x}}, width=1.1\linewidth, height=3.0cm, outer sep=-3pt, ymin=0,ymax=1,xmin=0, ylabel= \footnotesize CDF, ylabel near ticks, ylabel shift=-2.pt, xlabel shift=-1.pt, title={\footnotesize Random}, legend style={at={(0.50, 0.65)}, anchor= north west}, xtick pos=left, ytick pos=left, grid, thick]
\addplot+[mark repeat={5}] table[x=violation__random_greedy_x,y=violation__random_greedy_y] from \datatable ;
\addplot+[mark repeat={5}] table[x=violation__random_primaldual_x, y=violation__random_primaldual_y] from \datatable ;
\legend{\scriptsize greedy, \scriptsize LDM}
\end{axis}
\end{tikzpicture}
\end{subfigure}\\

\begin{subfigure}[c]{0.47\textwidth}
\centering
\begin{tikzpicture}
\begin{axis}[xlabel = \footnotesize optimality gap, xlabel near ticks, cycle list={{color=black, densely dotted, semithick}, {color=red, mark=square}, {color = blue, mark=x}}, width=1.1\linewidth, height=3.0cm, outer sep=-3pt, ymin=0,ymax=1,xmin=0 , ylabel= \footnotesize CDF, ylabel near ticks, ylabel shift=-2.pt, xlabel shift=-1.pt, legend style={at={(0.50, 0.65)}, anchor= north west}, xtick pos=left, ytick pos=left, grid, thick]
\addplot+[mark repeat={5}] table[x=optimality_loss__knn_greedy_x,y=optimality_loss__knn_greedy_y] from \datatable ;
\addplot+[ mark repeat={5}] table[x=optimality_loss__knn_primaldual_x, y=optimality_loss__knn_primaldual_y] from \datatable ;
\legend{\scriptsize greedy, \scriptsize LDM}
\end{axis}
\end{tikzpicture}
\end{subfigure}
~
\begin{subfigure}[c]{0.47\textwidth}
\centering
\begin{tikzpicture}
\begin{axis}[xlabel = \footnotesize optimality gap, xlabel near ticks, cycle list={{color=black, densely dotted, semithick}, {color=red, mark=square}, {color = blue, mark=x}}, width=1.1\linewidth, height=3.0cm, outer sep=-3pt, ymin=0,ymax=1,xmin=0, ylabel= \footnotesize CDF, ylabel near ticks, ylabel shift=-2.pt, xlabel shift=-1.pt, legend style={at={(0.50, 0.65)}, anchor= north west}, xtick pos=left, ytick pos=left, grid, thick]
\addplot+[mark repeat={5}] table[x=optimality_loss__random_greedy_x,y=optimality_loss__random_greedy_y] from \datatable ;
\addplot+[mark repeat={5}] table[x=optimality_loss__random_primaldual_x, y=optimality_loss__random_primaldual_y] from \datatable ;
\legend{\scriptsize greedy, \scriptsize LDM}
\end{axis}
\end{tikzpicture}
\end{subfigure}\\
\vspace{-12pt}
\caption{\small Optimality of OCS-mapping algorithms, using nearest neighbor (left col) and random (right col) TMs.}
\label{fig:combined_ocs_benchmark}
\end{figure}

Although LDM is motivated by convex optimization theories, our problem requires integer solutions and is thus not convex. Therefore, LDM cannot guarantee optimality. That said, we found via simulation that LDM shows superior performance.

We generated 900 DCN instances with pod-counts between 12 and 66. Each DCN instance is heterogeneous, containing pods with a mixture of 256, 512, and 1024 ports, interconnected via 128-port OCSs. The greedy algorithm described in Helios~\cite{farrington2011helios} acts as a baseline. All 900 instances are tested using: 1) nearest-neighbor, and 2) random permutation TMs. For nearest-neighbor TM, each pod sends traffic only to pods within $\rho$-units of circular index distance, where $\rho$ is $\sim\frac{1}{8}th$ the fabric size; this imitates skewed, neighbor-intensive traffic. Random TM is generated by treating each off-diagonal entry as a uniform random variable.

Next, we compute a logical topology \emph{w.r.t.} to its TM.  We use two solution-optimality metrics: 1) soft-constraint violation ratio, and 2) optimality loss. Soft-constraint violations count the number of $(i, j)$ pairs where (\ref{con_constraint}) is violated. Optimality loss measures the throughput loss/gap as we approximate the fractional topology with an integer one. This is measured as $1 - \frac{\mu_{int}^*}{\mu_{frac}^*}$; $\mu_{int}^*$ and $\mu_{frac}^*$ denote the throughputs under the integer and fractional logical topologies. Fig. \ref{fig:combined_ocs_benchmark} shows that LDM outperforms the greedy method in both metrics.

\end{appendices}

\end{document}

%% file: plot_data/facebook_data/path_len_total_capacity.pgf
\begingroup%
\makeatletter%
\begin{pgfpicture}%
\pgfpathrectangle{\pgfpointorigin}{\pgfqpoint{2.190000in}{1.400000in}}%
\pgfusepath{use as bounding box, clip}%
\begin{pgfscope}%
\pgfsetbuttcap%
\pgfsetmiterjoin%
\definecolor{currentfill}{rgb}{1.000000,1.000000,1.000000}%
\pgfsetfillcolor{currentfill}%
\pgfsetlinewidth{0.000000pt}%
\definecolor{currentstroke}{rgb}{1.000000,1.000000,1.000000}%
\pgfsetstrokecolor{currentstroke}%
\pgfsetdash{}{0pt}%
\pgfpathmoveto{\pgfqpoint{0.000000in}{0.000000in}}%
\pgfpathlineto{\pgfqpoint{2.190000in}{0.000000in}}%
\pgfpathlineto{\pgfqpoint{2.190000in}{1.400000in}}%
\pgfpathlineto{\pgfqpoint{0.000000in}{1.400000in}}%
\pgfpathclose%
\pgfusepath{fill}%
\end{pgfscope}%
\begin{pgfscope}%
\pgfsetbuttcap%
\pgfsetmiterjoin%
\definecolor{currentfill}{rgb}{1.000000,1.000000,1.000000}%
\pgfsetfillcolor{currentfill}%
\pgfsetlinewidth{0.000000pt}%
\definecolor{currentstroke}{rgb}{0.000000,0.000000,0.000000}%
\pgfsetstrokecolor{currentstroke}%
\pgfsetstrokeopacity{0.000000}%
\pgfsetdash{}{0pt}%
\pgfpathmoveto{\pgfqpoint{0.394200in}{0.322000in}}%
\pgfpathlineto{\pgfqpoint{2.146200in}{0.322000in}}%
\pgfpathlineto{\pgfqpoint{2.146200in}{1.358000in}}%
\pgfpathlineto{\pgfqpoint{0.394200in}{1.358000in}}%
\pgfpathclose%
\pgfusepath{fill}%
\end{pgfscope}%
\begin{pgfscope}%
\pgfsetbuttcap%
\pgfsetroundjoin%
\definecolor{currentfill}{rgb}{0.000000,0.000000,0.000000}%
\pgfsetfillcolor{currentfill}%
\pgfsetlinewidth{0.803000pt}%
\definecolor{currentstroke}{rgb}{0.000000,0.000000,0.000000}%
\pgfsetstrokecolor{currentstroke}%
\pgfsetdash{}{0pt}%
\pgfsys@defobject{currentmarker}{\pgfqpoint{0.000000in}{-0.048611in}}{\pgfqpoint{0.000000in}{0.000000in}}{%
\pgfpathmoveto{\pgfqpoint{0.000000in}{0.000000in}}%
\pgfpathlineto{\pgfqpoint{0.000000in}{-0.048611in}}%
\pgfusepath{stroke,fill}%
}%
\begin{pgfscope}%
\pgfsys@transformshift{0.613200in}{0.322000in}%
\pgfsys@useobject{currentmarker}{}%
\end{pgfscope}%
\end{pgfscope}%
\begin{pgfscope}%
\pgftext[x=0.613200in,y=0.224778in,,top]{\rmfamily\fontsize{6.400000}{7.680000}\selectfont 1}%
\end{pgfscope}%
\begin{pgfscope}%
\pgfsetbuttcap%
\pgfsetroundjoin%
\definecolor{currentfill}{rgb}{0.000000,0.000000,0.000000}%
\pgfsetfillcolor{currentfill}%
\pgfsetlinewidth{0.803000pt}%
\definecolor{currentstroke}{rgb}{0.000000,0.000000,0.000000}%
\pgfsetstrokecolor{currentstroke}%
\pgfsetdash{}{0pt}%
\pgfsys@defobject{currentmarker}{\pgfqpoint{0.000000in}{-0.048611in}}{\pgfqpoint{0.000000in}{0.000000in}}{%
\pgfpathmoveto{\pgfqpoint{0.000000in}{0.000000in}}%
\pgfpathlineto{\pgfqpoint{0.000000in}{-0.048611in}}%
\pgfusepath{stroke,fill}%
}%
\begin{pgfscope}%
\pgfsys@transformshift{1.051200in}{0.322000in}%
\pgfsys@useobject{currentmarker}{}%
\end{pgfscope}%
\end{pgfscope}%
\begin{pgfscope}%
\pgftext[x=1.051200in,y=0.224778in,,top]{\rmfamily\fontsize{6.400000}{7.680000}\selectfont 2}%
\end{pgfscope}%
\begin{pgfscope}%
\pgfsetbuttcap%
\pgfsetroundjoin%
\definecolor{currentfill}{rgb}{0.000000,0.000000,0.000000}%
\pgfsetfillcolor{currentfill}%
\pgfsetlinewidth{0.803000pt}%
\definecolor{currentstroke}{rgb}{0.000000,0.000000,0.000000}%
\pgfsetstrokecolor{currentstroke}%
\pgfsetdash{}{0pt}%
\pgfsys@defobject{currentmarker}{\pgfqpoint{0.000000in}{-0.048611in}}{\pgfqpoint{0.000000in}{0.000000in}}{%
\pgfpathmoveto{\pgfqpoint{0.000000in}{0.000000in}}%
\pgfpathlineto{\pgfqpoint{0.000000in}{-0.048611in}}%
\pgfusepath{stroke,fill}%
}%
\begin{pgfscope}%
\pgfsys@transformshift{1.489200in}{0.322000in}%
\pgfsys@useobject{currentmarker}{}%
\end{pgfscope}%
\end{pgfscope}%
\begin{pgfscope}%
\pgftext[x=1.489200in,y=0.224778in,,top]{\rmfamily\fontsize{6.400000}{7.680000}\selectfont 3}%
\end{pgfscope}%
\begin{pgfscope}%
\pgfsetbuttcap%
\pgfsetroundjoin%
\definecolor{currentfill}{rgb}{0.000000,0.000000,0.000000}%
\pgfsetfillcolor{currentfill}%
\pgfsetlinewidth{0.803000pt}%
\definecolor{currentstroke}{rgb}{0.000000,0.000000,0.000000}%
\pgfsetstrokecolor{currentstroke}%
\pgfsetdash{}{0pt}%
\pgfsys@defobject{currentmarker}{\pgfqpoint{0.000000in}{-0.048611in}}{\pgfqpoint{0.000000in}{0.000000in}}{%
\pgfpathmoveto{\pgfqpoint{0.000000in}{0.000000in}}%
\pgfpathlineto{\pgfqpoint{0.000000in}{-0.048611in}}%
\pgfusepath{stroke,fill}%
}%
\begin{pgfscope}%
\pgfsys@transformshift{1.927200in}{0.322000in}%
\pgfsys@useobject{currentmarker}{}%
\end{pgfscope}%
\end{pgfscope}%
\begin{pgfscope}%
\pgftext[x=1.927200in,y=0.224778in,,top]{\rmfamily\fontsize{6.400000}{7.680000}\selectfont 4}%
\end{pgfscope}%
\begin{pgfscope}%
\pgftext[x=1.270200in,y=0.136815in,,top]{\rmfamily\fontsize{7.200000}{8.640000}\selectfont Max. Path Length}%
\end{pgfscope}%
\begin{pgfscope}%
\pgfpathrectangle{\pgfqpoint{0.394200in}{0.322000in}}{\pgfqpoint{1.752000in}{1.036000in}}%
\pgfusepath{clip}%
\pgfsetbuttcap%
\pgfsetroundjoin%
\pgfsetlinewidth{0.702625pt}%
\definecolor{currentstroke}{rgb}{0.690196,0.690196,0.690196}%
\pgfsetstrokecolor{currentstroke}%
\pgfsetdash{{4.480000pt}{1.120000pt}{0.700000pt}{1.120000pt}}{0.000000pt}%
\pgfpathmoveto{\pgfqpoint{0.394200in}{0.322000in}}%
\pgfpathlineto{\pgfqpoint{2.146200in}{0.322000in}}%
\pgfusepath{stroke}%
\end{pgfscope}%
\begin{pgfscope}%
\pgfsetbuttcap%
\pgfsetroundjoin%
\definecolor{currentfill}{rgb}{0.000000,0.000000,0.000000}%
\pgfsetfillcolor{currentfill}%
\pgfsetlinewidth{0.803000pt}%
\definecolor{currentstroke}{rgb}{0.000000,0.000000,0.000000}%
\pgfsetstrokecolor{currentstroke}%
\pgfsetdash{}{0pt}%
\pgfsys@defobject{currentmarker}{\pgfqpoint{-0.048611in}{0.000000in}}{\pgfqpoint{0.000000in}{0.000000in}}{%
\pgfpathmoveto{\pgfqpoint{0.000000in}{0.000000in}}%
\pgfpathlineto{\pgfqpoint{-0.048611in}{0.000000in}}%
\pgfusepath{stroke,fill}%
}%
\begin{pgfscope}%
\pgfsys@transformshift{0.394200in}{0.322000in}%
\pgfsys@useobject{currentmarker}{}%
\end{pgfscope}%
\end{pgfscope}%
\begin{pgfscope}%
\pgftext[x=0.246053in,y=0.293065in,left,base]{\rmfamily\fontsize{6.400000}{7.680000}\selectfont \(\displaystyle 0\)}%
\end{pgfscope}%
\begin{pgfscope}%
\pgfpathrectangle{\pgfqpoint{0.394200in}{0.322000in}}{\pgfqpoint{1.752000in}{1.036000in}}%
\pgfusepath{clip}%
\pgfsetbuttcap%
\pgfsetroundjoin%
\pgfsetlinewidth{0.702625pt}%
\definecolor{currentstroke}{rgb}{0.690196,0.690196,0.690196}%
\pgfsetstrokecolor{currentstroke}%
\pgfsetdash{{4.480000pt}{1.120000pt}{0.700000pt}{1.120000pt}}{0.000000pt}%
\pgfpathmoveto{\pgfqpoint{0.394200in}{0.710425in}}%
\pgfpathlineto{\pgfqpoint{2.146200in}{0.710425in}}%
\pgfusepath{stroke}%
\end{pgfscope}%
\begin{pgfscope}%
\pgfsetbuttcap%
\pgfsetroundjoin%
\definecolor{currentfill}{rgb}{0.000000,0.000000,0.000000}%
\pgfsetfillcolor{currentfill}%
\pgfsetlinewidth{0.803000pt}%
\definecolor{currentstroke}{rgb}{0.000000,0.000000,0.000000}%
\pgfsetstrokecolor{currentstroke}%
\pgfsetdash{}{0pt}%
\pgfsys@defobject{currentmarker}{\pgfqpoint{-0.048611in}{0.000000in}}{\pgfqpoint{0.000000in}{0.000000in}}{%
\pgfpathmoveto{\pgfqpoint{0.000000in}{0.000000in}}%
\pgfpathlineto{\pgfqpoint{-0.048611in}{0.000000in}}%
\pgfusepath{stroke,fill}%
}%
\begin{pgfscope}%
\pgfsys@transformshift{0.394200in}{0.710425in}%
\pgfsys@useobject{currentmarker}{}%
\end{pgfscope}%
\end{pgfscope}%
\begin{pgfscope}%
\pgftext[x=0.195128in,y=0.681490in,left,base]{\rmfamily\fontsize{6.400000}{7.680000}\selectfont \(\displaystyle 50\)}%
\end{pgfscope}%
\begin{pgfscope}%
\pgfpathrectangle{\pgfqpoint{0.394200in}{0.322000in}}{\pgfqpoint{1.752000in}{1.036000in}}%
\pgfusepath{clip}%
\pgfsetbuttcap%
\pgfsetroundjoin%
\pgfsetlinewidth{0.702625pt}%
\definecolor{currentstroke}{rgb}{0.690196,0.690196,0.690196}%
\pgfsetstrokecolor{currentstroke}%
\pgfsetdash{{4.480000pt}{1.120000pt}{0.700000pt}{1.120000pt}}{0.000000pt}%
\pgfpathmoveto{\pgfqpoint{0.394200in}{1.098851in}}%
\pgfpathlineto{\pgfqpoint{2.146200in}{1.098851in}}%
\pgfusepath{stroke}%
\end{pgfscope}%
\begin{pgfscope}%
\pgfsetbuttcap%
\pgfsetroundjoin%
\definecolor{currentfill}{rgb}{0.000000,0.000000,0.000000}%
\pgfsetfillcolor{currentfill}%
\pgfsetlinewidth{0.803000pt}%
\definecolor{currentstroke}{rgb}{0.000000,0.000000,0.000000}%
\pgfsetstrokecolor{currentstroke}%
\pgfsetdash{}{0pt}%
\pgfsys@defobject{currentmarker}{\pgfqpoint{-0.048611in}{0.000000in}}{\pgfqpoint{0.000000in}{0.000000in}}{%
\pgfpathmoveto{\pgfqpoint{0.000000in}{0.000000in}}%
\pgfpathlineto{\pgfqpoint{-0.048611in}{0.000000in}}%
\pgfusepath{stroke,fill}%
}%
\begin{pgfscope}%
\pgfsys@transformshift{0.394200in}{1.098851in}%
\pgfsys@useobject{currentmarker}{}%
\end{pgfscope}%
\end{pgfscope}%
\begin{pgfscope}%
\pgftext[x=0.144202in,y=1.069916in,left,base]{\rmfamily\fontsize{6.400000}{7.680000}\selectfont \(\displaystyle 100\)}%
\end{pgfscope}%
\begin{pgfscope}%
\pgftext[x=0.130313in,y=0.840000in,,bottom,rotate=90.000000]{\rmfamily\fontsize{7.200000}{8.640000}\selectfont Path Capacity}%
\end{pgfscope}%
\begin{pgfscope}%
\pgfpathrectangle{\pgfqpoint{0.394200in}{0.322000in}}{\pgfqpoint{1.752000in}{1.036000in}}%
\pgfusepath{clip}%
\pgfsetbuttcap%
\pgfsetroundjoin%
\pgfsetlinewidth{1.204500pt}%
\definecolor{currentstroke}{rgb}{0.200000,0.200000,0.200000}%
\pgfsetstrokecolor{currentstroke}%
\pgfsetdash{{1.200000pt}{1.980000pt}}{0.000000pt}%
\pgfpathmoveto{\pgfqpoint{0.613200in}{0.463692in}}%
\pgfpathlineto{\pgfqpoint{1.051200in}{1.221285in}}%
\pgfpathlineto{\pgfqpoint{1.489200in}{1.305578in}}%
\pgfpathlineto{\pgfqpoint{1.927200in}{1.315414in}}%
\pgfusepath{stroke}%
\end{pgfscope}%
\begin{pgfscope}%
\pgfpathrectangle{\pgfqpoint{0.394200in}{0.322000in}}{\pgfqpoint{1.752000in}{1.036000in}}%
\pgfusepath{clip}%
\pgfsetbuttcap%
\pgfsetroundjoin%
\definecolor{currentfill}{rgb}{0.200000,0.200000,0.200000}%
\pgfsetfillcolor{currentfill}%
\pgfsetlinewidth{1.003750pt}%
\definecolor{currentstroke}{rgb}{0.200000,0.200000,0.200000}%
\pgfsetstrokecolor{currentstroke}%
\pgfsetdash{}{0pt}%
\pgfsys@defobject{currentmarker}{\pgfqpoint{-0.034722in}{-0.034722in}}{\pgfqpoint{0.034722in}{0.034722in}}{%
\pgfpathmoveto{\pgfqpoint{-0.034722in}{-0.034722in}}%
\pgfpathlineto{\pgfqpoint{0.034722in}{0.034722in}}%
\pgfpathmoveto{\pgfqpoint{-0.034722in}{0.034722in}}%
\pgfpathlineto{\pgfqpoint{0.034722in}{-0.034722in}}%
\pgfusepath{stroke,fill}%
}%
\begin{pgfscope}%
\pgfsys@transformshift{0.613200in}{0.463692in}%
\pgfsys@useobject{currentmarker}{}%
\end{pgfscope}%
\begin{pgfscope}%
\pgfsys@transformshift{1.051200in}{1.221285in}%
\pgfsys@useobject{currentmarker}{}%
\end{pgfscope}%
\begin{pgfscope}%
\pgfsys@transformshift{1.489200in}{1.305578in}%
\pgfsys@useobject{currentmarker}{}%
\end{pgfscope}%
\begin{pgfscope}%
\pgfsys@transformshift{1.927200in}{1.315414in}%
\pgfsys@useobject{currentmarker}{}%
\end{pgfscope}%
\end{pgfscope}%
\begin{pgfscope}%
\pgfsetrectcap%
\pgfsetmiterjoin%
\pgfsetlinewidth{0.803000pt}%
\definecolor{currentstroke}{rgb}{0.000000,0.000000,0.000000}%
\pgfsetstrokecolor{currentstroke}%
\pgfsetdash{}{0pt}%
\pgfpathmoveto{\pgfqpoint{0.394200in}{0.322000in}}%
\pgfpathlineto{\pgfqpoint{0.394200in}{1.358000in}}%
\pgfusepath{stroke}%
\end{pgfscope}%
\begin{pgfscope}%
\pgfsetrectcap%
\pgfsetmiterjoin%
\pgfsetlinewidth{0.803000pt}%
\definecolor{currentstroke}{rgb}{0.000000,0.000000,0.000000}%
\pgfsetstrokecolor{currentstroke}%
\pgfsetdash{}{0pt}%
\pgfpathmoveto{\pgfqpoint{2.146200in}{0.322000in}}%
\pgfpathlineto{\pgfqpoint{2.146200in}{1.358000in}}%
\pgfusepath{stroke}%
\end{pgfscope}%
\begin{pgfscope}%
\pgfsetrectcap%
\pgfsetmiterjoin%
\pgfsetlinewidth{0.803000pt}%
\definecolor{currentstroke}{rgb}{0.000000,0.000000,0.000000}%
\pgfsetstrokecolor{currentstroke}%
\pgfsetdash{}{0pt}%
\pgfpathmoveto{\pgfqpoint{0.394200in}{0.322000in}}%
\pgfpathlineto{\pgfqpoint{2.146200in}{0.322000in}}%
\pgfusepath{stroke}%
\end{pgfscope}%
\begin{pgfscope}%
\pgfsetrectcap%
\pgfsetmiterjoin%
\pgfsetlinewidth{0.803000pt}%
\definecolor{currentstroke}{rgb}{0.000000,0.000000,0.000000}%
\pgfsetstrokecolor{currentstroke}%
\pgfsetdash{}{0pt}%
\pgfpathmoveto{\pgfqpoint{0.394200in}{1.358000in}}%
\pgfpathlineto{\pgfqpoint{2.146200in}{1.358000in}}%
\pgfusepath{stroke}%
\end{pgfscope}%
\end{pgfpicture}%
\makeatother%
\endgroup%

%% file: plot_data/facebook_data/sensitivity_pdf.pgf
\begingroup%
\makeatletter%
\begin{pgfpicture}%
\pgfpathrectangle{\pgfpointorigin}{\pgfqpoint{1.640000in}{1.600000in}}%
\pgfusepath{use as bounding box, clip}%
\begin{pgfscope}%
\pgfsetbuttcap%
\pgfsetmiterjoin%
\definecolor{currentfill}{rgb}{1.000000,1.000000,1.000000}%
\pgfsetfillcolor{currentfill}%
\pgfsetlinewidth{0.000000pt}%
\definecolor{currentstroke}{rgb}{1.000000,1.000000,1.000000}%
\pgfsetstrokecolor{currentstroke}%
\pgfsetdash{}{0pt}%
\pgfpathmoveto{\pgfqpoint{0.000000in}{0.000000in}}%
\pgfpathlineto{\pgfqpoint{1.640000in}{0.000000in}}%
\pgfpathlineto{\pgfqpoint{1.640000in}{1.600000in}}%
\pgfpathlineto{\pgfqpoint{0.000000in}{1.600000in}}%
\pgfpathclose%
\pgfusepath{fill}%
\end{pgfscope}%
\begin{pgfscope}%
\pgfsetbuttcap%
\pgfsetmiterjoin%
\definecolor{currentfill}{rgb}{1.000000,1.000000,1.000000}%
\pgfsetfillcolor{currentfill}%
\pgfsetlinewidth{0.000000pt}%
\definecolor{currentstroke}{rgb}{0.000000,0.000000,0.000000}%
\pgfsetstrokecolor{currentstroke}%
\pgfsetstrokeopacity{0.000000}%
\pgfsetdash{}{0pt}%
\pgfpathmoveto{\pgfqpoint{0.410000in}{0.938667in}}%
\pgfpathlineto{\pgfqpoint{1.541600in}{0.938667in}}%
\pgfpathlineto{\pgfqpoint{1.541600in}{1.472000in}}%
\pgfpathlineto{\pgfqpoint{0.410000in}{1.472000in}}%
\pgfpathclose%
\pgfusepath{fill}%
\end{pgfscope}%
\begin{pgfscope}%
\pgfpathrectangle{\pgfqpoint{0.410000in}{0.938667in}}{\pgfqpoint{1.131600in}{0.533333in}}%
\pgfusepath{clip}%
\pgfsetbuttcap%
\pgfsetmiterjoin%
\definecolor{currentfill}{rgb}{0.200000,0.400000,0.600000}%
\pgfsetfillcolor{currentfill}%
\pgfsetlinewidth{0.000000pt}%
\definecolor{currentstroke}{rgb}{0.000000,0.000000,0.000000}%
\pgfsetstrokecolor{currentstroke}%
\pgfsetstrokeopacity{0.000000}%
\pgfsetdash{}{0pt}%
\pgfpathmoveto{\pgfqpoint{0.410000in}{0.938667in}}%
\pgfpathlineto{\pgfqpoint{0.451019in}{0.938667in}}%
\pgfpathlineto{\pgfqpoint{0.451019in}{0.984843in}}%
\pgfpathlineto{\pgfqpoint{0.410000in}{0.984843in}}%
\pgfpathclose%
\pgfusepath{fill}%
\end{pgfscope}%
\begin{pgfscope}%
\pgfpathrectangle{\pgfqpoint{0.410000in}{0.938667in}}{\pgfqpoint{1.131600in}{0.533333in}}%
\pgfusepath{clip}%
\pgfsetbuttcap%
\pgfsetmiterjoin%
\definecolor{currentfill}{rgb}{0.200000,0.400000,0.600000}%
\pgfsetfillcolor{currentfill}%
\pgfsetlinewidth{0.000000pt}%
\definecolor{currentstroke}{rgb}{0.000000,0.000000,0.000000}%
\pgfsetstrokecolor{currentstroke}%
\pgfsetstrokeopacity{0.000000}%
\pgfsetdash{}{0pt}%
\pgfpathmoveto{\pgfqpoint{0.451019in}{0.938667in}}%
\pgfpathlineto{\pgfqpoint{0.492037in}{0.938667in}}%
\pgfpathlineto{\pgfqpoint{0.492037in}{0.938667in}}%
\pgfpathlineto{\pgfqpoint{0.451019in}{0.938667in}}%
\pgfpathclose%
\pgfusepath{fill}%
\end{pgfscope}%
\begin{pgfscope}%
\pgfpathrectangle{\pgfqpoint{0.410000in}{0.938667in}}{\pgfqpoint{1.131600in}{0.533333in}}%
\pgfusepath{clip}%
\pgfsetbuttcap%
\pgfsetmiterjoin%
\definecolor{currentfill}{rgb}{0.200000,0.400000,0.600000}%
\pgfsetfillcolor{currentfill}%
\pgfsetlinewidth{0.000000pt}%
\definecolor{currentstroke}{rgb}{0.000000,0.000000,0.000000}%
\pgfsetstrokecolor{currentstroke}%
\pgfsetstrokeopacity{0.000000}%
\pgfsetdash{}{0pt}%
\pgfpathmoveto{\pgfqpoint{0.492037in}{0.938667in}}%
\pgfpathlineto{\pgfqpoint{0.533056in}{0.938667in}}%
\pgfpathlineto{\pgfqpoint{0.533056in}{0.938667in}}%
\pgfpathlineto{\pgfqpoint{0.492037in}{0.938667in}}%
\pgfpathclose%
\pgfusepath{fill}%
\end{pgfscope}%
\begin{pgfscope}%
\pgfpathrectangle{\pgfqpoint{0.410000in}{0.938667in}}{\pgfqpoint{1.131600in}{0.533333in}}%
\pgfusepath{clip}%
\pgfsetbuttcap%
\pgfsetmiterjoin%
\definecolor{currentfill}{rgb}{0.200000,0.400000,0.600000}%
\pgfsetfillcolor{currentfill}%
\pgfsetlinewidth{0.000000pt}%
\definecolor{currentstroke}{rgb}{0.000000,0.000000,0.000000}%
\pgfsetstrokecolor{currentstroke}%
\pgfsetstrokeopacity{0.000000}%
\pgfsetdash{}{0pt}%
\pgfpathmoveto{\pgfqpoint{0.533056in}{0.938667in}}%
\pgfpathlineto{\pgfqpoint{0.574074in}{0.938667in}}%
\pgfpathlineto{\pgfqpoint{0.574074in}{0.938667in}}%
\pgfpathlineto{\pgfqpoint{0.533056in}{0.938667in}}%
\pgfpathclose%
\pgfusepath{fill}%
\end{pgfscope}%
\begin{pgfscope}%
\pgfpathrectangle{\pgfqpoint{0.410000in}{0.938667in}}{\pgfqpoint{1.131600in}{0.533333in}}%
\pgfusepath{clip}%
\pgfsetbuttcap%
\pgfsetmiterjoin%
\definecolor{currentfill}{rgb}{0.200000,0.400000,0.600000}%
\pgfsetfillcolor{currentfill}%
\pgfsetlinewidth{0.000000pt}%
\definecolor{currentstroke}{rgb}{0.000000,0.000000,0.000000}%
\pgfsetstrokecolor{currentstroke}%
\pgfsetstrokeopacity{0.000000}%
\pgfsetdash{}{0pt}%
\pgfpathmoveto{\pgfqpoint{0.574074in}{0.938667in}}%
\pgfpathlineto{\pgfqpoint{0.615093in}{0.938667in}}%
\pgfpathlineto{\pgfqpoint{0.615093in}{0.938667in}}%
\pgfpathlineto{\pgfqpoint{0.574074in}{0.938667in}}%
\pgfpathclose%
\pgfusepath{fill}%
\end{pgfscope}%
\begin{pgfscope}%
\pgfpathrectangle{\pgfqpoint{0.410000in}{0.938667in}}{\pgfqpoint{1.131600in}{0.533333in}}%
\pgfusepath{clip}%
\pgfsetbuttcap%
\pgfsetmiterjoin%
\definecolor{currentfill}{rgb}{0.200000,0.400000,0.600000}%
\pgfsetfillcolor{currentfill}%
\pgfsetlinewidth{0.000000pt}%
\definecolor{currentstroke}{rgb}{0.000000,0.000000,0.000000}%
\pgfsetstrokecolor{currentstroke}%
\pgfsetstrokeopacity{0.000000}%
\pgfsetdash{}{0pt}%
\pgfpathmoveto{\pgfqpoint{0.615093in}{0.938667in}}%
\pgfpathlineto{\pgfqpoint{0.656111in}{0.938667in}}%
\pgfpathlineto{\pgfqpoint{0.656111in}{0.984843in}}%
\pgfpathlineto{\pgfqpoint{0.615093in}{0.984843in}}%
\pgfpathclose%
\pgfusepath{fill}%
\end{pgfscope}%
\begin{pgfscope}%
\pgfpathrectangle{\pgfqpoint{0.410000in}{0.938667in}}{\pgfqpoint{1.131600in}{0.533333in}}%
\pgfusepath{clip}%
\pgfsetbuttcap%
\pgfsetmiterjoin%
\definecolor{currentfill}{rgb}{0.200000,0.400000,0.600000}%
\pgfsetfillcolor{currentfill}%
\pgfsetlinewidth{0.000000pt}%
\definecolor{currentstroke}{rgb}{0.000000,0.000000,0.000000}%
\pgfsetstrokecolor{currentstroke}%
\pgfsetstrokeopacity{0.000000}%
\pgfsetdash{}{0pt}%
\pgfpathmoveto{\pgfqpoint{0.656111in}{0.938667in}}%
\pgfpathlineto{\pgfqpoint{0.697130in}{0.938667in}}%
\pgfpathlineto{\pgfqpoint{0.697130in}{1.031019in}}%
\pgfpathlineto{\pgfqpoint{0.656111in}{1.031019in}}%
\pgfpathclose%
\pgfusepath{fill}%
\end{pgfscope}%
\begin{pgfscope}%
\pgfpathrectangle{\pgfqpoint{0.410000in}{0.938667in}}{\pgfqpoint{1.131600in}{0.533333in}}%
\pgfusepath{clip}%
\pgfsetbuttcap%
\pgfsetmiterjoin%
\definecolor{currentfill}{rgb}{0.200000,0.400000,0.600000}%
\pgfsetfillcolor{currentfill}%
\pgfsetlinewidth{0.000000pt}%
\definecolor{currentstroke}{rgb}{0.000000,0.000000,0.000000}%
\pgfsetstrokecolor{currentstroke}%
\pgfsetstrokeopacity{0.000000}%
\pgfsetdash{}{0pt}%
\pgfpathmoveto{\pgfqpoint{0.697130in}{0.938667in}}%
\pgfpathlineto{\pgfqpoint{0.738148in}{0.938667in}}%
\pgfpathlineto{\pgfqpoint{0.738148in}{1.077195in}}%
\pgfpathlineto{\pgfqpoint{0.697130in}{1.077195in}}%
\pgfpathclose%
\pgfusepath{fill}%
\end{pgfscope}%
\begin{pgfscope}%
\pgfpathrectangle{\pgfqpoint{0.410000in}{0.938667in}}{\pgfqpoint{1.131600in}{0.533333in}}%
\pgfusepath{clip}%
\pgfsetbuttcap%
\pgfsetmiterjoin%
\definecolor{currentfill}{rgb}{0.200000,0.400000,0.600000}%
\pgfsetfillcolor{currentfill}%
\pgfsetlinewidth{0.000000pt}%
\definecolor{currentstroke}{rgb}{0.000000,0.000000,0.000000}%
\pgfsetstrokecolor{currentstroke}%
\pgfsetstrokeopacity{0.000000}%
\pgfsetdash{}{0pt}%
\pgfpathmoveto{\pgfqpoint{0.738148in}{0.938667in}}%
\pgfpathlineto{\pgfqpoint{0.779167in}{0.938667in}}%
\pgfpathlineto{\pgfqpoint{0.779167in}{1.308075in}}%
\pgfpathlineto{\pgfqpoint{0.738148in}{1.308075in}}%
\pgfpathclose%
\pgfusepath{fill}%
\end{pgfscope}%
\begin{pgfscope}%
\pgfpathrectangle{\pgfqpoint{0.410000in}{0.938667in}}{\pgfqpoint{1.131600in}{0.533333in}}%
\pgfusepath{clip}%
\pgfsetbuttcap%
\pgfsetmiterjoin%
\definecolor{currentfill}{rgb}{0.200000,0.400000,0.600000}%
\pgfsetfillcolor{currentfill}%
\pgfsetlinewidth{0.000000pt}%
\definecolor{currentstroke}{rgb}{0.000000,0.000000,0.000000}%
\pgfsetstrokecolor{currentstroke}%
\pgfsetstrokeopacity{0.000000}%
\pgfsetdash{}{0pt}%
\pgfpathmoveto{\pgfqpoint{0.779167in}{0.938667in}}%
\pgfpathlineto{\pgfqpoint{0.820185in}{0.938667in}}%
\pgfpathlineto{\pgfqpoint{0.820185in}{1.446603in}}%
\pgfpathlineto{\pgfqpoint{0.779167in}{1.446603in}}%
\pgfpathclose%
\pgfusepath{fill}%
\end{pgfscope}%
\begin{pgfscope}%
\pgfpathrectangle{\pgfqpoint{0.410000in}{0.938667in}}{\pgfqpoint{1.131600in}{0.533333in}}%
\pgfusepath{clip}%
\pgfsetbuttcap%
\pgfsetmiterjoin%
\definecolor{currentfill}{rgb}{0.200000,0.400000,0.600000}%
\pgfsetfillcolor{currentfill}%
\pgfsetlinewidth{0.000000pt}%
\definecolor{currentstroke}{rgb}{0.000000,0.000000,0.000000}%
\pgfsetstrokecolor{currentstroke}%
\pgfsetstrokeopacity{0.000000}%
\pgfsetdash{}{0pt}%
\pgfpathmoveto{\pgfqpoint{0.820185in}{0.938667in}}%
\pgfpathlineto{\pgfqpoint{0.861204in}{0.938667in}}%
\pgfpathlineto{\pgfqpoint{0.861204in}{1.446603in}}%
\pgfpathlineto{\pgfqpoint{0.820185in}{1.446603in}}%
\pgfpathclose%
\pgfusepath{fill}%
\end{pgfscope}%
\begin{pgfscope}%
\pgfpathrectangle{\pgfqpoint{0.410000in}{0.938667in}}{\pgfqpoint{1.131600in}{0.533333in}}%
\pgfusepath{clip}%
\pgfsetbuttcap%
\pgfsetmiterjoin%
\definecolor{currentfill}{rgb}{0.200000,0.400000,0.600000}%
\pgfsetfillcolor{currentfill}%
\pgfsetlinewidth{0.000000pt}%
\definecolor{currentstroke}{rgb}{0.000000,0.000000,0.000000}%
\pgfsetstrokecolor{currentstroke}%
\pgfsetstrokeopacity{0.000000}%
\pgfsetdash{}{0pt}%
\pgfpathmoveto{\pgfqpoint{0.861204in}{0.938667in}}%
\pgfpathlineto{\pgfqpoint{0.902222in}{0.938667in}}%
\pgfpathlineto{\pgfqpoint{0.902222in}{1.400427in}}%
\pgfpathlineto{\pgfqpoint{0.861204in}{1.400427in}}%
\pgfpathclose%
\pgfusepath{fill}%
\end{pgfscope}%
\begin{pgfscope}%
\pgfpathrectangle{\pgfqpoint{0.410000in}{0.938667in}}{\pgfqpoint{1.131600in}{0.533333in}}%
\pgfusepath{clip}%
\pgfsetbuttcap%
\pgfsetmiterjoin%
\definecolor{currentfill}{rgb}{0.200000,0.400000,0.600000}%
\pgfsetfillcolor{currentfill}%
\pgfsetlinewidth{0.000000pt}%
\definecolor{currentstroke}{rgb}{0.000000,0.000000,0.000000}%
\pgfsetstrokecolor{currentstroke}%
\pgfsetstrokeopacity{0.000000}%
\pgfsetdash{}{0pt}%
\pgfpathmoveto{\pgfqpoint{0.902222in}{0.938667in}}%
\pgfpathlineto{\pgfqpoint{0.943241in}{0.938667in}}%
\pgfpathlineto{\pgfqpoint{0.943241in}{1.308075in}}%
\pgfpathlineto{\pgfqpoint{0.902222in}{1.308075in}}%
\pgfpathclose%
\pgfusepath{fill}%
\end{pgfscope}%
\begin{pgfscope}%
\pgfpathrectangle{\pgfqpoint{0.410000in}{0.938667in}}{\pgfqpoint{1.131600in}{0.533333in}}%
\pgfusepath{clip}%
\pgfsetbuttcap%
\pgfsetmiterjoin%
\definecolor{currentfill}{rgb}{0.200000,0.400000,0.600000}%
\pgfsetfillcolor{currentfill}%
\pgfsetlinewidth{0.000000pt}%
\definecolor{currentstroke}{rgb}{0.000000,0.000000,0.000000}%
\pgfsetstrokecolor{currentstroke}%
\pgfsetstrokeopacity{0.000000}%
\pgfsetdash{}{0pt}%
\pgfpathmoveto{\pgfqpoint{0.943241in}{0.938667in}}%
\pgfpathlineto{\pgfqpoint{0.984259in}{0.938667in}}%
\pgfpathlineto{\pgfqpoint{0.984259in}{1.123371in}}%
\pgfpathlineto{\pgfqpoint{0.943241in}{1.123371in}}%
\pgfpathclose%
\pgfusepath{fill}%
\end{pgfscope}%
\begin{pgfscope}%
\pgfpathrectangle{\pgfqpoint{0.410000in}{0.938667in}}{\pgfqpoint{1.131600in}{0.533333in}}%
\pgfusepath{clip}%
\pgfsetbuttcap%
\pgfsetmiterjoin%
\definecolor{currentfill}{rgb}{0.200000,0.400000,0.600000}%
\pgfsetfillcolor{currentfill}%
\pgfsetlinewidth{0.000000pt}%
\definecolor{currentstroke}{rgb}{0.000000,0.000000,0.000000}%
\pgfsetstrokecolor{currentstroke}%
\pgfsetstrokeopacity{0.000000}%
\pgfsetdash{}{0pt}%
\pgfpathmoveto{\pgfqpoint{0.984259in}{0.938667in}}%
\pgfpathlineto{\pgfqpoint{1.025278in}{0.938667in}}%
\pgfpathlineto{\pgfqpoint{1.025278in}{1.077195in}}%
\pgfpathlineto{\pgfqpoint{0.984259in}{1.077195in}}%
\pgfpathclose%
\pgfusepath{fill}%
\end{pgfscope}%
\begin{pgfscope}%
\pgfpathrectangle{\pgfqpoint{0.410000in}{0.938667in}}{\pgfqpoint{1.131600in}{0.533333in}}%
\pgfusepath{clip}%
\pgfsetbuttcap%
\pgfsetmiterjoin%
\definecolor{currentfill}{rgb}{0.200000,0.400000,0.600000}%
\pgfsetfillcolor{currentfill}%
\pgfsetlinewidth{0.000000pt}%
\definecolor{currentstroke}{rgb}{0.000000,0.000000,0.000000}%
\pgfsetstrokecolor{currentstroke}%
\pgfsetstrokeopacity{0.000000}%
\pgfsetdash{}{0pt}%
\pgfpathmoveto{\pgfqpoint{1.025278in}{0.938667in}}%
\pgfpathlineto{\pgfqpoint{1.066296in}{0.938667in}}%
\pgfpathlineto{\pgfqpoint{1.066296in}{0.938667in}}%
\pgfpathlineto{\pgfqpoint{1.025278in}{0.938667in}}%
\pgfpathclose%
\pgfusepath{fill}%
\end{pgfscope}%
\begin{pgfscope}%
\pgfpathrectangle{\pgfqpoint{0.410000in}{0.938667in}}{\pgfqpoint{1.131600in}{0.533333in}}%
\pgfusepath{clip}%
\pgfsetbuttcap%
\pgfsetmiterjoin%
\definecolor{currentfill}{rgb}{0.200000,0.400000,0.600000}%
\pgfsetfillcolor{currentfill}%
\pgfsetlinewidth{0.000000pt}%
\definecolor{currentstroke}{rgb}{0.000000,0.000000,0.000000}%
\pgfsetstrokecolor{currentstroke}%
\pgfsetstrokeopacity{0.000000}%
\pgfsetdash{}{0pt}%
\pgfpathmoveto{\pgfqpoint{1.066296in}{0.938667in}}%
\pgfpathlineto{\pgfqpoint{1.107315in}{0.938667in}}%
\pgfpathlineto{\pgfqpoint{1.107315in}{0.984843in}}%
\pgfpathlineto{\pgfqpoint{1.066296in}{0.984843in}}%
\pgfpathclose%
\pgfusepath{fill}%
\end{pgfscope}%
\begin{pgfscope}%
\pgfpathrectangle{\pgfqpoint{0.410000in}{0.938667in}}{\pgfqpoint{1.131600in}{0.533333in}}%
\pgfusepath{clip}%
\pgfsetbuttcap%
\pgfsetmiterjoin%
\definecolor{currentfill}{rgb}{0.200000,0.400000,0.600000}%
\pgfsetfillcolor{currentfill}%
\pgfsetlinewidth{0.000000pt}%
\definecolor{currentstroke}{rgb}{0.000000,0.000000,0.000000}%
\pgfsetstrokecolor{currentstroke}%
\pgfsetstrokeopacity{0.000000}%
\pgfsetdash{}{0pt}%
\pgfpathmoveto{\pgfqpoint{1.107315in}{0.938667in}}%
\pgfpathlineto{\pgfqpoint{1.148333in}{0.938667in}}%
\pgfpathlineto{\pgfqpoint{1.148333in}{1.031019in}}%
\pgfpathlineto{\pgfqpoint{1.107315in}{1.031019in}}%
\pgfpathclose%
\pgfusepath{fill}%
\end{pgfscope}%
\begin{pgfscope}%
\pgfpathrectangle{\pgfqpoint{0.410000in}{0.938667in}}{\pgfqpoint{1.131600in}{0.533333in}}%
\pgfusepath{clip}%
\pgfsetbuttcap%
\pgfsetmiterjoin%
\definecolor{currentfill}{rgb}{0.200000,0.400000,0.600000}%
\pgfsetfillcolor{currentfill}%
\pgfsetlinewidth{0.000000pt}%
\definecolor{currentstroke}{rgb}{0.000000,0.000000,0.000000}%
\pgfsetstrokecolor{currentstroke}%
\pgfsetstrokeopacity{0.000000}%
\pgfsetdash{}{0pt}%
\pgfpathmoveto{\pgfqpoint{1.148333in}{0.938667in}}%
\pgfpathlineto{\pgfqpoint{1.189352in}{0.938667in}}%
\pgfpathlineto{\pgfqpoint{1.189352in}{1.031019in}}%
\pgfpathlineto{\pgfqpoint{1.148333in}{1.031019in}}%
\pgfpathclose%
\pgfusepath{fill}%
\end{pgfscope}%
\begin{pgfscope}%
\pgfpathrectangle{\pgfqpoint{0.410000in}{0.938667in}}{\pgfqpoint{1.131600in}{0.533333in}}%
\pgfusepath{clip}%
\pgfsetbuttcap%
\pgfsetmiterjoin%
\definecolor{currentfill}{rgb}{0.200000,0.400000,0.600000}%
\pgfsetfillcolor{currentfill}%
\pgfsetlinewidth{0.000000pt}%
\definecolor{currentstroke}{rgb}{0.000000,0.000000,0.000000}%
\pgfsetstrokecolor{currentstroke}%
\pgfsetstrokeopacity{0.000000}%
\pgfsetdash{}{0pt}%
\pgfpathmoveto{\pgfqpoint{1.189352in}{0.938667in}}%
\pgfpathlineto{\pgfqpoint{1.230370in}{0.938667in}}%
\pgfpathlineto{\pgfqpoint{1.230370in}{1.031019in}}%
\pgfpathlineto{\pgfqpoint{1.189352in}{1.031019in}}%
\pgfpathclose%
\pgfusepath{fill}%
\end{pgfscope}%
\begin{pgfscope}%
\pgfpathrectangle{\pgfqpoint{0.410000in}{0.938667in}}{\pgfqpoint{1.131600in}{0.533333in}}%
\pgfusepath{clip}%
\pgfsetbuttcap%
\pgfsetmiterjoin%
\definecolor{currentfill}{rgb}{0.200000,0.400000,0.600000}%
\pgfsetfillcolor{currentfill}%
\pgfsetlinewidth{0.000000pt}%
\definecolor{currentstroke}{rgb}{0.000000,0.000000,0.000000}%
\pgfsetstrokecolor{currentstroke}%
\pgfsetstrokeopacity{0.000000}%
\pgfsetdash{}{0pt}%
\pgfpathmoveto{\pgfqpoint{1.230370in}{0.938667in}}%
\pgfpathlineto{\pgfqpoint{1.271389in}{0.938667in}}%
\pgfpathlineto{\pgfqpoint{1.271389in}{0.938667in}}%
\pgfpathlineto{\pgfqpoint{1.230370in}{0.938667in}}%
\pgfpathclose%
\pgfusepath{fill}%
\end{pgfscope}%
\begin{pgfscope}%
\pgfpathrectangle{\pgfqpoint{0.410000in}{0.938667in}}{\pgfqpoint{1.131600in}{0.533333in}}%
\pgfusepath{clip}%
\pgfsetbuttcap%
\pgfsetmiterjoin%
\definecolor{currentfill}{rgb}{0.200000,0.400000,0.600000}%
\pgfsetfillcolor{currentfill}%
\pgfsetlinewidth{0.000000pt}%
\definecolor{currentstroke}{rgb}{0.000000,0.000000,0.000000}%
\pgfsetstrokecolor{currentstroke}%
\pgfsetstrokeopacity{0.000000}%
\pgfsetdash{}{0pt}%
\pgfpathmoveto{\pgfqpoint{1.271389in}{0.938667in}}%
\pgfpathlineto{\pgfqpoint{1.312407in}{0.938667in}}%
\pgfpathlineto{\pgfqpoint{1.312407in}{0.984843in}}%
\pgfpathlineto{\pgfqpoint{1.271389in}{0.984843in}}%
\pgfpathclose%
\pgfusepath{fill}%
\end{pgfscope}%
\begin{pgfscope}%
\pgfpathrectangle{\pgfqpoint{0.410000in}{0.938667in}}{\pgfqpoint{1.131600in}{0.533333in}}%
\pgfusepath{clip}%
\pgfsetbuttcap%
\pgfsetmiterjoin%
\definecolor{currentfill}{rgb}{0.200000,0.400000,0.600000}%
\pgfsetfillcolor{currentfill}%
\pgfsetlinewidth{0.000000pt}%
\definecolor{currentstroke}{rgb}{0.000000,0.000000,0.000000}%
\pgfsetstrokecolor{currentstroke}%
\pgfsetstrokeopacity{0.000000}%
\pgfsetdash{}{0pt}%
\pgfpathmoveto{\pgfqpoint{1.312407in}{0.938667in}}%
\pgfpathlineto{\pgfqpoint{1.353426in}{0.938667in}}%
\pgfpathlineto{\pgfqpoint{1.353426in}{0.938667in}}%
\pgfpathlineto{\pgfqpoint{1.312407in}{0.938667in}}%
\pgfpathclose%
\pgfusepath{fill}%
\end{pgfscope}%
\begin{pgfscope}%
\pgfpathrectangle{\pgfqpoint{0.410000in}{0.938667in}}{\pgfqpoint{1.131600in}{0.533333in}}%
\pgfusepath{clip}%
\pgfsetbuttcap%
\pgfsetmiterjoin%
\definecolor{currentfill}{rgb}{0.200000,0.400000,0.600000}%
\pgfsetfillcolor{currentfill}%
\pgfsetlinewidth{0.000000pt}%
\definecolor{currentstroke}{rgb}{0.000000,0.000000,0.000000}%
\pgfsetstrokecolor{currentstroke}%
\pgfsetstrokeopacity{0.000000}%
\pgfsetdash{}{0pt}%
\pgfpathmoveto{\pgfqpoint{1.353426in}{0.938667in}}%
\pgfpathlineto{\pgfqpoint{1.394444in}{0.938667in}}%
\pgfpathlineto{\pgfqpoint{1.394444in}{0.984843in}}%
\pgfpathlineto{\pgfqpoint{1.353426in}{0.984843in}}%
\pgfpathclose%
\pgfusepath{fill}%
\end{pgfscope}%
\begin{pgfscope}%
\pgfpathrectangle{\pgfqpoint{0.410000in}{0.938667in}}{\pgfqpoint{1.131600in}{0.533333in}}%
\pgfusepath{clip}%
\pgfsetbuttcap%
\pgfsetmiterjoin%
\definecolor{currentfill}{rgb}{0.200000,0.400000,0.600000}%
\pgfsetfillcolor{currentfill}%
\pgfsetlinewidth{0.000000pt}%
\definecolor{currentstroke}{rgb}{0.000000,0.000000,0.000000}%
\pgfsetstrokecolor{currentstroke}%
\pgfsetstrokeopacity{0.000000}%
\pgfsetdash{}{0pt}%
\pgfpathmoveto{\pgfqpoint{1.394444in}{0.938667in}}%
\pgfpathlineto{\pgfqpoint{1.435463in}{0.938667in}}%
\pgfpathlineto{\pgfqpoint{1.435463in}{0.984843in}}%
\pgfpathlineto{\pgfqpoint{1.394444in}{0.984843in}}%
\pgfpathclose%
\pgfusepath{fill}%
\end{pgfscope}%
\begin{pgfscope}%
\pgfsetbuttcap%
\pgfsetroundjoin%
\definecolor{currentfill}{rgb}{0.000000,0.000000,0.000000}%
\pgfsetfillcolor{currentfill}%
\pgfsetlinewidth{0.803000pt}%
\definecolor{currentstroke}{rgb}{0.000000,0.000000,0.000000}%
\pgfsetstrokecolor{currentstroke}%
\pgfsetdash{}{0pt}%
\pgfsys@defobject{currentmarker}{\pgfqpoint{-0.048611in}{0.000000in}}{\pgfqpoint{0.000000in}{0.000000in}}{%
\pgfpathmoveto{\pgfqpoint{0.000000in}{0.000000in}}%
\pgfpathlineto{\pgfqpoint{-0.048611in}{0.000000in}}%
\pgfusepath{stroke,fill}%
}%
\begin{pgfscope}%
\pgfsys@transformshift{0.410000in}{0.938667in}%
\pgfsys@useobject{currentmarker}{}%
\end{pgfscope}%
\end{pgfscope}%
\begin{pgfscope}%
\pgftext[x=0.128367in,y=0.909731in,left,base]{\rmfamily\fontsize{6.000000}{7.200000}\selectfont \(\displaystyle 0.00\)}%
\end{pgfscope}%
\begin{pgfscope}%
\pgfsetbuttcap%
\pgfsetroundjoin%
\definecolor{currentfill}{rgb}{0.000000,0.000000,0.000000}%
\pgfsetfillcolor{currentfill}%
\pgfsetlinewidth{0.803000pt}%
\definecolor{currentstroke}{rgb}{0.000000,0.000000,0.000000}%
\pgfsetstrokecolor{currentstroke}%
\pgfsetdash{}{0pt}%
\pgfsys@defobject{currentmarker}{\pgfqpoint{-0.048611in}{0.000000in}}{\pgfqpoint{0.000000in}{0.000000in}}{%
\pgfpathmoveto{\pgfqpoint{0.000000in}{0.000000in}}%
\pgfpathlineto{\pgfqpoint{-0.048611in}{0.000000in}}%
\pgfusepath{stroke,fill}%
}%
\begin{pgfscope}%
\pgfsys@transformshift{0.410000in}{1.104900in}%
\pgfsys@useobject{currentmarker}{}%
\end{pgfscope}%
\end{pgfscope}%
\begin{pgfscope}%
\pgftext[x=0.128367in,y=1.075965in,left,base]{\rmfamily\fontsize{6.000000}{7.200000}\selectfont \(\displaystyle 0.05\)}%
\end{pgfscope}%
\begin{pgfscope}%
\pgfsetbuttcap%
\pgfsetroundjoin%
\definecolor{currentfill}{rgb}{0.000000,0.000000,0.000000}%
\pgfsetfillcolor{currentfill}%
\pgfsetlinewidth{0.803000pt}%
\definecolor{currentstroke}{rgb}{0.000000,0.000000,0.000000}%
\pgfsetstrokecolor{currentstroke}%
\pgfsetdash{}{0pt}%
\pgfsys@defobject{currentmarker}{\pgfqpoint{-0.048611in}{0.000000in}}{\pgfqpoint{0.000000in}{0.000000in}}{%
\pgfpathmoveto{\pgfqpoint{0.000000in}{0.000000in}}%
\pgfpathlineto{\pgfqpoint{-0.048611in}{0.000000in}}%
\pgfusepath{stroke,fill}%
}%
\begin{pgfscope}%
\pgfsys@transformshift{0.410000in}{1.271134in}%
\pgfsys@useobject{currentmarker}{}%
\end{pgfscope}%
\end{pgfscope}%
\begin{pgfscope}%
\pgftext[x=0.128367in,y=1.242199in,left,base]{\rmfamily\fontsize{6.000000}{7.200000}\selectfont \(\displaystyle 0.10\)}%
\end{pgfscope}%
\begin{pgfscope}%
\pgfsetbuttcap%
\pgfsetroundjoin%
\definecolor{currentfill}{rgb}{0.000000,0.000000,0.000000}%
\pgfsetfillcolor{currentfill}%
\pgfsetlinewidth{0.803000pt}%
\definecolor{currentstroke}{rgb}{0.000000,0.000000,0.000000}%
\pgfsetstrokecolor{currentstroke}%
\pgfsetdash{}{0pt}%
\pgfsys@defobject{currentmarker}{\pgfqpoint{-0.048611in}{0.000000in}}{\pgfqpoint{0.000000in}{0.000000in}}{%
\pgfpathmoveto{\pgfqpoint{0.000000in}{0.000000in}}%
\pgfpathlineto{\pgfqpoint{-0.048611in}{0.000000in}}%
\pgfusepath{stroke,fill}%
}%
\begin{pgfscope}%
\pgfsys@transformshift{0.410000in}{1.437368in}%
\pgfsys@useobject{currentmarker}{}%
\end{pgfscope}%
\end{pgfscope}%
\begin{pgfscope}%
\pgftext[x=0.128367in,y=1.408433in,left,base]{\rmfamily\fontsize{6.000000}{7.200000}\selectfont \(\displaystyle 0.15\)}%
\end{pgfscope}%
\begin{pgfscope}%
\pgftext[x=0.113089in,y=1.205333in,,bottom,rotate=90.000000]{\rmfamily\fontsize{6.700000}{8.040000}\selectfont PDF}%
\end{pgfscope}%
\begin{pgfscope}%
\pgfsetrectcap%
\pgfsetmiterjoin%
\pgfsetlinewidth{0.803000pt}%
\definecolor{currentstroke}{rgb}{0.000000,0.000000,0.000000}%
\pgfsetstrokecolor{currentstroke}%
\pgfsetdash{}{0pt}%
\pgfpathmoveto{\pgfqpoint{0.410000in}{0.938667in}}%
\pgfpathlineto{\pgfqpoint{0.410000in}{1.472000in}}%
\pgfusepath{stroke}%
\end{pgfscope}%
\begin{pgfscope}%
\pgfsetrectcap%
\pgfsetmiterjoin%
\pgfsetlinewidth{0.803000pt}%
\definecolor{currentstroke}{rgb}{0.000000,0.000000,0.000000}%
\pgfsetstrokecolor{currentstroke}%
\pgfsetdash{}{0pt}%
\pgfpathmoveto{\pgfqpoint{1.541600in}{0.938667in}}%
\pgfpathlineto{\pgfqpoint{1.541600in}{1.472000in}}%
\pgfusepath{stroke}%
\end{pgfscope}%
\begin{pgfscope}%
\pgfsetrectcap%
\pgfsetmiterjoin%
\pgfsetlinewidth{0.803000pt}%
\definecolor{currentstroke}{rgb}{0.000000,0.000000,0.000000}%
\pgfsetstrokecolor{currentstroke}%
\pgfsetdash{}{0pt}%
\pgfpathmoveto{\pgfqpoint{0.410000in}{0.938667in}}%
\pgfpathlineto{\pgfqpoint{1.541600in}{0.938667in}}%
\pgfusepath{stroke}%
\end{pgfscope}%
\begin{pgfscope}%
\pgfsetrectcap%
\pgfsetmiterjoin%
\pgfsetlinewidth{0.803000pt}%
\definecolor{currentstroke}{rgb}{0.000000,0.000000,0.000000}%
\pgfsetstrokecolor{currentstroke}%
\pgfsetdash{}{0pt}%
\pgfpathmoveto{\pgfqpoint{0.410000in}{1.472000in}}%
\pgfpathlineto{\pgfqpoint{1.541600in}{1.472000in}}%
\pgfusepath{stroke}%
\end{pgfscope}%
\begin{pgfscope}%
\pgftext[x=0.975800in,y=1.499778in,,base]{\rmfamily\fontsize{6.800000}{8.160000}\selectfont Not Desensitized}%
\end{pgfscope}%
\begin{pgfscope}%
\pgfsetbuttcap%
\pgfsetmiterjoin%
\definecolor{currentfill}{rgb}{1.000000,1.000000,1.000000}%
\pgfsetfillcolor{currentfill}%
\pgfsetlinewidth{0.000000pt}%
\definecolor{currentstroke}{rgb}{0.000000,0.000000,0.000000}%
\pgfsetstrokecolor{currentstroke}%
\pgfsetstrokeopacity{0.000000}%
\pgfsetdash{}{0pt}%
\pgfpathmoveto{\pgfqpoint{0.410000in}{0.288000in}}%
\pgfpathlineto{\pgfqpoint{1.541600in}{0.288000in}}%
\pgfpathlineto{\pgfqpoint{1.541600in}{0.821333in}}%
\pgfpathlineto{\pgfqpoint{0.410000in}{0.821333in}}%
\pgfpathclose%
\pgfusepath{fill}%
\end{pgfscope}%
\begin{pgfscope}%
\pgfpathrectangle{\pgfqpoint{0.410000in}{0.288000in}}{\pgfqpoint{1.131600in}{0.533333in}}%
\pgfusepath{clip}%
\pgfsetbuttcap%
\pgfsetmiterjoin%
\definecolor{currentfill}{rgb}{0.200000,0.400000,0.600000}%
\pgfsetfillcolor{currentfill}%
\pgfsetlinewidth{0.000000pt}%
\definecolor{currentstroke}{rgb}{0.000000,0.000000,0.000000}%
\pgfsetstrokecolor{currentstroke}%
\pgfsetstrokeopacity{0.000000}%
\pgfsetdash{}{0pt}%
\pgfpathmoveto{\pgfqpoint{0.410000in}{0.288000in}}%
\pgfpathlineto{\pgfqpoint{0.451019in}{0.288000in}}%
\pgfpathlineto{\pgfqpoint{0.451019in}{0.288000in}}%
\pgfpathlineto{\pgfqpoint{0.410000in}{0.288000in}}%
\pgfpathclose%
\pgfusepath{fill}%
\end{pgfscope}%
\begin{pgfscope}%
\pgfpathrectangle{\pgfqpoint{0.410000in}{0.288000in}}{\pgfqpoint{1.131600in}{0.533333in}}%
\pgfusepath{clip}%
\pgfsetbuttcap%
\pgfsetmiterjoin%
\definecolor{currentfill}{rgb}{0.200000,0.400000,0.600000}%
\pgfsetfillcolor{currentfill}%
\pgfsetlinewidth{0.000000pt}%
\definecolor{currentstroke}{rgb}{0.000000,0.000000,0.000000}%
\pgfsetstrokecolor{currentstroke}%
\pgfsetstrokeopacity{0.000000}%
\pgfsetdash{}{0pt}%
\pgfpathmoveto{\pgfqpoint{0.451019in}{0.288000in}}%
\pgfpathlineto{\pgfqpoint{0.492037in}{0.288000in}}%
\pgfpathlineto{\pgfqpoint{0.492037in}{0.288000in}}%
\pgfpathlineto{\pgfqpoint{0.451019in}{0.288000in}}%
\pgfpathclose%
\pgfusepath{fill}%
\end{pgfscope}%
\begin{pgfscope}%
\pgfpathrectangle{\pgfqpoint{0.410000in}{0.288000in}}{\pgfqpoint{1.131600in}{0.533333in}}%
\pgfusepath{clip}%
\pgfsetbuttcap%
\pgfsetmiterjoin%
\definecolor{currentfill}{rgb}{0.200000,0.400000,0.600000}%
\pgfsetfillcolor{currentfill}%
\pgfsetlinewidth{0.000000pt}%
\definecolor{currentstroke}{rgb}{0.000000,0.000000,0.000000}%
\pgfsetstrokecolor{currentstroke}%
\pgfsetstrokeopacity{0.000000}%
\pgfsetdash{}{0pt}%
\pgfpathmoveto{\pgfqpoint{0.492037in}{0.288000in}}%
\pgfpathlineto{\pgfqpoint{0.533056in}{0.288000in}}%
\pgfpathlineto{\pgfqpoint{0.533056in}{0.288000in}}%
\pgfpathlineto{\pgfqpoint{0.492037in}{0.288000in}}%
\pgfpathclose%
\pgfusepath{fill}%
\end{pgfscope}%
\begin{pgfscope}%
\pgfpathrectangle{\pgfqpoint{0.410000in}{0.288000in}}{\pgfqpoint{1.131600in}{0.533333in}}%
\pgfusepath{clip}%
\pgfsetbuttcap%
\pgfsetmiterjoin%
\definecolor{currentfill}{rgb}{0.200000,0.400000,0.600000}%
\pgfsetfillcolor{currentfill}%
\pgfsetlinewidth{0.000000pt}%
\definecolor{currentstroke}{rgb}{0.000000,0.000000,0.000000}%
\pgfsetstrokecolor{currentstroke}%
\pgfsetstrokeopacity{0.000000}%
\pgfsetdash{}{0pt}%
\pgfpathmoveto{\pgfqpoint{0.533056in}{0.288000in}}%
\pgfpathlineto{\pgfqpoint{0.574074in}{0.288000in}}%
\pgfpathlineto{\pgfqpoint{0.574074in}{0.288000in}}%
\pgfpathlineto{\pgfqpoint{0.533056in}{0.288000in}}%
\pgfpathclose%
\pgfusepath{fill}%
\end{pgfscope}%
\begin{pgfscope}%
\pgfpathrectangle{\pgfqpoint{0.410000in}{0.288000in}}{\pgfqpoint{1.131600in}{0.533333in}}%
\pgfusepath{clip}%
\pgfsetbuttcap%
\pgfsetmiterjoin%
\definecolor{currentfill}{rgb}{0.200000,0.400000,0.600000}%
\pgfsetfillcolor{currentfill}%
\pgfsetlinewidth{0.000000pt}%
\definecolor{currentstroke}{rgb}{0.000000,0.000000,0.000000}%
\pgfsetstrokecolor{currentstroke}%
\pgfsetstrokeopacity{0.000000}%
\pgfsetdash{}{0pt}%
\pgfpathmoveto{\pgfqpoint{0.574074in}{0.288000in}}%
\pgfpathlineto{\pgfqpoint{0.615093in}{0.288000in}}%
\pgfpathlineto{\pgfqpoint{0.615093in}{0.288000in}}%
\pgfpathlineto{\pgfqpoint{0.574074in}{0.288000in}}%
\pgfpathclose%
\pgfusepath{fill}%
\end{pgfscope}%
\begin{pgfscope}%
\pgfpathrectangle{\pgfqpoint{0.410000in}{0.288000in}}{\pgfqpoint{1.131600in}{0.533333in}}%
\pgfusepath{clip}%
\pgfsetbuttcap%
\pgfsetmiterjoin%
\definecolor{currentfill}{rgb}{0.200000,0.400000,0.600000}%
\pgfsetfillcolor{currentfill}%
\pgfsetlinewidth{0.000000pt}%
\definecolor{currentstroke}{rgb}{0.000000,0.000000,0.000000}%
\pgfsetstrokecolor{currentstroke}%
\pgfsetstrokeopacity{0.000000}%
\pgfsetdash{}{0pt}%
\pgfpathmoveto{\pgfqpoint{0.615093in}{0.288000in}}%
\pgfpathlineto{\pgfqpoint{0.656111in}{0.288000in}}%
\pgfpathlineto{\pgfqpoint{0.656111in}{0.288000in}}%
\pgfpathlineto{\pgfqpoint{0.615093in}{0.288000in}}%
\pgfpathclose%
\pgfusepath{fill}%
\end{pgfscope}%
\begin{pgfscope}%
\pgfpathrectangle{\pgfqpoint{0.410000in}{0.288000in}}{\pgfqpoint{1.131600in}{0.533333in}}%
\pgfusepath{clip}%
\pgfsetbuttcap%
\pgfsetmiterjoin%
\definecolor{currentfill}{rgb}{0.200000,0.400000,0.600000}%
\pgfsetfillcolor{currentfill}%
\pgfsetlinewidth{0.000000pt}%
\definecolor{currentstroke}{rgb}{0.000000,0.000000,0.000000}%
\pgfsetstrokecolor{currentstroke}%
\pgfsetstrokeopacity{0.000000}%
\pgfsetdash{}{0pt}%
\pgfpathmoveto{\pgfqpoint{0.656111in}{0.288000in}}%
\pgfpathlineto{\pgfqpoint{0.697130in}{0.288000in}}%
\pgfpathlineto{\pgfqpoint{0.697130in}{0.288000in}}%
\pgfpathlineto{\pgfqpoint{0.656111in}{0.288000in}}%
\pgfpathclose%
\pgfusepath{fill}%
\end{pgfscope}%
\begin{pgfscope}%
\pgfpathrectangle{\pgfqpoint{0.410000in}{0.288000in}}{\pgfqpoint{1.131600in}{0.533333in}}%
\pgfusepath{clip}%
\pgfsetbuttcap%
\pgfsetmiterjoin%
\definecolor{currentfill}{rgb}{0.200000,0.400000,0.600000}%
\pgfsetfillcolor{currentfill}%
\pgfsetlinewidth{0.000000pt}%
\definecolor{currentstroke}{rgb}{0.000000,0.000000,0.000000}%
\pgfsetstrokecolor{currentstroke}%
\pgfsetstrokeopacity{0.000000}%
\pgfsetdash{}{0pt}%
\pgfpathmoveto{\pgfqpoint{0.697130in}{0.288000in}}%
\pgfpathlineto{\pgfqpoint{0.738148in}{0.288000in}}%
\pgfpathlineto{\pgfqpoint{0.738148in}{0.302939in}}%
\pgfpathlineto{\pgfqpoint{0.697130in}{0.302939in}}%
\pgfpathclose%
\pgfusepath{fill}%
\end{pgfscope}%
\begin{pgfscope}%
\pgfpathrectangle{\pgfqpoint{0.410000in}{0.288000in}}{\pgfqpoint{1.131600in}{0.533333in}}%
\pgfusepath{clip}%
\pgfsetbuttcap%
\pgfsetmiterjoin%
\definecolor{currentfill}{rgb}{0.200000,0.400000,0.600000}%
\pgfsetfillcolor{currentfill}%
\pgfsetlinewidth{0.000000pt}%
\definecolor{currentstroke}{rgb}{0.000000,0.000000,0.000000}%
\pgfsetstrokecolor{currentstroke}%
\pgfsetstrokeopacity{0.000000}%
\pgfsetdash{}{0pt}%
\pgfpathmoveto{\pgfqpoint{0.738148in}{0.288000in}}%
\pgfpathlineto{\pgfqpoint{0.779167in}{0.288000in}}%
\pgfpathlineto{\pgfqpoint{0.779167in}{0.302939in}}%
\pgfpathlineto{\pgfqpoint{0.738148in}{0.302939in}}%
\pgfpathclose%
\pgfusepath{fill}%
\end{pgfscope}%
\begin{pgfscope}%
\pgfpathrectangle{\pgfqpoint{0.410000in}{0.288000in}}{\pgfqpoint{1.131600in}{0.533333in}}%
\pgfusepath{clip}%
\pgfsetbuttcap%
\pgfsetmiterjoin%
\definecolor{currentfill}{rgb}{0.200000,0.400000,0.600000}%
\pgfsetfillcolor{currentfill}%
\pgfsetlinewidth{0.000000pt}%
\definecolor{currentstroke}{rgb}{0.000000,0.000000,0.000000}%
\pgfsetstrokecolor{currentstroke}%
\pgfsetstrokeopacity{0.000000}%
\pgfsetdash{}{0pt}%
\pgfpathmoveto{\pgfqpoint{0.779167in}{0.288000in}}%
\pgfpathlineto{\pgfqpoint{0.820185in}{0.288000in}}%
\pgfpathlineto{\pgfqpoint{0.820185in}{0.795937in}}%
\pgfpathlineto{\pgfqpoint{0.779167in}{0.795937in}}%
\pgfpathclose%
\pgfusepath{fill}%
\end{pgfscope}%
\begin{pgfscope}%
\pgfpathrectangle{\pgfqpoint{0.410000in}{0.288000in}}{\pgfqpoint{1.131600in}{0.533333in}}%
\pgfusepath{clip}%
\pgfsetbuttcap%
\pgfsetmiterjoin%
\definecolor{currentfill}{rgb}{0.200000,0.400000,0.600000}%
\pgfsetfillcolor{currentfill}%
\pgfsetlinewidth{0.000000pt}%
\definecolor{currentstroke}{rgb}{0.000000,0.000000,0.000000}%
\pgfsetstrokecolor{currentstroke}%
\pgfsetstrokeopacity{0.000000}%
\pgfsetdash{}{0pt}%
\pgfpathmoveto{\pgfqpoint{0.820185in}{0.288000in}}%
\pgfpathlineto{\pgfqpoint{0.861204in}{0.288000in}}%
\pgfpathlineto{\pgfqpoint{0.861204in}{0.288000in}}%
\pgfpathlineto{\pgfqpoint{0.820185in}{0.288000in}}%
\pgfpathclose%
\pgfusepath{fill}%
\end{pgfscope}%
\begin{pgfscope}%
\pgfpathrectangle{\pgfqpoint{0.410000in}{0.288000in}}{\pgfqpoint{1.131600in}{0.533333in}}%
\pgfusepath{clip}%
\pgfsetbuttcap%
\pgfsetmiterjoin%
\definecolor{currentfill}{rgb}{0.200000,0.400000,0.600000}%
\pgfsetfillcolor{currentfill}%
\pgfsetlinewidth{0.000000pt}%
\definecolor{currentstroke}{rgb}{0.000000,0.000000,0.000000}%
\pgfsetstrokecolor{currentstroke}%
\pgfsetstrokeopacity{0.000000}%
\pgfsetdash{}{0pt}%
\pgfpathmoveto{\pgfqpoint{0.861204in}{0.288000in}}%
\pgfpathlineto{\pgfqpoint{0.902222in}{0.288000in}}%
\pgfpathlineto{\pgfqpoint{0.902222in}{0.288000in}}%
\pgfpathlineto{\pgfqpoint{0.861204in}{0.288000in}}%
\pgfpathclose%
\pgfusepath{fill}%
\end{pgfscope}%
\begin{pgfscope}%
\pgfpathrectangle{\pgfqpoint{0.410000in}{0.288000in}}{\pgfqpoint{1.131600in}{0.533333in}}%
\pgfusepath{clip}%
\pgfsetbuttcap%
\pgfsetmiterjoin%
\definecolor{currentfill}{rgb}{0.200000,0.400000,0.600000}%
\pgfsetfillcolor{currentfill}%
\pgfsetlinewidth{0.000000pt}%
\definecolor{currentstroke}{rgb}{0.000000,0.000000,0.000000}%
\pgfsetstrokecolor{currentstroke}%
\pgfsetstrokeopacity{0.000000}%
\pgfsetdash{}{0pt}%
\pgfpathmoveto{\pgfqpoint{0.902222in}{0.288000in}}%
\pgfpathlineto{\pgfqpoint{0.943241in}{0.288000in}}%
\pgfpathlineto{\pgfqpoint{0.943241in}{0.288000in}}%
\pgfpathlineto{\pgfqpoint{0.902222in}{0.288000in}}%
\pgfpathclose%
\pgfusepath{fill}%
\end{pgfscope}%
\begin{pgfscope}%
\pgfpathrectangle{\pgfqpoint{0.410000in}{0.288000in}}{\pgfqpoint{1.131600in}{0.533333in}}%
\pgfusepath{clip}%
\pgfsetbuttcap%
\pgfsetmiterjoin%
\definecolor{currentfill}{rgb}{0.200000,0.400000,0.600000}%
\pgfsetfillcolor{currentfill}%
\pgfsetlinewidth{0.000000pt}%
\definecolor{currentstroke}{rgb}{0.000000,0.000000,0.000000}%
\pgfsetstrokecolor{currentstroke}%
\pgfsetstrokeopacity{0.000000}%
\pgfsetdash{}{0pt}%
\pgfpathmoveto{\pgfqpoint{0.943241in}{0.288000in}}%
\pgfpathlineto{\pgfqpoint{0.984259in}{0.288000in}}%
\pgfpathlineto{\pgfqpoint{0.984259in}{0.288000in}}%
\pgfpathlineto{\pgfqpoint{0.943241in}{0.288000in}}%
\pgfpathclose%
\pgfusepath{fill}%
\end{pgfscope}%
\begin{pgfscope}%
\pgfpathrectangle{\pgfqpoint{0.410000in}{0.288000in}}{\pgfqpoint{1.131600in}{0.533333in}}%
\pgfusepath{clip}%
\pgfsetbuttcap%
\pgfsetmiterjoin%
\definecolor{currentfill}{rgb}{0.200000,0.400000,0.600000}%
\pgfsetfillcolor{currentfill}%
\pgfsetlinewidth{0.000000pt}%
\definecolor{currentstroke}{rgb}{0.000000,0.000000,0.000000}%
\pgfsetstrokecolor{currentstroke}%
\pgfsetstrokeopacity{0.000000}%
\pgfsetdash{}{0pt}%
\pgfpathmoveto{\pgfqpoint{0.984259in}{0.288000in}}%
\pgfpathlineto{\pgfqpoint{1.025278in}{0.288000in}}%
\pgfpathlineto{\pgfqpoint{1.025278in}{0.288000in}}%
\pgfpathlineto{\pgfqpoint{0.984259in}{0.288000in}}%
\pgfpathclose%
\pgfusepath{fill}%
\end{pgfscope}%
\begin{pgfscope}%
\pgfpathrectangle{\pgfqpoint{0.410000in}{0.288000in}}{\pgfqpoint{1.131600in}{0.533333in}}%
\pgfusepath{clip}%
\pgfsetbuttcap%
\pgfsetmiterjoin%
\definecolor{currentfill}{rgb}{0.200000,0.400000,0.600000}%
\pgfsetfillcolor{currentfill}%
\pgfsetlinewidth{0.000000pt}%
\definecolor{currentstroke}{rgb}{0.000000,0.000000,0.000000}%
\pgfsetstrokecolor{currentstroke}%
\pgfsetstrokeopacity{0.000000}%
\pgfsetdash{}{0pt}%
\pgfpathmoveto{\pgfqpoint{1.025278in}{0.288000in}}%
\pgfpathlineto{\pgfqpoint{1.066296in}{0.288000in}}%
\pgfpathlineto{\pgfqpoint{1.066296in}{0.288000in}}%
\pgfpathlineto{\pgfqpoint{1.025278in}{0.288000in}}%
\pgfpathclose%
\pgfusepath{fill}%
\end{pgfscope}%
\begin{pgfscope}%
\pgfpathrectangle{\pgfqpoint{0.410000in}{0.288000in}}{\pgfqpoint{1.131600in}{0.533333in}}%
\pgfusepath{clip}%
\pgfsetbuttcap%
\pgfsetmiterjoin%
\definecolor{currentfill}{rgb}{0.200000,0.400000,0.600000}%
\pgfsetfillcolor{currentfill}%
\pgfsetlinewidth{0.000000pt}%
\definecolor{currentstroke}{rgb}{0.000000,0.000000,0.000000}%
\pgfsetstrokecolor{currentstroke}%
\pgfsetstrokeopacity{0.000000}%
\pgfsetdash{}{0pt}%
\pgfpathmoveto{\pgfqpoint{1.066296in}{0.288000in}}%
\pgfpathlineto{\pgfqpoint{1.107315in}{0.288000in}}%
\pgfpathlineto{\pgfqpoint{1.107315in}{0.288000in}}%
\pgfpathlineto{\pgfqpoint{1.066296in}{0.288000in}}%
\pgfpathclose%
\pgfusepath{fill}%
\end{pgfscope}%
\begin{pgfscope}%
\pgfpathrectangle{\pgfqpoint{0.410000in}{0.288000in}}{\pgfqpoint{1.131600in}{0.533333in}}%
\pgfusepath{clip}%
\pgfsetbuttcap%
\pgfsetmiterjoin%
\definecolor{currentfill}{rgb}{0.200000,0.400000,0.600000}%
\pgfsetfillcolor{currentfill}%
\pgfsetlinewidth{0.000000pt}%
\definecolor{currentstroke}{rgb}{0.000000,0.000000,0.000000}%
\pgfsetstrokecolor{currentstroke}%
\pgfsetstrokeopacity{0.000000}%
\pgfsetdash{}{0pt}%
\pgfpathmoveto{\pgfqpoint{1.107315in}{0.288000in}}%
\pgfpathlineto{\pgfqpoint{1.148333in}{0.288000in}}%
\pgfpathlineto{\pgfqpoint{1.148333in}{0.288000in}}%
\pgfpathlineto{\pgfqpoint{1.107315in}{0.288000in}}%
\pgfpathclose%
\pgfusepath{fill}%
\end{pgfscope}%
\begin{pgfscope}%
\pgfpathrectangle{\pgfqpoint{0.410000in}{0.288000in}}{\pgfqpoint{1.131600in}{0.533333in}}%
\pgfusepath{clip}%
\pgfsetbuttcap%
\pgfsetmiterjoin%
\definecolor{currentfill}{rgb}{0.200000,0.400000,0.600000}%
\pgfsetfillcolor{currentfill}%
\pgfsetlinewidth{0.000000pt}%
\definecolor{currentstroke}{rgb}{0.000000,0.000000,0.000000}%
\pgfsetstrokecolor{currentstroke}%
\pgfsetstrokeopacity{0.000000}%
\pgfsetdash{}{0pt}%
\pgfpathmoveto{\pgfqpoint{1.148333in}{0.288000in}}%
\pgfpathlineto{\pgfqpoint{1.189352in}{0.288000in}}%
\pgfpathlineto{\pgfqpoint{1.189352in}{0.288000in}}%
\pgfpathlineto{\pgfqpoint{1.148333in}{0.288000in}}%
\pgfpathclose%
\pgfusepath{fill}%
\end{pgfscope}%
\begin{pgfscope}%
\pgfpathrectangle{\pgfqpoint{0.410000in}{0.288000in}}{\pgfqpoint{1.131600in}{0.533333in}}%
\pgfusepath{clip}%
\pgfsetbuttcap%
\pgfsetmiterjoin%
\definecolor{currentfill}{rgb}{0.200000,0.400000,0.600000}%
\pgfsetfillcolor{currentfill}%
\pgfsetlinewidth{0.000000pt}%
\definecolor{currentstroke}{rgb}{0.000000,0.000000,0.000000}%
\pgfsetstrokecolor{currentstroke}%
\pgfsetstrokeopacity{0.000000}%
\pgfsetdash{}{0pt}%
\pgfpathmoveto{\pgfqpoint{1.189352in}{0.288000in}}%
\pgfpathlineto{\pgfqpoint{1.230370in}{0.288000in}}%
\pgfpathlineto{\pgfqpoint{1.230370in}{0.288000in}}%
\pgfpathlineto{\pgfqpoint{1.189352in}{0.288000in}}%
\pgfpathclose%
\pgfusepath{fill}%
\end{pgfscope}%
\begin{pgfscope}%
\pgfpathrectangle{\pgfqpoint{0.410000in}{0.288000in}}{\pgfqpoint{1.131600in}{0.533333in}}%
\pgfusepath{clip}%
\pgfsetbuttcap%
\pgfsetmiterjoin%
\definecolor{currentfill}{rgb}{0.200000,0.400000,0.600000}%
\pgfsetfillcolor{currentfill}%
\pgfsetlinewidth{0.000000pt}%
\definecolor{currentstroke}{rgb}{0.000000,0.000000,0.000000}%
\pgfsetstrokecolor{currentstroke}%
\pgfsetstrokeopacity{0.000000}%
\pgfsetdash{}{0pt}%
\pgfpathmoveto{\pgfqpoint{1.230370in}{0.288000in}}%
\pgfpathlineto{\pgfqpoint{1.271389in}{0.288000in}}%
\pgfpathlineto{\pgfqpoint{1.271389in}{0.288000in}}%
\pgfpathlineto{\pgfqpoint{1.230370in}{0.288000in}}%
\pgfpathclose%
\pgfusepath{fill}%
\end{pgfscope}%
\begin{pgfscope}%
\pgfpathrectangle{\pgfqpoint{0.410000in}{0.288000in}}{\pgfqpoint{1.131600in}{0.533333in}}%
\pgfusepath{clip}%
\pgfsetbuttcap%
\pgfsetmiterjoin%
\definecolor{currentfill}{rgb}{0.200000,0.400000,0.600000}%
\pgfsetfillcolor{currentfill}%
\pgfsetlinewidth{0.000000pt}%
\definecolor{currentstroke}{rgb}{0.000000,0.000000,0.000000}%
\pgfsetstrokecolor{currentstroke}%
\pgfsetstrokeopacity{0.000000}%
\pgfsetdash{}{0pt}%
\pgfpathmoveto{\pgfqpoint{1.271389in}{0.288000in}}%
\pgfpathlineto{\pgfqpoint{1.312407in}{0.288000in}}%
\pgfpathlineto{\pgfqpoint{1.312407in}{0.288000in}}%
\pgfpathlineto{\pgfqpoint{1.271389in}{0.288000in}}%
\pgfpathclose%
\pgfusepath{fill}%
\end{pgfscope}%
\begin{pgfscope}%
\pgfpathrectangle{\pgfqpoint{0.410000in}{0.288000in}}{\pgfqpoint{1.131600in}{0.533333in}}%
\pgfusepath{clip}%
\pgfsetbuttcap%
\pgfsetmiterjoin%
\definecolor{currentfill}{rgb}{0.200000,0.400000,0.600000}%
\pgfsetfillcolor{currentfill}%
\pgfsetlinewidth{0.000000pt}%
\definecolor{currentstroke}{rgb}{0.000000,0.000000,0.000000}%
\pgfsetstrokecolor{currentstroke}%
\pgfsetstrokeopacity{0.000000}%
\pgfsetdash{}{0pt}%
\pgfpathmoveto{\pgfqpoint{1.312407in}{0.288000in}}%
\pgfpathlineto{\pgfqpoint{1.353426in}{0.288000in}}%
\pgfpathlineto{\pgfqpoint{1.353426in}{0.288000in}}%
\pgfpathlineto{\pgfqpoint{1.312407in}{0.288000in}}%
\pgfpathclose%
\pgfusepath{fill}%
\end{pgfscope}%
\begin{pgfscope}%
\pgfpathrectangle{\pgfqpoint{0.410000in}{0.288000in}}{\pgfqpoint{1.131600in}{0.533333in}}%
\pgfusepath{clip}%
\pgfsetbuttcap%
\pgfsetmiterjoin%
\definecolor{currentfill}{rgb}{0.200000,0.400000,0.600000}%
\pgfsetfillcolor{currentfill}%
\pgfsetlinewidth{0.000000pt}%
\definecolor{currentstroke}{rgb}{0.000000,0.000000,0.000000}%
\pgfsetstrokecolor{currentstroke}%
\pgfsetstrokeopacity{0.000000}%
\pgfsetdash{}{0pt}%
\pgfpathmoveto{\pgfqpoint{1.353426in}{0.288000in}}%
\pgfpathlineto{\pgfqpoint{1.394444in}{0.288000in}}%
\pgfpathlineto{\pgfqpoint{1.394444in}{0.288000in}}%
\pgfpathlineto{\pgfqpoint{1.353426in}{0.288000in}}%
\pgfpathclose%
\pgfusepath{fill}%
\end{pgfscope}%
\begin{pgfscope}%
\pgfpathrectangle{\pgfqpoint{0.410000in}{0.288000in}}{\pgfqpoint{1.131600in}{0.533333in}}%
\pgfusepath{clip}%
\pgfsetbuttcap%
\pgfsetmiterjoin%
\definecolor{currentfill}{rgb}{0.200000,0.400000,0.600000}%
\pgfsetfillcolor{currentfill}%
\pgfsetlinewidth{0.000000pt}%
\definecolor{currentstroke}{rgb}{0.000000,0.000000,0.000000}%
\pgfsetstrokecolor{currentstroke}%
\pgfsetstrokeopacity{0.000000}%
\pgfsetdash{}{0pt}%
\pgfpathmoveto{\pgfqpoint{1.394444in}{0.288000in}}%
\pgfpathlineto{\pgfqpoint{1.435463in}{0.288000in}}%
\pgfpathlineto{\pgfqpoint{1.435463in}{0.288000in}}%
\pgfpathlineto{\pgfqpoint{1.394444in}{0.288000in}}%
\pgfpathclose%
\pgfusepath{fill}%
\end{pgfscope}%
\begin{pgfscope}%
\pgfsetbuttcap%
\pgfsetroundjoin%
\definecolor{currentfill}{rgb}{0.000000,0.000000,0.000000}%
\pgfsetfillcolor{currentfill}%
\pgfsetlinewidth{0.803000pt}%
\definecolor{currentstroke}{rgb}{0.000000,0.000000,0.000000}%
\pgfsetstrokecolor{currentstroke}%
\pgfsetdash{}{0pt}%
\pgfsys@defobject{currentmarker}{\pgfqpoint{0.000000in}{-0.048611in}}{\pgfqpoint{0.000000in}{0.000000in}}{%
\pgfpathmoveto{\pgfqpoint{0.000000in}{0.000000in}}%
\pgfpathlineto{\pgfqpoint{0.000000in}{-0.048611in}}%
\pgfusepath{stroke,fill}%
}%
\begin{pgfscope}%
\pgfsys@transformshift{0.410000in}{0.288000in}%
\pgfsys@useobject{currentmarker}{}%
\end{pgfscope}%
\end{pgfscope}%
\begin{pgfscope}%
\pgftext[x=0.410000in,y=0.190778in,,top]{\rmfamily\fontsize{6.000000}{7.200000}\selectfont \(\displaystyle 0.00\)}%
\end{pgfscope}%
\begin{pgfscope}%
\pgfsetbuttcap%
\pgfsetroundjoin%
\definecolor{currentfill}{rgb}{0.000000,0.000000,0.000000}%
\pgfsetfillcolor{currentfill}%
\pgfsetlinewidth{0.803000pt}%
\definecolor{currentstroke}{rgb}{0.000000,0.000000,0.000000}%
\pgfsetstrokecolor{currentstroke}%
\pgfsetdash{}{0pt}%
\pgfsys@defobject{currentmarker}{\pgfqpoint{0.000000in}{-0.048611in}}{\pgfqpoint{0.000000in}{0.000000in}}{%
\pgfpathmoveto{\pgfqpoint{0.000000in}{0.000000in}}%
\pgfpathlineto{\pgfqpoint{0.000000in}{-0.048611in}}%
\pgfusepath{stroke,fill}%
}%
\begin{pgfscope}%
\pgfsys@transformshift{0.787200in}{0.288000in}%
\pgfsys@useobject{currentmarker}{}%
\end{pgfscope}%
\end{pgfscope}%
\begin{pgfscope}%
\pgftext[x=0.787200in,y=0.190778in,,top]{\rmfamily\fontsize{6.000000}{7.200000}\selectfont \(\displaystyle 0.05\)}%
\end{pgfscope}%
\begin{pgfscope}%
\pgfsetbuttcap%
\pgfsetroundjoin%
\definecolor{currentfill}{rgb}{0.000000,0.000000,0.000000}%
\pgfsetfillcolor{currentfill}%
\pgfsetlinewidth{0.803000pt}%
\definecolor{currentstroke}{rgb}{0.000000,0.000000,0.000000}%
\pgfsetstrokecolor{currentstroke}%
\pgfsetdash{}{0pt}%
\pgfsys@defobject{currentmarker}{\pgfqpoint{0.000000in}{-0.048611in}}{\pgfqpoint{0.000000in}{0.000000in}}{%
\pgfpathmoveto{\pgfqpoint{0.000000in}{0.000000in}}%
\pgfpathlineto{\pgfqpoint{0.000000in}{-0.048611in}}%
\pgfusepath{stroke,fill}%
}%
\begin{pgfscope}%
\pgfsys@transformshift{1.164400in}{0.288000in}%
\pgfsys@useobject{currentmarker}{}%
\end{pgfscope}%
\end{pgfscope}%
\begin{pgfscope}%
\pgftext[x=1.164400in,y=0.190778in,,top]{\rmfamily\fontsize{6.000000}{7.200000}\selectfont \(\displaystyle 0.10\)}%
\end{pgfscope}%
\begin{pgfscope}%
\pgfsetbuttcap%
\pgfsetroundjoin%
\definecolor{currentfill}{rgb}{0.000000,0.000000,0.000000}%
\pgfsetfillcolor{currentfill}%
\pgfsetlinewidth{0.803000pt}%
\definecolor{currentstroke}{rgb}{0.000000,0.000000,0.000000}%
\pgfsetstrokecolor{currentstroke}%
\pgfsetdash{}{0pt}%
\pgfsys@defobject{currentmarker}{\pgfqpoint{0.000000in}{-0.048611in}}{\pgfqpoint{0.000000in}{0.000000in}}{%
\pgfpathmoveto{\pgfqpoint{0.000000in}{0.000000in}}%
\pgfpathlineto{\pgfqpoint{0.000000in}{-0.048611in}}%
\pgfusepath{stroke,fill}%
}%
\begin{pgfscope}%
\pgfsys@transformshift{1.541600in}{0.288000in}%
\pgfsys@useobject{currentmarker}{}%
\end{pgfscope}%
\end{pgfscope}%
\begin{pgfscope}%
\pgftext[x=1.541600in,y=0.190778in,,top]{\rmfamily\fontsize{6.000000}{7.200000}\selectfont \(\displaystyle 0.15\)}%
\end{pgfscope}%
\begin{pgfscope}%
\pgftext[x=0.975800in,y=0.101426in,,top]{\rmfamily\fontsize{6.700000}{8.040000}\selectfont Sensitivity}%
\end{pgfscope}%
\begin{pgfscope}%
\pgfsetbuttcap%
\pgfsetroundjoin%
\definecolor{currentfill}{rgb}{0.000000,0.000000,0.000000}%
\pgfsetfillcolor{currentfill}%
\pgfsetlinewidth{0.803000pt}%
\definecolor{currentstroke}{rgb}{0.000000,0.000000,0.000000}%
\pgfsetstrokecolor{currentstroke}%
\pgfsetdash{}{0pt}%
\pgfsys@defobject{currentmarker}{\pgfqpoint{-0.048611in}{0.000000in}}{\pgfqpoint{0.000000in}{0.000000in}}{%
\pgfpathmoveto{\pgfqpoint{0.000000in}{0.000000in}}%
\pgfpathlineto{\pgfqpoint{-0.048611in}{0.000000in}}%
\pgfusepath{stroke,fill}%
}%
\begin{pgfscope}%
\pgfsys@transformshift{0.410000in}{0.288000in}%
\pgfsys@useobject{currentmarker}{}%
\end{pgfscope}%
\end{pgfscope}%
\begin{pgfscope}%
\pgftext[x=0.179292in,y=0.259065in,left,base]{\rmfamily\fontsize{6.000000}{7.200000}\selectfont \(\displaystyle 0.0\)}%
\end{pgfscope}%
\begin{pgfscope}%
\pgfsetbuttcap%
\pgfsetroundjoin%
\definecolor{currentfill}{rgb}{0.000000,0.000000,0.000000}%
\pgfsetfillcolor{currentfill}%
\pgfsetlinewidth{0.803000pt}%
\definecolor{currentstroke}{rgb}{0.000000,0.000000,0.000000}%
\pgfsetstrokecolor{currentstroke}%
\pgfsetdash{}{0pt}%
\pgfsys@defobject{currentmarker}{\pgfqpoint{-0.048611in}{0.000000in}}{\pgfqpoint{0.000000in}{0.000000in}}{%
\pgfpathmoveto{\pgfqpoint{0.000000in}{0.000000in}}%
\pgfpathlineto{\pgfqpoint{-0.048611in}{0.000000in}}%
\pgfusepath{stroke,fill}%
}%
\begin{pgfscope}%
\pgfsys@transformshift{0.410000in}{0.556908in}%
\pgfsys@useobject{currentmarker}{}%
\end{pgfscope}%
\end{pgfscope}%
\begin{pgfscope}%
\pgftext[x=0.179292in,y=0.527972in,left,base]{\rmfamily\fontsize{6.000000}{7.200000}\selectfont \(\displaystyle 0.5\)}%
\end{pgfscope}%
\begin{pgfscope}%
\pgfsetbuttcap%
\pgfsetroundjoin%
\definecolor{currentfill}{rgb}{0.000000,0.000000,0.000000}%
\pgfsetfillcolor{currentfill}%
\pgfsetlinewidth{0.803000pt}%
\definecolor{currentstroke}{rgb}{0.000000,0.000000,0.000000}%
\pgfsetstrokecolor{currentstroke}%
\pgfsetdash{}{0pt}%
\pgfsys@defobject{currentmarker}{\pgfqpoint{-0.048611in}{0.000000in}}{\pgfqpoint{0.000000in}{0.000000in}}{%
\pgfpathmoveto{\pgfqpoint{0.000000in}{0.000000in}}%
\pgfpathlineto{\pgfqpoint{-0.048611in}{0.000000in}}%
\pgfusepath{stroke,fill}%
}%
\begin{pgfscope}%
\pgfsys@transformshift{0.410000in}{0.825815in}%
\pgfsys@useobject{currentmarker}{}%
\end{pgfscope}%
\end{pgfscope}%
\begin{pgfscope}%
\pgftext[x=0.179292in,y=0.796880in,left,base]{\rmfamily\fontsize{6.000000}{7.200000}\selectfont \(\displaystyle 1.0\)}%
\end{pgfscope}%
\begin{pgfscope}%
\pgftext[x=0.164014in,y=0.554667in,,bottom,rotate=90.000000]{\rmfamily\fontsize{6.700000}{8.040000}\selectfont PDF}%
\end{pgfscope}%
\begin{pgfscope}%
\pgfsetrectcap%
\pgfsetmiterjoin%
\pgfsetlinewidth{0.803000pt}%
\definecolor{currentstroke}{rgb}{0.000000,0.000000,0.000000}%
\pgfsetstrokecolor{currentstroke}%
\pgfsetdash{}{0pt}%
\pgfpathmoveto{\pgfqpoint{0.410000in}{0.288000in}}%
\pgfpathlineto{\pgfqpoint{0.410000in}{0.821333in}}%
\pgfusepath{stroke}%
\end{pgfscope}%
\begin{pgfscope}%
\pgfsetrectcap%
\pgfsetmiterjoin%
\pgfsetlinewidth{0.803000pt}%
\definecolor{currentstroke}{rgb}{0.000000,0.000000,0.000000}%
\pgfsetstrokecolor{currentstroke}%
\pgfsetdash{}{0pt}%
\pgfpathmoveto{\pgfqpoint{1.541600in}{0.288000in}}%
\pgfpathlineto{\pgfqpoint{1.541600in}{0.821333in}}%
\pgfusepath{stroke}%
\end{pgfscope}%
\begin{pgfscope}%
\pgfsetrectcap%
\pgfsetmiterjoin%
\pgfsetlinewidth{0.803000pt}%
\definecolor{currentstroke}{rgb}{0.000000,0.000000,0.000000}%
\pgfsetstrokecolor{currentstroke}%
\pgfsetdash{}{0pt}%
\pgfpathmoveto{\pgfqpoint{0.410000in}{0.288000in}}%
\pgfpathlineto{\pgfqpoint{1.541600in}{0.288000in}}%
\pgfusepath{stroke}%
\end{pgfscope}%
\begin{pgfscope}%
\pgfsetrectcap%
\pgfsetmiterjoin%
\pgfsetlinewidth{0.803000pt}%
\definecolor{currentstroke}{rgb}{0.000000,0.000000,0.000000}%
\pgfsetstrokecolor{currentstroke}%
\pgfsetdash{}{0pt}%
\pgfpathmoveto{\pgfqpoint{0.410000in}{0.821333in}}%
\pgfpathlineto{\pgfqpoint{1.541600in}{0.821333in}}%
\pgfusepath{stroke}%
\end{pgfscope}%
\begin{pgfscope}%
\pgftext[x=0.975800in,y=0.849111in,,base]{\rmfamily\fontsize{6.800000}{8.160000}\selectfont Desensitized}%
\end{pgfscope}%
\end{pgfpicture}%
\makeatother%
\endgroup%

%% file: plot_data/facebook_data/fb_cluster_combined/combined_fabric_diff_k_ahc.pgf
\begingroup%
\makeatletter%
\begin{pgfpicture}%
\pgfpathrectangle{\pgfpointorigin}{\pgfqpoint{2.620000in}{1.600000in}}%
\pgfusepath{use as bounding box, clip}%
\begin{pgfscope}%
\pgfsetbuttcap%
\pgfsetmiterjoin%
\definecolor{currentfill}{rgb}{1.000000,1.000000,1.000000}%
\pgfsetfillcolor{currentfill}%
\pgfsetlinewidth{0.000000pt}%
\definecolor{currentstroke}{rgb}{1.000000,1.000000,1.000000}%
\pgfsetstrokecolor{currentstroke}%
\pgfsetdash{}{0pt}%
\pgfpathmoveto{\pgfqpoint{0.000000in}{0.000000in}}%
\pgfpathlineto{\pgfqpoint{2.620000in}{0.000000in}}%
\pgfpathlineto{\pgfqpoint{2.620000in}{1.600000in}}%
\pgfpathlineto{\pgfqpoint{0.000000in}{1.600000in}}%
\pgfpathclose%
\pgfusepath{fill}%
\end{pgfscope}%
\begin{pgfscope}%
\pgfsetbuttcap%
\pgfsetmiterjoin%
\definecolor{currentfill}{rgb}{1.000000,1.000000,1.000000}%
\pgfsetfillcolor{currentfill}%
\pgfsetlinewidth{0.000000pt}%
\definecolor{currentstroke}{rgb}{0.000000,0.000000,0.000000}%
\pgfsetstrokecolor{currentstroke}%
\pgfsetstrokeopacity{0.000000}%
\pgfsetdash{}{0pt}%
\pgfpathmoveto{\pgfqpoint{0.393000in}{0.320000in}}%
\pgfpathlineto{\pgfqpoint{2.593800in}{0.320000in}}%
\pgfpathlineto{\pgfqpoint{2.593800in}{1.408000in}}%
\pgfpathlineto{\pgfqpoint{0.393000in}{1.408000in}}%
\pgfpathclose%
\pgfusepath{fill}%
\end{pgfscope}%
\begin{pgfscope}%
\pgfsetbuttcap%
\pgfsetroundjoin%
\definecolor{currentfill}{rgb}{0.000000,0.000000,0.000000}%
\pgfsetfillcolor{currentfill}%
\pgfsetlinewidth{0.803000pt}%
\definecolor{currentstroke}{rgb}{0.000000,0.000000,0.000000}%
\pgfsetstrokecolor{currentstroke}%
\pgfsetdash{}{0pt}%
\pgfsys@defobject{currentmarker}{\pgfqpoint{0.000000in}{-0.048611in}}{\pgfqpoint{0.000000in}{0.000000in}}{%
\pgfpathmoveto{\pgfqpoint{0.000000in}{0.000000in}}%
\pgfpathlineto{\pgfqpoint{0.000000in}{-0.048611in}}%
\pgfusepath{stroke,fill}%
}%
\begin{pgfscope}%
\pgfsys@transformshift{0.613080in}{0.320000in}%
\pgfsys@useobject{currentmarker}{}%
\end{pgfscope}%
\end{pgfscope}%
\begin{pgfscope}%
\pgftext[x=0.613080in,y=0.222778in,,top]{\rmfamily\fontsize{6.500000}{7.800000}\selectfont 1}%
\end{pgfscope}%
\begin{pgfscope}%
\pgfsetbuttcap%
\pgfsetroundjoin%
\definecolor{currentfill}{rgb}{0.000000,0.000000,0.000000}%
\pgfsetfillcolor{currentfill}%
\pgfsetlinewidth{0.803000pt}%
\definecolor{currentstroke}{rgb}{0.000000,0.000000,0.000000}%
\pgfsetstrokecolor{currentstroke}%
\pgfsetdash{}{0pt}%
\pgfsys@defobject{currentmarker}{\pgfqpoint{0.000000in}{-0.048611in}}{\pgfqpoint{0.000000in}{0.000000in}}{%
\pgfpathmoveto{\pgfqpoint{0.000000in}{0.000000in}}%
\pgfpathlineto{\pgfqpoint{0.000000in}{-0.048611in}}%
\pgfusepath{stroke,fill}%
}%
\begin{pgfscope}%
\pgfsys@transformshift{1.053240in}{0.320000in}%
\pgfsys@useobject{currentmarker}{}%
\end{pgfscope}%
\end{pgfscope}%
\begin{pgfscope}%
\pgftext[x=1.053240in,y=0.222778in,,top]{\rmfamily\fontsize{6.500000}{7.800000}\selectfont 3}%
\end{pgfscope}%
\begin{pgfscope}%
\pgfsetbuttcap%
\pgfsetroundjoin%
\definecolor{currentfill}{rgb}{0.000000,0.000000,0.000000}%
\pgfsetfillcolor{currentfill}%
\pgfsetlinewidth{0.803000pt}%
\definecolor{currentstroke}{rgb}{0.000000,0.000000,0.000000}%
\pgfsetstrokecolor{currentstroke}%
\pgfsetdash{}{0pt}%
\pgfsys@defobject{currentmarker}{\pgfqpoint{0.000000in}{-0.048611in}}{\pgfqpoint{0.000000in}{0.000000in}}{%
\pgfpathmoveto{\pgfqpoint{0.000000in}{0.000000in}}%
\pgfpathlineto{\pgfqpoint{0.000000in}{-0.048611in}}%
\pgfusepath{stroke,fill}%
}%
\begin{pgfscope}%
\pgfsys@transformshift{1.493400in}{0.320000in}%
\pgfsys@useobject{currentmarker}{}%
\end{pgfscope}%
\end{pgfscope}%
\begin{pgfscope}%
\pgftext[x=1.493400in,y=0.222778in,,top]{\rmfamily\fontsize{6.500000}{7.800000}\selectfont 5}%
\end{pgfscope}%
\begin{pgfscope}%
\pgfsetbuttcap%
\pgfsetroundjoin%
\definecolor{currentfill}{rgb}{0.000000,0.000000,0.000000}%
\pgfsetfillcolor{currentfill}%
\pgfsetlinewidth{0.803000pt}%
\definecolor{currentstroke}{rgb}{0.000000,0.000000,0.000000}%
\pgfsetstrokecolor{currentstroke}%
\pgfsetdash{}{0pt}%
\pgfsys@defobject{currentmarker}{\pgfqpoint{0.000000in}{-0.048611in}}{\pgfqpoint{0.000000in}{0.000000in}}{%
\pgfpathmoveto{\pgfqpoint{0.000000in}{0.000000in}}%
\pgfpathlineto{\pgfqpoint{0.000000in}{-0.048611in}}%
\pgfusepath{stroke,fill}%
}%
\begin{pgfscope}%
\pgfsys@transformshift{1.933560in}{0.320000in}%
\pgfsys@useobject{currentmarker}{}%
\end{pgfscope}%
\end{pgfscope}%
\begin{pgfscope}%
\pgftext[x=1.933560in,y=0.222778in,,top]{\rmfamily\fontsize{6.500000}{7.800000}\selectfont 7}%
\end{pgfscope}%
\begin{pgfscope}%
\pgfsetbuttcap%
\pgfsetroundjoin%
\definecolor{currentfill}{rgb}{0.000000,0.000000,0.000000}%
\pgfsetfillcolor{currentfill}%
\pgfsetlinewidth{0.803000pt}%
\definecolor{currentstroke}{rgb}{0.000000,0.000000,0.000000}%
\pgfsetstrokecolor{currentstroke}%
\pgfsetdash{}{0pt}%
\pgfsys@defobject{currentmarker}{\pgfqpoint{0.000000in}{-0.048611in}}{\pgfqpoint{0.000000in}{0.000000in}}{%
\pgfpathmoveto{\pgfqpoint{0.000000in}{0.000000in}}%
\pgfpathlineto{\pgfqpoint{0.000000in}{-0.048611in}}%
\pgfusepath{stroke,fill}%
}%
\begin{pgfscope}%
\pgfsys@transformshift{2.373720in}{0.320000in}%
\pgfsys@useobject{currentmarker}{}%
\end{pgfscope}%
\end{pgfscope}%
\begin{pgfscope}%
\pgftext[x=2.373720in,y=0.222778in,,top]{\rmfamily\fontsize{6.500000}{7.800000}\selectfont 9}%
\end{pgfscope}%
\begin{pgfscope}%
\pgftext[x=1.493400in,y=0.093148in,,top]{\rmfamily\fontsize{7.400000}{8.880000}\selectfont Num. of Critical TMs}%
\end{pgfscope}%
\begin{pgfscope}%
\pgfpathrectangle{\pgfqpoint{0.393000in}{0.320000in}}{\pgfqpoint{2.200800in}{1.088000in}}%
\pgfusepath{clip}%
\pgfsetbuttcap%
\pgfsetroundjoin%
\pgfsetlinewidth{0.702625pt}%
\definecolor{currentstroke}{rgb}{0.690196,0.690196,0.690196}%
\pgfsetstrokecolor{currentstroke}%
\pgfsetdash{{4.480000pt}{1.120000pt}{0.700000pt}{1.120000pt}}{0.000000pt}%
\pgfpathmoveto{\pgfqpoint{0.393000in}{0.495295in}}%
\pgfpathlineto{\pgfqpoint{2.593800in}{0.495295in}}%
\pgfusepath{stroke}%
\end{pgfscope}%
\begin{pgfscope}%
\pgfsetbuttcap%
\pgfsetroundjoin%
\definecolor{currentfill}{rgb}{0.000000,0.000000,0.000000}%
\pgfsetfillcolor{currentfill}%
\pgfsetlinewidth{0.803000pt}%
\definecolor{currentstroke}{rgb}{0.000000,0.000000,0.000000}%
\pgfsetstrokecolor{currentstroke}%
\pgfsetdash{}{0pt}%
\pgfsys@defobject{currentmarker}{\pgfqpoint{0.000000in}{0.000000in}}{\pgfqpoint{0.048611in}{0.000000in}}{%
\pgfpathmoveto{\pgfqpoint{0.000000in}{0.000000in}}%
\pgfpathlineto{\pgfqpoint{0.048611in}{0.000000in}}%
\pgfusepath{stroke,fill}%
}%
\begin{pgfscope}%
\pgfsys@transformshift{0.393000in}{0.495295in}%
\pgfsys@useobject{currentmarker}{}%
\end{pgfscope}%
\end{pgfscope}%
\begin{pgfscope}%
\pgftext[x=0.210903in,y=0.466360in,left,base]{\rmfamily\fontsize{6.500000}{7.800000}\selectfont \(\displaystyle 1.4\)}%
\end{pgfscope}%
\begin{pgfscope}%
\pgfpathrectangle{\pgfqpoint{0.393000in}{0.320000in}}{\pgfqpoint{2.200800in}{1.088000in}}%
\pgfusepath{clip}%
\pgfsetbuttcap%
\pgfsetroundjoin%
\pgfsetlinewidth{0.702625pt}%
\definecolor{currentstroke}{rgb}{0.690196,0.690196,0.690196}%
\pgfsetstrokecolor{currentstroke}%
\pgfsetdash{{4.480000pt}{1.120000pt}{0.700000pt}{1.120000pt}}{0.000000pt}%
\pgfpathmoveto{\pgfqpoint{0.393000in}{0.936066in}}%
\pgfpathlineto{\pgfqpoint{2.593800in}{0.936066in}}%
\pgfusepath{stroke}%
\end{pgfscope}%
\begin{pgfscope}%
\pgfsetbuttcap%
\pgfsetroundjoin%
\definecolor{currentfill}{rgb}{0.000000,0.000000,0.000000}%
\pgfsetfillcolor{currentfill}%
\pgfsetlinewidth{0.803000pt}%
\definecolor{currentstroke}{rgb}{0.000000,0.000000,0.000000}%
\pgfsetstrokecolor{currentstroke}%
\pgfsetdash{}{0pt}%
\pgfsys@defobject{currentmarker}{\pgfqpoint{0.000000in}{0.000000in}}{\pgfqpoint{0.048611in}{0.000000in}}{%
\pgfpathmoveto{\pgfqpoint{0.000000in}{0.000000in}}%
\pgfpathlineto{\pgfqpoint{0.048611in}{0.000000in}}%
\pgfusepath{stroke,fill}%
}%
\begin{pgfscope}%
\pgfsys@transformshift{0.393000in}{0.936066in}%
\pgfsys@useobject{currentmarker}{}%
\end{pgfscope}%
\end{pgfscope}%
\begin{pgfscope}%
\pgftext[x=0.210903in,y=0.907131in,left,base]{\rmfamily\fontsize{6.500000}{7.800000}\selectfont \(\displaystyle 1.6\)}%
\end{pgfscope}%
\begin{pgfscope}%
\pgfpathrectangle{\pgfqpoint{0.393000in}{0.320000in}}{\pgfqpoint{2.200800in}{1.088000in}}%
\pgfusepath{clip}%
\pgfsetbuttcap%
\pgfsetroundjoin%
\pgfsetlinewidth{0.702625pt}%
\definecolor{currentstroke}{rgb}{0.690196,0.690196,0.690196}%
\pgfsetstrokecolor{currentstroke}%
\pgfsetdash{{4.480000pt}{1.120000pt}{0.700000pt}{1.120000pt}}{0.000000pt}%
\pgfpathmoveto{\pgfqpoint{0.393000in}{1.376837in}}%
\pgfpathlineto{\pgfqpoint{2.593800in}{1.376837in}}%
\pgfusepath{stroke}%
\end{pgfscope}%
\begin{pgfscope}%
\pgfsetbuttcap%
\pgfsetroundjoin%
\definecolor{currentfill}{rgb}{0.000000,0.000000,0.000000}%
\pgfsetfillcolor{currentfill}%
\pgfsetlinewidth{0.803000pt}%
\definecolor{currentstroke}{rgb}{0.000000,0.000000,0.000000}%
\pgfsetstrokecolor{currentstroke}%
\pgfsetdash{}{0pt}%
\pgfsys@defobject{currentmarker}{\pgfqpoint{0.000000in}{0.000000in}}{\pgfqpoint{0.048611in}{0.000000in}}{%
\pgfpathmoveto{\pgfqpoint{0.000000in}{0.000000in}}%
\pgfpathlineto{\pgfqpoint{0.048611in}{0.000000in}}%
\pgfusepath{stroke,fill}%
}%
\begin{pgfscope}%
\pgfsys@transformshift{0.393000in}{1.376837in}%
\pgfsys@useobject{currentmarker}{}%
\end{pgfscope}%
\end{pgfscope}%
\begin{pgfscope}%
\pgftext[x=0.210903in,y=1.347902in,left,base]{\rmfamily\fontsize{6.500000}{7.800000}\selectfont \(\displaystyle 1.8\)}%
\end{pgfscope}%
\begin{pgfscope}%
\pgfsetbuttcap%
\pgfsetroundjoin%
\definecolor{currentfill}{rgb}{0.000000,0.000000,0.000000}%
\pgfsetfillcolor{currentfill}%
\pgfsetlinewidth{0.602250pt}%
\definecolor{currentstroke}{rgb}{0.000000,0.000000,0.000000}%
\pgfsetstrokecolor{currentstroke}%
\pgfsetdash{}{0pt}%
\pgfsys@defobject{currentmarker}{\pgfqpoint{0.000000in}{0.000000in}}{\pgfqpoint{0.027778in}{0.000000in}}{%
\pgfpathmoveto{\pgfqpoint{0.000000in}{0.000000in}}%
\pgfpathlineto{\pgfqpoint{0.027778in}{0.000000in}}%
\pgfusepath{stroke,fill}%
}%
\begin{pgfscope}%
\pgfsys@transformshift{0.393000in}{0.385102in}%
\pgfsys@useobject{currentmarker}{}%
\end{pgfscope}%
\end{pgfscope}%
\begin{pgfscope}%
\pgfsetbuttcap%
\pgfsetroundjoin%
\definecolor{currentfill}{rgb}{0.000000,0.000000,0.000000}%
\pgfsetfillcolor{currentfill}%
\pgfsetlinewidth{0.602250pt}%
\definecolor{currentstroke}{rgb}{0.000000,0.000000,0.000000}%
\pgfsetstrokecolor{currentstroke}%
\pgfsetdash{}{0pt}%
\pgfsys@defobject{currentmarker}{\pgfqpoint{0.000000in}{0.000000in}}{\pgfqpoint{0.027778in}{0.000000in}}{%
\pgfpathmoveto{\pgfqpoint{0.000000in}{0.000000in}}%
\pgfpathlineto{\pgfqpoint{0.027778in}{0.000000in}}%
\pgfusepath{stroke,fill}%
}%
\begin{pgfscope}%
\pgfsys@transformshift{0.393000in}{0.605488in}%
\pgfsys@useobject{currentmarker}{}%
\end{pgfscope}%
\end{pgfscope}%
\begin{pgfscope}%
\pgfsetbuttcap%
\pgfsetroundjoin%
\definecolor{currentfill}{rgb}{0.000000,0.000000,0.000000}%
\pgfsetfillcolor{currentfill}%
\pgfsetlinewidth{0.602250pt}%
\definecolor{currentstroke}{rgb}{0.000000,0.000000,0.000000}%
\pgfsetstrokecolor{currentstroke}%
\pgfsetdash{}{0pt}%
\pgfsys@defobject{currentmarker}{\pgfqpoint{0.000000in}{0.000000in}}{\pgfqpoint{0.027778in}{0.000000in}}{%
\pgfpathmoveto{\pgfqpoint{0.000000in}{0.000000in}}%
\pgfpathlineto{\pgfqpoint{0.027778in}{0.000000in}}%
\pgfusepath{stroke,fill}%
}%
\begin{pgfscope}%
\pgfsys@transformshift{0.393000in}{0.715680in}%
\pgfsys@useobject{currentmarker}{}%
\end{pgfscope}%
\end{pgfscope}%
\begin{pgfscope}%
\pgfsetbuttcap%
\pgfsetroundjoin%
\definecolor{currentfill}{rgb}{0.000000,0.000000,0.000000}%
\pgfsetfillcolor{currentfill}%
\pgfsetlinewidth{0.602250pt}%
\definecolor{currentstroke}{rgb}{0.000000,0.000000,0.000000}%
\pgfsetstrokecolor{currentstroke}%
\pgfsetdash{}{0pt}%
\pgfsys@defobject{currentmarker}{\pgfqpoint{0.000000in}{0.000000in}}{\pgfqpoint{0.027778in}{0.000000in}}{%
\pgfpathmoveto{\pgfqpoint{0.000000in}{0.000000in}}%
\pgfpathlineto{\pgfqpoint{0.027778in}{0.000000in}}%
\pgfusepath{stroke,fill}%
}%
\begin{pgfscope}%
\pgfsys@transformshift{0.393000in}{0.825873in}%
\pgfsys@useobject{currentmarker}{}%
\end{pgfscope}%
\end{pgfscope}%
\begin{pgfscope}%
\pgfsetbuttcap%
\pgfsetroundjoin%
\definecolor{currentfill}{rgb}{0.000000,0.000000,0.000000}%
\pgfsetfillcolor{currentfill}%
\pgfsetlinewidth{0.602250pt}%
\definecolor{currentstroke}{rgb}{0.000000,0.000000,0.000000}%
\pgfsetstrokecolor{currentstroke}%
\pgfsetdash{}{0pt}%
\pgfsys@defobject{currentmarker}{\pgfqpoint{0.000000in}{0.000000in}}{\pgfqpoint{0.027778in}{0.000000in}}{%
\pgfpathmoveto{\pgfqpoint{0.000000in}{0.000000in}}%
\pgfpathlineto{\pgfqpoint{0.027778in}{0.000000in}}%
\pgfusepath{stroke,fill}%
}%
\begin{pgfscope}%
\pgfsys@transformshift{0.393000in}{1.046259in}%
\pgfsys@useobject{currentmarker}{}%
\end{pgfscope}%
\end{pgfscope}%
\begin{pgfscope}%
\pgfsetbuttcap%
\pgfsetroundjoin%
\definecolor{currentfill}{rgb}{0.000000,0.000000,0.000000}%
\pgfsetfillcolor{currentfill}%
\pgfsetlinewidth{0.602250pt}%
\definecolor{currentstroke}{rgb}{0.000000,0.000000,0.000000}%
\pgfsetstrokecolor{currentstroke}%
\pgfsetdash{}{0pt}%
\pgfsys@defobject{currentmarker}{\pgfqpoint{0.000000in}{0.000000in}}{\pgfqpoint{0.027778in}{0.000000in}}{%
\pgfpathmoveto{\pgfqpoint{0.000000in}{0.000000in}}%
\pgfpathlineto{\pgfqpoint{0.027778in}{0.000000in}}%
\pgfusepath{stroke,fill}%
}%
\begin{pgfscope}%
\pgfsys@transformshift{0.393000in}{1.156452in}%
\pgfsys@useobject{currentmarker}{}%
\end{pgfscope}%
\end{pgfscope}%
\begin{pgfscope}%
\pgfsetbuttcap%
\pgfsetroundjoin%
\definecolor{currentfill}{rgb}{0.000000,0.000000,0.000000}%
\pgfsetfillcolor{currentfill}%
\pgfsetlinewidth{0.602250pt}%
\definecolor{currentstroke}{rgb}{0.000000,0.000000,0.000000}%
\pgfsetstrokecolor{currentstroke}%
\pgfsetdash{}{0pt}%
\pgfsys@defobject{currentmarker}{\pgfqpoint{0.000000in}{0.000000in}}{\pgfqpoint{0.027778in}{0.000000in}}{%
\pgfpathmoveto{\pgfqpoint{0.000000in}{0.000000in}}%
\pgfpathlineto{\pgfqpoint{0.027778in}{0.000000in}}%
\pgfusepath{stroke,fill}%
}%
\begin{pgfscope}%
\pgfsys@transformshift{0.393000in}{1.266645in}%
\pgfsys@useobject{currentmarker}{}%
\end{pgfscope}%
\end{pgfscope}%
\begin{pgfscope}%
\pgftext[x=0.155347in,y=0.864000in,,bottom,rotate=90.000000]{\rmfamily\fontsize{7.400000}{8.880000}\selectfont Avg. Hop Count}%
\end{pgfscope}%
\begin{pgfscope}%
\pgfpathrectangle{\pgfqpoint{0.393000in}{0.320000in}}{\pgfqpoint{2.200800in}{1.088000in}}%
\pgfusepath{clip}%
\pgfsetbuttcap%
\pgfsetmiterjoin%
\definecolor{currentfill}{rgb}{0.194608,0.453431,0.632843}%
\pgfsetfillcolor{currentfill}%
\pgfsetlinewidth{0.803000pt}%
\definecolor{currentstroke}{rgb}{0.247059,0.247059,0.247059}%
\pgfsetstrokecolor{currentstroke}%
\pgfsetdash{}{0pt}%
\pgfpathmoveto{\pgfqpoint{0.438777in}{0.518314in}}%
\pgfpathlineto{\pgfqpoint{0.611319in}{0.518314in}}%
\pgfpathlineto{\pgfqpoint{0.611319in}{0.604477in}}%
\pgfpathlineto{\pgfqpoint{0.438777in}{0.604477in}}%
\pgfpathlineto{\pgfqpoint{0.438777in}{0.518314in}}%
\pgfpathclose%
\pgfusepath{stroke,fill}%
\end{pgfscope}%
\begin{pgfscope}%
\pgfpathrectangle{\pgfqpoint{0.393000in}{0.320000in}}{\pgfqpoint{2.200800in}{1.088000in}}%
\pgfusepath{clip}%
\pgfsetbuttcap%
\pgfsetmiterjoin%
\definecolor{currentfill}{rgb}{0.881863,0.505392,0.173039}%
\pgfsetfillcolor{currentfill}%
\pgfsetlinewidth{0.803000pt}%
\definecolor{currentstroke}{rgb}{0.247059,0.247059,0.247059}%
\pgfsetstrokecolor{currentstroke}%
\pgfsetdash{}{0pt}%
\pgfpathmoveto{\pgfqpoint{0.614841in}{1.174565in}}%
\pgfpathlineto{\pgfqpoint{0.787383in}{1.174565in}}%
\pgfpathlineto{\pgfqpoint{0.787383in}{1.270837in}}%
\pgfpathlineto{\pgfqpoint{0.614841in}{1.270837in}}%
\pgfpathlineto{\pgfqpoint{0.614841in}{1.174565in}}%
\pgfpathclose%
\pgfusepath{stroke,fill}%
\end{pgfscope}%
\begin{pgfscope}%
\pgfpathrectangle{\pgfqpoint{0.393000in}{0.320000in}}{\pgfqpoint{2.200800in}{1.088000in}}%
\pgfusepath{clip}%
\pgfsetbuttcap%
\pgfsetmiterjoin%
\definecolor{currentfill}{rgb}{0.194608,0.453431,0.632843}%
\pgfsetfillcolor{currentfill}%
\pgfsetlinewidth{0.803000pt}%
\definecolor{currentstroke}{rgb}{0.247059,0.247059,0.247059}%
\pgfsetstrokecolor{currentstroke}%
\pgfsetdash{}{0pt}%
\pgfpathmoveto{\pgfqpoint{0.878937in}{0.528481in}}%
\pgfpathlineto{\pgfqpoint{1.051479in}{0.528481in}}%
\pgfpathlineto{\pgfqpoint{1.051479in}{0.630264in}}%
\pgfpathlineto{\pgfqpoint{0.878937in}{0.630264in}}%
\pgfpathlineto{\pgfqpoint{0.878937in}{0.528481in}}%
\pgfpathclose%
\pgfusepath{stroke,fill}%
\end{pgfscope}%
\begin{pgfscope}%
\pgfpathrectangle{\pgfqpoint{0.393000in}{0.320000in}}{\pgfqpoint{2.200800in}{1.088000in}}%
\pgfusepath{clip}%
\pgfsetbuttcap%
\pgfsetmiterjoin%
\definecolor{currentfill}{rgb}{0.881863,0.505392,0.173039}%
\pgfsetfillcolor{currentfill}%
\pgfsetlinewidth{0.803000pt}%
\definecolor{currentstroke}{rgb}{0.247059,0.247059,0.247059}%
\pgfsetstrokecolor{currentstroke}%
\pgfsetdash{}{0pt}%
\pgfpathmoveto{\pgfqpoint{1.055001in}{1.191337in}}%
\pgfpathlineto{\pgfqpoint{1.227543in}{1.191337in}}%
\pgfpathlineto{\pgfqpoint{1.227543in}{1.271136in}}%
\pgfpathlineto{\pgfqpoint{1.055001in}{1.271136in}}%
\pgfpathlineto{\pgfqpoint{1.055001in}{1.191337in}}%
\pgfpathclose%
\pgfusepath{stroke,fill}%
\end{pgfscope}%
\begin{pgfscope}%
\pgfpathrectangle{\pgfqpoint{0.393000in}{0.320000in}}{\pgfqpoint{2.200800in}{1.088000in}}%
\pgfusepath{clip}%
\pgfsetbuttcap%
\pgfsetmiterjoin%
\definecolor{currentfill}{rgb}{0.194608,0.453431,0.632843}%
\pgfsetfillcolor{currentfill}%
\pgfsetlinewidth{0.803000pt}%
\definecolor{currentstroke}{rgb}{0.247059,0.247059,0.247059}%
\pgfsetstrokecolor{currentstroke}%
\pgfsetdash{}{0pt}%
\pgfpathmoveto{\pgfqpoint{1.319097in}{0.467901in}}%
\pgfpathlineto{\pgfqpoint{1.491639in}{0.467901in}}%
\pgfpathlineto{\pgfqpoint{1.491639in}{0.555957in}}%
\pgfpathlineto{\pgfqpoint{1.319097in}{0.555957in}}%
\pgfpathlineto{\pgfqpoint{1.319097in}{0.467901in}}%
\pgfpathclose%
\pgfusepath{stroke,fill}%
\end{pgfscope}%
\begin{pgfscope}%
\pgfpathrectangle{\pgfqpoint{0.393000in}{0.320000in}}{\pgfqpoint{2.200800in}{1.088000in}}%
\pgfusepath{clip}%
\pgfsetbuttcap%
\pgfsetmiterjoin%
\definecolor{currentfill}{rgb}{0.881863,0.505392,0.173039}%
\pgfsetfillcolor{currentfill}%
\pgfsetlinewidth{0.803000pt}%
\definecolor{currentstroke}{rgb}{0.247059,0.247059,0.247059}%
\pgfsetstrokecolor{currentstroke}%
\pgfsetdash{}{0pt}%
\pgfpathmoveto{\pgfqpoint{1.495161in}{1.141876in}}%
\pgfpathlineto{\pgfqpoint{1.667703in}{1.141876in}}%
\pgfpathlineto{\pgfqpoint{1.667703in}{1.235406in}}%
\pgfpathlineto{\pgfqpoint{1.495161in}{1.235406in}}%
\pgfpathlineto{\pgfqpoint{1.495161in}{1.141876in}}%
\pgfpathclose%
\pgfusepath{stroke,fill}%
\end{pgfscope}%
\begin{pgfscope}%
\pgfpathrectangle{\pgfqpoint{0.393000in}{0.320000in}}{\pgfqpoint{2.200800in}{1.088000in}}%
\pgfusepath{clip}%
\pgfsetbuttcap%
\pgfsetmiterjoin%
\definecolor{currentfill}{rgb}{0.194608,0.453431,0.632843}%
\pgfsetfillcolor{currentfill}%
\pgfsetlinewidth{0.803000pt}%
\definecolor{currentstroke}{rgb}{0.247059,0.247059,0.247059}%
\pgfsetstrokecolor{currentstroke}%
\pgfsetdash{}{0pt}%
\pgfpathmoveto{\pgfqpoint{1.759257in}{0.479695in}}%
\pgfpathlineto{\pgfqpoint{1.931799in}{0.479695in}}%
\pgfpathlineto{\pgfqpoint{1.931799in}{0.577581in}}%
\pgfpathlineto{\pgfqpoint{1.759257in}{0.577581in}}%
\pgfpathlineto{\pgfqpoint{1.759257in}{0.479695in}}%
\pgfpathclose%
\pgfusepath{stroke,fill}%
\end{pgfscope}%
\begin{pgfscope}%
\pgfpathrectangle{\pgfqpoint{0.393000in}{0.320000in}}{\pgfqpoint{2.200800in}{1.088000in}}%
\pgfusepath{clip}%
\pgfsetbuttcap%
\pgfsetmiterjoin%
\definecolor{currentfill}{rgb}{0.881863,0.505392,0.173039}%
\pgfsetfillcolor{currentfill}%
\pgfsetlinewidth{0.803000pt}%
\definecolor{currentstroke}{rgb}{0.247059,0.247059,0.247059}%
\pgfsetstrokecolor{currentstroke}%
\pgfsetdash{}{0pt}%
\pgfpathmoveto{\pgfqpoint{1.935321in}{1.153722in}}%
\pgfpathlineto{\pgfqpoint{2.107863in}{1.153722in}}%
\pgfpathlineto{\pgfqpoint{2.107863in}{1.254038in}}%
\pgfpathlineto{\pgfqpoint{1.935321in}{1.254038in}}%
\pgfpathlineto{\pgfqpoint{1.935321in}{1.153722in}}%
\pgfpathclose%
\pgfusepath{stroke,fill}%
\end{pgfscope}%
\begin{pgfscope}%
\pgfpathrectangle{\pgfqpoint{0.393000in}{0.320000in}}{\pgfqpoint{2.200800in}{1.088000in}}%
\pgfusepath{clip}%
\pgfsetbuttcap%
\pgfsetmiterjoin%
\definecolor{currentfill}{rgb}{0.194608,0.453431,0.632843}%
\pgfsetfillcolor{currentfill}%
\pgfsetlinewidth{0.803000pt}%
\definecolor{currentstroke}{rgb}{0.247059,0.247059,0.247059}%
\pgfsetstrokecolor{currentstroke}%
\pgfsetdash{}{0pt}%
\pgfpathmoveto{\pgfqpoint{2.199417in}{0.459473in}}%
\pgfpathlineto{\pgfqpoint{2.371959in}{0.459473in}}%
\pgfpathlineto{\pgfqpoint{2.371959in}{0.547567in}}%
\pgfpathlineto{\pgfqpoint{2.199417in}{0.547567in}}%
\pgfpathlineto{\pgfqpoint{2.199417in}{0.459473in}}%
\pgfpathclose%
\pgfusepath{stroke,fill}%
\end{pgfscope}%
\begin{pgfscope}%
\pgfpathrectangle{\pgfqpoint{0.393000in}{0.320000in}}{\pgfqpoint{2.200800in}{1.088000in}}%
\pgfusepath{clip}%
\pgfsetbuttcap%
\pgfsetmiterjoin%
\definecolor{currentfill}{rgb}{0.881863,0.505392,0.173039}%
\pgfsetfillcolor{currentfill}%
\pgfsetlinewidth{0.803000pt}%
\definecolor{currentstroke}{rgb}{0.247059,0.247059,0.247059}%
\pgfsetstrokecolor{currentstroke}%
\pgfsetdash{}{0pt}%
\pgfpathmoveto{\pgfqpoint{2.375481in}{1.183131in}}%
\pgfpathlineto{\pgfqpoint{2.548023in}{1.183131in}}%
\pgfpathlineto{\pgfqpoint{2.548023in}{1.251061in}}%
\pgfpathlineto{\pgfqpoint{2.375481in}{1.251061in}}%
\pgfpathlineto{\pgfqpoint{2.375481in}{1.183131in}}%
\pgfpathclose%
\pgfusepath{stroke,fill}%
\end{pgfscope}%
\begin{pgfscope}%
\pgfpathrectangle{\pgfqpoint{0.393000in}{0.320000in}}{\pgfqpoint{2.200800in}{1.088000in}}%
\pgfusepath{clip}%
\pgfsetbuttcap%
\pgfsetmiterjoin%
\definecolor{currentfill}{rgb}{0.194608,0.453431,0.632843}%
\pgfsetfillcolor{currentfill}%
\pgfsetlinewidth{0.401500pt}%
\definecolor{currentstroke}{rgb}{0.247059,0.247059,0.247059}%
\pgfsetstrokecolor{currentstroke}%
\pgfsetdash{}{0pt}%
\pgfpathmoveto{\pgfqpoint{0.613080in}{-2.590105in}}%
\pgfpathlineto{\pgfqpoint{0.613080in}{-2.590105in}}%
\pgfpathlineto{\pgfqpoint{0.613080in}{-2.590105in}}%
\pgfpathlineto{\pgfqpoint{0.613080in}{-2.590105in}}%
\pgfpathclose%
\pgfusepath{stroke,fill}%
\end{pgfscope}%
\begin{pgfscope}%
\pgfpathrectangle{\pgfqpoint{0.393000in}{0.320000in}}{\pgfqpoint{2.200800in}{1.088000in}}%
\pgfusepath{clip}%
\pgfsetbuttcap%
\pgfsetmiterjoin%
\definecolor{currentfill}{rgb}{0.881863,0.505392,0.173039}%
\pgfsetfillcolor{currentfill}%
\pgfsetlinewidth{0.401500pt}%
\definecolor{currentstroke}{rgb}{0.247059,0.247059,0.247059}%
\pgfsetstrokecolor{currentstroke}%
\pgfsetdash{}{0pt}%
\pgfpathmoveto{\pgfqpoint{0.613080in}{-2.590105in}}%
\pgfpathlineto{\pgfqpoint{0.613080in}{-2.590105in}}%
\pgfpathlineto{\pgfqpoint{0.613080in}{-2.590105in}}%
\pgfpathlineto{\pgfqpoint{0.613080in}{-2.590105in}}%
\pgfpathclose%
\pgfusepath{stroke,fill}%
\end{pgfscope}%
\begin{pgfscope}%
\pgfpathrectangle{\pgfqpoint{0.393000in}{0.320000in}}{\pgfqpoint{2.200800in}{1.088000in}}%
\pgfusepath{clip}%
\pgfsetrectcap%
\pgfsetroundjoin%
\pgfsetlinewidth{0.803000pt}%
\definecolor{currentstroke}{rgb}{0.247059,0.247059,0.247059}%
\pgfsetstrokecolor{currentstroke}%
\pgfsetdash{}{0pt}%
\pgfpathmoveto{\pgfqpoint{0.525048in}{0.518314in}}%
\pgfpathlineto{\pgfqpoint{0.525048in}{0.427762in}}%
\pgfusepath{stroke}%
\end{pgfscope}%
\begin{pgfscope}%
\pgfpathrectangle{\pgfqpoint{0.393000in}{0.320000in}}{\pgfqpoint{2.200800in}{1.088000in}}%
\pgfusepath{clip}%
\pgfsetrectcap%
\pgfsetroundjoin%
\pgfsetlinewidth{0.803000pt}%
\definecolor{currentstroke}{rgb}{0.247059,0.247059,0.247059}%
\pgfsetstrokecolor{currentstroke}%
\pgfsetdash{}{0pt}%
\pgfpathmoveto{\pgfqpoint{0.525048in}{0.604477in}}%
\pgfpathlineto{\pgfqpoint{0.525048in}{0.702067in}}%
\pgfusepath{stroke}%
\end{pgfscope}%
\begin{pgfscope}%
\pgfpathrectangle{\pgfqpoint{0.393000in}{0.320000in}}{\pgfqpoint{2.200800in}{1.088000in}}%
\pgfusepath{clip}%
\pgfsetrectcap%
\pgfsetroundjoin%
\pgfsetlinewidth{0.803000pt}%
\definecolor{currentstroke}{rgb}{0.247059,0.247059,0.247059}%
\pgfsetstrokecolor{currentstroke}%
\pgfsetdash{}{0pt}%
\pgfpathmoveto{\pgfqpoint{0.481912in}{0.427762in}}%
\pgfpathlineto{\pgfqpoint{0.568184in}{0.427762in}}%
\pgfusepath{stroke}%
\end{pgfscope}%
\begin{pgfscope}%
\pgfpathrectangle{\pgfqpoint{0.393000in}{0.320000in}}{\pgfqpoint{2.200800in}{1.088000in}}%
\pgfusepath{clip}%
\pgfsetrectcap%
\pgfsetroundjoin%
\pgfsetlinewidth{0.803000pt}%
\definecolor{currentstroke}{rgb}{0.247059,0.247059,0.247059}%
\pgfsetstrokecolor{currentstroke}%
\pgfsetdash{}{0pt}%
\pgfpathmoveto{\pgfqpoint{0.481912in}{0.702067in}}%
\pgfpathlineto{\pgfqpoint{0.568184in}{0.702067in}}%
\pgfusepath{stroke}%
\end{pgfscope}%
\begin{pgfscope}%
\pgfpathrectangle{\pgfqpoint{0.393000in}{0.320000in}}{\pgfqpoint{2.200800in}{1.088000in}}%
\pgfusepath{clip}%
\pgfsetrectcap%
\pgfsetroundjoin%
\pgfsetlinewidth{0.803000pt}%
\definecolor{currentstroke}{rgb}{0.247059,0.247059,0.247059}%
\pgfsetstrokecolor{currentstroke}%
\pgfsetdash{}{0pt}%
\pgfpathmoveto{\pgfqpoint{0.701112in}{1.174565in}}%
\pgfpathlineto{\pgfqpoint{0.701112in}{1.084915in}}%
\pgfusepath{stroke}%
\end{pgfscope}%
\begin{pgfscope}%
\pgfpathrectangle{\pgfqpoint{0.393000in}{0.320000in}}{\pgfqpoint{2.200800in}{1.088000in}}%
\pgfusepath{clip}%
\pgfsetrectcap%
\pgfsetroundjoin%
\pgfsetlinewidth{0.803000pt}%
\definecolor{currentstroke}{rgb}{0.247059,0.247059,0.247059}%
\pgfsetstrokecolor{currentstroke}%
\pgfsetdash{}{0pt}%
\pgfpathmoveto{\pgfqpoint{0.701112in}{1.270837in}}%
\pgfpathlineto{\pgfqpoint{0.701112in}{1.358545in}}%
\pgfusepath{stroke}%
\end{pgfscope}%
\begin{pgfscope}%
\pgfpathrectangle{\pgfqpoint{0.393000in}{0.320000in}}{\pgfqpoint{2.200800in}{1.088000in}}%
\pgfusepath{clip}%
\pgfsetrectcap%
\pgfsetroundjoin%
\pgfsetlinewidth{0.803000pt}%
\definecolor{currentstroke}{rgb}{0.247059,0.247059,0.247059}%
\pgfsetstrokecolor{currentstroke}%
\pgfsetdash{}{0pt}%
\pgfpathmoveto{\pgfqpoint{0.657976in}{1.084915in}}%
\pgfpathlineto{\pgfqpoint{0.744248in}{1.084915in}}%
\pgfusepath{stroke}%
\end{pgfscope}%
\begin{pgfscope}%
\pgfpathrectangle{\pgfqpoint{0.393000in}{0.320000in}}{\pgfqpoint{2.200800in}{1.088000in}}%
\pgfusepath{clip}%
\pgfsetrectcap%
\pgfsetroundjoin%
\pgfsetlinewidth{0.803000pt}%
\definecolor{currentstroke}{rgb}{0.247059,0.247059,0.247059}%
\pgfsetstrokecolor{currentstroke}%
\pgfsetdash{}{0pt}%
\pgfpathmoveto{\pgfqpoint{0.657976in}{1.358545in}}%
\pgfpathlineto{\pgfqpoint{0.744248in}{1.358545in}}%
\pgfusepath{stroke}%
\end{pgfscope}%
\begin{pgfscope}%
\pgfpathrectangle{\pgfqpoint{0.393000in}{0.320000in}}{\pgfqpoint{2.200800in}{1.088000in}}%
\pgfusepath{clip}%
\pgfsetrectcap%
\pgfsetroundjoin%
\pgfsetlinewidth{0.803000pt}%
\definecolor{currentstroke}{rgb}{0.247059,0.247059,0.247059}%
\pgfsetstrokecolor{currentstroke}%
\pgfsetdash{}{0pt}%
\pgfpathmoveto{\pgfqpoint{0.965208in}{0.528481in}}%
\pgfpathlineto{\pgfqpoint{0.965208in}{0.427813in}}%
\pgfusepath{stroke}%
\end{pgfscope}%
\begin{pgfscope}%
\pgfpathrectangle{\pgfqpoint{0.393000in}{0.320000in}}{\pgfqpoint{2.200800in}{1.088000in}}%
\pgfusepath{clip}%
\pgfsetrectcap%
\pgfsetroundjoin%
\pgfsetlinewidth{0.803000pt}%
\definecolor{currentstroke}{rgb}{0.247059,0.247059,0.247059}%
\pgfsetstrokecolor{currentstroke}%
\pgfsetdash{}{0pt}%
\pgfpathmoveto{\pgfqpoint{0.965208in}{0.630264in}}%
\pgfpathlineto{\pgfqpoint{0.965208in}{0.725377in}}%
\pgfusepath{stroke}%
\end{pgfscope}%
\begin{pgfscope}%
\pgfpathrectangle{\pgfqpoint{0.393000in}{0.320000in}}{\pgfqpoint{2.200800in}{1.088000in}}%
\pgfusepath{clip}%
\pgfsetrectcap%
\pgfsetroundjoin%
\pgfsetlinewidth{0.803000pt}%
\definecolor{currentstroke}{rgb}{0.247059,0.247059,0.247059}%
\pgfsetstrokecolor{currentstroke}%
\pgfsetdash{}{0pt}%
\pgfpathmoveto{\pgfqpoint{0.922072in}{0.427813in}}%
\pgfpathlineto{\pgfqpoint{1.008344in}{0.427813in}}%
\pgfusepath{stroke}%
\end{pgfscope}%
\begin{pgfscope}%
\pgfpathrectangle{\pgfqpoint{0.393000in}{0.320000in}}{\pgfqpoint{2.200800in}{1.088000in}}%
\pgfusepath{clip}%
\pgfsetrectcap%
\pgfsetroundjoin%
\pgfsetlinewidth{0.803000pt}%
\definecolor{currentstroke}{rgb}{0.247059,0.247059,0.247059}%
\pgfsetstrokecolor{currentstroke}%
\pgfsetdash{}{0pt}%
\pgfpathmoveto{\pgfqpoint{0.922072in}{0.725377in}}%
\pgfpathlineto{\pgfqpoint{1.008344in}{0.725377in}}%
\pgfusepath{stroke}%
\end{pgfscope}%
\begin{pgfscope}%
\pgfpathrectangle{\pgfqpoint{0.393000in}{0.320000in}}{\pgfqpoint{2.200800in}{1.088000in}}%
\pgfusepath{clip}%
\pgfsetrectcap%
\pgfsetroundjoin%
\pgfsetlinewidth{0.803000pt}%
\definecolor{currentstroke}{rgb}{0.247059,0.247059,0.247059}%
\pgfsetstrokecolor{currentstroke}%
\pgfsetdash{}{0pt}%
\pgfpathmoveto{\pgfqpoint{1.141272in}{1.191337in}}%
\pgfpathlineto{\pgfqpoint{1.141272in}{1.120970in}}%
\pgfusepath{stroke}%
\end{pgfscope}%
\begin{pgfscope}%
\pgfpathrectangle{\pgfqpoint{0.393000in}{0.320000in}}{\pgfqpoint{2.200800in}{1.088000in}}%
\pgfusepath{clip}%
\pgfsetrectcap%
\pgfsetroundjoin%
\pgfsetlinewidth{0.803000pt}%
\definecolor{currentstroke}{rgb}{0.247059,0.247059,0.247059}%
\pgfsetstrokecolor{currentstroke}%
\pgfsetdash{}{0pt}%
\pgfpathmoveto{\pgfqpoint{1.141272in}{1.271136in}}%
\pgfpathlineto{\pgfqpoint{1.141272in}{1.341135in}}%
\pgfusepath{stroke}%
\end{pgfscope}%
\begin{pgfscope}%
\pgfpathrectangle{\pgfqpoint{0.393000in}{0.320000in}}{\pgfqpoint{2.200800in}{1.088000in}}%
\pgfusepath{clip}%
\pgfsetrectcap%
\pgfsetroundjoin%
\pgfsetlinewidth{0.803000pt}%
\definecolor{currentstroke}{rgb}{0.247059,0.247059,0.247059}%
\pgfsetstrokecolor{currentstroke}%
\pgfsetdash{}{0pt}%
\pgfpathmoveto{\pgfqpoint{1.098136in}{1.120970in}}%
\pgfpathlineto{\pgfqpoint{1.184408in}{1.120970in}}%
\pgfusepath{stroke}%
\end{pgfscope}%
\begin{pgfscope}%
\pgfpathrectangle{\pgfqpoint{0.393000in}{0.320000in}}{\pgfqpoint{2.200800in}{1.088000in}}%
\pgfusepath{clip}%
\pgfsetrectcap%
\pgfsetroundjoin%
\pgfsetlinewidth{0.803000pt}%
\definecolor{currentstroke}{rgb}{0.247059,0.247059,0.247059}%
\pgfsetstrokecolor{currentstroke}%
\pgfsetdash{}{0pt}%
\pgfpathmoveto{\pgfqpoint{1.098136in}{1.341135in}}%
\pgfpathlineto{\pgfqpoint{1.184408in}{1.341135in}}%
\pgfusepath{stroke}%
\end{pgfscope}%
\begin{pgfscope}%
\pgfpathrectangle{\pgfqpoint{0.393000in}{0.320000in}}{\pgfqpoint{2.200800in}{1.088000in}}%
\pgfusepath{clip}%
\pgfsetrectcap%
\pgfsetroundjoin%
\pgfsetlinewidth{0.803000pt}%
\definecolor{currentstroke}{rgb}{0.247059,0.247059,0.247059}%
\pgfsetstrokecolor{currentstroke}%
\pgfsetdash{}{0pt}%
\pgfpathmoveto{\pgfqpoint{1.405368in}{0.467901in}}%
\pgfpathlineto{\pgfqpoint{1.405368in}{0.377609in}}%
\pgfusepath{stroke}%
\end{pgfscope}%
\begin{pgfscope}%
\pgfpathrectangle{\pgfqpoint{0.393000in}{0.320000in}}{\pgfqpoint{2.200800in}{1.088000in}}%
\pgfusepath{clip}%
\pgfsetrectcap%
\pgfsetroundjoin%
\pgfsetlinewidth{0.803000pt}%
\definecolor{currentstroke}{rgb}{0.247059,0.247059,0.247059}%
\pgfsetstrokecolor{currentstroke}%
\pgfsetdash{}{0pt}%
\pgfpathmoveto{\pgfqpoint{1.405368in}{0.555957in}}%
\pgfpathlineto{\pgfqpoint{1.405368in}{0.635608in}}%
\pgfusepath{stroke}%
\end{pgfscope}%
\begin{pgfscope}%
\pgfpathrectangle{\pgfqpoint{0.393000in}{0.320000in}}{\pgfqpoint{2.200800in}{1.088000in}}%
\pgfusepath{clip}%
\pgfsetrectcap%
\pgfsetroundjoin%
\pgfsetlinewidth{0.803000pt}%
\definecolor{currentstroke}{rgb}{0.247059,0.247059,0.247059}%
\pgfsetstrokecolor{currentstroke}%
\pgfsetdash{}{0pt}%
\pgfpathmoveto{\pgfqpoint{1.362232in}{0.377609in}}%
\pgfpathlineto{\pgfqpoint{1.448504in}{0.377609in}}%
\pgfusepath{stroke}%
\end{pgfscope}%
\begin{pgfscope}%
\pgfpathrectangle{\pgfqpoint{0.393000in}{0.320000in}}{\pgfqpoint{2.200800in}{1.088000in}}%
\pgfusepath{clip}%
\pgfsetrectcap%
\pgfsetroundjoin%
\pgfsetlinewidth{0.803000pt}%
\definecolor{currentstroke}{rgb}{0.247059,0.247059,0.247059}%
\pgfsetstrokecolor{currentstroke}%
\pgfsetdash{}{0pt}%
\pgfpathmoveto{\pgfqpoint{1.362232in}{0.635608in}}%
\pgfpathlineto{\pgfqpoint{1.448504in}{0.635608in}}%
\pgfusepath{stroke}%
\end{pgfscope}%
\begin{pgfscope}%
\pgfpathrectangle{\pgfqpoint{0.393000in}{0.320000in}}{\pgfqpoint{2.200800in}{1.088000in}}%
\pgfusepath{clip}%
\pgfsetrectcap%
\pgfsetroundjoin%
\pgfsetlinewidth{0.803000pt}%
\definecolor{currentstroke}{rgb}{0.247059,0.247059,0.247059}%
\pgfsetstrokecolor{currentstroke}%
\pgfsetdash{}{0pt}%
\pgfpathmoveto{\pgfqpoint{1.581432in}{1.141876in}}%
\pgfpathlineto{\pgfqpoint{1.581432in}{1.044496in}}%
\pgfusepath{stroke}%
\end{pgfscope}%
\begin{pgfscope}%
\pgfpathrectangle{\pgfqpoint{0.393000in}{0.320000in}}{\pgfqpoint{2.200800in}{1.088000in}}%
\pgfusepath{clip}%
\pgfsetrectcap%
\pgfsetroundjoin%
\pgfsetlinewidth{0.803000pt}%
\definecolor{currentstroke}{rgb}{0.247059,0.247059,0.247059}%
\pgfsetstrokecolor{currentstroke}%
\pgfsetdash{}{0pt}%
\pgfpathmoveto{\pgfqpoint{1.581432in}{1.235406in}}%
\pgfpathlineto{\pgfqpoint{1.581432in}{1.331218in}}%
\pgfusepath{stroke}%
\end{pgfscope}%
\begin{pgfscope}%
\pgfpathrectangle{\pgfqpoint{0.393000in}{0.320000in}}{\pgfqpoint{2.200800in}{1.088000in}}%
\pgfusepath{clip}%
\pgfsetrectcap%
\pgfsetroundjoin%
\pgfsetlinewidth{0.803000pt}%
\definecolor{currentstroke}{rgb}{0.247059,0.247059,0.247059}%
\pgfsetstrokecolor{currentstroke}%
\pgfsetdash{}{0pt}%
\pgfpathmoveto{\pgfqpoint{1.538296in}{1.044496in}}%
\pgfpathlineto{\pgfqpoint{1.624568in}{1.044496in}}%
\pgfusepath{stroke}%
\end{pgfscope}%
\begin{pgfscope}%
\pgfpathrectangle{\pgfqpoint{0.393000in}{0.320000in}}{\pgfqpoint{2.200800in}{1.088000in}}%
\pgfusepath{clip}%
\pgfsetrectcap%
\pgfsetroundjoin%
\pgfsetlinewidth{0.803000pt}%
\definecolor{currentstroke}{rgb}{0.247059,0.247059,0.247059}%
\pgfsetstrokecolor{currentstroke}%
\pgfsetdash{}{0pt}%
\pgfpathmoveto{\pgfqpoint{1.538296in}{1.331218in}}%
\pgfpathlineto{\pgfqpoint{1.624568in}{1.331218in}}%
\pgfusepath{stroke}%
\end{pgfscope}%
\begin{pgfscope}%
\pgfpathrectangle{\pgfqpoint{0.393000in}{0.320000in}}{\pgfqpoint{2.200800in}{1.088000in}}%
\pgfusepath{clip}%
\pgfsetrectcap%
\pgfsetroundjoin%
\pgfsetlinewidth{0.803000pt}%
\definecolor{currentstroke}{rgb}{0.247059,0.247059,0.247059}%
\pgfsetstrokecolor{currentstroke}%
\pgfsetdash{}{0pt}%
\pgfpathmoveto{\pgfqpoint{1.845528in}{0.479695in}}%
\pgfpathlineto{\pgfqpoint{1.845528in}{0.384661in}}%
\pgfusepath{stroke}%
\end{pgfscope}%
\begin{pgfscope}%
\pgfpathrectangle{\pgfqpoint{0.393000in}{0.320000in}}{\pgfqpoint{2.200800in}{1.088000in}}%
\pgfusepath{clip}%
\pgfsetrectcap%
\pgfsetroundjoin%
\pgfsetlinewidth{0.803000pt}%
\definecolor{currentstroke}{rgb}{0.247059,0.247059,0.247059}%
\pgfsetstrokecolor{currentstroke}%
\pgfsetdash{}{0pt}%
\pgfpathmoveto{\pgfqpoint{1.845528in}{0.577581in}}%
\pgfpathlineto{\pgfqpoint{1.845528in}{0.667857in}}%
\pgfusepath{stroke}%
\end{pgfscope}%
\begin{pgfscope}%
\pgfpathrectangle{\pgfqpoint{0.393000in}{0.320000in}}{\pgfqpoint{2.200800in}{1.088000in}}%
\pgfusepath{clip}%
\pgfsetrectcap%
\pgfsetroundjoin%
\pgfsetlinewidth{0.803000pt}%
\definecolor{currentstroke}{rgb}{0.247059,0.247059,0.247059}%
\pgfsetstrokecolor{currentstroke}%
\pgfsetdash{}{0pt}%
\pgfpathmoveto{\pgfqpoint{1.802392in}{0.384661in}}%
\pgfpathlineto{\pgfqpoint{1.888664in}{0.384661in}}%
\pgfusepath{stroke}%
\end{pgfscope}%
\begin{pgfscope}%
\pgfpathrectangle{\pgfqpoint{0.393000in}{0.320000in}}{\pgfqpoint{2.200800in}{1.088000in}}%
\pgfusepath{clip}%
\pgfsetrectcap%
\pgfsetroundjoin%
\pgfsetlinewidth{0.803000pt}%
\definecolor{currentstroke}{rgb}{0.247059,0.247059,0.247059}%
\pgfsetstrokecolor{currentstroke}%
\pgfsetdash{}{0pt}%
\pgfpathmoveto{\pgfqpoint{1.802392in}{0.667857in}}%
\pgfpathlineto{\pgfqpoint{1.888664in}{0.667857in}}%
\pgfusepath{stroke}%
\end{pgfscope}%
\begin{pgfscope}%
\pgfpathrectangle{\pgfqpoint{0.393000in}{0.320000in}}{\pgfqpoint{2.200800in}{1.088000in}}%
\pgfusepath{clip}%
\pgfsetrectcap%
\pgfsetroundjoin%
\pgfsetlinewidth{0.803000pt}%
\definecolor{currentstroke}{rgb}{0.247059,0.247059,0.247059}%
\pgfsetstrokecolor{currentstroke}%
\pgfsetdash{}{0pt}%
\pgfpathmoveto{\pgfqpoint{2.021592in}{1.153722in}}%
\pgfpathlineto{\pgfqpoint{2.021592in}{1.051989in}}%
\pgfusepath{stroke}%
\end{pgfscope}%
\begin{pgfscope}%
\pgfpathrectangle{\pgfqpoint{0.393000in}{0.320000in}}{\pgfqpoint{2.200800in}{1.088000in}}%
\pgfusepath{clip}%
\pgfsetrectcap%
\pgfsetroundjoin%
\pgfsetlinewidth{0.803000pt}%
\definecolor{currentstroke}{rgb}{0.247059,0.247059,0.247059}%
\pgfsetstrokecolor{currentstroke}%
\pgfsetdash{}{0pt}%
\pgfpathmoveto{\pgfqpoint{2.021592in}{1.254038in}}%
\pgfpathlineto{\pgfqpoint{2.021592in}{1.350391in}}%
\pgfusepath{stroke}%
\end{pgfscope}%
\begin{pgfscope}%
\pgfpathrectangle{\pgfqpoint{0.393000in}{0.320000in}}{\pgfqpoint{2.200800in}{1.088000in}}%
\pgfusepath{clip}%
\pgfsetrectcap%
\pgfsetroundjoin%
\pgfsetlinewidth{0.803000pt}%
\definecolor{currentstroke}{rgb}{0.247059,0.247059,0.247059}%
\pgfsetstrokecolor{currentstroke}%
\pgfsetdash{}{0pt}%
\pgfpathmoveto{\pgfqpoint{1.978456in}{1.051989in}}%
\pgfpathlineto{\pgfqpoint{2.064728in}{1.051989in}}%
\pgfusepath{stroke}%
\end{pgfscope}%
\begin{pgfscope}%
\pgfpathrectangle{\pgfqpoint{0.393000in}{0.320000in}}{\pgfqpoint{2.200800in}{1.088000in}}%
\pgfusepath{clip}%
\pgfsetrectcap%
\pgfsetroundjoin%
\pgfsetlinewidth{0.803000pt}%
\definecolor{currentstroke}{rgb}{0.247059,0.247059,0.247059}%
\pgfsetstrokecolor{currentstroke}%
\pgfsetdash{}{0pt}%
\pgfpathmoveto{\pgfqpoint{1.978456in}{1.350391in}}%
\pgfpathlineto{\pgfqpoint{2.064728in}{1.350391in}}%
\pgfusepath{stroke}%
\end{pgfscope}%
\begin{pgfscope}%
\pgfpathrectangle{\pgfqpoint{0.393000in}{0.320000in}}{\pgfqpoint{2.200800in}{1.088000in}}%
\pgfusepath{clip}%
\pgfsetrectcap%
\pgfsetroundjoin%
\pgfsetlinewidth{0.803000pt}%
\definecolor{currentstroke}{rgb}{0.247059,0.247059,0.247059}%
\pgfsetstrokecolor{currentstroke}%
\pgfsetdash{}{0pt}%
\pgfpathmoveto{\pgfqpoint{2.285688in}{0.459473in}}%
\pgfpathlineto{\pgfqpoint{2.285688in}{0.369455in}}%
\pgfusepath{stroke}%
\end{pgfscope}%
\begin{pgfscope}%
\pgfpathrectangle{\pgfqpoint{0.393000in}{0.320000in}}{\pgfqpoint{2.200800in}{1.088000in}}%
\pgfusepath{clip}%
\pgfsetrectcap%
\pgfsetroundjoin%
\pgfsetlinewidth{0.803000pt}%
\definecolor{currentstroke}{rgb}{0.247059,0.247059,0.247059}%
\pgfsetstrokecolor{currentstroke}%
\pgfsetdash{}{0pt}%
\pgfpathmoveto{\pgfqpoint{2.285688in}{0.547567in}}%
\pgfpathlineto{\pgfqpoint{2.285688in}{0.633477in}}%
\pgfusepath{stroke}%
\end{pgfscope}%
\begin{pgfscope}%
\pgfpathrectangle{\pgfqpoint{0.393000in}{0.320000in}}{\pgfqpoint{2.200800in}{1.088000in}}%
\pgfusepath{clip}%
\pgfsetrectcap%
\pgfsetroundjoin%
\pgfsetlinewidth{0.803000pt}%
\definecolor{currentstroke}{rgb}{0.247059,0.247059,0.247059}%
\pgfsetstrokecolor{currentstroke}%
\pgfsetdash{}{0pt}%
\pgfpathmoveto{\pgfqpoint{2.242552in}{0.369455in}}%
\pgfpathlineto{\pgfqpoint{2.328824in}{0.369455in}}%
\pgfusepath{stroke}%
\end{pgfscope}%
\begin{pgfscope}%
\pgfpathrectangle{\pgfqpoint{0.393000in}{0.320000in}}{\pgfqpoint{2.200800in}{1.088000in}}%
\pgfusepath{clip}%
\pgfsetrectcap%
\pgfsetroundjoin%
\pgfsetlinewidth{0.803000pt}%
\definecolor{currentstroke}{rgb}{0.247059,0.247059,0.247059}%
\pgfsetstrokecolor{currentstroke}%
\pgfsetdash{}{0pt}%
\pgfpathmoveto{\pgfqpoint{2.242552in}{0.633477in}}%
\pgfpathlineto{\pgfqpoint{2.328824in}{0.633477in}}%
\pgfusepath{stroke}%
\end{pgfscope}%
\begin{pgfscope}%
\pgfpathrectangle{\pgfqpoint{0.393000in}{0.320000in}}{\pgfqpoint{2.200800in}{1.088000in}}%
\pgfusepath{clip}%
\pgfsetrectcap%
\pgfsetroundjoin%
\pgfsetlinewidth{0.803000pt}%
\definecolor{currentstroke}{rgb}{0.247059,0.247059,0.247059}%
\pgfsetstrokecolor{currentstroke}%
\pgfsetdash{}{0pt}%
\pgfpathmoveto{\pgfqpoint{2.461752in}{1.183131in}}%
\pgfpathlineto{\pgfqpoint{2.461752in}{1.111934in}}%
\pgfusepath{stroke}%
\end{pgfscope}%
\begin{pgfscope}%
\pgfpathrectangle{\pgfqpoint{0.393000in}{0.320000in}}{\pgfqpoint{2.200800in}{1.088000in}}%
\pgfusepath{clip}%
\pgfsetrectcap%
\pgfsetroundjoin%
\pgfsetlinewidth{0.803000pt}%
\definecolor{currentstroke}{rgb}{0.247059,0.247059,0.247059}%
\pgfsetstrokecolor{currentstroke}%
\pgfsetdash{}{0pt}%
\pgfpathmoveto{\pgfqpoint{2.461752in}{1.251061in}}%
\pgfpathlineto{\pgfqpoint{2.461752in}{1.321587in}}%
\pgfusepath{stroke}%
\end{pgfscope}%
\begin{pgfscope}%
\pgfpathrectangle{\pgfqpoint{0.393000in}{0.320000in}}{\pgfqpoint{2.200800in}{1.088000in}}%
\pgfusepath{clip}%
\pgfsetrectcap%
\pgfsetroundjoin%
\pgfsetlinewidth{0.803000pt}%
\definecolor{currentstroke}{rgb}{0.247059,0.247059,0.247059}%
\pgfsetstrokecolor{currentstroke}%
\pgfsetdash{}{0pt}%
\pgfpathmoveto{\pgfqpoint{2.418616in}{1.111934in}}%
\pgfpathlineto{\pgfqpoint{2.504888in}{1.111934in}}%
\pgfusepath{stroke}%
\end{pgfscope}%
\begin{pgfscope}%
\pgfpathrectangle{\pgfqpoint{0.393000in}{0.320000in}}{\pgfqpoint{2.200800in}{1.088000in}}%
\pgfusepath{clip}%
\pgfsetrectcap%
\pgfsetroundjoin%
\pgfsetlinewidth{0.803000pt}%
\definecolor{currentstroke}{rgb}{0.247059,0.247059,0.247059}%
\pgfsetstrokecolor{currentstroke}%
\pgfsetdash{}{0pt}%
\pgfpathmoveto{\pgfqpoint{2.418616in}{1.321587in}}%
\pgfpathlineto{\pgfqpoint{2.504888in}{1.321587in}}%
\pgfusepath{stroke}%
\end{pgfscope}%
\begin{pgfscope}%
\pgfpathrectangle{\pgfqpoint{0.393000in}{0.320000in}}{\pgfqpoint{2.200800in}{1.088000in}}%
\pgfusepath{clip}%
\pgfsetrectcap%
\pgfsetroundjoin%
\pgfsetlinewidth{0.803000pt}%
\definecolor{currentstroke}{rgb}{0.247059,0.247059,0.247059}%
\pgfsetstrokecolor{currentstroke}%
\pgfsetdash{}{0pt}%
\pgfpathmoveto{\pgfqpoint{0.438777in}{0.563061in}}%
\pgfpathlineto{\pgfqpoint{0.611319in}{0.563061in}}%
\pgfusepath{stroke}%
\end{pgfscope}%
\begin{pgfscope}%
\pgfpathrectangle{\pgfqpoint{0.393000in}{0.320000in}}{\pgfqpoint{2.200800in}{1.088000in}}%
\pgfusepath{clip}%
\pgfsetrectcap%
\pgfsetroundjoin%
\pgfsetlinewidth{0.803000pt}%
\definecolor{currentstroke}{rgb}{0.247059,0.247059,0.247059}%
\pgfsetstrokecolor{currentstroke}%
\pgfsetdash{}{0pt}%
\pgfpathmoveto{\pgfqpoint{0.614841in}{1.222678in}}%
\pgfpathlineto{\pgfqpoint{0.787383in}{1.222678in}}%
\pgfusepath{stroke}%
\end{pgfscope}%
\begin{pgfscope}%
\pgfpathrectangle{\pgfqpoint{0.393000in}{0.320000in}}{\pgfqpoint{2.200800in}{1.088000in}}%
\pgfusepath{clip}%
\pgfsetrectcap%
\pgfsetroundjoin%
\pgfsetlinewidth{0.803000pt}%
\definecolor{currentstroke}{rgb}{0.247059,0.247059,0.247059}%
\pgfsetstrokecolor{currentstroke}%
\pgfsetdash{}{0pt}%
\pgfpathmoveto{\pgfqpoint{0.878937in}{0.581310in}}%
\pgfpathlineto{\pgfqpoint{1.051479in}{0.581310in}}%
\pgfusepath{stroke}%
\end{pgfscope}%
\begin{pgfscope}%
\pgfpathrectangle{\pgfqpoint{0.393000in}{0.320000in}}{\pgfqpoint{2.200800in}{1.088000in}}%
\pgfusepath{clip}%
\pgfsetrectcap%
\pgfsetroundjoin%
\pgfsetlinewidth{0.803000pt}%
\definecolor{currentstroke}{rgb}{0.247059,0.247059,0.247059}%
\pgfsetstrokecolor{currentstroke}%
\pgfsetdash{}{0pt}%
\pgfpathmoveto{\pgfqpoint{1.055001in}{1.230954in}}%
\pgfpathlineto{\pgfqpoint{1.227543in}{1.230954in}}%
\pgfusepath{stroke}%
\end{pgfscope}%
\begin{pgfscope}%
\pgfpathrectangle{\pgfqpoint{0.393000in}{0.320000in}}{\pgfqpoint{2.200800in}{1.088000in}}%
\pgfusepath{clip}%
\pgfsetrectcap%
\pgfsetroundjoin%
\pgfsetlinewidth{0.803000pt}%
\definecolor{currentstroke}{rgb}{0.247059,0.247059,0.247059}%
\pgfsetstrokecolor{currentstroke}%
\pgfsetdash{}{0pt}%
\pgfpathmoveto{\pgfqpoint{1.319097in}{0.509527in}}%
\pgfpathlineto{\pgfqpoint{1.491639in}{0.509527in}}%
\pgfusepath{stroke}%
\end{pgfscope}%
\begin{pgfscope}%
\pgfpathrectangle{\pgfqpoint{0.393000in}{0.320000in}}{\pgfqpoint{2.200800in}{1.088000in}}%
\pgfusepath{clip}%
\pgfsetrectcap%
\pgfsetroundjoin%
\pgfsetlinewidth{0.803000pt}%
\definecolor{currentstroke}{rgb}{0.247059,0.247059,0.247059}%
\pgfsetstrokecolor{currentstroke}%
\pgfsetdash{}{0pt}%
\pgfpathmoveto{\pgfqpoint{1.495161in}{1.188225in}}%
\pgfpathlineto{\pgfqpoint{1.667703in}{1.188225in}}%
\pgfusepath{stroke}%
\end{pgfscope}%
\begin{pgfscope}%
\pgfpathrectangle{\pgfqpoint{0.393000in}{0.320000in}}{\pgfqpoint{2.200800in}{1.088000in}}%
\pgfusepath{clip}%
\pgfsetrectcap%
\pgfsetroundjoin%
\pgfsetlinewidth{0.803000pt}%
\definecolor{currentstroke}{rgb}{0.247059,0.247059,0.247059}%
\pgfsetstrokecolor{currentstroke}%
\pgfsetdash{}{0pt}%
\pgfpathmoveto{\pgfqpoint{1.759257in}{0.531310in}}%
\pgfpathlineto{\pgfqpoint{1.931799in}{0.531310in}}%
\pgfusepath{stroke}%
\end{pgfscope}%
\begin{pgfscope}%
\pgfpathrectangle{\pgfqpoint{0.393000in}{0.320000in}}{\pgfqpoint{2.200800in}{1.088000in}}%
\pgfusepath{clip}%
\pgfsetrectcap%
\pgfsetroundjoin%
\pgfsetlinewidth{0.803000pt}%
\definecolor{currentstroke}{rgb}{0.247059,0.247059,0.247059}%
\pgfsetstrokecolor{currentstroke}%
\pgfsetdash{}{0pt}%
\pgfpathmoveto{\pgfqpoint{1.935321in}{1.203783in}}%
\pgfpathlineto{\pgfqpoint{2.107863in}{1.203783in}}%
\pgfusepath{stroke}%
\end{pgfscope}%
\begin{pgfscope}%
\pgfpathrectangle{\pgfqpoint{0.393000in}{0.320000in}}{\pgfqpoint{2.200800in}{1.088000in}}%
\pgfusepath{clip}%
\pgfsetrectcap%
\pgfsetroundjoin%
\pgfsetlinewidth{0.803000pt}%
\definecolor{currentstroke}{rgb}{0.247059,0.247059,0.247059}%
\pgfsetstrokecolor{currentstroke}%
\pgfsetdash{}{0pt}%
\pgfpathmoveto{\pgfqpoint{2.199417in}{0.501046in}}%
\pgfpathlineto{\pgfqpoint{2.371959in}{0.501046in}}%
\pgfusepath{stroke}%
\end{pgfscope}%
\begin{pgfscope}%
\pgfpathrectangle{\pgfqpoint{0.393000in}{0.320000in}}{\pgfqpoint{2.200800in}{1.088000in}}%
\pgfusepath{clip}%
\pgfsetrectcap%
\pgfsetroundjoin%
\pgfsetlinewidth{0.803000pt}%
\definecolor{currentstroke}{rgb}{0.247059,0.247059,0.247059}%
\pgfsetstrokecolor{currentstroke}%
\pgfsetdash{}{0pt}%
\pgfpathmoveto{\pgfqpoint{2.375481in}{1.215533in}}%
\pgfpathlineto{\pgfqpoint{2.548023in}{1.215533in}}%
\pgfusepath{stroke}%
\end{pgfscope}%
\begin{pgfscope}%
\pgfsetrectcap%
\pgfsetmiterjoin%
\pgfsetlinewidth{0.803000pt}%
\definecolor{currentstroke}{rgb}{0.000000,0.000000,0.000000}%
\pgfsetstrokecolor{currentstroke}%
\pgfsetdash{}{0pt}%
\pgfpathmoveto{\pgfqpoint{0.393000in}{0.320000in}}%
\pgfpathlineto{\pgfqpoint{0.393000in}{1.408000in}}%
\pgfusepath{stroke}%
\end{pgfscope}%
\begin{pgfscope}%
\pgfsetrectcap%
\pgfsetmiterjoin%
\pgfsetlinewidth{0.803000pt}%
\definecolor{currentstroke}{rgb}{0.000000,0.000000,0.000000}%
\pgfsetstrokecolor{currentstroke}%
\pgfsetdash{}{0pt}%
\pgfpathmoveto{\pgfqpoint{2.593800in}{0.320000in}}%
\pgfpathlineto{\pgfqpoint{2.593800in}{1.408000in}}%
\pgfusepath{stroke}%
\end{pgfscope}%
\begin{pgfscope}%
\pgfsetrectcap%
\pgfsetmiterjoin%
\pgfsetlinewidth{0.803000pt}%
\definecolor{currentstroke}{rgb}{0.000000,0.000000,0.000000}%
\pgfsetstrokecolor{currentstroke}%
\pgfsetdash{}{0pt}%
\pgfpathmoveto{\pgfqpoint{0.393000in}{0.320000in}}%
\pgfpathlineto{\pgfqpoint{2.593800in}{0.320000in}}%
\pgfusepath{stroke}%
\end{pgfscope}%
\begin{pgfscope}%
\pgfsetrectcap%
\pgfsetmiterjoin%
\pgfsetlinewidth{0.803000pt}%
\definecolor{currentstroke}{rgb}{0.000000,0.000000,0.000000}%
\pgfsetstrokecolor{currentstroke}%
\pgfsetdash{}{0pt}%
\pgfpathmoveto{\pgfqpoint{0.393000in}{1.408000in}}%
\pgfpathlineto{\pgfqpoint{2.593800in}{1.408000in}}%
\pgfusepath{stroke}%
\end{pgfscope}%
\begin{pgfscope}%
\pgfsetbuttcap%
\pgfsetmiterjoin%
\definecolor{currentfill}{rgb}{1.000000,1.000000,1.000000}%
\pgfsetfillcolor{currentfill}%
\pgfsetfillopacity{0.800000}%
\pgfsetlinewidth{1.003750pt}%
\definecolor{currentstroke}{rgb}{0.800000,0.800000,0.800000}%
\pgfsetstrokecolor{currentstroke}%
\pgfsetstrokeopacity{0.800000}%
\pgfsetdash{}{0pt}%
\pgfpathmoveto{\pgfqpoint{0.517366in}{1.430684in}}%
\pgfpathlineto{\pgfqpoint{2.469434in}{1.430684in}}%
\pgfpathquadraticcurveto{\pgfqpoint{2.487212in}{1.430684in}}{\pgfqpoint{2.487212in}{1.448462in}}%
\pgfpathlineto{\pgfqpoint{2.487212in}{1.562444in}}%
\pgfpathquadraticcurveto{\pgfqpoint{2.487212in}{1.580222in}}{\pgfqpoint{2.469434in}{1.580222in}}%
\pgfpathlineto{\pgfqpoint{0.517366in}{1.580222in}}%
\pgfpathquadraticcurveto{\pgfqpoint{0.499588in}{1.580222in}}{\pgfqpoint{0.499588in}{1.562444in}}%
\pgfpathlineto{\pgfqpoint{0.499588in}{1.448462in}}%
\pgfpathquadraticcurveto{\pgfqpoint{0.499588in}{1.430684in}}{\pgfqpoint{0.517366in}{1.430684in}}%
\pgfpathclose%
\pgfusepath{stroke,fill}%
\end{pgfscope}%
\begin{pgfscope}%
\pgfsetbuttcap%
\pgfsetmiterjoin%
\definecolor{currentfill}{rgb}{0.194608,0.453431,0.632843}%
\pgfsetfillcolor{currentfill}%
\pgfsetlinewidth{0.401500pt}%
\definecolor{currentstroke}{rgb}{0.247059,0.247059,0.247059}%
\pgfsetstrokecolor{currentstroke}%
\pgfsetdash{}{0pt}%
\pgfpathmoveto{\pgfqpoint{0.535144in}{1.482444in}}%
\pgfpathlineto{\pgfqpoint{0.712922in}{1.482444in}}%
\pgfpathlineto{\pgfqpoint{0.712922in}{1.544666in}}%
\pgfpathlineto{\pgfqpoint{0.535144in}{1.544666in}}%
\pgfpathclose%
\pgfusepath{stroke,fill}%
\end{pgfscope}%
\begin{pgfscope}%
\pgftext[x=0.784033in,y=1.482444in,left,base]{\rmfamily\fontsize{6.400000}{7.680000}\selectfont COUDER}%
\end{pgfscope}%
\begin{pgfscope}%
\pgfsetbuttcap%
\pgfsetmiterjoin%
\definecolor{currentfill}{rgb}{0.881863,0.505392,0.173039}%
\pgfsetfillcolor{currentfill}%
\pgfsetlinewidth{0.401500pt}%
\definecolor{currentstroke}{rgb}{0.247059,0.247059,0.247059}%
\pgfsetstrokecolor{currentstroke}%
\pgfsetdash{}{0pt}%
\pgfpathmoveto{\pgfqpoint{1.405092in}{1.482444in}}%
\pgfpathlineto{\pgfqpoint{1.582870in}{1.482444in}}%
\pgfpathlineto{\pgfqpoint{1.582870in}{1.544666in}}%
\pgfpathlineto{\pgfqpoint{1.405092in}{1.544666in}}%
\pgfpathclose%
\pgfusepath{stroke,fill}%
\end{pgfscope}%
\begin{pgfscope}%
\pgftext[x=1.653981in,y=1.482444in,left,base]{\rmfamily\fontsize{6.400000}{7.680000}\selectfont Static Unif. Mesh}%
\end{pgfscope}%
\end{pgfpicture}%
\makeatother%
\endgroup%

%% file: plot_data/facebook_data/reconfiguration_leftover_dcn_capacity.pgf
\begingroup%
\makeatletter%
\begin{pgfpicture}%
\pgfpathrectangle{\pgfpointorigin}{\pgfqpoint{1.260000in}{1.600000in}}%
\pgfusepath{use as bounding box, clip}%
\begin{pgfscope}%
\pgfsetbuttcap%
\pgfsetmiterjoin%
\definecolor{currentfill}{rgb}{1.000000,1.000000,1.000000}%
\pgfsetfillcolor{currentfill}%
\pgfsetlinewidth{0.000000pt}%
\definecolor{currentstroke}{rgb}{1.000000,1.000000,1.000000}%
\pgfsetstrokecolor{currentstroke}%
\pgfsetdash{}{0pt}%
\pgfpathmoveto{\pgfqpoint{0.000000in}{0.000000in}}%
\pgfpathlineto{\pgfqpoint{1.260000in}{0.000000in}}%
\pgfpathlineto{\pgfqpoint{1.260000in}{1.600000in}}%
\pgfpathlineto{\pgfqpoint{0.000000in}{1.600000in}}%
\pgfpathclose%
\pgfusepath{fill}%
\end{pgfscope}%
\begin{pgfscope}%
\pgfsetbuttcap%
\pgfsetmiterjoin%
\definecolor{currentfill}{rgb}{1.000000,1.000000,1.000000}%
\pgfsetfillcolor{currentfill}%
\pgfsetlinewidth{0.000000pt}%
\definecolor{currentstroke}{rgb}{0.000000,0.000000,0.000000}%
\pgfsetstrokecolor{currentstroke}%
\pgfsetstrokeopacity{0.000000}%
\pgfsetdash{}{0pt}%
\pgfpathmoveto{\pgfqpoint{0.346500in}{0.320000in}}%
\pgfpathlineto{\pgfqpoint{1.222200in}{0.320000in}}%
\pgfpathlineto{\pgfqpoint{1.222200in}{1.568000in}}%
\pgfpathlineto{\pgfqpoint{0.346500in}{1.568000in}}%
\pgfpathclose%
\pgfusepath{fill}%
\end{pgfscope}%
\begin{pgfscope}%
\pgfpathrectangle{\pgfqpoint{0.346500in}{0.320000in}}{\pgfqpoint{0.875700in}{1.248000in}}%
\pgfusepath{clip}%
\pgfsetbuttcap%
\pgfsetroundjoin%
\definecolor{currentfill}{rgb}{0.500000,0.200000,0.100000}%
\pgfsetfillcolor{currentfill}%
\pgfsetfillopacity{0.500000}%
\pgfsetlinewidth{0.401500pt}%
\definecolor{currentstroke}{rgb}{0.500000,0.200000,0.100000}%
\pgfsetstrokecolor{currentstroke}%
\pgfsetstrokeopacity{0.500000}%
\pgfsetdash{}{0pt}%
\pgfpathmoveto{\pgfqpoint{0.784350in}{1.551579in}}%
\pgfpathlineto{\pgfqpoint{0.784350in}{1.141053in}}%
\pgfpathlineto{\pgfqpoint{1.152144in}{1.141053in}}%
\pgfpathlineto{\pgfqpoint{1.152144in}{1.551579in}}%
\pgfpathlineto{\pgfqpoint{1.152144in}{1.551579in}}%
\pgfpathlineto{\pgfqpoint{0.784350in}{1.551579in}}%
\pgfpathclose%
\pgfusepath{stroke,fill}%
\end{pgfscope}%
\begin{pgfscope}%
\pgfpathrectangle{\pgfqpoint{0.346500in}{0.320000in}}{\pgfqpoint{0.875700in}{1.248000in}}%
\pgfusepath{clip}%
\pgfsetbuttcap%
\pgfsetroundjoin%
\definecolor{currentfill}{rgb}{0.100000,0.050000,0.400000}%
\pgfsetfillcolor{currentfill}%
\pgfsetfillopacity{0.100000}%
\pgfsetlinewidth{1.003750pt}%
\definecolor{currentstroke}{rgb}{0.100000,0.050000,0.400000}%
\pgfsetstrokecolor{currentstroke}%
\pgfsetstrokeopacity{0.100000}%
\pgfsetdash{}{0pt}%
\pgfpathmoveto{\pgfqpoint{0.346500in}{1.551579in}}%
\pgfpathlineto{\pgfqpoint{0.346500in}{1.551579in}}%
\pgfpathlineto{\pgfqpoint{0.416556in}{1.551579in}}%
\pgfpathlineto{\pgfqpoint{0.416556in}{0.730526in}}%
\pgfpathlineto{\pgfqpoint{0.600453in}{0.730526in}}%
\pgfpathlineto{\pgfqpoint{0.784350in}{0.730526in}}%
\pgfpathlineto{\pgfqpoint{0.784350in}{1.551579in}}%
\pgfpathlineto{\pgfqpoint{1.152144in}{1.551579in}}%
\pgfpathlineto{\pgfqpoint{1.152144in}{1.551579in}}%
\pgfpathlineto{\pgfqpoint{1.222200in}{1.551579in}}%
\pgfpathlineto{\pgfqpoint{1.222200in}{1.551579in}}%
\pgfpathlineto{\pgfqpoint{1.222200in}{1.551579in}}%
\pgfpathlineto{\pgfqpoint{1.152144in}{1.551579in}}%
\pgfpathlineto{\pgfqpoint{1.152144in}{1.551579in}}%
\pgfpathlineto{\pgfqpoint{0.784350in}{1.551579in}}%
\pgfpathlineto{\pgfqpoint{0.784350in}{1.551579in}}%
\pgfpathlineto{\pgfqpoint{0.600453in}{1.551579in}}%
\pgfpathlineto{\pgfqpoint{0.416556in}{1.551579in}}%
\pgfpathlineto{\pgfqpoint{0.416556in}{1.551579in}}%
\pgfpathlineto{\pgfqpoint{0.346500in}{1.551579in}}%
\pgfpathclose%
\pgfusepath{stroke,fill}%
\end{pgfscope}%
\begin{pgfscope}%
\pgfsetbuttcap%
\pgfsetroundjoin%
\definecolor{currentfill}{rgb}{0.000000,0.000000,0.000000}%
\pgfsetfillcolor{currentfill}%
\pgfsetlinewidth{0.803000pt}%
\definecolor{currentstroke}{rgb}{0.000000,0.000000,0.000000}%
\pgfsetstrokecolor{currentstroke}%
\pgfsetdash{}{0pt}%
\pgfsys@defobject{currentmarker}{\pgfqpoint{0.000000in}{-0.048611in}}{\pgfqpoint{0.000000in}{0.000000in}}{%
\pgfpathmoveto{\pgfqpoint{0.000000in}{0.000000in}}%
\pgfpathlineto{\pgfqpoint{0.000000in}{-0.048611in}}%
\pgfusepath{stroke,fill}%
}%
\begin{pgfscope}%
\pgfsys@transformshift{0.416556in}{0.320000in}%
\pgfsys@useobject{currentmarker}{}%
\end{pgfscope}%
\end{pgfscope}%
\begin{pgfscope}%
\pgftext[x=0.416556in,y=0.222778in,,top]{\rmfamily\fontsize{6.200000}{7.440000}\selectfont t0}%
\end{pgfscope}%
\begin{pgfscope}%
\pgfsetbuttcap%
\pgfsetroundjoin%
\definecolor{currentfill}{rgb}{0.000000,0.000000,0.000000}%
\pgfsetfillcolor{currentfill}%
\pgfsetlinewidth{0.803000pt}%
\definecolor{currentstroke}{rgb}{0.000000,0.000000,0.000000}%
\pgfsetstrokecolor{currentstroke}%
\pgfsetdash{}{0pt}%
\pgfsys@defobject{currentmarker}{\pgfqpoint{0.000000in}{-0.048611in}}{\pgfqpoint{0.000000in}{0.000000in}}{%
\pgfpathmoveto{\pgfqpoint{0.000000in}{0.000000in}}%
\pgfpathlineto{\pgfqpoint{0.000000in}{-0.048611in}}%
\pgfusepath{stroke,fill}%
}%
\begin{pgfscope}%
\pgfsys@transformshift{0.600453in}{0.320000in}%
\pgfsys@useobject{currentmarker}{}%
\end{pgfscope}%
\end{pgfscope}%
\begin{pgfscope}%
\pgftext[x=0.600453in,y=0.222778in,,top]{\rmfamily\fontsize{6.200000}{7.440000}\selectfont t1}%
\end{pgfscope}%
\begin{pgfscope}%
\pgfsetbuttcap%
\pgfsetroundjoin%
\definecolor{currentfill}{rgb}{0.000000,0.000000,0.000000}%
\pgfsetfillcolor{currentfill}%
\pgfsetlinewidth{0.803000pt}%
\definecolor{currentstroke}{rgb}{0.000000,0.000000,0.000000}%
\pgfsetstrokecolor{currentstroke}%
\pgfsetdash{}{0pt}%
\pgfsys@defobject{currentmarker}{\pgfqpoint{0.000000in}{-0.048611in}}{\pgfqpoint{0.000000in}{0.000000in}}{%
\pgfpathmoveto{\pgfqpoint{0.000000in}{0.000000in}}%
\pgfpathlineto{\pgfqpoint{0.000000in}{-0.048611in}}%
\pgfusepath{stroke,fill}%
}%
\begin{pgfscope}%
\pgfsys@transformshift{0.784350in}{0.320000in}%
\pgfsys@useobject{currentmarker}{}%
\end{pgfscope}%
\end{pgfscope}%
\begin{pgfscope}%
\pgftext[x=0.784350in,y=0.222778in,,top]{\rmfamily\fontsize{6.200000}{7.440000}\selectfont t2}%
\end{pgfscope}%
\begin{pgfscope}%
\pgfsetbuttcap%
\pgfsetroundjoin%
\definecolor{currentfill}{rgb}{0.000000,0.000000,0.000000}%
\pgfsetfillcolor{currentfill}%
\pgfsetlinewidth{0.803000pt}%
\definecolor{currentstroke}{rgb}{0.000000,0.000000,0.000000}%
\pgfsetstrokecolor{currentstroke}%
\pgfsetdash{}{0pt}%
\pgfsys@defobject{currentmarker}{\pgfqpoint{0.000000in}{-0.048611in}}{\pgfqpoint{0.000000in}{0.000000in}}{%
\pgfpathmoveto{\pgfqpoint{0.000000in}{0.000000in}}%
\pgfpathlineto{\pgfqpoint{0.000000in}{-0.048611in}}%
\pgfusepath{stroke,fill}%
}%
\begin{pgfscope}%
\pgfsys@transformshift{0.968247in}{0.320000in}%
\pgfsys@useobject{currentmarker}{}%
\end{pgfscope}%
\end{pgfscope}%
\begin{pgfscope}%
\pgftext[x=0.968247in,y=0.222778in,,top]{\rmfamily\fontsize{6.200000}{7.440000}\selectfont t3}%
\end{pgfscope}%
\begin{pgfscope}%
\pgfsetbuttcap%
\pgfsetroundjoin%
\definecolor{currentfill}{rgb}{0.000000,0.000000,0.000000}%
\pgfsetfillcolor{currentfill}%
\pgfsetlinewidth{0.803000pt}%
\definecolor{currentstroke}{rgb}{0.000000,0.000000,0.000000}%
\pgfsetstrokecolor{currentstroke}%
\pgfsetdash{}{0pt}%
\pgfsys@defobject{currentmarker}{\pgfqpoint{0.000000in}{-0.048611in}}{\pgfqpoint{0.000000in}{0.000000in}}{%
\pgfpathmoveto{\pgfqpoint{0.000000in}{0.000000in}}%
\pgfpathlineto{\pgfqpoint{0.000000in}{-0.048611in}}%
\pgfusepath{stroke,fill}%
}%
\begin{pgfscope}%
\pgfsys@transformshift{1.152144in}{0.320000in}%
\pgfsys@useobject{currentmarker}{}%
\end{pgfscope}%
\end{pgfscope}%
\begin{pgfscope}%
\pgftext[x=1.152144in,y=0.222778in,,top]{\rmfamily\fontsize{6.200000}{7.440000}\selectfont t4}%
\end{pgfscope}%
\begin{pgfscope}%
\pgftext[x=0.784350in,y=0.130648in,,top]{\rmfamily\fontsize{6.600000}{7.920000}\selectfont Time}%
\end{pgfscope}%
\begin{pgfscope}%
\pgfsetbuttcap%
\pgfsetroundjoin%
\definecolor{currentfill}{rgb}{0.000000,0.000000,0.000000}%
\pgfsetfillcolor{currentfill}%
\pgfsetlinewidth{0.803000pt}%
\definecolor{currentstroke}{rgb}{0.000000,0.000000,0.000000}%
\pgfsetstrokecolor{currentstroke}%
\pgfsetdash{}{0pt}%
\pgfsys@defobject{currentmarker}{\pgfqpoint{-0.048611in}{0.000000in}}{\pgfqpoint{0.000000in}{0.000000in}}{%
\pgfpathmoveto{\pgfqpoint{0.000000in}{0.000000in}}%
\pgfpathlineto{\pgfqpoint{-0.048611in}{0.000000in}}%
\pgfusepath{stroke,fill}%
}%
\begin{pgfscope}%
\pgfsys@transformshift{0.346500in}{0.730526in}%
\pgfsys@useobject{currentmarker}{}%
\end{pgfscope}%
\end{pgfscope}%
\begin{pgfscope}%
\pgftext[x=0.118107in,y=0.701591in,left,base]{\rmfamily\fontsize{6.200000}{7.440000}\selectfont 0.5}%
\end{pgfscope}%
\begin{pgfscope}%
\pgfsetbuttcap%
\pgfsetroundjoin%
\definecolor{currentfill}{rgb}{0.000000,0.000000,0.000000}%
\pgfsetfillcolor{currentfill}%
\pgfsetlinewidth{0.803000pt}%
\definecolor{currentstroke}{rgb}{0.000000,0.000000,0.000000}%
\pgfsetstrokecolor{currentstroke}%
\pgfsetdash{}{0pt}%
\pgfsys@defobject{currentmarker}{\pgfqpoint{-0.048611in}{0.000000in}}{\pgfqpoint{0.000000in}{0.000000in}}{%
\pgfpathmoveto{\pgfqpoint{0.000000in}{0.000000in}}%
\pgfpathlineto{\pgfqpoint{-0.048611in}{0.000000in}}%
\pgfusepath{stroke,fill}%
}%
\begin{pgfscope}%
\pgfsys@transformshift{0.346500in}{1.551579in}%
\pgfsys@useobject{currentmarker}{}%
\end{pgfscope}%
\end{pgfscope}%
\begin{pgfscope}%
\pgftext[x=0.198353in,y=1.522644in,left,base]{\rmfamily\fontsize{6.200000}{7.440000}\selectfont 1}%
\end{pgfscope}%
\begin{pgfscope}%
\pgftext[x=0.111163in,y=0.944000in,,bottom,rotate=90.000000]{\rmfamily\fontsize{6.600000}{7.920000}\selectfont DCN Capacity}%
\end{pgfscope}%
\begin{pgfscope}%
\pgfpathrectangle{\pgfqpoint{0.346500in}{0.320000in}}{\pgfqpoint{0.875700in}{1.248000in}}%
\pgfusepath{clip}%
\pgfsetrectcap%
\pgfsetroundjoin%
\pgfsetlinewidth{1.204500pt}%
\definecolor{currentstroke}{rgb}{0.100000,0.100000,0.100000}%
\pgfsetstrokecolor{currentstroke}%
\pgfsetdash{}{0pt}%
\pgfpathmoveto{\pgfqpoint{0.346500in}{1.551579in}}%
\pgfpathlineto{\pgfqpoint{0.416556in}{1.551579in}}%
\pgfpathlineto{\pgfqpoint{0.416556in}{1.141053in}}%
\pgfpathlineto{\pgfqpoint{0.600453in}{1.141053in}}%
\pgfpathlineto{\pgfqpoint{0.784350in}{1.141053in}}%
\pgfpathlineto{\pgfqpoint{0.968247in}{1.141053in}}%
\pgfpathlineto{\pgfqpoint{1.152144in}{1.141053in}}%
\pgfpathlineto{\pgfqpoint{1.152144in}{1.551579in}}%
\pgfpathlineto{\pgfqpoint{1.222200in}{1.551579in}}%
\pgfusepath{stroke}%
\end{pgfscope}%
\begin{pgfscope}%
\pgfpathrectangle{\pgfqpoint{0.346500in}{0.320000in}}{\pgfqpoint{0.875700in}{1.248000in}}%
\pgfusepath{clip}%
\pgfsetrectcap%
\pgfsetroundjoin%
\pgfsetlinewidth{1.204500pt}%
\definecolor{currentstroke}{rgb}{0.900000,0.200000,0.200000}%
\pgfsetstrokecolor{currentstroke}%
\pgfsetdash{}{0pt}%
\pgfpathmoveto{\pgfqpoint{0.346500in}{1.551579in}}%
\pgfpathlineto{\pgfqpoint{0.416556in}{1.551579in}}%
\pgfpathlineto{\pgfqpoint{0.416556in}{0.730526in}}%
\pgfpathlineto{\pgfqpoint{0.600453in}{0.730526in}}%
\pgfpathlineto{\pgfqpoint{0.784350in}{0.730526in}}%
\pgfpathlineto{\pgfqpoint{0.784350in}{1.551579in}}%
\pgfpathlineto{\pgfqpoint{1.152144in}{1.551579in}}%
\pgfpathlineto{\pgfqpoint{1.152144in}{1.551579in}}%
\pgfpathlineto{\pgfqpoint{1.222200in}{1.551579in}}%
\pgfusepath{stroke}%
\end{pgfscope}%
\begin{pgfscope}%
\pgfpathrectangle{\pgfqpoint{0.346500in}{0.320000in}}{\pgfqpoint{0.875700in}{1.248000in}}%
\pgfusepath{clip}%
\pgfsetbuttcap%
\pgfsetroundjoin%
\pgfsetlinewidth{0.702625pt}%
\definecolor{currentstroke}{rgb}{0.800000,0.800000,0.800000}%
\pgfsetstrokecolor{currentstroke}%
\pgfsetdash{{2.590000pt}{1.120000pt}}{0.000000pt}%
\pgfpathmoveto{\pgfqpoint{0.416556in}{0.320000in}}%
\pgfpathlineto{\pgfqpoint{0.416556in}{1.568000in}}%
\pgfusepath{stroke}%
\end{pgfscope}%
\begin{pgfscope}%
\pgfpathrectangle{\pgfqpoint{0.346500in}{0.320000in}}{\pgfqpoint{0.875700in}{1.248000in}}%
\pgfusepath{clip}%
\pgfsetbuttcap%
\pgfsetroundjoin%
\pgfsetlinewidth{0.702625pt}%
\definecolor{currentstroke}{rgb}{0.800000,0.800000,0.800000}%
\pgfsetstrokecolor{currentstroke}%
\pgfsetdash{{2.590000pt}{1.120000pt}}{0.000000pt}%
\pgfpathmoveto{\pgfqpoint{0.600453in}{0.320000in}}%
\pgfpathlineto{\pgfqpoint{0.600453in}{1.568000in}}%
\pgfusepath{stroke}%
\end{pgfscope}%
\begin{pgfscope}%
\pgfpathrectangle{\pgfqpoint{0.346500in}{0.320000in}}{\pgfqpoint{0.875700in}{1.248000in}}%
\pgfusepath{clip}%
\pgfsetbuttcap%
\pgfsetroundjoin%
\pgfsetlinewidth{0.702625pt}%
\definecolor{currentstroke}{rgb}{0.800000,0.800000,0.800000}%
\pgfsetstrokecolor{currentstroke}%
\pgfsetdash{{2.590000pt}{1.120000pt}}{0.000000pt}%
\pgfpathmoveto{\pgfqpoint{0.784350in}{0.320000in}}%
\pgfpathlineto{\pgfqpoint{0.784350in}{1.568000in}}%
\pgfusepath{stroke}%
\end{pgfscope}%
\begin{pgfscope}%
\pgfpathrectangle{\pgfqpoint{0.346500in}{0.320000in}}{\pgfqpoint{0.875700in}{1.248000in}}%
\pgfusepath{clip}%
\pgfsetbuttcap%
\pgfsetroundjoin%
\pgfsetlinewidth{0.702625pt}%
\definecolor{currentstroke}{rgb}{0.800000,0.800000,0.800000}%
\pgfsetstrokecolor{currentstroke}%
\pgfsetdash{{2.590000pt}{1.120000pt}}{0.000000pt}%
\pgfpathmoveto{\pgfqpoint{0.968247in}{0.320000in}}%
\pgfpathlineto{\pgfqpoint{0.968247in}{1.568000in}}%
\pgfusepath{stroke}%
\end{pgfscope}%
\begin{pgfscope}%
\pgfpathrectangle{\pgfqpoint{0.346500in}{0.320000in}}{\pgfqpoint{0.875700in}{1.248000in}}%
\pgfusepath{clip}%
\pgfsetbuttcap%
\pgfsetroundjoin%
\pgfsetlinewidth{0.702625pt}%
\definecolor{currentstroke}{rgb}{0.800000,0.800000,0.800000}%
\pgfsetstrokecolor{currentstroke}%
\pgfsetdash{{2.590000pt}{1.120000pt}}{0.000000pt}%
\pgfpathmoveto{\pgfqpoint{1.152144in}{0.320000in}}%
\pgfpathlineto{\pgfqpoint{1.152144in}{1.568000in}}%
\pgfusepath{stroke}%
\end{pgfscope}%
\begin{pgfscope}%
\pgfsetrectcap%
\pgfsetmiterjoin%
\pgfsetlinewidth{0.803000pt}%
\definecolor{currentstroke}{rgb}{0.000000,0.000000,0.000000}%
\pgfsetstrokecolor{currentstroke}%
\pgfsetdash{}{0pt}%
\pgfpathmoveto{\pgfqpoint{0.346500in}{0.320000in}}%
\pgfpathlineto{\pgfqpoint{0.346500in}{1.568000in}}%
\pgfusepath{stroke}%
\end{pgfscope}%
\begin{pgfscope}%
\pgfsetrectcap%
\pgfsetmiterjoin%
\pgfsetlinewidth{0.803000pt}%
\definecolor{currentstroke}{rgb}{0.000000,0.000000,0.000000}%
\pgfsetstrokecolor{currentstroke}%
\pgfsetdash{}{0pt}%
\pgfpathmoveto{\pgfqpoint{1.222200in}{0.320000in}}%
\pgfpathlineto{\pgfqpoint{1.222200in}{1.568000in}}%
\pgfusepath{stroke}%
\end{pgfscope}%
\begin{pgfscope}%
\pgfsetrectcap%
\pgfsetmiterjoin%
\pgfsetlinewidth{0.803000pt}%
\definecolor{currentstroke}{rgb}{0.000000,0.000000,0.000000}%
\pgfsetstrokecolor{currentstroke}%
\pgfsetdash{}{0pt}%
\pgfpathmoveto{\pgfqpoint{0.346500in}{0.320000in}}%
\pgfpathlineto{\pgfqpoint{1.222200in}{0.320000in}}%
\pgfusepath{stroke}%
\end{pgfscope}%
\begin{pgfscope}%
\pgfsetrectcap%
\pgfsetmiterjoin%
\pgfsetlinewidth{0.803000pt}%
\definecolor{currentstroke}{rgb}{0.000000,0.000000,0.000000}%
\pgfsetstrokecolor{currentstroke}%
\pgfsetdash{}{0pt}%
\pgfpathmoveto{\pgfqpoint{0.346500in}{1.568000in}}%
\pgfpathlineto{\pgfqpoint{1.222200in}{1.568000in}}%
\pgfusepath{stroke}%
\end{pgfscope}%
\begin{pgfscope}%
\pgfsetbuttcap%
\pgfsetmiterjoin%
\pgfsetlinewidth{0.000000pt}%
\definecolor{currentstroke}{rgb}{0.800000,0.800000,0.800000}%
\pgfsetstrokecolor{currentstroke}%
\pgfsetstrokeopacity{0.000000}%
\pgfsetdash{}{0pt}%
\pgfpathmoveto{\pgfqpoint{0.404833in}{0.361667in}}%
\pgfpathlineto{\pgfqpoint{1.242635in}{0.361667in}}%
\pgfpathquadraticcurveto{\pgfqpoint{1.259301in}{0.361667in}}{\pgfqpoint{1.259301in}{0.378333in}}%
\pgfpathlineto{\pgfqpoint{1.259301in}{0.602407in}}%
\pgfpathquadraticcurveto{\pgfqpoint{1.259301in}{0.619074in}}{\pgfqpoint{1.242635in}{0.619074in}}%
\pgfpathlineto{\pgfqpoint{0.404833in}{0.619074in}}%
\pgfpathquadraticcurveto{\pgfqpoint{0.388167in}{0.619074in}}{\pgfqpoint{0.388167in}{0.602407in}}%
\pgfpathlineto{\pgfqpoint{0.388167in}{0.378333in}}%
\pgfpathquadraticcurveto{\pgfqpoint{0.388167in}{0.361667in}}{\pgfqpoint{0.404833in}{0.361667in}}%
\pgfpathclose%
\pgfusepath{}%
\end{pgfscope}%
\begin{pgfscope}%
\pgfsetrectcap%
\pgfsetroundjoin%
\pgfsetlinewidth{1.204500pt}%
\definecolor{currentstroke}{rgb}{0.100000,0.100000,0.100000}%
\pgfsetstrokecolor{currentstroke}%
\pgfsetdash{}{0pt}%
\pgfpathmoveto{\pgfqpoint{0.421500in}{0.556574in}}%
\pgfpathlineto{\pgfqpoint{0.588167in}{0.556574in}}%
\pgfusepath{stroke}%
\end{pgfscope}%
\begin{pgfscope}%
\pgftext[x=0.654833in,y=0.527407in,left,base]{\rmfamily\fontsize{6.000000}{7.200000}\selectfont Conservative}%
\end{pgfscope}%
\begin{pgfscope}%
\pgfsetrectcap%
\pgfsetroundjoin%
\pgfsetlinewidth{1.204500pt}%
\definecolor{currentstroke}{rgb}{0.900000,0.200000,0.200000}%
\pgfsetstrokecolor{currentstroke}%
\pgfsetdash{}{0pt}%
\pgfpathmoveto{\pgfqpoint{0.421500in}{0.440370in}}%
\pgfpathlineto{\pgfqpoint{0.588167in}{0.440370in}}%
\pgfusepath{stroke}%
\end{pgfscope}%
\begin{pgfscope}%
\pgftext[x=0.654833in,y=0.411204in,left,base]{\rmfamily\fontsize{6.000000}{7.200000}\selectfont Aggressive}%
\end{pgfscope}%
\end{pgfpicture}%
\makeatother%
\endgroup%

%% file: plot_data/facebook_data/fb_cluster_combined/p999_nosense_reconfig_frequency_latency_combined.pgf
\begingroup%
\makeatletter%
\begin{pgfpicture}%
\pgfpathrectangle{\pgfpointorigin}{\pgfqpoint{1.430000in}{1.600000in}}%
\pgfusepath{use as bounding box, clip}%
\begin{pgfscope}%
\pgfsetbuttcap%
\pgfsetmiterjoin%
\definecolor{currentfill}{rgb}{1.000000,1.000000,1.000000}%
\pgfsetfillcolor{currentfill}%
\pgfsetlinewidth{0.000000pt}%
\definecolor{currentstroke}{rgb}{1.000000,1.000000,1.000000}%
\pgfsetstrokecolor{currentstroke}%
\pgfsetdash{}{0pt}%
\pgfpathmoveto{\pgfqpoint{0.000000in}{0.000000in}}%
\pgfpathlineto{\pgfqpoint{1.430000in}{0.000000in}}%
\pgfpathlineto{\pgfqpoint{1.430000in}{1.600000in}}%
\pgfpathlineto{\pgfqpoint{0.000000in}{1.600000in}}%
\pgfpathclose%
\pgfusepath{fill}%
\end{pgfscope}%
\begin{pgfscope}%
\pgfsetbuttcap%
\pgfsetmiterjoin%
\definecolor{currentfill}{rgb}{1.000000,1.000000,1.000000}%
\pgfsetfillcolor{currentfill}%
\pgfsetlinewidth{0.000000pt}%
\definecolor{currentstroke}{rgb}{0.000000,0.000000,0.000000}%
\pgfsetstrokecolor{currentstroke}%
\pgfsetstrokeopacity{0.000000}%
\pgfsetdash{}{0pt}%
\pgfpathmoveto{\pgfqpoint{0.386100in}{0.336000in}}%
\pgfpathlineto{\pgfqpoint{1.401400in}{0.336000in}}%
\pgfpathlineto{\pgfqpoint{1.401400in}{1.568000in}}%
\pgfpathlineto{\pgfqpoint{0.386100in}{1.568000in}}%
\pgfpathclose%
\pgfusepath{fill}%
\end{pgfscope}%
\begin{pgfscope}%
\pgfsetbuttcap%
\pgfsetroundjoin%
\definecolor{currentfill}{rgb}{0.000000,0.000000,0.000000}%
\pgfsetfillcolor{currentfill}%
\pgfsetlinewidth{0.803000pt}%
\definecolor{currentstroke}{rgb}{0.000000,0.000000,0.000000}%
\pgfsetstrokecolor{currentstroke}%
\pgfsetdash{}{0pt}%
\pgfsys@defobject{currentmarker}{\pgfqpoint{0.000000in}{-0.048611in}}{\pgfqpoint{0.000000in}{0.000000in}}{%
\pgfpathmoveto{\pgfqpoint{0.000000in}{0.000000in}}%
\pgfpathlineto{\pgfqpoint{0.000000in}{-0.048611in}}%
\pgfusepath{stroke,fill}%
}%
\begin{pgfscope}%
\pgfsys@transformshift{0.517015in}{0.336000in}%
\pgfsys@useobject{currentmarker}{}%
\end{pgfscope}%
\end{pgfscope}%
\begin{pgfscope}%
\pgftext[x=0.517015in,y=0.238778in,,top]{\rmfamily\fontsize{6.500000}{7.800000}\selectfont 30s}%
\end{pgfscope}%
\begin{pgfscope}%
\pgfsetbuttcap%
\pgfsetroundjoin%
\definecolor{currentfill}{rgb}{0.000000,0.000000,0.000000}%
\pgfsetfillcolor{currentfill}%
\pgfsetlinewidth{0.803000pt}%
\definecolor{currentstroke}{rgb}{0.000000,0.000000,0.000000}%
\pgfsetstrokecolor{currentstroke}%
\pgfsetdash{}{0pt}%
\pgfsys@defobject{currentmarker}{\pgfqpoint{0.000000in}{-0.048611in}}{\pgfqpoint{0.000000in}{0.000000in}}{%
\pgfpathmoveto{\pgfqpoint{0.000000in}{0.000000in}}%
\pgfpathlineto{\pgfqpoint{0.000000in}{-0.048611in}}%
\pgfusepath{stroke,fill}%
}%
\begin{pgfscope}%
\pgfsys@transformshift{0.768172in}{0.336000in}%
\pgfsys@useobject{currentmarker}{}%
\end{pgfscope}%
\end{pgfscope}%
\begin{pgfscope}%
\pgftext[x=0.768172in,y=0.238778in,,top]{\rmfamily\fontsize{6.500000}{7.800000}\selectfont 300s}%
\end{pgfscope}%
\begin{pgfscope}%
\pgfsetbuttcap%
\pgfsetroundjoin%
\definecolor{currentfill}{rgb}{0.000000,0.000000,0.000000}%
\pgfsetfillcolor{currentfill}%
\pgfsetlinewidth{0.803000pt}%
\definecolor{currentstroke}{rgb}{0.000000,0.000000,0.000000}%
\pgfsetstrokecolor{currentstroke}%
\pgfsetdash{}{0pt}%
\pgfsys@defobject{currentmarker}{\pgfqpoint{0.000000in}{-0.048611in}}{\pgfqpoint{0.000000in}{0.000000in}}{%
\pgfpathmoveto{\pgfqpoint{0.000000in}{0.000000in}}%
\pgfpathlineto{\pgfqpoint{0.000000in}{-0.048611in}}%
\pgfusepath{stroke,fill}%
}%
\begin{pgfscope}%
\pgfsys@transformshift{1.019328in}{0.336000in}%
\pgfsys@useobject{currentmarker}{}%
\end{pgfscope}%
\end{pgfscope}%
\begin{pgfscope}%
\pgftext[x=1.019328in,y=0.238778in,,top]{\rmfamily\fontsize{6.500000}{7.800000}\selectfont 3600s}%
\end{pgfscope}%
\begin{pgfscope}%
\pgfsetbuttcap%
\pgfsetroundjoin%
\definecolor{currentfill}{rgb}{0.000000,0.000000,0.000000}%
\pgfsetfillcolor{currentfill}%
\pgfsetlinewidth{0.803000pt}%
\definecolor{currentstroke}{rgb}{0.000000,0.000000,0.000000}%
\pgfsetstrokecolor{currentstroke}%
\pgfsetdash{}{0pt}%
\pgfsys@defobject{currentmarker}{\pgfqpoint{0.000000in}{-0.048611in}}{\pgfqpoint{0.000000in}{0.000000in}}{%
\pgfpathmoveto{\pgfqpoint{0.000000in}{0.000000in}}%
\pgfpathlineto{\pgfqpoint{0.000000in}{-0.048611in}}%
\pgfusepath{stroke,fill}%
}%
\begin{pgfscope}%
\pgfsys@transformshift{1.270485in}{0.336000in}%
\pgfsys@useobject{currentmarker}{}%
\end{pgfscope}%
\end{pgfscope}%
\begin{pgfscope}%
\pgftext[x=1.270485in,y=0.238778in,,top]{\rmfamily\fontsize{6.500000}{7.800000}\selectfont 24H}%
\end{pgfscope}%
\begin{pgfscope}%
\pgftext[x=0.893750in,y=0.150815in,,top]{\rmfamily\fontsize{7.400000}{8.880000}\selectfont Reconfig. Freq.}%
\end{pgfscope}%
\begin{pgfscope}%
\pgfpathrectangle{\pgfqpoint{0.386100in}{0.336000in}}{\pgfqpoint{1.015300in}{1.232000in}}%
\pgfusepath{clip}%
\pgfsetbuttcap%
\pgfsetroundjoin%
\pgfsetlinewidth{0.702625pt}%
\definecolor{currentstroke}{rgb}{0.690196,0.690196,0.690196}%
\pgfsetstrokecolor{currentstroke}%
\pgfsetdash{{4.480000pt}{1.120000pt}{0.700000pt}{1.120000pt}}{0.000000pt}%
\pgfpathmoveto{\pgfqpoint{0.386100in}{0.525538in}}%
\pgfpathlineto{\pgfqpoint{1.401400in}{0.525538in}}%
\pgfusepath{stroke}%
\end{pgfscope}%
\begin{pgfscope}%
\pgfsetbuttcap%
\pgfsetroundjoin%
\definecolor{currentfill}{rgb}{0.000000,0.000000,0.000000}%
\pgfsetfillcolor{currentfill}%
\pgfsetlinewidth{0.803000pt}%
\definecolor{currentstroke}{rgb}{0.000000,0.000000,0.000000}%
\pgfsetstrokecolor{currentstroke}%
\pgfsetdash{}{0pt}%
\pgfsys@defobject{currentmarker}{\pgfqpoint{0.000000in}{0.000000in}}{\pgfqpoint{0.048611in}{0.000000in}}{%
\pgfpathmoveto{\pgfqpoint{0.000000in}{0.000000in}}%
\pgfpathlineto{\pgfqpoint{0.048611in}{0.000000in}}%
\pgfusepath{stroke,fill}%
}%
\begin{pgfscope}%
\pgfsys@transformshift{0.386100in}{0.525538in}%
\pgfsys@useobject{currentmarker}{}%
\end{pgfscope}%
\end{pgfscope}%
\begin{pgfscope}%
\pgftext[x=0.153078in,y=0.496603in,left,base]{\rmfamily\fontsize{6.500000}{7.800000}\selectfont \(\displaystyle 1.00\)}%
\end{pgfscope}%
\begin{pgfscope}%
\pgfpathrectangle{\pgfqpoint{0.386100in}{0.336000in}}{\pgfqpoint{1.015300in}{1.232000in}}%
\pgfusepath{clip}%
\pgfsetbuttcap%
\pgfsetroundjoin%
\pgfsetlinewidth{0.702625pt}%
\definecolor{currentstroke}{rgb}{0.690196,0.690196,0.690196}%
\pgfsetstrokecolor{currentstroke}%
\pgfsetdash{{4.480000pt}{1.120000pt}{0.700000pt}{1.120000pt}}{0.000000pt}%
\pgfpathmoveto{\pgfqpoint{0.386100in}{0.762462in}}%
\pgfpathlineto{\pgfqpoint{1.401400in}{0.762462in}}%
\pgfusepath{stroke}%
\end{pgfscope}%
\begin{pgfscope}%
\pgfsetbuttcap%
\pgfsetroundjoin%
\definecolor{currentfill}{rgb}{0.000000,0.000000,0.000000}%
\pgfsetfillcolor{currentfill}%
\pgfsetlinewidth{0.803000pt}%
\definecolor{currentstroke}{rgb}{0.000000,0.000000,0.000000}%
\pgfsetstrokecolor{currentstroke}%
\pgfsetdash{}{0pt}%
\pgfsys@defobject{currentmarker}{\pgfqpoint{0.000000in}{0.000000in}}{\pgfqpoint{0.048611in}{0.000000in}}{%
\pgfpathmoveto{\pgfqpoint{0.000000in}{0.000000in}}%
\pgfpathlineto{\pgfqpoint{0.048611in}{0.000000in}}%
\pgfusepath{stroke,fill}%
}%
\begin{pgfscope}%
\pgfsys@transformshift{0.386100in}{0.762462in}%
\pgfsys@useobject{currentmarker}{}%
\end{pgfscope}%
\end{pgfscope}%
\begin{pgfscope}%
\pgftext[x=0.153078in,y=0.733526in,left,base]{\rmfamily\fontsize{6.500000}{7.800000}\selectfont \(\displaystyle 1.25\)}%
\end{pgfscope}%
\begin{pgfscope}%
\pgfpathrectangle{\pgfqpoint{0.386100in}{0.336000in}}{\pgfqpoint{1.015300in}{1.232000in}}%
\pgfusepath{clip}%
\pgfsetbuttcap%
\pgfsetroundjoin%
\pgfsetlinewidth{0.702625pt}%
\definecolor{currentstroke}{rgb}{0.690196,0.690196,0.690196}%
\pgfsetstrokecolor{currentstroke}%
\pgfsetdash{{4.480000pt}{1.120000pt}{0.700000pt}{1.120000pt}}{0.000000pt}%
\pgfpathmoveto{\pgfqpoint{0.386100in}{0.999385in}}%
\pgfpathlineto{\pgfqpoint{1.401400in}{0.999385in}}%
\pgfusepath{stroke}%
\end{pgfscope}%
\begin{pgfscope}%
\pgfsetbuttcap%
\pgfsetroundjoin%
\definecolor{currentfill}{rgb}{0.000000,0.000000,0.000000}%
\pgfsetfillcolor{currentfill}%
\pgfsetlinewidth{0.803000pt}%
\definecolor{currentstroke}{rgb}{0.000000,0.000000,0.000000}%
\pgfsetstrokecolor{currentstroke}%
\pgfsetdash{}{0pt}%
\pgfsys@defobject{currentmarker}{\pgfqpoint{0.000000in}{0.000000in}}{\pgfqpoint{0.048611in}{0.000000in}}{%
\pgfpathmoveto{\pgfqpoint{0.000000in}{0.000000in}}%
\pgfpathlineto{\pgfqpoint{0.048611in}{0.000000in}}%
\pgfusepath{stroke,fill}%
}%
\begin{pgfscope}%
\pgfsys@transformshift{0.386100in}{0.999385in}%
\pgfsys@useobject{currentmarker}{}%
\end{pgfscope}%
\end{pgfscope}%
\begin{pgfscope}%
\pgftext[x=0.153078in,y=0.970449in,left,base]{\rmfamily\fontsize{6.500000}{7.800000}\selectfont \(\displaystyle 1.50\)}%
\end{pgfscope}%
\begin{pgfscope}%
\pgfpathrectangle{\pgfqpoint{0.386100in}{0.336000in}}{\pgfqpoint{1.015300in}{1.232000in}}%
\pgfusepath{clip}%
\pgfsetbuttcap%
\pgfsetroundjoin%
\pgfsetlinewidth{0.702625pt}%
\definecolor{currentstroke}{rgb}{0.690196,0.690196,0.690196}%
\pgfsetstrokecolor{currentstroke}%
\pgfsetdash{{4.480000pt}{1.120000pt}{0.700000pt}{1.120000pt}}{0.000000pt}%
\pgfpathmoveto{\pgfqpoint{0.386100in}{1.236308in}}%
\pgfpathlineto{\pgfqpoint{1.401400in}{1.236308in}}%
\pgfusepath{stroke}%
\end{pgfscope}%
\begin{pgfscope}%
\pgfsetbuttcap%
\pgfsetroundjoin%
\definecolor{currentfill}{rgb}{0.000000,0.000000,0.000000}%
\pgfsetfillcolor{currentfill}%
\pgfsetlinewidth{0.803000pt}%
\definecolor{currentstroke}{rgb}{0.000000,0.000000,0.000000}%
\pgfsetstrokecolor{currentstroke}%
\pgfsetdash{}{0pt}%
\pgfsys@defobject{currentmarker}{\pgfqpoint{0.000000in}{0.000000in}}{\pgfqpoint{0.048611in}{0.000000in}}{%
\pgfpathmoveto{\pgfqpoint{0.000000in}{0.000000in}}%
\pgfpathlineto{\pgfqpoint{0.048611in}{0.000000in}}%
\pgfusepath{stroke,fill}%
}%
\begin{pgfscope}%
\pgfsys@transformshift{0.386100in}{1.236308in}%
\pgfsys@useobject{currentmarker}{}%
\end{pgfscope}%
\end{pgfscope}%
\begin{pgfscope}%
\pgftext[x=0.153078in,y=1.207372in,left,base]{\rmfamily\fontsize{6.500000}{7.800000}\selectfont \(\displaystyle 1.75\)}%
\end{pgfscope}%
\begin{pgfscope}%
\pgfpathrectangle{\pgfqpoint{0.386100in}{0.336000in}}{\pgfqpoint{1.015300in}{1.232000in}}%
\pgfusepath{clip}%
\pgfsetbuttcap%
\pgfsetroundjoin%
\pgfsetlinewidth{0.702625pt}%
\definecolor{currentstroke}{rgb}{0.690196,0.690196,0.690196}%
\pgfsetstrokecolor{currentstroke}%
\pgfsetdash{{4.480000pt}{1.120000pt}{0.700000pt}{1.120000pt}}{0.000000pt}%
\pgfpathmoveto{\pgfqpoint{0.386100in}{1.473231in}}%
\pgfpathlineto{\pgfqpoint{1.401400in}{1.473231in}}%
\pgfusepath{stroke}%
\end{pgfscope}%
\begin{pgfscope}%
\pgfsetbuttcap%
\pgfsetroundjoin%
\definecolor{currentfill}{rgb}{0.000000,0.000000,0.000000}%
\pgfsetfillcolor{currentfill}%
\pgfsetlinewidth{0.803000pt}%
\definecolor{currentstroke}{rgb}{0.000000,0.000000,0.000000}%
\pgfsetstrokecolor{currentstroke}%
\pgfsetdash{}{0pt}%
\pgfsys@defobject{currentmarker}{\pgfqpoint{0.000000in}{0.000000in}}{\pgfqpoint{0.048611in}{0.000000in}}{%
\pgfpathmoveto{\pgfqpoint{0.000000in}{0.000000in}}%
\pgfpathlineto{\pgfqpoint{0.048611in}{0.000000in}}%
\pgfusepath{stroke,fill}%
}%
\begin{pgfscope}%
\pgfsys@transformshift{0.386100in}{1.473231in}%
\pgfsys@useobject{currentmarker}{}%
\end{pgfscope}%
\end{pgfscope}%
\begin{pgfscope}%
\pgftext[x=0.153078in,y=1.444296in,left,base]{\rmfamily\fontsize{6.500000}{7.800000}\selectfont \(\displaystyle 2.00\)}%
\end{pgfscope}%
\begin{pgfscope}%
\pgfsetbuttcap%
\pgfsetroundjoin%
\definecolor{currentfill}{rgb}{0.000000,0.000000,0.000000}%
\pgfsetfillcolor{currentfill}%
\pgfsetlinewidth{0.602250pt}%
\definecolor{currentstroke}{rgb}{0.000000,0.000000,0.000000}%
\pgfsetstrokecolor{currentstroke}%
\pgfsetdash{}{0pt}%
\pgfsys@defobject{currentmarker}{\pgfqpoint{0.000000in}{0.000000in}}{\pgfqpoint{0.027778in}{0.000000in}}{%
\pgfpathmoveto{\pgfqpoint{0.000000in}{0.000000in}}%
\pgfpathlineto{\pgfqpoint{0.027778in}{0.000000in}}%
\pgfusepath{stroke,fill}%
}%
\begin{pgfscope}%
\pgfsys@transformshift{0.386100in}{0.347846in}%
\pgfsys@useobject{currentmarker}{}%
\end{pgfscope}%
\end{pgfscope}%
\begin{pgfscope}%
\pgfsetbuttcap%
\pgfsetroundjoin%
\definecolor{currentfill}{rgb}{0.000000,0.000000,0.000000}%
\pgfsetfillcolor{currentfill}%
\pgfsetlinewidth{0.602250pt}%
\definecolor{currentstroke}{rgb}{0.000000,0.000000,0.000000}%
\pgfsetstrokecolor{currentstroke}%
\pgfsetdash{}{0pt}%
\pgfsys@defobject{currentmarker}{\pgfqpoint{0.000000in}{0.000000in}}{\pgfqpoint{0.027778in}{0.000000in}}{%
\pgfpathmoveto{\pgfqpoint{0.000000in}{0.000000in}}%
\pgfpathlineto{\pgfqpoint{0.027778in}{0.000000in}}%
\pgfusepath{stroke,fill}%
}%
\begin{pgfscope}%
\pgfsys@transformshift{0.386100in}{0.407077in}%
\pgfsys@useobject{currentmarker}{}%
\end{pgfscope}%
\end{pgfscope}%
\begin{pgfscope}%
\pgfsetbuttcap%
\pgfsetroundjoin%
\definecolor{currentfill}{rgb}{0.000000,0.000000,0.000000}%
\pgfsetfillcolor{currentfill}%
\pgfsetlinewidth{0.602250pt}%
\definecolor{currentstroke}{rgb}{0.000000,0.000000,0.000000}%
\pgfsetstrokecolor{currentstroke}%
\pgfsetdash{}{0pt}%
\pgfsys@defobject{currentmarker}{\pgfqpoint{0.000000in}{0.000000in}}{\pgfqpoint{0.027778in}{0.000000in}}{%
\pgfpathmoveto{\pgfqpoint{0.000000in}{0.000000in}}%
\pgfpathlineto{\pgfqpoint{0.027778in}{0.000000in}}%
\pgfusepath{stroke,fill}%
}%
\begin{pgfscope}%
\pgfsys@transformshift{0.386100in}{0.466308in}%
\pgfsys@useobject{currentmarker}{}%
\end{pgfscope}%
\end{pgfscope}%
\begin{pgfscope}%
\pgfsetbuttcap%
\pgfsetroundjoin%
\definecolor{currentfill}{rgb}{0.000000,0.000000,0.000000}%
\pgfsetfillcolor{currentfill}%
\pgfsetlinewidth{0.602250pt}%
\definecolor{currentstroke}{rgb}{0.000000,0.000000,0.000000}%
\pgfsetstrokecolor{currentstroke}%
\pgfsetdash{}{0pt}%
\pgfsys@defobject{currentmarker}{\pgfqpoint{0.000000in}{0.000000in}}{\pgfqpoint{0.027778in}{0.000000in}}{%
\pgfpathmoveto{\pgfqpoint{0.000000in}{0.000000in}}%
\pgfpathlineto{\pgfqpoint{0.027778in}{0.000000in}}%
\pgfusepath{stroke,fill}%
}%
\begin{pgfscope}%
\pgfsys@transformshift{0.386100in}{0.584769in}%
\pgfsys@useobject{currentmarker}{}%
\end{pgfscope}%
\end{pgfscope}%
\begin{pgfscope}%
\pgfsetbuttcap%
\pgfsetroundjoin%
\definecolor{currentfill}{rgb}{0.000000,0.000000,0.000000}%
\pgfsetfillcolor{currentfill}%
\pgfsetlinewidth{0.602250pt}%
\definecolor{currentstroke}{rgb}{0.000000,0.000000,0.000000}%
\pgfsetstrokecolor{currentstroke}%
\pgfsetdash{}{0pt}%
\pgfsys@defobject{currentmarker}{\pgfqpoint{0.000000in}{0.000000in}}{\pgfqpoint{0.027778in}{0.000000in}}{%
\pgfpathmoveto{\pgfqpoint{0.000000in}{0.000000in}}%
\pgfpathlineto{\pgfqpoint{0.027778in}{0.000000in}}%
\pgfusepath{stroke,fill}%
}%
\begin{pgfscope}%
\pgfsys@transformshift{0.386100in}{0.644000in}%
\pgfsys@useobject{currentmarker}{}%
\end{pgfscope}%
\end{pgfscope}%
\begin{pgfscope}%
\pgfsetbuttcap%
\pgfsetroundjoin%
\definecolor{currentfill}{rgb}{0.000000,0.000000,0.000000}%
\pgfsetfillcolor{currentfill}%
\pgfsetlinewidth{0.602250pt}%
\definecolor{currentstroke}{rgb}{0.000000,0.000000,0.000000}%
\pgfsetstrokecolor{currentstroke}%
\pgfsetdash{}{0pt}%
\pgfsys@defobject{currentmarker}{\pgfqpoint{0.000000in}{0.000000in}}{\pgfqpoint{0.027778in}{0.000000in}}{%
\pgfpathmoveto{\pgfqpoint{0.000000in}{0.000000in}}%
\pgfpathlineto{\pgfqpoint{0.027778in}{0.000000in}}%
\pgfusepath{stroke,fill}%
}%
\begin{pgfscope}%
\pgfsys@transformshift{0.386100in}{0.703231in}%
\pgfsys@useobject{currentmarker}{}%
\end{pgfscope}%
\end{pgfscope}%
\begin{pgfscope}%
\pgfsetbuttcap%
\pgfsetroundjoin%
\definecolor{currentfill}{rgb}{0.000000,0.000000,0.000000}%
\pgfsetfillcolor{currentfill}%
\pgfsetlinewidth{0.602250pt}%
\definecolor{currentstroke}{rgb}{0.000000,0.000000,0.000000}%
\pgfsetstrokecolor{currentstroke}%
\pgfsetdash{}{0pt}%
\pgfsys@defobject{currentmarker}{\pgfqpoint{0.000000in}{0.000000in}}{\pgfqpoint{0.027778in}{0.000000in}}{%
\pgfpathmoveto{\pgfqpoint{0.000000in}{0.000000in}}%
\pgfpathlineto{\pgfqpoint{0.027778in}{0.000000in}}%
\pgfusepath{stroke,fill}%
}%
\begin{pgfscope}%
\pgfsys@transformshift{0.386100in}{0.821692in}%
\pgfsys@useobject{currentmarker}{}%
\end{pgfscope}%
\end{pgfscope}%
\begin{pgfscope}%
\pgfsetbuttcap%
\pgfsetroundjoin%
\definecolor{currentfill}{rgb}{0.000000,0.000000,0.000000}%
\pgfsetfillcolor{currentfill}%
\pgfsetlinewidth{0.602250pt}%
\definecolor{currentstroke}{rgb}{0.000000,0.000000,0.000000}%
\pgfsetstrokecolor{currentstroke}%
\pgfsetdash{}{0pt}%
\pgfsys@defobject{currentmarker}{\pgfqpoint{0.000000in}{0.000000in}}{\pgfqpoint{0.027778in}{0.000000in}}{%
\pgfpathmoveto{\pgfqpoint{0.000000in}{0.000000in}}%
\pgfpathlineto{\pgfqpoint{0.027778in}{0.000000in}}%
\pgfusepath{stroke,fill}%
}%
\begin{pgfscope}%
\pgfsys@transformshift{0.386100in}{0.880923in}%
\pgfsys@useobject{currentmarker}{}%
\end{pgfscope}%
\end{pgfscope}%
\begin{pgfscope}%
\pgfsetbuttcap%
\pgfsetroundjoin%
\definecolor{currentfill}{rgb}{0.000000,0.000000,0.000000}%
\pgfsetfillcolor{currentfill}%
\pgfsetlinewidth{0.602250pt}%
\definecolor{currentstroke}{rgb}{0.000000,0.000000,0.000000}%
\pgfsetstrokecolor{currentstroke}%
\pgfsetdash{}{0pt}%
\pgfsys@defobject{currentmarker}{\pgfqpoint{0.000000in}{0.000000in}}{\pgfqpoint{0.027778in}{0.000000in}}{%
\pgfpathmoveto{\pgfqpoint{0.000000in}{0.000000in}}%
\pgfpathlineto{\pgfqpoint{0.027778in}{0.000000in}}%
\pgfusepath{stroke,fill}%
}%
\begin{pgfscope}%
\pgfsys@transformshift{0.386100in}{0.940154in}%
\pgfsys@useobject{currentmarker}{}%
\end{pgfscope}%
\end{pgfscope}%
\begin{pgfscope}%
\pgfsetbuttcap%
\pgfsetroundjoin%
\definecolor{currentfill}{rgb}{0.000000,0.000000,0.000000}%
\pgfsetfillcolor{currentfill}%
\pgfsetlinewidth{0.602250pt}%
\definecolor{currentstroke}{rgb}{0.000000,0.000000,0.000000}%
\pgfsetstrokecolor{currentstroke}%
\pgfsetdash{}{0pt}%
\pgfsys@defobject{currentmarker}{\pgfqpoint{0.000000in}{0.000000in}}{\pgfqpoint{0.027778in}{0.000000in}}{%
\pgfpathmoveto{\pgfqpoint{0.000000in}{0.000000in}}%
\pgfpathlineto{\pgfqpoint{0.027778in}{0.000000in}}%
\pgfusepath{stroke,fill}%
}%
\begin{pgfscope}%
\pgfsys@transformshift{0.386100in}{1.058615in}%
\pgfsys@useobject{currentmarker}{}%
\end{pgfscope}%
\end{pgfscope}%
\begin{pgfscope}%
\pgfsetbuttcap%
\pgfsetroundjoin%
\definecolor{currentfill}{rgb}{0.000000,0.000000,0.000000}%
\pgfsetfillcolor{currentfill}%
\pgfsetlinewidth{0.602250pt}%
\definecolor{currentstroke}{rgb}{0.000000,0.000000,0.000000}%
\pgfsetstrokecolor{currentstroke}%
\pgfsetdash{}{0pt}%
\pgfsys@defobject{currentmarker}{\pgfqpoint{0.000000in}{0.000000in}}{\pgfqpoint{0.027778in}{0.000000in}}{%
\pgfpathmoveto{\pgfqpoint{0.000000in}{0.000000in}}%
\pgfpathlineto{\pgfqpoint{0.027778in}{0.000000in}}%
\pgfusepath{stroke,fill}%
}%
\begin{pgfscope}%
\pgfsys@transformshift{0.386100in}{1.117846in}%
\pgfsys@useobject{currentmarker}{}%
\end{pgfscope}%
\end{pgfscope}%
\begin{pgfscope}%
\pgfsetbuttcap%
\pgfsetroundjoin%
\definecolor{currentfill}{rgb}{0.000000,0.000000,0.000000}%
\pgfsetfillcolor{currentfill}%
\pgfsetlinewidth{0.602250pt}%
\definecolor{currentstroke}{rgb}{0.000000,0.000000,0.000000}%
\pgfsetstrokecolor{currentstroke}%
\pgfsetdash{}{0pt}%
\pgfsys@defobject{currentmarker}{\pgfqpoint{0.000000in}{0.000000in}}{\pgfqpoint{0.027778in}{0.000000in}}{%
\pgfpathmoveto{\pgfqpoint{0.000000in}{0.000000in}}%
\pgfpathlineto{\pgfqpoint{0.027778in}{0.000000in}}%
\pgfusepath{stroke,fill}%
}%
\begin{pgfscope}%
\pgfsys@transformshift{0.386100in}{1.177077in}%
\pgfsys@useobject{currentmarker}{}%
\end{pgfscope}%
\end{pgfscope}%
\begin{pgfscope}%
\pgfsetbuttcap%
\pgfsetroundjoin%
\definecolor{currentfill}{rgb}{0.000000,0.000000,0.000000}%
\pgfsetfillcolor{currentfill}%
\pgfsetlinewidth{0.602250pt}%
\definecolor{currentstroke}{rgb}{0.000000,0.000000,0.000000}%
\pgfsetstrokecolor{currentstroke}%
\pgfsetdash{}{0pt}%
\pgfsys@defobject{currentmarker}{\pgfqpoint{0.000000in}{0.000000in}}{\pgfqpoint{0.027778in}{0.000000in}}{%
\pgfpathmoveto{\pgfqpoint{0.000000in}{0.000000in}}%
\pgfpathlineto{\pgfqpoint{0.027778in}{0.000000in}}%
\pgfusepath{stroke,fill}%
}%
\begin{pgfscope}%
\pgfsys@transformshift{0.386100in}{1.295538in}%
\pgfsys@useobject{currentmarker}{}%
\end{pgfscope}%
\end{pgfscope}%
\begin{pgfscope}%
\pgfsetbuttcap%
\pgfsetroundjoin%
\definecolor{currentfill}{rgb}{0.000000,0.000000,0.000000}%
\pgfsetfillcolor{currentfill}%
\pgfsetlinewidth{0.602250pt}%
\definecolor{currentstroke}{rgb}{0.000000,0.000000,0.000000}%
\pgfsetstrokecolor{currentstroke}%
\pgfsetdash{}{0pt}%
\pgfsys@defobject{currentmarker}{\pgfqpoint{0.000000in}{0.000000in}}{\pgfqpoint{0.027778in}{0.000000in}}{%
\pgfpathmoveto{\pgfqpoint{0.000000in}{0.000000in}}%
\pgfpathlineto{\pgfqpoint{0.027778in}{0.000000in}}%
\pgfusepath{stroke,fill}%
}%
\begin{pgfscope}%
\pgfsys@transformshift{0.386100in}{1.354769in}%
\pgfsys@useobject{currentmarker}{}%
\end{pgfscope}%
\end{pgfscope}%
\begin{pgfscope}%
\pgfsetbuttcap%
\pgfsetroundjoin%
\definecolor{currentfill}{rgb}{0.000000,0.000000,0.000000}%
\pgfsetfillcolor{currentfill}%
\pgfsetlinewidth{0.602250pt}%
\definecolor{currentstroke}{rgb}{0.000000,0.000000,0.000000}%
\pgfsetstrokecolor{currentstroke}%
\pgfsetdash{}{0pt}%
\pgfsys@defobject{currentmarker}{\pgfqpoint{0.000000in}{0.000000in}}{\pgfqpoint{0.027778in}{0.000000in}}{%
\pgfpathmoveto{\pgfqpoint{0.000000in}{0.000000in}}%
\pgfpathlineto{\pgfqpoint{0.027778in}{0.000000in}}%
\pgfusepath{stroke,fill}%
}%
\begin{pgfscope}%
\pgfsys@transformshift{0.386100in}{1.414000in}%
\pgfsys@useobject{currentmarker}{}%
\end{pgfscope}%
\end{pgfscope}%
\begin{pgfscope}%
\pgfsetbuttcap%
\pgfsetroundjoin%
\definecolor{currentfill}{rgb}{0.000000,0.000000,0.000000}%
\pgfsetfillcolor{currentfill}%
\pgfsetlinewidth{0.602250pt}%
\definecolor{currentstroke}{rgb}{0.000000,0.000000,0.000000}%
\pgfsetstrokecolor{currentstroke}%
\pgfsetdash{}{0pt}%
\pgfsys@defobject{currentmarker}{\pgfqpoint{0.000000in}{0.000000in}}{\pgfqpoint{0.027778in}{0.000000in}}{%
\pgfpathmoveto{\pgfqpoint{0.000000in}{0.000000in}}%
\pgfpathlineto{\pgfqpoint{0.027778in}{0.000000in}}%
\pgfusepath{stroke,fill}%
}%
\begin{pgfscope}%
\pgfsys@transformshift{0.386100in}{1.532462in}%
\pgfsys@useobject{currentmarker}{}%
\end{pgfscope}%
\end{pgfscope}%
\begin{pgfscope}%
\pgftext[x=0.139189in,y=0.952000in,,bottom,rotate=90.000000]{\rmfamily\fontsize{7.400000}{8.880000}\selectfont \(\displaystyle 99.9^{th}\) \%-tile MLU}%
\end{pgfscope}%
\begin{pgfscope}%
\pgfpathrectangle{\pgfqpoint{0.386100in}{0.336000in}}{\pgfqpoint{1.015300in}{1.232000in}}%
\pgfusepath{clip}%
\pgfsetbuttcap%
\pgfsetmiterjoin%
\definecolor{currentfill}{rgb}{0.900000,0.100000,0.100000}%
\pgfsetfillcolor{currentfill}%
\pgfsetlinewidth{0.000000pt}%
\definecolor{currentstroke}{rgb}{0.000000,0.000000,0.000000}%
\pgfsetstrokecolor{currentstroke}%
\pgfsetstrokeopacity{0.000000}%
\pgfsetdash{}{0pt}%
\pgfpathmoveto{\pgfqpoint{0.432250in}{-0.422154in}}%
\pgfpathlineto{\pgfqpoint{0.476202in}{-0.422154in}}%
\pgfpathlineto{\pgfqpoint{0.476202in}{0.553969in}}%
\pgfpathlineto{\pgfqpoint{0.432250in}{0.553969in}}%
\pgfpathclose%
\pgfusepath{fill}%
\end{pgfscope}%
\begin{pgfscope}%
\pgfpathrectangle{\pgfqpoint{0.386100in}{0.336000in}}{\pgfqpoint{1.015300in}{1.232000in}}%
\pgfusepath{clip}%
\pgfsetbuttcap%
\pgfsetmiterjoin%
\definecolor{currentfill}{rgb}{0.900000,0.100000,0.100000}%
\pgfsetfillcolor{currentfill}%
\pgfsetlinewidth{0.000000pt}%
\definecolor{currentstroke}{rgb}{0.000000,0.000000,0.000000}%
\pgfsetstrokecolor{currentstroke}%
\pgfsetstrokeopacity{0.000000}%
\pgfsetdash{}{0pt}%
\pgfpathmoveto{\pgfqpoint{0.683406in}{-0.422154in}}%
\pgfpathlineto{\pgfqpoint{0.727359in}{-0.422154in}}%
\pgfpathlineto{\pgfqpoint{0.727359in}{0.724554in}}%
\pgfpathlineto{\pgfqpoint{0.683406in}{0.724554in}}%
\pgfpathclose%
\pgfusepath{fill}%
\end{pgfscope}%
\begin{pgfscope}%
\pgfpathrectangle{\pgfqpoint{0.386100in}{0.336000in}}{\pgfqpoint{1.015300in}{1.232000in}}%
\pgfusepath{clip}%
\pgfsetbuttcap%
\pgfsetmiterjoin%
\definecolor{currentfill}{rgb}{0.900000,0.100000,0.100000}%
\pgfsetfillcolor{currentfill}%
\pgfsetlinewidth{0.000000pt}%
\definecolor{currentstroke}{rgb}{0.000000,0.000000,0.000000}%
\pgfsetstrokecolor{currentstroke}%
\pgfsetstrokeopacity{0.000000}%
\pgfsetdash{}{0pt}%
\pgfpathmoveto{\pgfqpoint{0.934563in}{-0.422154in}}%
\pgfpathlineto{\pgfqpoint{0.978515in}{-0.422154in}}%
\pgfpathlineto{\pgfqpoint{0.978515in}{1.008862in}}%
\pgfpathlineto{\pgfqpoint{0.934563in}{1.008862in}}%
\pgfpathclose%
\pgfusepath{fill}%
\end{pgfscope}%
\begin{pgfscope}%
\pgfpathrectangle{\pgfqpoint{0.386100in}{0.336000in}}{\pgfqpoint{1.015300in}{1.232000in}}%
\pgfusepath{clip}%
\pgfsetbuttcap%
\pgfsetmiterjoin%
\definecolor{currentfill}{rgb}{0.900000,0.100000,0.100000}%
\pgfsetfillcolor{currentfill}%
\pgfsetlinewidth{0.000000pt}%
\definecolor{currentstroke}{rgb}{0.000000,0.000000,0.000000}%
\pgfsetstrokecolor{currentstroke}%
\pgfsetstrokeopacity{0.000000}%
\pgfsetdash{}{0pt}%
\pgfpathmoveto{\pgfqpoint{1.185719in}{-0.422154in}}%
\pgfpathlineto{\pgfqpoint{1.229672in}{-0.422154in}}%
\pgfpathlineto{\pgfqpoint{1.229672in}{1.245785in}}%
\pgfpathlineto{\pgfqpoint{1.185719in}{1.245785in}}%
\pgfpathclose%
\pgfusepath{fill}%
\end{pgfscope}%
\begin{pgfscope}%
\pgfpathrectangle{\pgfqpoint{0.386100in}{0.336000in}}{\pgfqpoint{1.015300in}{1.232000in}}%
\pgfusepath{clip}%
\pgfsetbuttcap%
\pgfsetmiterjoin%
\definecolor{currentfill}{rgb}{0.300000,0.800000,0.200000}%
\pgfsetfillcolor{currentfill}%
\pgfsetlinewidth{0.000000pt}%
\definecolor{currentstroke}{rgb}{0.000000,0.000000,0.000000}%
\pgfsetstrokecolor{currentstroke}%
\pgfsetstrokeopacity{0.000000}%
\pgfsetdash{}{0pt}%
\pgfpathmoveto{\pgfqpoint{0.495039in}{-0.422154in}}%
\pgfpathlineto{\pgfqpoint{0.538991in}{-0.422154in}}%
\pgfpathlineto{\pgfqpoint{0.538991in}{0.724554in}}%
\pgfpathlineto{\pgfqpoint{0.495039in}{0.724554in}}%
\pgfpathclose%
\pgfusepath{fill}%
\end{pgfscope}%
\begin{pgfscope}%
\pgfpathrectangle{\pgfqpoint{0.386100in}{0.336000in}}{\pgfqpoint{1.015300in}{1.232000in}}%
\pgfusepath{clip}%
\pgfsetbuttcap%
\pgfsetmiterjoin%
\definecolor{currentfill}{rgb}{0.300000,0.800000,0.200000}%
\pgfsetfillcolor{currentfill}%
\pgfsetlinewidth{0.000000pt}%
\definecolor{currentstroke}{rgb}{0.000000,0.000000,0.000000}%
\pgfsetstrokecolor{currentstroke}%
\pgfsetstrokeopacity{0.000000}%
\pgfsetdash{}{0pt}%
\pgfpathmoveto{\pgfqpoint{0.746196in}{-0.422154in}}%
\pgfpathlineto{\pgfqpoint{0.790148in}{-0.422154in}}%
\pgfpathlineto{\pgfqpoint{0.790148in}{0.800369in}}%
\pgfpathlineto{\pgfqpoint{0.746196in}{0.800369in}}%
\pgfpathclose%
\pgfusepath{fill}%
\end{pgfscope}%
\begin{pgfscope}%
\pgfpathrectangle{\pgfqpoint{0.386100in}{0.336000in}}{\pgfqpoint{1.015300in}{1.232000in}}%
\pgfusepath{clip}%
\pgfsetbuttcap%
\pgfsetmiterjoin%
\definecolor{currentfill}{rgb}{0.300000,0.800000,0.200000}%
\pgfsetfillcolor{currentfill}%
\pgfsetlinewidth{0.000000pt}%
\definecolor{currentstroke}{rgb}{0.000000,0.000000,0.000000}%
\pgfsetstrokecolor{currentstroke}%
\pgfsetstrokeopacity{0.000000}%
\pgfsetdash{}{0pt}%
\pgfpathmoveto{\pgfqpoint{0.997352in}{-0.422154in}}%
\pgfpathlineto{\pgfqpoint{1.041304in}{-0.422154in}}%
\pgfpathlineto{\pgfqpoint{1.041304in}{1.019286in}}%
\pgfpathlineto{\pgfqpoint{0.997352in}{1.019286in}}%
\pgfpathclose%
\pgfusepath{fill}%
\end{pgfscope}%
\begin{pgfscope}%
\pgfpathrectangle{\pgfqpoint{0.386100in}{0.336000in}}{\pgfqpoint{1.015300in}{1.232000in}}%
\pgfusepath{clip}%
\pgfsetbuttcap%
\pgfsetmiterjoin%
\definecolor{currentfill}{rgb}{0.300000,0.800000,0.200000}%
\pgfsetfillcolor{currentfill}%
\pgfsetlinewidth{0.000000pt}%
\definecolor{currentstroke}{rgb}{0.000000,0.000000,0.000000}%
\pgfsetstrokecolor{currentstroke}%
\pgfsetstrokeopacity{0.000000}%
\pgfsetdash{}{0pt}%
\pgfpathmoveto{\pgfqpoint{1.248509in}{-0.422154in}}%
\pgfpathlineto{\pgfqpoint{1.292461in}{-0.422154in}}%
\pgfpathlineto{\pgfqpoint{1.292461in}{1.246164in}}%
\pgfpathlineto{\pgfqpoint{1.248509in}{1.246164in}}%
\pgfpathclose%
\pgfusepath{fill}%
\end{pgfscope}%
\begin{pgfscope}%
\pgfpathrectangle{\pgfqpoint{0.386100in}{0.336000in}}{\pgfqpoint{1.015300in}{1.232000in}}%
\pgfusepath{clip}%
\pgfsetbuttcap%
\pgfsetmiterjoin%
\definecolor{currentfill}{rgb}{0.050000,0.050000,0.700000}%
\pgfsetfillcolor{currentfill}%
\pgfsetlinewidth{0.000000pt}%
\definecolor{currentstroke}{rgb}{0.000000,0.000000,0.000000}%
\pgfsetstrokecolor{currentstroke}%
\pgfsetstrokeopacity{0.000000}%
\pgfsetdash{}{0pt}%
\pgfpathmoveto{\pgfqpoint{0.557828in}{-0.422154in}}%
\pgfpathlineto{\pgfqpoint{0.601781in}{-0.422154in}}%
\pgfpathlineto{\pgfqpoint{0.601781in}{1.160492in}}%
\pgfpathlineto{\pgfqpoint{0.557828in}{1.160492in}}%
\pgfpathclose%
\pgfusepath{fill}%
\end{pgfscope}%
\begin{pgfscope}%
\pgfpathrectangle{\pgfqpoint{0.386100in}{0.336000in}}{\pgfqpoint{1.015300in}{1.232000in}}%
\pgfusepath{clip}%
\pgfsetbuttcap%
\pgfsetmiterjoin%
\definecolor{currentfill}{rgb}{0.050000,0.050000,0.700000}%
\pgfsetfillcolor{currentfill}%
\pgfsetlinewidth{0.000000pt}%
\definecolor{currentstroke}{rgb}{0.000000,0.000000,0.000000}%
\pgfsetstrokecolor{currentstroke}%
\pgfsetstrokeopacity{0.000000}%
\pgfsetdash{}{0pt}%
\pgfpathmoveto{\pgfqpoint{0.808985in}{-0.422154in}}%
\pgfpathlineto{\pgfqpoint{0.852937in}{-0.422154in}}%
\pgfpathlineto{\pgfqpoint{0.852937in}{1.018338in}}%
\pgfpathlineto{\pgfqpoint{0.808985in}{1.018338in}}%
\pgfpathclose%
\pgfusepath{fill}%
\end{pgfscope}%
\begin{pgfscope}%
\pgfpathrectangle{\pgfqpoint{0.386100in}{0.336000in}}{\pgfqpoint{1.015300in}{1.232000in}}%
\pgfusepath{clip}%
\pgfsetbuttcap%
\pgfsetmiterjoin%
\definecolor{currentfill}{rgb}{0.050000,0.050000,0.700000}%
\pgfsetfillcolor{currentfill}%
\pgfsetlinewidth{0.000000pt}%
\definecolor{currentstroke}{rgb}{0.000000,0.000000,0.000000}%
\pgfsetstrokecolor{currentstroke}%
\pgfsetstrokeopacity{0.000000}%
\pgfsetdash{}{0pt}%
\pgfpathmoveto{\pgfqpoint{1.060141in}{-0.422154in}}%
\pgfpathlineto{\pgfqpoint{1.104094in}{-0.422154in}}%
\pgfpathlineto{\pgfqpoint{1.104094in}{1.029711in}}%
\pgfpathlineto{\pgfqpoint{1.060141in}{1.029711in}}%
\pgfpathclose%
\pgfusepath{fill}%
\end{pgfscope}%
\begin{pgfscope}%
\pgfpathrectangle{\pgfqpoint{0.386100in}{0.336000in}}{\pgfqpoint{1.015300in}{1.232000in}}%
\pgfusepath{clip}%
\pgfsetbuttcap%
\pgfsetmiterjoin%
\definecolor{currentfill}{rgb}{0.050000,0.050000,0.700000}%
\pgfsetfillcolor{currentfill}%
\pgfsetlinewidth{0.000000pt}%
\definecolor{currentstroke}{rgb}{0.000000,0.000000,0.000000}%
\pgfsetstrokecolor{currentstroke}%
\pgfsetstrokeopacity{0.000000}%
\pgfsetdash{}{0pt}%
\pgfpathmoveto{\pgfqpoint{1.311298in}{-0.422154in}}%
\pgfpathlineto{\pgfqpoint{1.355250in}{-0.422154in}}%
\pgfpathlineto{\pgfqpoint{1.355250in}{1.247680in}}%
\pgfpathlineto{\pgfqpoint{1.311298in}{1.247680in}}%
\pgfpathclose%
\pgfusepath{fill}%
\end{pgfscope}%
\begin{pgfscope}%
\pgfsetrectcap%
\pgfsetmiterjoin%
\pgfsetlinewidth{0.803000pt}%
\definecolor{currentstroke}{rgb}{0.000000,0.000000,0.000000}%
\pgfsetstrokecolor{currentstroke}%
\pgfsetdash{}{0pt}%
\pgfpathmoveto{\pgfqpoint{0.386100in}{0.336000in}}%
\pgfpathlineto{\pgfqpoint{0.386100in}{1.568000in}}%
\pgfusepath{stroke}%
\end{pgfscope}%
\begin{pgfscope}%
\pgfsetrectcap%
\pgfsetmiterjoin%
\pgfsetlinewidth{0.803000pt}%
\definecolor{currentstroke}{rgb}{0.000000,0.000000,0.000000}%
\pgfsetstrokecolor{currentstroke}%
\pgfsetdash{}{0pt}%
\pgfpathmoveto{\pgfqpoint{1.401400in}{0.336000in}}%
\pgfpathlineto{\pgfqpoint{1.401400in}{1.568000in}}%
\pgfusepath{stroke}%
\end{pgfscope}%
\begin{pgfscope}%
\pgfsetrectcap%
\pgfsetmiterjoin%
\pgfsetlinewidth{0.803000pt}%
\definecolor{currentstroke}{rgb}{0.000000,0.000000,0.000000}%
\pgfsetstrokecolor{currentstroke}%
\pgfsetdash{}{0pt}%
\pgfpathmoveto{\pgfqpoint{0.386100in}{0.336000in}}%
\pgfpathlineto{\pgfqpoint{1.401400in}{0.336000in}}%
\pgfusepath{stroke}%
\end{pgfscope}%
\begin{pgfscope}%
\pgfsetrectcap%
\pgfsetmiterjoin%
\pgfsetlinewidth{0.803000pt}%
\definecolor{currentstroke}{rgb}{0.000000,0.000000,0.000000}%
\pgfsetstrokecolor{currentstroke}%
\pgfsetdash{}{0pt}%
\pgfpathmoveto{\pgfqpoint{0.386100in}{1.568000in}}%
\pgfpathlineto{\pgfqpoint{1.401400in}{1.568000in}}%
\pgfusepath{stroke}%
\end{pgfscope}%
\begin{pgfscope}%
\pgfsetbuttcap%
\pgfsetmiterjoin%
\pgfsetlinewidth{0.000000pt}%
\definecolor{currentstroke}{rgb}{0.800000,0.800000,0.800000}%
\pgfsetstrokecolor{currentstroke}%
\pgfsetstrokeopacity{0.000000}%
\pgfsetdash{}{0pt}%
\pgfpathmoveto{\pgfqpoint{0.613142in}{1.128278in}}%
\pgfpathlineto{\pgfqpoint{1.174358in}{1.128278in}}%
\pgfpathquadraticcurveto{\pgfqpoint{1.192135in}{1.128278in}}{\pgfqpoint{1.192135in}{1.146055in}}%
\pgfpathlineto{\pgfqpoint{1.192135in}{1.505778in}}%
\pgfpathquadraticcurveto{\pgfqpoint{1.192135in}{1.523556in}}{\pgfqpoint{1.174358in}{1.523556in}}%
\pgfpathlineto{\pgfqpoint{0.613142in}{1.523556in}}%
\pgfpathquadraticcurveto{\pgfqpoint{0.595365in}{1.523556in}}{\pgfqpoint{0.595365in}{1.505778in}}%
\pgfpathlineto{\pgfqpoint{0.595365in}{1.146055in}}%
\pgfpathquadraticcurveto{\pgfqpoint{0.595365in}{1.128278in}}{\pgfqpoint{0.613142in}{1.128278in}}%
\pgfpathclose%
\pgfusepath{}%
\end{pgfscope}%
\begin{pgfscope}%
\pgfsetbuttcap%
\pgfsetmiterjoin%
\definecolor{currentfill}{rgb}{0.900000,0.100000,0.100000}%
\pgfsetfillcolor{currentfill}%
\pgfsetlinewidth{0.000000pt}%
\definecolor{currentstroke}{rgb}{0.000000,0.000000,0.000000}%
\pgfsetstrokecolor{currentstroke}%
\pgfsetstrokeopacity{0.000000}%
\pgfsetdash{}{0pt}%
\pgfpathmoveto{\pgfqpoint{0.630920in}{1.425778in}}%
\pgfpathlineto{\pgfqpoint{0.808698in}{1.425778in}}%
\pgfpathlineto{\pgfqpoint{0.808698in}{1.488000in}}%
\pgfpathlineto{\pgfqpoint{0.630920in}{1.488000in}}%
\pgfpathclose%
\pgfusepath{fill}%
\end{pgfscope}%
\begin{pgfscope}%
\pgftext[x=0.879809in,y=1.425778in,left,base]{\rmfamily\fontsize{6.400000}{7.680000}\selectfont 0s}%
\end{pgfscope}%
\begin{pgfscope}%
\pgfsetbuttcap%
\pgfsetmiterjoin%
\definecolor{currentfill}{rgb}{0.300000,0.800000,0.200000}%
\pgfsetfillcolor{currentfill}%
\pgfsetlinewidth{0.000000pt}%
\definecolor{currentstroke}{rgb}{0.000000,0.000000,0.000000}%
\pgfsetstrokecolor{currentstroke}%
\pgfsetstrokeopacity{0.000000}%
\pgfsetdash{}{0pt}%
\pgfpathmoveto{\pgfqpoint{0.630920in}{1.302907in}}%
\pgfpathlineto{\pgfqpoint{0.808698in}{1.302907in}}%
\pgfpathlineto{\pgfqpoint{0.808698in}{1.365130in}}%
\pgfpathlineto{\pgfqpoint{0.630920in}{1.365130in}}%
\pgfpathclose%
\pgfusepath{fill}%
\end{pgfscope}%
\begin{pgfscope}%
\pgftext[x=0.879809in,y=1.302907in,left,base]{\rmfamily\fontsize{6.400000}{7.680000}\selectfont 500\(\displaystyle \mu\)s}%
\end{pgfscope}%
\begin{pgfscope}%
\pgfsetbuttcap%
\pgfsetmiterjoin%
\definecolor{currentfill}{rgb}{0.050000,0.050000,0.700000}%
\pgfsetfillcolor{currentfill}%
\pgfsetlinewidth{0.000000pt}%
\definecolor{currentstroke}{rgb}{0.000000,0.000000,0.000000}%
\pgfsetstrokecolor{currentstroke}%
\pgfsetstrokeopacity{0.000000}%
\pgfsetdash{}{0pt}%
\pgfpathmoveto{\pgfqpoint{0.630920in}{1.180037in}}%
\pgfpathlineto{\pgfqpoint{0.808698in}{1.180037in}}%
\pgfpathlineto{\pgfqpoint{0.808698in}{1.242259in}}%
\pgfpathlineto{\pgfqpoint{0.630920in}{1.242259in}}%
\pgfpathclose%
\pgfusepath{fill}%
\end{pgfscope}%
\begin{pgfscope}%
\pgftext[x=0.879809in,y=1.180037in,left,base]{\rmfamily\fontsize{6.400000}{7.680000}\selectfont 500ms}%
\end{pgfscope}%
\end{pgfpicture}%
\makeatother%
\endgroup%

%% file: plot_data/facebook_data/fb_cluster_combined/p999_reconfig_frequency_latency_combined.pgf
\begingroup%
\makeatletter%
\begin{pgfpicture}%
\pgfpathrectangle{\pgfpointorigin}{\pgfqpoint{1.360000in}{1.600000in}}%
\pgfusepath{use as bounding box, clip}%
\begin{pgfscope}%
\pgfsetbuttcap%
\pgfsetmiterjoin%
\definecolor{currentfill}{rgb}{1.000000,1.000000,1.000000}%
\pgfsetfillcolor{currentfill}%
\pgfsetlinewidth{0.000000pt}%
\definecolor{currentstroke}{rgb}{1.000000,1.000000,1.000000}%
\pgfsetstrokecolor{currentstroke}%
\pgfsetdash{}{0pt}%
\pgfpathmoveto{\pgfqpoint{0.000000in}{0.000000in}}%
\pgfpathlineto{\pgfqpoint{1.360000in}{0.000000in}}%
\pgfpathlineto{\pgfqpoint{1.360000in}{1.600000in}}%
\pgfpathlineto{\pgfqpoint{0.000000in}{1.600000in}}%
\pgfpathclose%
\pgfusepath{fill}%
\end{pgfscope}%
\begin{pgfscope}%
\pgfsetbuttcap%
\pgfsetmiterjoin%
\definecolor{currentfill}{rgb}{1.000000,1.000000,1.000000}%
\pgfsetfillcolor{currentfill}%
\pgfsetlinewidth{0.000000pt}%
\definecolor{currentstroke}{rgb}{0.000000,0.000000,0.000000}%
\pgfsetstrokecolor{currentstroke}%
\pgfsetstrokeopacity{0.000000}%
\pgfsetdash{}{0pt}%
\pgfpathmoveto{\pgfqpoint{0.340000in}{0.336000in}}%
\pgfpathlineto{\pgfqpoint{1.332800in}{0.336000in}}%
\pgfpathlineto{\pgfqpoint{1.332800in}{1.584000in}}%
\pgfpathlineto{\pgfqpoint{0.340000in}{1.584000in}}%
\pgfpathclose%
\pgfusepath{fill}%
\end{pgfscope}%
\begin{pgfscope}%
\pgfsetbuttcap%
\pgfsetroundjoin%
\definecolor{currentfill}{rgb}{0.000000,0.000000,0.000000}%
\pgfsetfillcolor{currentfill}%
\pgfsetlinewidth{0.803000pt}%
\definecolor{currentstroke}{rgb}{0.000000,0.000000,0.000000}%
\pgfsetstrokecolor{currentstroke}%
\pgfsetdash{}{0pt}%
\pgfsys@defobject{currentmarker}{\pgfqpoint{0.000000in}{-0.048611in}}{\pgfqpoint{0.000000in}{0.000000in}}{%
\pgfpathmoveto{\pgfqpoint{0.000000in}{0.000000in}}%
\pgfpathlineto{\pgfqpoint{0.000000in}{-0.048611in}}%
\pgfusepath{stroke,fill}%
}%
\begin{pgfscope}%
\pgfsys@transformshift{0.468014in}{0.336000in}%
\pgfsys@useobject{currentmarker}{}%
\end{pgfscope}%
\end{pgfscope}%
\begin{pgfscope}%
\pgftext[x=0.468014in,y=0.238778in,,top]{\rmfamily\fontsize{6.500000}{7.800000}\selectfont 30s}%
\end{pgfscope}%
\begin{pgfscope}%
\pgfsetbuttcap%
\pgfsetroundjoin%
\definecolor{currentfill}{rgb}{0.000000,0.000000,0.000000}%
\pgfsetfillcolor{currentfill}%
\pgfsetlinewidth{0.803000pt}%
\definecolor{currentstroke}{rgb}{0.000000,0.000000,0.000000}%
\pgfsetstrokecolor{currentstroke}%
\pgfsetdash{}{0pt}%
\pgfsys@defobject{currentmarker}{\pgfqpoint{0.000000in}{-0.048611in}}{\pgfqpoint{0.000000in}{0.000000in}}{%
\pgfpathmoveto{\pgfqpoint{0.000000in}{0.000000in}}%
\pgfpathlineto{\pgfqpoint{0.000000in}{-0.048611in}}%
\pgfusepath{stroke,fill}%
}%
\begin{pgfscope}%
\pgfsys@transformshift{0.713605in}{0.336000in}%
\pgfsys@useobject{currentmarker}{}%
\end{pgfscope}%
\end{pgfscope}%
\begin{pgfscope}%
\pgftext[x=0.713605in,y=0.238778in,,top]{\rmfamily\fontsize{6.500000}{7.800000}\selectfont 300s}%
\end{pgfscope}%
\begin{pgfscope}%
\pgfsetbuttcap%
\pgfsetroundjoin%
\definecolor{currentfill}{rgb}{0.000000,0.000000,0.000000}%
\pgfsetfillcolor{currentfill}%
\pgfsetlinewidth{0.803000pt}%
\definecolor{currentstroke}{rgb}{0.000000,0.000000,0.000000}%
\pgfsetstrokecolor{currentstroke}%
\pgfsetdash{}{0pt}%
\pgfsys@defobject{currentmarker}{\pgfqpoint{0.000000in}{-0.048611in}}{\pgfqpoint{0.000000in}{0.000000in}}{%
\pgfpathmoveto{\pgfqpoint{0.000000in}{0.000000in}}%
\pgfpathlineto{\pgfqpoint{0.000000in}{-0.048611in}}%
\pgfusepath{stroke,fill}%
}%
\begin{pgfscope}%
\pgfsys@transformshift{0.959195in}{0.336000in}%
\pgfsys@useobject{currentmarker}{}%
\end{pgfscope}%
\end{pgfscope}%
\begin{pgfscope}%
\pgftext[x=0.959195in,y=0.238778in,,top]{\rmfamily\fontsize{6.500000}{7.800000}\selectfont 3600s}%
\end{pgfscope}%
\begin{pgfscope}%
\pgfsetbuttcap%
\pgfsetroundjoin%
\definecolor{currentfill}{rgb}{0.000000,0.000000,0.000000}%
\pgfsetfillcolor{currentfill}%
\pgfsetlinewidth{0.803000pt}%
\definecolor{currentstroke}{rgb}{0.000000,0.000000,0.000000}%
\pgfsetstrokecolor{currentstroke}%
\pgfsetdash{}{0pt}%
\pgfsys@defobject{currentmarker}{\pgfqpoint{0.000000in}{-0.048611in}}{\pgfqpoint{0.000000in}{0.000000in}}{%
\pgfpathmoveto{\pgfqpoint{0.000000in}{0.000000in}}%
\pgfpathlineto{\pgfqpoint{0.000000in}{-0.048611in}}%
\pgfusepath{stroke,fill}%
}%
\begin{pgfscope}%
\pgfsys@transformshift{1.204786in}{0.336000in}%
\pgfsys@useobject{currentmarker}{}%
\end{pgfscope}%
\end{pgfscope}%
\begin{pgfscope}%
\pgftext[x=1.204786in,y=0.238778in,,top]{\rmfamily\fontsize{6.500000}{7.800000}\selectfont 24H}%
\end{pgfscope}%
\begin{pgfscope}%
\pgftext[x=0.836400in,y=0.150815in,,top]{\rmfamily\fontsize{7.400000}{8.880000}\selectfont Reconfig. Freq.}%
\end{pgfscope}%
\begin{pgfscope}%
\pgfpathrectangle{\pgfqpoint{0.340000in}{0.336000in}}{\pgfqpoint{0.992800in}{1.248000in}}%
\pgfusepath{clip}%
\pgfsetbuttcap%
\pgfsetroundjoin%
\pgfsetlinewidth{0.702625pt}%
\definecolor{currentstroke}{rgb}{0.690196,0.690196,0.690196}%
\pgfsetstrokecolor{currentstroke}%
\pgfsetdash{{4.480000pt}{1.120000pt}{0.700000pt}{1.120000pt}}{0.000000pt}%
\pgfpathmoveto{\pgfqpoint{0.340000in}{0.336000in}}%
\pgfpathlineto{\pgfqpoint{1.332800in}{0.336000in}}%
\pgfusepath{stroke}%
\end{pgfscope}%
\begin{pgfscope}%
\pgfsetbuttcap%
\pgfsetroundjoin%
\definecolor{currentfill}{rgb}{0.000000,0.000000,0.000000}%
\pgfsetfillcolor{currentfill}%
\pgfsetlinewidth{0.803000pt}%
\definecolor{currentstroke}{rgb}{0.000000,0.000000,0.000000}%
\pgfsetstrokecolor{currentstroke}%
\pgfsetdash{}{0pt}%
\pgfsys@defobject{currentmarker}{\pgfqpoint{0.000000in}{0.000000in}}{\pgfqpoint{0.048611in}{0.000000in}}{%
\pgfpathmoveto{\pgfqpoint{0.000000in}{0.000000in}}%
\pgfpathlineto{\pgfqpoint{0.048611in}{0.000000in}}%
\pgfusepath{stroke,fill}%
}%
\begin{pgfscope}%
\pgfsys@transformshift{0.340000in}{0.336000in}%
\pgfsys@useobject{currentmarker}{}%
\end{pgfscope}%
\end{pgfscope}%
\begin{pgfscope}%
\pgftext[x=0.157903in,y=0.307065in,left,base]{\rmfamily\fontsize{6.500000}{7.800000}\selectfont \(\displaystyle 0.5\)}%
\end{pgfscope}%
\begin{pgfscope}%
\pgfpathrectangle{\pgfqpoint{0.340000in}{0.336000in}}{\pgfqpoint{0.992800in}{1.248000in}}%
\pgfusepath{clip}%
\pgfsetbuttcap%
\pgfsetroundjoin%
\pgfsetlinewidth{0.702625pt}%
\definecolor{currentstroke}{rgb}{0.690196,0.690196,0.690196}%
\pgfsetstrokecolor{currentstroke}%
\pgfsetdash{{4.480000pt}{1.120000pt}{0.700000pt}{1.120000pt}}{0.000000pt}%
\pgfpathmoveto{\pgfqpoint{0.340000in}{0.714182in}}%
\pgfpathlineto{\pgfqpoint{1.332800in}{0.714182in}}%
\pgfusepath{stroke}%
\end{pgfscope}%
\begin{pgfscope}%
\pgfsetbuttcap%
\pgfsetroundjoin%
\definecolor{currentfill}{rgb}{0.000000,0.000000,0.000000}%
\pgfsetfillcolor{currentfill}%
\pgfsetlinewidth{0.803000pt}%
\definecolor{currentstroke}{rgb}{0.000000,0.000000,0.000000}%
\pgfsetstrokecolor{currentstroke}%
\pgfsetdash{}{0pt}%
\pgfsys@defobject{currentmarker}{\pgfqpoint{0.000000in}{0.000000in}}{\pgfqpoint{0.048611in}{0.000000in}}{%
\pgfpathmoveto{\pgfqpoint{0.000000in}{0.000000in}}%
\pgfpathlineto{\pgfqpoint{0.048611in}{0.000000in}}%
\pgfusepath{stroke,fill}%
}%
\begin{pgfscope}%
\pgfsys@transformshift{0.340000in}{0.714182in}%
\pgfsys@useobject{currentmarker}{}%
\end{pgfscope}%
\end{pgfscope}%
\begin{pgfscope}%
\pgftext[x=0.157903in,y=0.685247in,left,base]{\rmfamily\fontsize{6.500000}{7.800000}\selectfont \(\displaystyle 0.6\)}%
\end{pgfscope}%
\begin{pgfscope}%
\pgfpathrectangle{\pgfqpoint{0.340000in}{0.336000in}}{\pgfqpoint{0.992800in}{1.248000in}}%
\pgfusepath{clip}%
\pgfsetbuttcap%
\pgfsetroundjoin%
\pgfsetlinewidth{0.702625pt}%
\definecolor{currentstroke}{rgb}{0.690196,0.690196,0.690196}%
\pgfsetstrokecolor{currentstroke}%
\pgfsetdash{{4.480000pt}{1.120000pt}{0.700000pt}{1.120000pt}}{0.000000pt}%
\pgfpathmoveto{\pgfqpoint{0.340000in}{1.092364in}}%
\pgfpathlineto{\pgfqpoint{1.332800in}{1.092364in}}%
\pgfusepath{stroke}%
\end{pgfscope}%
\begin{pgfscope}%
\pgfsetbuttcap%
\pgfsetroundjoin%
\definecolor{currentfill}{rgb}{0.000000,0.000000,0.000000}%
\pgfsetfillcolor{currentfill}%
\pgfsetlinewidth{0.803000pt}%
\definecolor{currentstroke}{rgb}{0.000000,0.000000,0.000000}%
\pgfsetstrokecolor{currentstroke}%
\pgfsetdash{}{0pt}%
\pgfsys@defobject{currentmarker}{\pgfqpoint{0.000000in}{0.000000in}}{\pgfqpoint{0.048611in}{0.000000in}}{%
\pgfpathmoveto{\pgfqpoint{0.000000in}{0.000000in}}%
\pgfpathlineto{\pgfqpoint{0.048611in}{0.000000in}}%
\pgfusepath{stroke,fill}%
}%
\begin{pgfscope}%
\pgfsys@transformshift{0.340000in}{1.092364in}%
\pgfsys@useobject{currentmarker}{}%
\end{pgfscope}%
\end{pgfscope}%
\begin{pgfscope}%
\pgftext[x=0.157903in,y=1.063428in,left,base]{\rmfamily\fontsize{6.500000}{7.800000}\selectfont \(\displaystyle 0.7\)}%
\end{pgfscope}%
\begin{pgfscope}%
\pgfpathrectangle{\pgfqpoint{0.340000in}{0.336000in}}{\pgfqpoint{0.992800in}{1.248000in}}%
\pgfusepath{clip}%
\pgfsetbuttcap%
\pgfsetroundjoin%
\pgfsetlinewidth{0.702625pt}%
\definecolor{currentstroke}{rgb}{0.690196,0.690196,0.690196}%
\pgfsetstrokecolor{currentstroke}%
\pgfsetdash{{4.480000pt}{1.120000pt}{0.700000pt}{1.120000pt}}{0.000000pt}%
\pgfpathmoveto{\pgfqpoint{0.340000in}{1.470545in}}%
\pgfpathlineto{\pgfqpoint{1.332800in}{1.470545in}}%
\pgfusepath{stroke}%
\end{pgfscope}%
\begin{pgfscope}%
\pgfsetbuttcap%
\pgfsetroundjoin%
\definecolor{currentfill}{rgb}{0.000000,0.000000,0.000000}%
\pgfsetfillcolor{currentfill}%
\pgfsetlinewidth{0.803000pt}%
\definecolor{currentstroke}{rgb}{0.000000,0.000000,0.000000}%
\pgfsetstrokecolor{currentstroke}%
\pgfsetdash{}{0pt}%
\pgfsys@defobject{currentmarker}{\pgfqpoint{0.000000in}{0.000000in}}{\pgfqpoint{0.048611in}{0.000000in}}{%
\pgfpathmoveto{\pgfqpoint{0.000000in}{0.000000in}}%
\pgfpathlineto{\pgfqpoint{0.048611in}{0.000000in}}%
\pgfusepath{stroke,fill}%
}%
\begin{pgfscope}%
\pgfsys@transformshift{0.340000in}{1.470545in}%
\pgfsys@useobject{currentmarker}{}%
\end{pgfscope}%
\end{pgfscope}%
\begin{pgfscope}%
\pgftext[x=0.157903in,y=1.441610in,left,base]{\rmfamily\fontsize{6.500000}{7.800000}\selectfont \(\displaystyle 0.8\)}%
\end{pgfscope}%
\begin{pgfscope}%
\pgfsetbuttcap%
\pgfsetroundjoin%
\definecolor{currentfill}{rgb}{0.000000,0.000000,0.000000}%
\pgfsetfillcolor{currentfill}%
\pgfsetlinewidth{0.602250pt}%
\definecolor{currentstroke}{rgb}{0.000000,0.000000,0.000000}%
\pgfsetstrokecolor{currentstroke}%
\pgfsetdash{}{0pt}%
\pgfsys@defobject{currentmarker}{\pgfqpoint{0.000000in}{0.000000in}}{\pgfqpoint{0.027778in}{0.000000in}}{%
\pgfpathmoveto{\pgfqpoint{0.000000in}{0.000000in}}%
\pgfpathlineto{\pgfqpoint{0.027778in}{0.000000in}}%
\pgfusepath{stroke,fill}%
}%
\begin{pgfscope}%
\pgfsys@transformshift{0.340000in}{0.411636in}%
\pgfsys@useobject{currentmarker}{}%
\end{pgfscope}%
\end{pgfscope}%
\begin{pgfscope}%
\pgfsetbuttcap%
\pgfsetroundjoin%
\definecolor{currentfill}{rgb}{0.000000,0.000000,0.000000}%
\pgfsetfillcolor{currentfill}%
\pgfsetlinewidth{0.602250pt}%
\definecolor{currentstroke}{rgb}{0.000000,0.000000,0.000000}%
\pgfsetstrokecolor{currentstroke}%
\pgfsetdash{}{0pt}%
\pgfsys@defobject{currentmarker}{\pgfqpoint{0.000000in}{0.000000in}}{\pgfqpoint{0.027778in}{0.000000in}}{%
\pgfpathmoveto{\pgfqpoint{0.000000in}{0.000000in}}%
\pgfpathlineto{\pgfqpoint{0.027778in}{0.000000in}}%
\pgfusepath{stroke,fill}%
}%
\begin{pgfscope}%
\pgfsys@transformshift{0.340000in}{0.487273in}%
\pgfsys@useobject{currentmarker}{}%
\end{pgfscope}%
\end{pgfscope}%
\begin{pgfscope}%
\pgfsetbuttcap%
\pgfsetroundjoin%
\definecolor{currentfill}{rgb}{0.000000,0.000000,0.000000}%
\pgfsetfillcolor{currentfill}%
\pgfsetlinewidth{0.602250pt}%
\definecolor{currentstroke}{rgb}{0.000000,0.000000,0.000000}%
\pgfsetstrokecolor{currentstroke}%
\pgfsetdash{}{0pt}%
\pgfsys@defobject{currentmarker}{\pgfqpoint{0.000000in}{0.000000in}}{\pgfqpoint{0.027778in}{0.000000in}}{%
\pgfpathmoveto{\pgfqpoint{0.000000in}{0.000000in}}%
\pgfpathlineto{\pgfqpoint{0.027778in}{0.000000in}}%
\pgfusepath{stroke,fill}%
}%
\begin{pgfscope}%
\pgfsys@transformshift{0.340000in}{0.562909in}%
\pgfsys@useobject{currentmarker}{}%
\end{pgfscope}%
\end{pgfscope}%
\begin{pgfscope}%
\pgfsetbuttcap%
\pgfsetroundjoin%
\definecolor{currentfill}{rgb}{0.000000,0.000000,0.000000}%
\pgfsetfillcolor{currentfill}%
\pgfsetlinewidth{0.602250pt}%
\definecolor{currentstroke}{rgb}{0.000000,0.000000,0.000000}%
\pgfsetstrokecolor{currentstroke}%
\pgfsetdash{}{0pt}%
\pgfsys@defobject{currentmarker}{\pgfqpoint{0.000000in}{0.000000in}}{\pgfqpoint{0.027778in}{0.000000in}}{%
\pgfpathmoveto{\pgfqpoint{0.000000in}{0.000000in}}%
\pgfpathlineto{\pgfqpoint{0.027778in}{0.000000in}}%
\pgfusepath{stroke,fill}%
}%
\begin{pgfscope}%
\pgfsys@transformshift{0.340000in}{0.638545in}%
\pgfsys@useobject{currentmarker}{}%
\end{pgfscope}%
\end{pgfscope}%
\begin{pgfscope}%
\pgfsetbuttcap%
\pgfsetroundjoin%
\definecolor{currentfill}{rgb}{0.000000,0.000000,0.000000}%
\pgfsetfillcolor{currentfill}%
\pgfsetlinewidth{0.602250pt}%
\definecolor{currentstroke}{rgb}{0.000000,0.000000,0.000000}%
\pgfsetstrokecolor{currentstroke}%
\pgfsetdash{}{0pt}%
\pgfsys@defobject{currentmarker}{\pgfqpoint{0.000000in}{0.000000in}}{\pgfqpoint{0.027778in}{0.000000in}}{%
\pgfpathmoveto{\pgfqpoint{0.000000in}{0.000000in}}%
\pgfpathlineto{\pgfqpoint{0.027778in}{0.000000in}}%
\pgfusepath{stroke,fill}%
}%
\begin{pgfscope}%
\pgfsys@transformshift{0.340000in}{0.789818in}%
\pgfsys@useobject{currentmarker}{}%
\end{pgfscope}%
\end{pgfscope}%
\begin{pgfscope}%
\pgfsetbuttcap%
\pgfsetroundjoin%
\definecolor{currentfill}{rgb}{0.000000,0.000000,0.000000}%
\pgfsetfillcolor{currentfill}%
\pgfsetlinewidth{0.602250pt}%
\definecolor{currentstroke}{rgb}{0.000000,0.000000,0.000000}%
\pgfsetstrokecolor{currentstroke}%
\pgfsetdash{}{0pt}%
\pgfsys@defobject{currentmarker}{\pgfqpoint{0.000000in}{0.000000in}}{\pgfqpoint{0.027778in}{0.000000in}}{%
\pgfpathmoveto{\pgfqpoint{0.000000in}{0.000000in}}%
\pgfpathlineto{\pgfqpoint{0.027778in}{0.000000in}}%
\pgfusepath{stroke,fill}%
}%
\begin{pgfscope}%
\pgfsys@transformshift{0.340000in}{0.865455in}%
\pgfsys@useobject{currentmarker}{}%
\end{pgfscope}%
\end{pgfscope}%
\begin{pgfscope}%
\pgfsetbuttcap%
\pgfsetroundjoin%
\definecolor{currentfill}{rgb}{0.000000,0.000000,0.000000}%
\pgfsetfillcolor{currentfill}%
\pgfsetlinewidth{0.602250pt}%
\definecolor{currentstroke}{rgb}{0.000000,0.000000,0.000000}%
\pgfsetstrokecolor{currentstroke}%
\pgfsetdash{}{0pt}%
\pgfsys@defobject{currentmarker}{\pgfqpoint{0.000000in}{0.000000in}}{\pgfqpoint{0.027778in}{0.000000in}}{%
\pgfpathmoveto{\pgfqpoint{0.000000in}{0.000000in}}%
\pgfpathlineto{\pgfqpoint{0.027778in}{0.000000in}}%
\pgfusepath{stroke,fill}%
}%
\begin{pgfscope}%
\pgfsys@transformshift{0.340000in}{0.941091in}%
\pgfsys@useobject{currentmarker}{}%
\end{pgfscope}%
\end{pgfscope}%
\begin{pgfscope}%
\pgfsetbuttcap%
\pgfsetroundjoin%
\definecolor{currentfill}{rgb}{0.000000,0.000000,0.000000}%
\pgfsetfillcolor{currentfill}%
\pgfsetlinewidth{0.602250pt}%
\definecolor{currentstroke}{rgb}{0.000000,0.000000,0.000000}%
\pgfsetstrokecolor{currentstroke}%
\pgfsetdash{}{0pt}%
\pgfsys@defobject{currentmarker}{\pgfqpoint{0.000000in}{0.000000in}}{\pgfqpoint{0.027778in}{0.000000in}}{%
\pgfpathmoveto{\pgfqpoint{0.000000in}{0.000000in}}%
\pgfpathlineto{\pgfqpoint{0.027778in}{0.000000in}}%
\pgfusepath{stroke,fill}%
}%
\begin{pgfscope}%
\pgfsys@transformshift{0.340000in}{1.016727in}%
\pgfsys@useobject{currentmarker}{}%
\end{pgfscope}%
\end{pgfscope}%
\begin{pgfscope}%
\pgfsetbuttcap%
\pgfsetroundjoin%
\definecolor{currentfill}{rgb}{0.000000,0.000000,0.000000}%
\pgfsetfillcolor{currentfill}%
\pgfsetlinewidth{0.602250pt}%
\definecolor{currentstroke}{rgb}{0.000000,0.000000,0.000000}%
\pgfsetstrokecolor{currentstroke}%
\pgfsetdash{}{0pt}%
\pgfsys@defobject{currentmarker}{\pgfqpoint{0.000000in}{0.000000in}}{\pgfqpoint{0.027778in}{0.000000in}}{%
\pgfpathmoveto{\pgfqpoint{0.000000in}{0.000000in}}%
\pgfpathlineto{\pgfqpoint{0.027778in}{0.000000in}}%
\pgfusepath{stroke,fill}%
}%
\begin{pgfscope}%
\pgfsys@transformshift{0.340000in}{1.168000in}%
\pgfsys@useobject{currentmarker}{}%
\end{pgfscope}%
\end{pgfscope}%
\begin{pgfscope}%
\pgfsetbuttcap%
\pgfsetroundjoin%
\definecolor{currentfill}{rgb}{0.000000,0.000000,0.000000}%
\pgfsetfillcolor{currentfill}%
\pgfsetlinewidth{0.602250pt}%
\definecolor{currentstroke}{rgb}{0.000000,0.000000,0.000000}%
\pgfsetstrokecolor{currentstroke}%
\pgfsetdash{}{0pt}%
\pgfsys@defobject{currentmarker}{\pgfqpoint{0.000000in}{0.000000in}}{\pgfqpoint{0.027778in}{0.000000in}}{%
\pgfpathmoveto{\pgfqpoint{0.000000in}{0.000000in}}%
\pgfpathlineto{\pgfqpoint{0.027778in}{0.000000in}}%
\pgfusepath{stroke,fill}%
}%
\begin{pgfscope}%
\pgfsys@transformshift{0.340000in}{1.243636in}%
\pgfsys@useobject{currentmarker}{}%
\end{pgfscope}%
\end{pgfscope}%
\begin{pgfscope}%
\pgfsetbuttcap%
\pgfsetroundjoin%
\definecolor{currentfill}{rgb}{0.000000,0.000000,0.000000}%
\pgfsetfillcolor{currentfill}%
\pgfsetlinewidth{0.602250pt}%
\definecolor{currentstroke}{rgb}{0.000000,0.000000,0.000000}%
\pgfsetstrokecolor{currentstroke}%
\pgfsetdash{}{0pt}%
\pgfsys@defobject{currentmarker}{\pgfqpoint{0.000000in}{0.000000in}}{\pgfqpoint{0.027778in}{0.000000in}}{%
\pgfpathmoveto{\pgfqpoint{0.000000in}{0.000000in}}%
\pgfpathlineto{\pgfqpoint{0.027778in}{0.000000in}}%
\pgfusepath{stroke,fill}%
}%
\begin{pgfscope}%
\pgfsys@transformshift{0.340000in}{1.319273in}%
\pgfsys@useobject{currentmarker}{}%
\end{pgfscope}%
\end{pgfscope}%
\begin{pgfscope}%
\pgfsetbuttcap%
\pgfsetroundjoin%
\definecolor{currentfill}{rgb}{0.000000,0.000000,0.000000}%
\pgfsetfillcolor{currentfill}%
\pgfsetlinewidth{0.602250pt}%
\definecolor{currentstroke}{rgb}{0.000000,0.000000,0.000000}%
\pgfsetstrokecolor{currentstroke}%
\pgfsetdash{}{0pt}%
\pgfsys@defobject{currentmarker}{\pgfqpoint{0.000000in}{0.000000in}}{\pgfqpoint{0.027778in}{0.000000in}}{%
\pgfpathmoveto{\pgfqpoint{0.000000in}{0.000000in}}%
\pgfpathlineto{\pgfqpoint{0.027778in}{0.000000in}}%
\pgfusepath{stroke,fill}%
}%
\begin{pgfscope}%
\pgfsys@transformshift{0.340000in}{1.394909in}%
\pgfsys@useobject{currentmarker}{}%
\end{pgfscope}%
\end{pgfscope}%
\begin{pgfscope}%
\pgfsetbuttcap%
\pgfsetroundjoin%
\definecolor{currentfill}{rgb}{0.000000,0.000000,0.000000}%
\pgfsetfillcolor{currentfill}%
\pgfsetlinewidth{0.602250pt}%
\definecolor{currentstroke}{rgb}{0.000000,0.000000,0.000000}%
\pgfsetstrokecolor{currentstroke}%
\pgfsetdash{}{0pt}%
\pgfsys@defobject{currentmarker}{\pgfqpoint{0.000000in}{0.000000in}}{\pgfqpoint{0.027778in}{0.000000in}}{%
\pgfpathmoveto{\pgfqpoint{0.000000in}{0.000000in}}%
\pgfpathlineto{\pgfqpoint{0.027778in}{0.000000in}}%
\pgfusepath{stroke,fill}%
}%
\begin{pgfscope}%
\pgfsys@transformshift{0.340000in}{1.546182in}%
\pgfsys@useobject{currentmarker}{}%
\end{pgfscope}%
\end{pgfscope}%
\begin{pgfscope}%
\pgftext[x=0.144014in,y=0.960000in,,bottom,rotate=90.000000]{\rmfamily\fontsize{7.400000}{8.880000}\selectfont \(\displaystyle 99.9^{th}\) \%-tile MLU}%
\end{pgfscope}%
\begin{pgfscope}%
\pgfpathrectangle{\pgfqpoint{0.340000in}{0.336000in}}{\pgfqpoint{0.992800in}{1.248000in}}%
\pgfusepath{clip}%
\pgfsetbuttcap%
\pgfsetmiterjoin%
\definecolor{currentfill}{rgb}{0.900000,0.100000,0.100000}%
\pgfsetfillcolor{currentfill}%
\pgfsetlinewidth{0.000000pt}%
\definecolor{currentstroke}{rgb}{0.000000,0.000000,0.000000}%
\pgfsetstrokecolor{currentstroke}%
\pgfsetstrokeopacity{0.000000}%
\pgfsetdash{}{0pt}%
\pgfpathmoveto{\pgfqpoint{0.385127in}{-1.554909in}}%
\pgfpathlineto{\pgfqpoint{0.428106in}{-1.554909in}}%
\pgfpathlineto{\pgfqpoint{0.428106in}{0.770891in}}%
\pgfpathlineto{\pgfqpoint{0.385127in}{0.770891in}}%
\pgfpathclose%
\pgfusepath{fill}%
\end{pgfscope}%
\begin{pgfscope}%
\pgfpathrectangle{\pgfqpoint{0.340000in}{0.336000in}}{\pgfqpoint{0.992800in}{1.248000in}}%
\pgfusepath{clip}%
\pgfsetbuttcap%
\pgfsetmiterjoin%
\definecolor{currentfill}{rgb}{0.900000,0.100000,0.100000}%
\pgfsetfillcolor{currentfill}%
\pgfsetlinewidth{0.000000pt}%
\definecolor{currentstroke}{rgb}{0.000000,0.000000,0.000000}%
\pgfsetstrokecolor{currentstroke}%
\pgfsetstrokeopacity{0.000000}%
\pgfsetdash{}{0pt}%
\pgfpathmoveto{\pgfqpoint{0.630718in}{-1.554909in}}%
\pgfpathlineto{\pgfqpoint{0.673696in}{-1.554909in}}%
\pgfpathlineto{\pgfqpoint{0.673696in}{0.780408in}}%
\pgfpathlineto{\pgfqpoint{0.630718in}{0.780408in}}%
\pgfpathclose%
\pgfusepath{fill}%
\end{pgfscope}%
\begin{pgfscope}%
\pgfpathrectangle{\pgfqpoint{0.340000in}{0.336000in}}{\pgfqpoint{0.992800in}{1.248000in}}%
\pgfusepath{clip}%
\pgfsetbuttcap%
\pgfsetmiterjoin%
\definecolor{currentfill}{rgb}{0.900000,0.100000,0.100000}%
\pgfsetfillcolor{currentfill}%
\pgfsetlinewidth{0.000000pt}%
\definecolor{currentstroke}{rgb}{0.000000,0.000000,0.000000}%
\pgfsetstrokecolor{currentstroke}%
\pgfsetstrokeopacity{0.000000}%
\pgfsetdash{}{0pt}%
\pgfpathmoveto{\pgfqpoint{0.876308in}{-1.554909in}}%
\pgfpathlineto{\pgfqpoint{0.919287in}{-1.554909in}}%
\pgfpathlineto{\pgfqpoint{0.919287in}{0.882246in}}%
\pgfpathlineto{\pgfqpoint{0.876308in}{0.882246in}}%
\pgfpathclose%
\pgfusepath{fill}%
\end{pgfscope}%
\begin{pgfscope}%
\pgfpathrectangle{\pgfqpoint{0.340000in}{0.336000in}}{\pgfqpoint{0.992800in}{1.248000in}}%
\pgfusepath{clip}%
\pgfsetbuttcap%
\pgfsetmiterjoin%
\definecolor{currentfill}{rgb}{0.900000,0.100000,0.100000}%
\pgfsetfillcolor{currentfill}%
\pgfsetlinewidth{0.000000pt}%
\definecolor{currentstroke}{rgb}{0.000000,0.000000,0.000000}%
\pgfsetstrokecolor{currentstroke}%
\pgfsetstrokeopacity{0.000000}%
\pgfsetdash{}{0pt}%
\pgfpathmoveto{\pgfqpoint{1.121899in}{-1.554909in}}%
\pgfpathlineto{\pgfqpoint{1.164877in}{-1.554909in}}%
\pgfpathlineto{\pgfqpoint{1.164877in}{0.904407in}}%
\pgfpathlineto{\pgfqpoint{1.121899in}{0.904407in}}%
\pgfpathclose%
\pgfusepath{fill}%
\end{pgfscope}%
\begin{pgfscope}%
\pgfpathrectangle{\pgfqpoint{0.340000in}{0.336000in}}{\pgfqpoint{0.992800in}{1.248000in}}%
\pgfusepath{clip}%
\pgfsetbuttcap%
\pgfsetmiterjoin%
\definecolor{currentfill}{rgb}{0.300000,0.800000,0.200000}%
\pgfsetfillcolor{currentfill}%
\pgfsetlinewidth{0.000000pt}%
\definecolor{currentstroke}{rgb}{0.000000,0.000000,0.000000}%
\pgfsetstrokecolor{currentstroke}%
\pgfsetstrokeopacity{0.000000}%
\pgfsetdash{}{0pt}%
\pgfpathmoveto{\pgfqpoint{0.446525in}{-1.554909in}}%
\pgfpathlineto{\pgfqpoint{0.489503in}{-1.554909in}}%
\pgfpathlineto{\pgfqpoint{0.489503in}{0.867799in}}%
\pgfpathlineto{\pgfqpoint{0.446525in}{0.867799in}}%
\pgfpathclose%
\pgfusepath{fill}%
\end{pgfscope}%
\begin{pgfscope}%
\pgfpathrectangle{\pgfqpoint{0.340000in}{0.336000in}}{\pgfqpoint{0.992800in}{1.248000in}}%
\pgfusepath{clip}%
\pgfsetbuttcap%
\pgfsetmiterjoin%
\definecolor{currentfill}{rgb}{0.300000,0.800000,0.200000}%
\pgfsetfillcolor{currentfill}%
\pgfsetlinewidth{0.000000pt}%
\definecolor{currentstroke}{rgb}{0.000000,0.000000,0.000000}%
\pgfsetstrokecolor{currentstroke}%
\pgfsetstrokeopacity{0.000000}%
\pgfsetdash{}{0pt}%
\pgfpathmoveto{\pgfqpoint{0.692116in}{-1.554909in}}%
\pgfpathlineto{\pgfqpoint{0.735094in}{-1.554909in}}%
\pgfpathlineto{\pgfqpoint{0.735094in}{0.852634in}}%
\pgfpathlineto{\pgfqpoint{0.692116in}{0.852634in}}%
\pgfpathclose%
\pgfusepath{fill}%
\end{pgfscope}%
\begin{pgfscope}%
\pgfpathrectangle{\pgfqpoint{0.340000in}{0.336000in}}{\pgfqpoint{0.992800in}{1.248000in}}%
\pgfusepath{clip}%
\pgfsetbuttcap%
\pgfsetmiterjoin%
\definecolor{currentfill}{rgb}{0.300000,0.800000,0.200000}%
\pgfsetfillcolor{currentfill}%
\pgfsetlinewidth{0.000000pt}%
\definecolor{currentstroke}{rgb}{0.000000,0.000000,0.000000}%
\pgfsetstrokecolor{currentstroke}%
\pgfsetstrokeopacity{0.000000}%
\pgfsetdash{}{0pt}%
\pgfpathmoveto{\pgfqpoint{0.937706in}{-1.554909in}}%
\pgfpathlineto{\pgfqpoint{0.980684in}{-1.554909in}}%
\pgfpathlineto{\pgfqpoint{0.980684in}{0.882246in}}%
\pgfpathlineto{\pgfqpoint{0.937706in}{0.882246in}}%
\pgfpathclose%
\pgfusepath{fill}%
\end{pgfscope}%
\begin{pgfscope}%
\pgfpathrectangle{\pgfqpoint{0.340000in}{0.336000in}}{\pgfqpoint{0.992800in}{1.248000in}}%
\pgfusepath{clip}%
\pgfsetbuttcap%
\pgfsetmiterjoin%
\definecolor{currentfill}{rgb}{0.300000,0.800000,0.200000}%
\pgfsetfillcolor{currentfill}%
\pgfsetlinewidth{0.000000pt}%
\definecolor{currentstroke}{rgb}{0.000000,0.000000,0.000000}%
\pgfsetstrokecolor{currentstroke}%
\pgfsetstrokeopacity{0.000000}%
\pgfsetdash{}{0pt}%
\pgfpathmoveto{\pgfqpoint{1.183297in}{-1.554909in}}%
\pgfpathlineto{\pgfqpoint{1.226275in}{-1.554909in}}%
\pgfpathlineto{\pgfqpoint{1.226275in}{0.904407in}}%
\pgfpathlineto{\pgfqpoint{1.183297in}{0.904407in}}%
\pgfpathclose%
\pgfusepath{fill}%
\end{pgfscope}%
\begin{pgfscope}%
\pgfpathrectangle{\pgfqpoint{0.340000in}{0.336000in}}{\pgfqpoint{0.992800in}{1.248000in}}%
\pgfusepath{clip}%
\pgfsetbuttcap%
\pgfsetmiterjoin%
\definecolor{currentfill}{rgb}{0.050000,0.050000,0.700000}%
\pgfsetfillcolor{currentfill}%
\pgfsetlinewidth{0.000000pt}%
\definecolor{currentstroke}{rgb}{0.000000,0.000000,0.000000}%
\pgfsetstrokecolor{currentstroke}%
\pgfsetstrokeopacity{0.000000}%
\pgfsetdash{}{0pt}%
\pgfpathmoveto{\pgfqpoint{0.507923in}{-1.554909in}}%
\pgfpathlineto{\pgfqpoint{0.550901in}{-1.554909in}}%
\pgfpathlineto{\pgfqpoint{0.550901in}{1.482761in}}%
\pgfpathlineto{\pgfqpoint{0.507923in}{1.482761in}}%
\pgfpathclose%
\pgfusepath{fill}%
\end{pgfscope}%
\begin{pgfscope}%
\pgfpathrectangle{\pgfqpoint{0.340000in}{0.336000in}}{\pgfqpoint{0.992800in}{1.248000in}}%
\pgfusepath{clip}%
\pgfsetbuttcap%
\pgfsetmiterjoin%
\definecolor{currentfill}{rgb}{0.050000,0.050000,0.700000}%
\pgfsetfillcolor{currentfill}%
\pgfsetlinewidth{0.000000pt}%
\definecolor{currentstroke}{rgb}{0.000000,0.000000,0.000000}%
\pgfsetstrokecolor{currentstroke}%
\pgfsetstrokeopacity{0.000000}%
\pgfsetdash{}{0pt}%
\pgfpathmoveto{\pgfqpoint{0.753513in}{-1.554909in}}%
\pgfpathlineto{\pgfqpoint{0.796492in}{-1.554909in}}%
\pgfpathlineto{\pgfqpoint{0.796492in}{1.016084in}}%
\pgfpathlineto{\pgfqpoint{0.753513in}{1.016084in}}%
\pgfpathclose%
\pgfusepath{fill}%
\end{pgfscope}%
\begin{pgfscope}%
\pgfpathrectangle{\pgfqpoint{0.340000in}{0.336000in}}{\pgfqpoint{0.992800in}{1.248000in}}%
\pgfusepath{clip}%
\pgfsetbuttcap%
\pgfsetmiterjoin%
\definecolor{currentfill}{rgb}{0.050000,0.050000,0.700000}%
\pgfsetfillcolor{currentfill}%
\pgfsetlinewidth{0.000000pt}%
\definecolor{currentstroke}{rgb}{0.000000,0.000000,0.000000}%
\pgfsetstrokecolor{currentstroke}%
\pgfsetstrokeopacity{0.000000}%
\pgfsetdash{}{0pt}%
\pgfpathmoveto{\pgfqpoint{0.999104in}{-1.554909in}}%
\pgfpathlineto{\pgfqpoint{1.042082in}{-1.554909in}}%
\pgfpathlineto{\pgfqpoint{1.042082in}{0.891549in}}%
\pgfpathlineto{\pgfqpoint{0.999104in}{0.891549in}}%
\pgfpathclose%
\pgfusepath{fill}%
\end{pgfscope}%
\begin{pgfscope}%
\pgfpathrectangle{\pgfqpoint{0.340000in}{0.336000in}}{\pgfqpoint{0.992800in}{1.248000in}}%
\pgfusepath{clip}%
\pgfsetbuttcap%
\pgfsetmiterjoin%
\definecolor{currentfill}{rgb}{0.050000,0.050000,0.700000}%
\pgfsetfillcolor{currentfill}%
\pgfsetlinewidth{0.000000pt}%
\definecolor{currentstroke}{rgb}{0.000000,0.000000,0.000000}%
\pgfsetstrokecolor{currentstroke}%
\pgfsetstrokeopacity{0.000000}%
\pgfsetdash{}{0pt}%
\pgfpathmoveto{\pgfqpoint{1.244694in}{-1.554909in}}%
\pgfpathlineto{\pgfqpoint{1.287673in}{-1.554909in}}%
\pgfpathlineto{\pgfqpoint{1.287673in}{0.904748in}}%
\pgfpathlineto{\pgfqpoint{1.244694in}{0.904748in}}%
\pgfpathclose%
\pgfusepath{fill}%
\end{pgfscope}%
\begin{pgfscope}%
\pgfsetrectcap%
\pgfsetmiterjoin%
\pgfsetlinewidth{0.803000pt}%
\definecolor{currentstroke}{rgb}{0.000000,0.000000,0.000000}%
\pgfsetstrokecolor{currentstroke}%
\pgfsetdash{}{0pt}%
\pgfpathmoveto{\pgfqpoint{0.340000in}{0.336000in}}%
\pgfpathlineto{\pgfqpoint{0.340000in}{1.584000in}}%
\pgfusepath{stroke}%
\end{pgfscope}%
\begin{pgfscope}%
\pgfsetrectcap%
\pgfsetmiterjoin%
\pgfsetlinewidth{0.803000pt}%
\definecolor{currentstroke}{rgb}{0.000000,0.000000,0.000000}%
\pgfsetstrokecolor{currentstroke}%
\pgfsetdash{}{0pt}%
\pgfpathmoveto{\pgfqpoint{1.332800in}{0.336000in}}%
\pgfpathlineto{\pgfqpoint{1.332800in}{1.584000in}}%
\pgfusepath{stroke}%
\end{pgfscope}%
\begin{pgfscope}%
\pgfsetrectcap%
\pgfsetmiterjoin%
\pgfsetlinewidth{0.803000pt}%
\definecolor{currentstroke}{rgb}{0.000000,0.000000,0.000000}%
\pgfsetstrokecolor{currentstroke}%
\pgfsetdash{}{0pt}%
\pgfpathmoveto{\pgfqpoint{0.340000in}{0.336000in}}%
\pgfpathlineto{\pgfqpoint{1.332800in}{0.336000in}}%
\pgfusepath{stroke}%
\end{pgfscope}%
\begin{pgfscope}%
\pgfsetrectcap%
\pgfsetmiterjoin%
\pgfsetlinewidth{0.803000pt}%
\definecolor{currentstroke}{rgb}{0.000000,0.000000,0.000000}%
\pgfsetstrokecolor{currentstroke}%
\pgfsetdash{}{0pt}%
\pgfpathmoveto{\pgfqpoint{0.340000in}{1.584000in}}%
\pgfpathlineto{\pgfqpoint{1.332800in}{1.584000in}}%
\pgfusepath{stroke}%
\end{pgfscope}%
\begin{pgfscope}%
\pgfsetbuttcap%
\pgfsetmiterjoin%
\pgfsetlinewidth{0.000000pt}%
\definecolor{currentstroke}{rgb}{0.800000,0.800000,0.800000}%
\pgfsetstrokecolor{currentstroke}%
\pgfsetstrokeopacity{0.000000}%
\pgfsetdash{}{0pt}%
\pgfpathmoveto{\pgfqpoint{0.709363in}{1.144278in}}%
\pgfpathlineto{\pgfqpoint{1.270578in}{1.144278in}}%
\pgfpathquadraticcurveto{\pgfqpoint{1.288356in}{1.144278in}}{\pgfqpoint{1.288356in}{1.162055in}}%
\pgfpathlineto{\pgfqpoint{1.288356in}{1.521778in}}%
\pgfpathquadraticcurveto{\pgfqpoint{1.288356in}{1.539556in}}{\pgfqpoint{1.270578in}{1.539556in}}%
\pgfpathlineto{\pgfqpoint{0.709363in}{1.539556in}}%
\pgfpathquadraticcurveto{\pgfqpoint{0.691585in}{1.539556in}}{\pgfqpoint{0.691585in}{1.521778in}}%
\pgfpathlineto{\pgfqpoint{0.691585in}{1.162055in}}%
\pgfpathquadraticcurveto{\pgfqpoint{0.691585in}{1.144278in}}{\pgfqpoint{0.709363in}{1.144278in}}%
\pgfpathclose%
\pgfusepath{}%
\end{pgfscope}%
\begin{pgfscope}%
\pgfsetbuttcap%
\pgfsetmiterjoin%
\definecolor{currentfill}{rgb}{0.900000,0.100000,0.100000}%
\pgfsetfillcolor{currentfill}%
\pgfsetlinewidth{0.000000pt}%
\definecolor{currentstroke}{rgb}{0.000000,0.000000,0.000000}%
\pgfsetstrokecolor{currentstroke}%
\pgfsetstrokeopacity{0.000000}%
\pgfsetdash{}{0pt}%
\pgfpathmoveto{\pgfqpoint{0.727140in}{1.441778in}}%
\pgfpathlineto{\pgfqpoint{0.904918in}{1.441778in}}%
\pgfpathlineto{\pgfqpoint{0.904918in}{1.504000in}}%
\pgfpathlineto{\pgfqpoint{0.727140in}{1.504000in}}%
\pgfpathclose%
\pgfusepath{fill}%
\end{pgfscope}%
\begin{pgfscope}%
\pgftext[x=0.976029in,y=1.441778in,left,base]{\rmfamily\fontsize{6.400000}{7.680000}\selectfont 0s}%
\end{pgfscope}%
\begin{pgfscope}%
\pgfsetbuttcap%
\pgfsetmiterjoin%
\definecolor{currentfill}{rgb}{0.300000,0.800000,0.200000}%
\pgfsetfillcolor{currentfill}%
\pgfsetlinewidth{0.000000pt}%
\definecolor{currentstroke}{rgb}{0.000000,0.000000,0.000000}%
\pgfsetstrokecolor{currentstroke}%
\pgfsetstrokeopacity{0.000000}%
\pgfsetdash{}{0pt}%
\pgfpathmoveto{\pgfqpoint{0.727140in}{1.318907in}}%
\pgfpathlineto{\pgfqpoint{0.904918in}{1.318907in}}%
\pgfpathlineto{\pgfqpoint{0.904918in}{1.381130in}}%
\pgfpathlineto{\pgfqpoint{0.727140in}{1.381130in}}%
\pgfpathclose%
\pgfusepath{fill}%
\end{pgfscope}%
\begin{pgfscope}%
\pgftext[x=0.976029in,y=1.318907in,left,base]{\rmfamily\fontsize{6.400000}{7.680000}\selectfont 500\(\displaystyle \mu\)s}%
\end{pgfscope}%
\begin{pgfscope}%
\pgfsetbuttcap%
\pgfsetmiterjoin%
\definecolor{currentfill}{rgb}{0.050000,0.050000,0.700000}%
\pgfsetfillcolor{currentfill}%
\pgfsetlinewidth{0.000000pt}%
\definecolor{currentstroke}{rgb}{0.000000,0.000000,0.000000}%
\pgfsetstrokecolor{currentstroke}%
\pgfsetstrokeopacity{0.000000}%
\pgfsetdash{}{0pt}%
\pgfpathmoveto{\pgfqpoint{0.727140in}{1.196037in}}%
\pgfpathlineto{\pgfqpoint{0.904918in}{1.196037in}}%
\pgfpathlineto{\pgfqpoint{0.904918in}{1.258259in}}%
\pgfpathlineto{\pgfqpoint{0.727140in}{1.258259in}}%
\pgfpathclose%
\pgfusepath{fill}%
\end{pgfscope}%
\begin{pgfscope}%
\pgftext[x=0.976029in,y=1.196037in,left,base]{\rmfamily\fontsize{6.400000}{7.680000}\selectfont 500ms}%
\end{pgfscope}%
\end{pgfpicture}%
\makeatother%
\endgroup%

%% file: plot_data/facebook_data/fb_cluster_combined/ahc_nosense_reconfig_frequency_latency_combined.pgf
\begingroup%
\makeatletter%
\begin{pgfpicture}%
\pgfpathrectangle{\pgfpointorigin}{\pgfqpoint{1.360000in}{1.600000in}}%
\pgfusepath{use as bounding box, clip}%
\begin{pgfscope}%
\pgfsetbuttcap%
\pgfsetmiterjoin%
\definecolor{currentfill}{rgb}{1.000000,1.000000,1.000000}%
\pgfsetfillcolor{currentfill}%
\pgfsetlinewidth{0.000000pt}%
\definecolor{currentstroke}{rgb}{1.000000,1.000000,1.000000}%
\pgfsetstrokecolor{currentstroke}%
\pgfsetdash{}{0pt}%
\pgfpathmoveto{\pgfqpoint{0.000000in}{0.000000in}}%
\pgfpathlineto{\pgfqpoint{1.360000in}{0.000000in}}%
\pgfpathlineto{\pgfqpoint{1.360000in}{1.600000in}}%
\pgfpathlineto{\pgfqpoint{0.000000in}{1.600000in}}%
\pgfpathclose%
\pgfusepath{fill}%
\end{pgfscope}%
\begin{pgfscope}%
\pgfsetbuttcap%
\pgfsetmiterjoin%
\definecolor{currentfill}{rgb}{1.000000,1.000000,1.000000}%
\pgfsetfillcolor{currentfill}%
\pgfsetlinewidth{0.000000pt}%
\definecolor{currentstroke}{rgb}{0.000000,0.000000,0.000000}%
\pgfsetstrokecolor{currentstroke}%
\pgfsetstrokeopacity{0.000000}%
\pgfsetdash{}{0pt}%
\pgfpathmoveto{\pgfqpoint{0.326400in}{0.336000in}}%
\pgfpathlineto{\pgfqpoint{1.278400in}{0.336000in}}%
\pgfpathlineto{\pgfqpoint{1.278400in}{1.568000in}}%
\pgfpathlineto{\pgfqpoint{0.326400in}{1.568000in}}%
\pgfpathclose%
\pgfusepath{fill}%
\end{pgfscope}%
\begin{pgfscope}%
\pgfsetbuttcap%
\pgfsetroundjoin%
\definecolor{currentfill}{rgb}{0.000000,0.000000,0.000000}%
\pgfsetfillcolor{currentfill}%
\pgfsetlinewidth{0.803000pt}%
\definecolor{currentstroke}{rgb}{0.000000,0.000000,0.000000}%
\pgfsetstrokecolor{currentstroke}%
\pgfsetdash{}{0pt}%
\pgfsys@defobject{currentmarker}{\pgfqpoint{0.000000in}{-0.048611in}}{\pgfqpoint{0.000000in}{0.000000in}}{%
\pgfpathmoveto{\pgfqpoint{0.000000in}{0.000000in}}%
\pgfpathlineto{\pgfqpoint{0.000000in}{-0.048611in}}%
\pgfusepath{stroke,fill}%
}%
\begin{pgfscope}%
\pgfsys@transformshift{0.326400in}{0.336000in}%
\pgfsys@useobject{currentmarker}{}%
\end{pgfscope}%
\end{pgfscope}%
\begin{pgfscope}%
\pgftext[x=0.326400in,y=0.238778in,,top]{\rmfamily\fontsize{6.500000}{7.800000}\selectfont \(\displaystyle 0\)}%
\end{pgfscope}%
\begin{pgfscope}%
\pgfsetbuttcap%
\pgfsetroundjoin%
\definecolor{currentfill}{rgb}{0.000000,0.000000,0.000000}%
\pgfsetfillcolor{currentfill}%
\pgfsetlinewidth{0.803000pt}%
\definecolor{currentstroke}{rgb}{0.000000,0.000000,0.000000}%
\pgfsetstrokecolor{currentstroke}%
\pgfsetdash{}{0pt}%
\pgfsys@defobject{currentmarker}{\pgfqpoint{0.000000in}{-0.048611in}}{\pgfqpoint{0.000000in}{0.000000in}}{%
\pgfpathmoveto{\pgfqpoint{0.000000in}{0.000000in}}%
\pgfpathlineto{\pgfqpoint{0.000000in}{-0.048611in}}%
\pgfusepath{stroke,fill}%
}%
\begin{pgfscope}%
\pgfsys@transformshift{0.802400in}{0.336000in}%
\pgfsys@useobject{currentmarker}{}%
\end{pgfscope}%
\end{pgfscope}%
\begin{pgfscope}%
\pgftext[x=0.802400in,y=0.238778in,,top]{\rmfamily\fontsize{6.500000}{7.800000}\selectfont \(\displaystyle 50\)}%
\end{pgfscope}%
\begin{pgfscope}%
\pgfsetbuttcap%
\pgfsetroundjoin%
\definecolor{currentfill}{rgb}{0.000000,0.000000,0.000000}%
\pgfsetfillcolor{currentfill}%
\pgfsetlinewidth{0.803000pt}%
\definecolor{currentstroke}{rgb}{0.000000,0.000000,0.000000}%
\pgfsetstrokecolor{currentstroke}%
\pgfsetdash{}{0pt}%
\pgfsys@defobject{currentmarker}{\pgfqpoint{0.000000in}{-0.048611in}}{\pgfqpoint{0.000000in}{0.000000in}}{%
\pgfpathmoveto{\pgfqpoint{0.000000in}{0.000000in}}%
\pgfpathlineto{\pgfqpoint{0.000000in}{-0.048611in}}%
\pgfusepath{stroke,fill}%
}%
\begin{pgfscope}%
\pgfsys@transformshift{1.278400in}{0.336000in}%
\pgfsys@useobject{currentmarker}{}%
\end{pgfscope}%
\end{pgfscope}%
\begin{pgfscope}%
\pgftext[x=1.278400in,y=0.238778in,,top]{\rmfamily\fontsize{6.500000}{7.800000}\selectfont \(\displaystyle 100\)}%
\end{pgfscope}%
\begin{pgfscope}%
\pgftext[x=0.802400in,y=0.150815in,,top]{\rmfamily\fontsize{7.400000}{8.880000}\selectfont Percentile}%
\end{pgfscope}%
\begin{pgfscope}%
\pgfpathrectangle{\pgfqpoint{0.326400in}{0.336000in}}{\pgfqpoint{0.952000in}{1.232000in}}%
\pgfusepath{clip}%
\pgfsetbuttcap%
\pgfsetroundjoin%
\pgfsetlinewidth{0.702625pt}%
\definecolor{currentstroke}{rgb}{0.690196,0.690196,0.690196}%
\pgfsetstrokecolor{currentstroke}%
\pgfsetdash{{4.480000pt}{1.120000pt}{0.700000pt}{1.120000pt}}{0.000000pt}%
\pgfpathmoveto{\pgfqpoint{0.326400in}{0.500267in}}%
\pgfpathlineto{\pgfqpoint{1.278400in}{0.500267in}}%
\pgfusepath{stroke}%
\end{pgfscope}%
\begin{pgfscope}%
\pgfsetbuttcap%
\pgfsetroundjoin%
\definecolor{currentfill}{rgb}{0.000000,0.000000,0.000000}%
\pgfsetfillcolor{currentfill}%
\pgfsetlinewidth{0.803000pt}%
\definecolor{currentstroke}{rgb}{0.000000,0.000000,0.000000}%
\pgfsetstrokecolor{currentstroke}%
\pgfsetdash{}{0pt}%
\pgfsys@defobject{currentmarker}{\pgfqpoint{0.000000in}{0.000000in}}{\pgfqpoint{0.048611in}{0.000000in}}{%
\pgfpathmoveto{\pgfqpoint{0.000000in}{0.000000in}}%
\pgfpathlineto{\pgfqpoint{0.048611in}{0.000000in}}%
\pgfusepath{stroke,fill}%
}%
\begin{pgfscope}%
\pgfsys@transformshift{0.326400in}{0.500267in}%
\pgfsys@useobject{currentmarker}{}%
\end{pgfscope}%
\end{pgfscope}%
\begin{pgfscope}%
\pgftext[x=0.144303in,y=0.471331in,left,base]{\rmfamily\fontsize{6.500000}{7.800000}\selectfont \(\displaystyle 1.2\)}%
\end{pgfscope}%
\begin{pgfscope}%
\pgfpathrectangle{\pgfqpoint{0.326400in}{0.336000in}}{\pgfqpoint{0.952000in}{1.232000in}}%
\pgfusepath{clip}%
\pgfsetbuttcap%
\pgfsetroundjoin%
\pgfsetlinewidth{0.702625pt}%
\definecolor{currentstroke}{rgb}{0.690196,0.690196,0.690196}%
\pgfsetstrokecolor{currentstroke}%
\pgfsetdash{{4.480000pt}{1.120000pt}{0.700000pt}{1.120000pt}}{0.000000pt}%
\pgfpathmoveto{\pgfqpoint{0.326400in}{0.828800in}}%
\pgfpathlineto{\pgfqpoint{1.278400in}{0.828800in}}%
\pgfusepath{stroke}%
\end{pgfscope}%
\begin{pgfscope}%
\pgfsetbuttcap%
\pgfsetroundjoin%
\definecolor{currentfill}{rgb}{0.000000,0.000000,0.000000}%
\pgfsetfillcolor{currentfill}%
\pgfsetlinewidth{0.803000pt}%
\definecolor{currentstroke}{rgb}{0.000000,0.000000,0.000000}%
\pgfsetstrokecolor{currentstroke}%
\pgfsetdash{}{0pt}%
\pgfsys@defobject{currentmarker}{\pgfqpoint{0.000000in}{0.000000in}}{\pgfqpoint{0.048611in}{0.000000in}}{%
\pgfpathmoveto{\pgfqpoint{0.000000in}{0.000000in}}%
\pgfpathlineto{\pgfqpoint{0.048611in}{0.000000in}}%
\pgfusepath{stroke,fill}%
}%
\begin{pgfscope}%
\pgfsys@transformshift{0.326400in}{0.828800in}%
\pgfsys@useobject{currentmarker}{}%
\end{pgfscope}%
\end{pgfscope}%
\begin{pgfscope}%
\pgftext[x=0.144303in,y=0.799865in,left,base]{\rmfamily\fontsize{6.500000}{7.800000}\selectfont \(\displaystyle 1.4\)}%
\end{pgfscope}%
\begin{pgfscope}%
\pgfpathrectangle{\pgfqpoint{0.326400in}{0.336000in}}{\pgfqpoint{0.952000in}{1.232000in}}%
\pgfusepath{clip}%
\pgfsetbuttcap%
\pgfsetroundjoin%
\pgfsetlinewidth{0.702625pt}%
\definecolor{currentstroke}{rgb}{0.690196,0.690196,0.690196}%
\pgfsetstrokecolor{currentstroke}%
\pgfsetdash{{4.480000pt}{1.120000pt}{0.700000pt}{1.120000pt}}{0.000000pt}%
\pgfpathmoveto{\pgfqpoint{0.326400in}{1.157333in}}%
\pgfpathlineto{\pgfqpoint{1.278400in}{1.157333in}}%
\pgfusepath{stroke}%
\end{pgfscope}%
\begin{pgfscope}%
\pgfsetbuttcap%
\pgfsetroundjoin%
\definecolor{currentfill}{rgb}{0.000000,0.000000,0.000000}%
\pgfsetfillcolor{currentfill}%
\pgfsetlinewidth{0.803000pt}%
\definecolor{currentstroke}{rgb}{0.000000,0.000000,0.000000}%
\pgfsetstrokecolor{currentstroke}%
\pgfsetdash{}{0pt}%
\pgfsys@defobject{currentmarker}{\pgfqpoint{0.000000in}{0.000000in}}{\pgfqpoint{0.048611in}{0.000000in}}{%
\pgfpathmoveto{\pgfqpoint{0.000000in}{0.000000in}}%
\pgfpathlineto{\pgfqpoint{0.048611in}{0.000000in}}%
\pgfusepath{stroke,fill}%
}%
\begin{pgfscope}%
\pgfsys@transformshift{0.326400in}{1.157333in}%
\pgfsys@useobject{currentmarker}{}%
\end{pgfscope}%
\end{pgfscope}%
\begin{pgfscope}%
\pgftext[x=0.144303in,y=1.128398in,left,base]{\rmfamily\fontsize{6.500000}{7.800000}\selectfont \(\displaystyle 1.6\)}%
\end{pgfscope}%
\begin{pgfscope}%
\pgfpathrectangle{\pgfqpoint{0.326400in}{0.336000in}}{\pgfqpoint{0.952000in}{1.232000in}}%
\pgfusepath{clip}%
\pgfsetbuttcap%
\pgfsetroundjoin%
\pgfsetlinewidth{0.702625pt}%
\definecolor{currentstroke}{rgb}{0.690196,0.690196,0.690196}%
\pgfsetstrokecolor{currentstroke}%
\pgfsetdash{{4.480000pt}{1.120000pt}{0.700000pt}{1.120000pt}}{0.000000pt}%
\pgfpathmoveto{\pgfqpoint{0.326400in}{1.485867in}}%
\pgfpathlineto{\pgfqpoint{1.278400in}{1.485867in}}%
\pgfusepath{stroke}%
\end{pgfscope}%
\begin{pgfscope}%
\pgfsetbuttcap%
\pgfsetroundjoin%
\definecolor{currentfill}{rgb}{0.000000,0.000000,0.000000}%
\pgfsetfillcolor{currentfill}%
\pgfsetlinewidth{0.803000pt}%
\definecolor{currentstroke}{rgb}{0.000000,0.000000,0.000000}%
\pgfsetstrokecolor{currentstroke}%
\pgfsetdash{}{0pt}%
\pgfsys@defobject{currentmarker}{\pgfqpoint{0.000000in}{0.000000in}}{\pgfqpoint{0.048611in}{0.000000in}}{%
\pgfpathmoveto{\pgfqpoint{0.000000in}{0.000000in}}%
\pgfpathlineto{\pgfqpoint{0.048611in}{0.000000in}}%
\pgfusepath{stroke,fill}%
}%
\begin{pgfscope}%
\pgfsys@transformshift{0.326400in}{1.485867in}%
\pgfsys@useobject{currentmarker}{}%
\end{pgfscope}%
\end{pgfscope}%
\begin{pgfscope}%
\pgftext[x=0.144303in,y=1.456931in,left,base]{\rmfamily\fontsize{6.500000}{7.800000}\selectfont \(\displaystyle 1.8\)}%
\end{pgfscope}%
\begin{pgfscope}%
\pgfsetbuttcap%
\pgfsetroundjoin%
\definecolor{currentfill}{rgb}{0.000000,0.000000,0.000000}%
\pgfsetfillcolor{currentfill}%
\pgfsetlinewidth{0.602250pt}%
\definecolor{currentstroke}{rgb}{0.000000,0.000000,0.000000}%
\pgfsetstrokecolor{currentstroke}%
\pgfsetdash{}{0pt}%
\pgfsys@defobject{currentmarker}{\pgfqpoint{0.000000in}{0.000000in}}{\pgfqpoint{0.027778in}{0.000000in}}{%
\pgfpathmoveto{\pgfqpoint{0.000000in}{0.000000in}}%
\pgfpathlineto{\pgfqpoint{0.027778in}{0.000000in}}%
\pgfusepath{stroke,fill}%
}%
\begin{pgfscope}%
\pgfsys@transformshift{0.326400in}{0.418133in}%
\pgfsys@useobject{currentmarker}{}%
\end{pgfscope}%
\end{pgfscope}%
\begin{pgfscope}%
\pgfsetbuttcap%
\pgfsetroundjoin%
\definecolor{currentfill}{rgb}{0.000000,0.000000,0.000000}%
\pgfsetfillcolor{currentfill}%
\pgfsetlinewidth{0.602250pt}%
\definecolor{currentstroke}{rgb}{0.000000,0.000000,0.000000}%
\pgfsetstrokecolor{currentstroke}%
\pgfsetdash{}{0pt}%
\pgfsys@defobject{currentmarker}{\pgfqpoint{0.000000in}{0.000000in}}{\pgfqpoint{0.027778in}{0.000000in}}{%
\pgfpathmoveto{\pgfqpoint{0.000000in}{0.000000in}}%
\pgfpathlineto{\pgfqpoint{0.027778in}{0.000000in}}%
\pgfusepath{stroke,fill}%
}%
\begin{pgfscope}%
\pgfsys@transformshift{0.326400in}{0.582400in}%
\pgfsys@useobject{currentmarker}{}%
\end{pgfscope}%
\end{pgfscope}%
\begin{pgfscope}%
\pgfsetbuttcap%
\pgfsetroundjoin%
\definecolor{currentfill}{rgb}{0.000000,0.000000,0.000000}%
\pgfsetfillcolor{currentfill}%
\pgfsetlinewidth{0.602250pt}%
\definecolor{currentstroke}{rgb}{0.000000,0.000000,0.000000}%
\pgfsetstrokecolor{currentstroke}%
\pgfsetdash{}{0pt}%
\pgfsys@defobject{currentmarker}{\pgfqpoint{0.000000in}{0.000000in}}{\pgfqpoint{0.027778in}{0.000000in}}{%
\pgfpathmoveto{\pgfqpoint{0.000000in}{0.000000in}}%
\pgfpathlineto{\pgfqpoint{0.027778in}{0.000000in}}%
\pgfusepath{stroke,fill}%
}%
\begin{pgfscope}%
\pgfsys@transformshift{0.326400in}{0.664533in}%
\pgfsys@useobject{currentmarker}{}%
\end{pgfscope}%
\end{pgfscope}%
\begin{pgfscope}%
\pgfsetbuttcap%
\pgfsetroundjoin%
\definecolor{currentfill}{rgb}{0.000000,0.000000,0.000000}%
\pgfsetfillcolor{currentfill}%
\pgfsetlinewidth{0.602250pt}%
\definecolor{currentstroke}{rgb}{0.000000,0.000000,0.000000}%
\pgfsetstrokecolor{currentstroke}%
\pgfsetdash{}{0pt}%
\pgfsys@defobject{currentmarker}{\pgfqpoint{0.000000in}{0.000000in}}{\pgfqpoint{0.027778in}{0.000000in}}{%
\pgfpathmoveto{\pgfqpoint{0.000000in}{0.000000in}}%
\pgfpathlineto{\pgfqpoint{0.027778in}{0.000000in}}%
\pgfusepath{stroke,fill}%
}%
\begin{pgfscope}%
\pgfsys@transformshift{0.326400in}{0.746667in}%
\pgfsys@useobject{currentmarker}{}%
\end{pgfscope}%
\end{pgfscope}%
\begin{pgfscope}%
\pgfsetbuttcap%
\pgfsetroundjoin%
\definecolor{currentfill}{rgb}{0.000000,0.000000,0.000000}%
\pgfsetfillcolor{currentfill}%
\pgfsetlinewidth{0.602250pt}%
\definecolor{currentstroke}{rgb}{0.000000,0.000000,0.000000}%
\pgfsetstrokecolor{currentstroke}%
\pgfsetdash{}{0pt}%
\pgfsys@defobject{currentmarker}{\pgfqpoint{0.000000in}{0.000000in}}{\pgfqpoint{0.027778in}{0.000000in}}{%
\pgfpathmoveto{\pgfqpoint{0.000000in}{0.000000in}}%
\pgfpathlineto{\pgfqpoint{0.027778in}{0.000000in}}%
\pgfusepath{stroke,fill}%
}%
\begin{pgfscope}%
\pgfsys@transformshift{0.326400in}{0.910933in}%
\pgfsys@useobject{currentmarker}{}%
\end{pgfscope}%
\end{pgfscope}%
\begin{pgfscope}%
\pgfsetbuttcap%
\pgfsetroundjoin%
\definecolor{currentfill}{rgb}{0.000000,0.000000,0.000000}%
\pgfsetfillcolor{currentfill}%
\pgfsetlinewidth{0.602250pt}%
\definecolor{currentstroke}{rgb}{0.000000,0.000000,0.000000}%
\pgfsetstrokecolor{currentstroke}%
\pgfsetdash{}{0pt}%
\pgfsys@defobject{currentmarker}{\pgfqpoint{0.000000in}{0.000000in}}{\pgfqpoint{0.027778in}{0.000000in}}{%
\pgfpathmoveto{\pgfqpoint{0.000000in}{0.000000in}}%
\pgfpathlineto{\pgfqpoint{0.027778in}{0.000000in}}%
\pgfusepath{stroke,fill}%
}%
\begin{pgfscope}%
\pgfsys@transformshift{0.326400in}{0.993067in}%
\pgfsys@useobject{currentmarker}{}%
\end{pgfscope}%
\end{pgfscope}%
\begin{pgfscope}%
\pgfsetbuttcap%
\pgfsetroundjoin%
\definecolor{currentfill}{rgb}{0.000000,0.000000,0.000000}%
\pgfsetfillcolor{currentfill}%
\pgfsetlinewidth{0.602250pt}%
\definecolor{currentstroke}{rgb}{0.000000,0.000000,0.000000}%
\pgfsetstrokecolor{currentstroke}%
\pgfsetdash{}{0pt}%
\pgfsys@defobject{currentmarker}{\pgfqpoint{0.000000in}{0.000000in}}{\pgfqpoint{0.027778in}{0.000000in}}{%
\pgfpathmoveto{\pgfqpoint{0.000000in}{0.000000in}}%
\pgfpathlineto{\pgfqpoint{0.027778in}{0.000000in}}%
\pgfusepath{stroke,fill}%
}%
\begin{pgfscope}%
\pgfsys@transformshift{0.326400in}{1.075200in}%
\pgfsys@useobject{currentmarker}{}%
\end{pgfscope}%
\end{pgfscope}%
\begin{pgfscope}%
\pgfsetbuttcap%
\pgfsetroundjoin%
\definecolor{currentfill}{rgb}{0.000000,0.000000,0.000000}%
\pgfsetfillcolor{currentfill}%
\pgfsetlinewidth{0.602250pt}%
\definecolor{currentstroke}{rgb}{0.000000,0.000000,0.000000}%
\pgfsetstrokecolor{currentstroke}%
\pgfsetdash{}{0pt}%
\pgfsys@defobject{currentmarker}{\pgfqpoint{0.000000in}{0.000000in}}{\pgfqpoint{0.027778in}{0.000000in}}{%
\pgfpathmoveto{\pgfqpoint{0.000000in}{0.000000in}}%
\pgfpathlineto{\pgfqpoint{0.027778in}{0.000000in}}%
\pgfusepath{stroke,fill}%
}%
\begin{pgfscope}%
\pgfsys@transformshift{0.326400in}{1.239467in}%
\pgfsys@useobject{currentmarker}{}%
\end{pgfscope}%
\end{pgfscope}%
\begin{pgfscope}%
\pgfsetbuttcap%
\pgfsetroundjoin%
\definecolor{currentfill}{rgb}{0.000000,0.000000,0.000000}%
\pgfsetfillcolor{currentfill}%
\pgfsetlinewidth{0.602250pt}%
\definecolor{currentstroke}{rgb}{0.000000,0.000000,0.000000}%
\pgfsetstrokecolor{currentstroke}%
\pgfsetdash{}{0pt}%
\pgfsys@defobject{currentmarker}{\pgfqpoint{0.000000in}{0.000000in}}{\pgfqpoint{0.027778in}{0.000000in}}{%
\pgfpathmoveto{\pgfqpoint{0.000000in}{0.000000in}}%
\pgfpathlineto{\pgfqpoint{0.027778in}{0.000000in}}%
\pgfusepath{stroke,fill}%
}%
\begin{pgfscope}%
\pgfsys@transformshift{0.326400in}{1.321600in}%
\pgfsys@useobject{currentmarker}{}%
\end{pgfscope}%
\end{pgfscope}%
\begin{pgfscope}%
\pgfsetbuttcap%
\pgfsetroundjoin%
\definecolor{currentfill}{rgb}{0.000000,0.000000,0.000000}%
\pgfsetfillcolor{currentfill}%
\pgfsetlinewidth{0.602250pt}%
\definecolor{currentstroke}{rgb}{0.000000,0.000000,0.000000}%
\pgfsetstrokecolor{currentstroke}%
\pgfsetdash{}{0pt}%
\pgfsys@defobject{currentmarker}{\pgfqpoint{0.000000in}{0.000000in}}{\pgfqpoint{0.027778in}{0.000000in}}{%
\pgfpathmoveto{\pgfqpoint{0.000000in}{0.000000in}}%
\pgfpathlineto{\pgfqpoint{0.027778in}{0.000000in}}%
\pgfusepath{stroke,fill}%
}%
\begin{pgfscope}%
\pgfsys@transformshift{0.326400in}{1.403733in}%
\pgfsys@useobject{currentmarker}{}%
\end{pgfscope}%
\end{pgfscope}%
\begin{pgfscope}%
\pgfsetbuttcap%
\pgfsetroundjoin%
\definecolor{currentfill}{rgb}{0.000000,0.000000,0.000000}%
\pgfsetfillcolor{currentfill}%
\pgfsetlinewidth{0.602250pt}%
\definecolor{currentstroke}{rgb}{0.000000,0.000000,0.000000}%
\pgfsetstrokecolor{currentstroke}%
\pgfsetdash{}{0pt}%
\pgfsys@defobject{currentmarker}{\pgfqpoint{0.000000in}{0.000000in}}{\pgfqpoint{0.027778in}{0.000000in}}{%
\pgfpathmoveto{\pgfqpoint{0.000000in}{0.000000in}}%
\pgfpathlineto{\pgfqpoint{0.027778in}{0.000000in}}%
\pgfusepath{stroke,fill}%
}%
\begin{pgfscope}%
\pgfsys@transformshift{0.326400in}{1.568000in}%
\pgfsys@useobject{currentmarker}{}%
\end{pgfscope}%
\end{pgfscope}%
\begin{pgfscope}%
\pgftext[x=0.130414in,y=0.952000in,,bottom,rotate=90.000000]{\rmfamily\fontsize{7.400000}{8.880000}\selectfont Average Hop Count}%
\end{pgfscope}%
\begin{pgfscope}%
\pgfpathrectangle{\pgfqpoint{0.326400in}{0.336000in}}{\pgfqpoint{0.952000in}{1.232000in}}%
\pgfusepath{clip}%
\pgfsetrectcap%
\pgfsetroundjoin%
\pgfsetlinewidth{1.204500pt}%
\definecolor{currentstroke}{rgb}{0.200000,0.200000,0.950000}%
\pgfsetstrokecolor{currentstroke}%
\pgfsetdash{}{0pt}%
\pgfpathmoveto{\pgfqpoint{0.326400in}{0.404385in}}%
\pgfpathlineto{\pgfqpoint{0.335920in}{0.488308in}}%
\pgfpathlineto{\pgfqpoint{0.345440in}{0.520498in}}%
\pgfpathlineto{\pgfqpoint{0.354960in}{0.533824in}}%
\pgfpathlineto{\pgfqpoint{0.364480in}{0.543110in}}%
\pgfpathlineto{\pgfqpoint{0.374000in}{0.550726in}}%
\pgfpathlineto{\pgfqpoint{0.383520in}{0.556422in}}%
\pgfpathlineto{\pgfqpoint{0.393040in}{0.560740in}}%
\pgfpathlineto{\pgfqpoint{0.402560in}{0.563920in}}%
\pgfpathlineto{\pgfqpoint{0.412080in}{0.566927in}}%
\pgfpathlineto{\pgfqpoint{0.421600in}{0.569809in}}%
\pgfpathlineto{\pgfqpoint{0.431120in}{0.572351in}}%
\pgfpathlineto{\pgfqpoint{0.440640in}{0.575036in}}%
\pgfpathlineto{\pgfqpoint{0.450160in}{0.578477in}}%
\pgfpathlineto{\pgfqpoint{0.459680in}{0.582725in}}%
\pgfpathlineto{\pgfqpoint{0.469200in}{0.587313in}}%
\pgfpathlineto{\pgfqpoint{0.478720in}{0.590812in}}%
\pgfpathlineto{\pgfqpoint{0.488240in}{0.593675in}}%
\pgfpathlineto{\pgfqpoint{0.497760in}{0.595678in}}%
\pgfpathlineto{\pgfqpoint{0.507280in}{0.597631in}}%
\pgfpathlineto{\pgfqpoint{0.516800in}{0.599415in}}%
\pgfpathlineto{\pgfqpoint{0.526320in}{0.601123in}}%
\pgfpathlineto{\pgfqpoint{0.535840in}{0.602791in}}%
\pgfpathlineto{\pgfqpoint{0.545360in}{0.604240in}}%
\pgfpathlineto{\pgfqpoint{0.554880in}{0.605523in}}%
\pgfpathlineto{\pgfqpoint{0.564400in}{0.606927in}}%
\pgfpathlineto{\pgfqpoint{0.573920in}{0.608218in}}%
\pgfpathlineto{\pgfqpoint{0.583440in}{0.609703in}}%
\pgfpathlineto{\pgfqpoint{0.592960in}{0.611250in}}%
\pgfpathlineto{\pgfqpoint{0.602480in}{0.612782in}}%
\pgfpathlineto{\pgfqpoint{0.612000in}{0.614436in}}%
\pgfpathlineto{\pgfqpoint{0.621520in}{0.616338in}}%
\pgfpathlineto{\pgfqpoint{0.631040in}{0.618333in}}%
\pgfpathlineto{\pgfqpoint{0.640560in}{0.620990in}}%
\pgfpathlineto{\pgfqpoint{0.650080in}{0.623981in}}%
\pgfpathlineto{\pgfqpoint{0.659600in}{0.627572in}}%
\pgfpathlineto{\pgfqpoint{0.669120in}{0.631667in}}%
\pgfpathlineto{\pgfqpoint{0.678640in}{0.636208in}}%
\pgfpathlineto{\pgfqpoint{0.688160in}{0.642360in}}%
\pgfpathlineto{\pgfqpoint{0.697680in}{0.651678in}}%
\pgfpathlineto{\pgfqpoint{0.707200in}{0.674565in}}%
\pgfpathlineto{\pgfqpoint{0.716720in}{0.680673in}}%
\pgfpathlineto{\pgfqpoint{0.726240in}{0.684684in}}%
\pgfpathlineto{\pgfqpoint{0.735760in}{0.687495in}}%
\pgfpathlineto{\pgfqpoint{0.745280in}{0.689617in}}%
\pgfpathlineto{\pgfqpoint{0.754800in}{0.691600in}}%
\pgfpathlineto{\pgfqpoint{0.764320in}{0.693464in}}%
\pgfpathlineto{\pgfqpoint{0.773840in}{0.695096in}}%
\pgfpathlineto{\pgfqpoint{0.783360in}{0.696544in}}%
\pgfpathlineto{\pgfqpoint{0.792880in}{0.697822in}}%
\pgfpathlineto{\pgfqpoint{0.802400in}{0.698968in}}%
\pgfpathlineto{\pgfqpoint{0.811920in}{0.699923in}}%
\pgfpathlineto{\pgfqpoint{0.821440in}{0.700938in}}%
\pgfpathlineto{\pgfqpoint{0.830960in}{0.701929in}}%
\pgfpathlineto{\pgfqpoint{0.840480in}{0.702901in}}%
\pgfpathlineto{\pgfqpoint{0.850000in}{0.703750in}}%
\pgfpathlineto{\pgfqpoint{0.859520in}{0.704563in}}%
\pgfpathlineto{\pgfqpoint{0.869040in}{0.705344in}}%
\pgfpathlineto{\pgfqpoint{0.878560in}{0.706122in}}%
\pgfpathlineto{\pgfqpoint{0.888080in}{0.706947in}}%
\pgfpathlineto{\pgfqpoint{0.897600in}{0.707604in}}%
\pgfpathlineto{\pgfqpoint{0.907120in}{0.708181in}}%
\pgfpathlineto{\pgfqpoint{0.916640in}{0.708745in}}%
\pgfpathlineto{\pgfqpoint{0.926160in}{0.709360in}}%
\pgfpathlineto{\pgfqpoint{0.935680in}{0.710014in}}%
\pgfpathlineto{\pgfqpoint{0.945200in}{0.710673in}}%
\pgfpathlineto{\pgfqpoint{0.954720in}{0.711337in}}%
\pgfpathlineto{\pgfqpoint{0.964240in}{0.711952in}}%
\pgfpathlineto{\pgfqpoint{0.973760in}{0.712633in}}%
\pgfpathlineto{\pgfqpoint{0.983280in}{0.713263in}}%
\pgfpathlineto{\pgfqpoint{0.992800in}{0.713987in}}%
\pgfpathlineto{\pgfqpoint{1.002320in}{0.714624in}}%
\pgfpathlineto{\pgfqpoint{1.011840in}{0.715241in}}%
\pgfpathlineto{\pgfqpoint{1.021360in}{0.715913in}}%
\pgfpathlineto{\pgfqpoint{1.030880in}{0.716654in}}%
\pgfpathlineto{\pgfqpoint{1.040400in}{0.717453in}}%
\pgfpathlineto{\pgfqpoint{1.049920in}{0.718133in}}%
\pgfpathlineto{\pgfqpoint{1.059440in}{0.719061in}}%
\pgfpathlineto{\pgfqpoint{1.068960in}{0.719910in}}%
\pgfpathlineto{\pgfqpoint{1.078480in}{0.720800in}}%
\pgfpathlineto{\pgfqpoint{1.088000in}{0.721683in}}%
\pgfpathlineto{\pgfqpoint{1.097520in}{0.722864in}}%
\pgfpathlineto{\pgfqpoint{1.107040in}{0.724167in}}%
\pgfpathlineto{\pgfqpoint{1.116560in}{0.726105in}}%
\pgfpathlineto{\pgfqpoint{1.126080in}{0.728861in}}%
\pgfpathlineto{\pgfqpoint{1.135600in}{0.733929in}}%
\pgfpathlineto{\pgfqpoint{1.145120in}{0.756283in}}%
\pgfpathlineto{\pgfqpoint{1.154640in}{0.769972in}}%
\pgfpathlineto{\pgfqpoint{1.164160in}{0.782139in}}%
\pgfpathlineto{\pgfqpoint{1.173680in}{0.789285in}}%
\pgfpathlineto{\pgfqpoint{1.183200in}{0.794190in}}%
\pgfpathlineto{\pgfqpoint{1.192720in}{0.797537in}}%
\pgfpathlineto{\pgfqpoint{1.202240in}{0.800105in}}%
\pgfpathlineto{\pgfqpoint{1.211760in}{0.802978in}}%
\pgfpathlineto{\pgfqpoint{1.221280in}{0.806011in}}%
\pgfpathlineto{\pgfqpoint{1.230800in}{0.808962in}}%
\pgfpathlineto{\pgfqpoint{1.240320in}{0.813262in}}%
\pgfpathlineto{\pgfqpoint{1.249840in}{0.823081in}}%
\pgfpathlineto{\pgfqpoint{1.259360in}{0.835440in}}%
\pgfpathlineto{\pgfqpoint{1.268880in}{0.853004in}}%
\pgfpathlineto{\pgfqpoint{1.278400in}{0.919841in}}%
\pgfusepath{stroke}%
\end{pgfscope}%
\begin{pgfscope}%
\pgfpathrectangle{\pgfqpoint{0.326400in}{0.336000in}}{\pgfqpoint{0.952000in}{1.232000in}}%
\pgfusepath{clip}%
\pgfsetrectcap%
\pgfsetroundjoin%
\pgfsetlinewidth{1.204500pt}%
\definecolor{currentstroke}{rgb}{0.900000,0.100000,0.100000}%
\pgfsetstrokecolor{currentstroke}%
\pgfsetdash{}{0pt}%
\pgfpathmoveto{\pgfqpoint{0.326400in}{0.494951in}}%
\pgfpathlineto{\pgfqpoint{0.335920in}{0.548426in}}%
\pgfpathlineto{\pgfqpoint{0.345440in}{0.555583in}}%
\pgfpathlineto{\pgfqpoint{0.354960in}{0.562136in}}%
\pgfpathlineto{\pgfqpoint{0.364480in}{0.569400in}}%
\pgfpathlineto{\pgfqpoint{0.374000in}{0.574959in}}%
\pgfpathlineto{\pgfqpoint{0.383520in}{0.579210in}}%
\pgfpathlineto{\pgfqpoint{0.393040in}{0.582940in}}%
\pgfpathlineto{\pgfqpoint{0.402560in}{0.585714in}}%
\pgfpathlineto{\pgfqpoint{0.412080in}{0.589042in}}%
\pgfpathlineto{\pgfqpoint{0.421600in}{0.591633in}}%
\pgfpathlineto{\pgfqpoint{0.431120in}{0.594363in}}%
\pgfpathlineto{\pgfqpoint{0.440640in}{0.597462in}}%
\pgfpathlineto{\pgfqpoint{0.450160in}{0.600763in}}%
\pgfpathlineto{\pgfqpoint{0.459680in}{0.604310in}}%
\pgfpathlineto{\pgfqpoint{0.469200in}{0.607730in}}%
\pgfpathlineto{\pgfqpoint{0.478720in}{0.610857in}}%
\pgfpathlineto{\pgfqpoint{0.488240in}{0.613657in}}%
\pgfpathlineto{\pgfqpoint{0.497760in}{0.615428in}}%
\pgfpathlineto{\pgfqpoint{0.507280in}{0.617271in}}%
\pgfpathlineto{\pgfqpoint{0.516800in}{0.618959in}}%
\pgfpathlineto{\pgfqpoint{0.526320in}{0.620554in}}%
\pgfpathlineto{\pgfqpoint{0.535840in}{0.622049in}}%
\pgfpathlineto{\pgfqpoint{0.545360in}{0.623552in}}%
\pgfpathlineto{\pgfqpoint{0.554880in}{0.624882in}}%
\pgfpathlineto{\pgfqpoint{0.564400in}{0.626233in}}%
\pgfpathlineto{\pgfqpoint{0.573920in}{0.627535in}}%
\pgfpathlineto{\pgfqpoint{0.583440in}{0.629040in}}%
\pgfpathlineto{\pgfqpoint{0.592960in}{0.630473in}}%
\pgfpathlineto{\pgfqpoint{0.602480in}{0.632246in}}%
\pgfpathlineto{\pgfqpoint{0.612000in}{0.633877in}}%
\pgfpathlineto{\pgfqpoint{0.621520in}{0.635835in}}%
\pgfpathlineto{\pgfqpoint{0.631040in}{0.638095in}}%
\pgfpathlineto{\pgfqpoint{0.640560in}{0.641260in}}%
\pgfpathlineto{\pgfqpoint{0.650080in}{0.644885in}}%
\pgfpathlineto{\pgfqpoint{0.659600in}{0.648847in}}%
\pgfpathlineto{\pgfqpoint{0.669120in}{0.652494in}}%
\pgfpathlineto{\pgfqpoint{0.678640in}{0.656492in}}%
\pgfpathlineto{\pgfqpoint{0.688160in}{0.662103in}}%
\pgfpathlineto{\pgfqpoint{0.697680in}{0.687782in}}%
\pgfpathlineto{\pgfqpoint{0.707200in}{0.697714in}}%
\pgfpathlineto{\pgfqpoint{0.716720in}{0.701758in}}%
\pgfpathlineto{\pgfqpoint{0.726240in}{0.704395in}}%
\pgfpathlineto{\pgfqpoint{0.735760in}{0.707099in}}%
\pgfpathlineto{\pgfqpoint{0.745280in}{0.708827in}}%
\pgfpathlineto{\pgfqpoint{0.754800in}{0.710756in}}%
\pgfpathlineto{\pgfqpoint{0.764320in}{0.712453in}}%
\pgfpathlineto{\pgfqpoint{0.773840in}{0.713821in}}%
\pgfpathlineto{\pgfqpoint{0.783360in}{0.715216in}}%
\pgfpathlineto{\pgfqpoint{0.792880in}{0.716353in}}%
\pgfpathlineto{\pgfqpoint{0.802400in}{0.717448in}}%
\pgfpathlineto{\pgfqpoint{0.811920in}{0.718475in}}%
\pgfpathlineto{\pgfqpoint{0.821440in}{0.719414in}}%
\pgfpathlineto{\pgfqpoint{0.830960in}{0.720395in}}%
\pgfpathlineto{\pgfqpoint{0.840480in}{0.721332in}}%
\pgfpathlineto{\pgfqpoint{0.850000in}{0.722162in}}%
\pgfpathlineto{\pgfqpoint{0.859520in}{0.723044in}}%
\pgfpathlineto{\pgfqpoint{0.869040in}{0.723802in}}%
\pgfpathlineto{\pgfqpoint{0.878560in}{0.724501in}}%
\pgfpathlineto{\pgfqpoint{0.888080in}{0.725250in}}%
\pgfpathlineto{\pgfqpoint{0.897600in}{0.725969in}}%
\pgfpathlineto{\pgfqpoint{0.907120in}{0.726601in}}%
\pgfpathlineto{\pgfqpoint{0.916640in}{0.727220in}}%
\pgfpathlineto{\pgfqpoint{0.926160in}{0.727886in}}%
\pgfpathlineto{\pgfqpoint{0.935680in}{0.728460in}}%
\pgfpathlineto{\pgfqpoint{0.945200in}{0.729085in}}%
\pgfpathlineto{\pgfqpoint{0.954720in}{0.729782in}}%
\pgfpathlineto{\pgfqpoint{0.964240in}{0.730421in}}%
\pgfpathlineto{\pgfqpoint{0.973760in}{0.731022in}}%
\pgfpathlineto{\pgfqpoint{0.983280in}{0.731609in}}%
\pgfpathlineto{\pgfqpoint{0.992800in}{0.732297in}}%
\pgfpathlineto{\pgfqpoint{1.002320in}{0.732947in}}%
\pgfpathlineto{\pgfqpoint{1.011840in}{0.733608in}}%
\pgfpathlineto{\pgfqpoint{1.021360in}{0.734227in}}%
\pgfpathlineto{\pgfqpoint{1.030880in}{0.734875in}}%
\pgfpathlineto{\pgfqpoint{1.040400in}{0.735666in}}%
\pgfpathlineto{\pgfqpoint{1.049920in}{0.736391in}}%
\pgfpathlineto{\pgfqpoint{1.059440in}{0.737137in}}%
\pgfpathlineto{\pgfqpoint{1.068960in}{0.737996in}}%
\pgfpathlineto{\pgfqpoint{1.078480in}{0.738923in}}%
\pgfpathlineto{\pgfqpoint{1.088000in}{0.739758in}}%
\pgfpathlineto{\pgfqpoint{1.097520in}{0.740732in}}%
\pgfpathlineto{\pgfqpoint{1.107040in}{0.741960in}}%
\pgfpathlineto{\pgfqpoint{1.116560in}{0.743360in}}%
\pgfpathlineto{\pgfqpoint{1.126080in}{0.745324in}}%
\pgfpathlineto{\pgfqpoint{1.135600in}{0.748299in}}%
\pgfpathlineto{\pgfqpoint{1.145120in}{0.756542in}}%
\pgfpathlineto{\pgfqpoint{1.154640in}{0.779171in}}%
\pgfpathlineto{\pgfqpoint{1.164160in}{0.799254in}}%
\pgfpathlineto{\pgfqpoint{1.173680in}{0.805776in}}%
\pgfpathlineto{\pgfqpoint{1.183200in}{0.810652in}}%
\pgfpathlineto{\pgfqpoint{1.192720in}{0.814804in}}%
\pgfpathlineto{\pgfqpoint{1.202240in}{0.817464in}}%
\pgfpathlineto{\pgfqpoint{1.211760in}{0.820484in}}%
\pgfpathlineto{\pgfqpoint{1.221280in}{0.823673in}}%
\pgfpathlineto{\pgfqpoint{1.230800in}{0.827183in}}%
\pgfpathlineto{\pgfqpoint{1.240320in}{0.831340in}}%
\pgfpathlineto{\pgfqpoint{1.249840in}{0.838156in}}%
\pgfpathlineto{\pgfqpoint{1.259360in}{0.844701in}}%
\pgfpathlineto{\pgfqpoint{1.268880in}{0.853395in}}%
\pgfpathlineto{\pgfqpoint{1.278400in}{0.879630in}}%
\pgfusepath{stroke}%
\end{pgfscope}%
\begin{pgfscope}%
\pgfpathrectangle{\pgfqpoint{0.326400in}{0.336000in}}{\pgfqpoint{0.952000in}{1.232000in}}%
\pgfusepath{clip}%
\pgfsetrectcap%
\pgfsetroundjoin%
\pgfsetlinewidth{1.204500pt}%
\definecolor{currentstroke}{rgb}{0.200000,0.800000,0.300000}%
\pgfsetstrokecolor{currentstroke}%
\pgfsetdash{}{0pt}%
\pgfpathmoveto{\pgfqpoint{0.326400in}{0.626614in}}%
\pgfpathlineto{\pgfqpoint{0.335920in}{0.641122in}}%
\pgfpathlineto{\pgfqpoint{0.345440in}{0.645407in}}%
\pgfpathlineto{\pgfqpoint{0.354960in}{0.648312in}}%
\pgfpathlineto{\pgfqpoint{0.364480in}{0.650576in}}%
\pgfpathlineto{\pgfqpoint{0.374000in}{0.653061in}}%
\pgfpathlineto{\pgfqpoint{0.383520in}{0.655158in}}%
\pgfpathlineto{\pgfqpoint{0.393040in}{0.657180in}}%
\pgfpathlineto{\pgfqpoint{0.402560in}{0.659183in}}%
\pgfpathlineto{\pgfqpoint{0.412080in}{0.661052in}}%
\pgfpathlineto{\pgfqpoint{0.421600in}{0.663124in}}%
\pgfpathlineto{\pgfqpoint{0.431120in}{0.664911in}}%
\pgfpathlineto{\pgfqpoint{0.440640in}{0.666640in}}%
\pgfpathlineto{\pgfqpoint{0.450160in}{0.669078in}}%
\pgfpathlineto{\pgfqpoint{0.459680in}{0.671633in}}%
\pgfpathlineto{\pgfqpoint{0.469200in}{0.674801in}}%
\pgfpathlineto{\pgfqpoint{0.478720in}{0.680337in}}%
\pgfpathlineto{\pgfqpoint{0.488240in}{0.688724in}}%
\pgfpathlineto{\pgfqpoint{0.497760in}{0.693393in}}%
\pgfpathlineto{\pgfqpoint{0.507280in}{0.696013in}}%
\pgfpathlineto{\pgfqpoint{0.516800in}{0.697846in}}%
\pgfpathlineto{\pgfqpoint{0.526320in}{0.699412in}}%
\pgfpathlineto{\pgfqpoint{0.535840in}{0.700795in}}%
\pgfpathlineto{\pgfqpoint{0.545360in}{0.702133in}}%
\pgfpathlineto{\pgfqpoint{0.554880in}{0.703306in}}%
\pgfpathlineto{\pgfqpoint{0.564400in}{0.704330in}}%
\pgfpathlineto{\pgfqpoint{0.573920in}{0.705511in}}%
\pgfpathlineto{\pgfqpoint{0.583440in}{0.706561in}}%
\pgfpathlineto{\pgfqpoint{0.592960in}{0.707600in}}%
\pgfpathlineto{\pgfqpoint{0.602480in}{0.708494in}}%
\pgfpathlineto{\pgfqpoint{0.612000in}{0.709304in}}%
\pgfpathlineto{\pgfqpoint{0.621520in}{0.710185in}}%
\pgfpathlineto{\pgfqpoint{0.631040in}{0.711078in}}%
\pgfpathlineto{\pgfqpoint{0.640560in}{0.711949in}}%
\pgfpathlineto{\pgfqpoint{0.650080in}{0.712846in}}%
\pgfpathlineto{\pgfqpoint{0.659600in}{0.713791in}}%
\pgfpathlineto{\pgfqpoint{0.669120in}{0.714730in}}%
\pgfpathlineto{\pgfqpoint{0.678640in}{0.715673in}}%
\pgfpathlineto{\pgfqpoint{0.688160in}{0.716648in}}%
\pgfpathlineto{\pgfqpoint{0.697680in}{0.717640in}}%
\pgfpathlineto{\pgfqpoint{0.707200in}{0.718814in}}%
\pgfpathlineto{\pgfqpoint{0.716720in}{0.719962in}}%
\pgfpathlineto{\pgfqpoint{0.726240in}{0.721167in}}%
\pgfpathlineto{\pgfqpoint{0.735760in}{0.722489in}}%
\pgfpathlineto{\pgfqpoint{0.745280in}{0.723573in}}%
\pgfpathlineto{\pgfqpoint{0.754800in}{0.724970in}}%
\pgfpathlineto{\pgfqpoint{0.764320in}{0.726817in}}%
\pgfpathlineto{\pgfqpoint{0.773840in}{0.728789in}}%
\pgfpathlineto{\pgfqpoint{0.783360in}{0.731596in}}%
\pgfpathlineto{\pgfqpoint{0.792880in}{0.735989in}}%
\pgfpathlineto{\pgfqpoint{0.802400in}{0.753253in}}%
\pgfpathlineto{\pgfqpoint{0.811920in}{0.778145in}}%
\pgfpathlineto{\pgfqpoint{0.821440in}{0.781451in}}%
\pgfpathlineto{\pgfqpoint{0.830960in}{0.783585in}}%
\pgfpathlineto{\pgfqpoint{0.840480in}{0.785631in}}%
\pgfpathlineto{\pgfqpoint{0.850000in}{0.787405in}}%
\pgfpathlineto{\pgfqpoint{0.859520in}{0.788913in}}%
\pgfpathlineto{\pgfqpoint{0.869040in}{0.790006in}}%
\pgfpathlineto{\pgfqpoint{0.878560in}{0.791169in}}%
\pgfpathlineto{\pgfqpoint{0.888080in}{0.792225in}}%
\pgfpathlineto{\pgfqpoint{0.897600in}{0.793308in}}%
\pgfpathlineto{\pgfqpoint{0.907120in}{0.794253in}}%
\pgfpathlineto{\pgfqpoint{0.916640in}{0.795126in}}%
\pgfpathlineto{\pgfqpoint{0.926160in}{0.795922in}}%
\pgfpathlineto{\pgfqpoint{0.935680in}{0.796736in}}%
\pgfpathlineto{\pgfqpoint{0.945200in}{0.797554in}}%
\pgfpathlineto{\pgfqpoint{0.954720in}{0.798284in}}%
\pgfpathlineto{\pgfqpoint{0.964240in}{0.799043in}}%
\pgfpathlineto{\pgfqpoint{0.973760in}{0.799781in}}%
\pgfpathlineto{\pgfqpoint{0.983280in}{0.800515in}}%
\pgfpathlineto{\pgfqpoint{0.992800in}{0.801203in}}%
\pgfpathlineto{\pgfqpoint{1.002320in}{0.801823in}}%
\pgfpathlineto{\pgfqpoint{1.011840in}{0.802435in}}%
\pgfpathlineto{\pgfqpoint{1.021360in}{0.803003in}}%
\pgfpathlineto{\pgfqpoint{1.030880in}{0.803589in}}%
\pgfpathlineto{\pgfqpoint{1.040400in}{0.804128in}}%
\pgfpathlineto{\pgfqpoint{1.049920in}{0.804765in}}%
\pgfpathlineto{\pgfqpoint{1.059440in}{0.805269in}}%
\pgfpathlineto{\pgfqpoint{1.068960in}{0.805840in}}%
\pgfpathlineto{\pgfqpoint{1.078480in}{0.806359in}}%
\pgfpathlineto{\pgfqpoint{1.088000in}{0.806971in}}%
\pgfpathlineto{\pgfqpoint{1.097520in}{0.807587in}}%
\pgfpathlineto{\pgfqpoint{1.107040in}{0.808188in}}%
\pgfpathlineto{\pgfqpoint{1.116560in}{0.808746in}}%
\pgfpathlineto{\pgfqpoint{1.126080in}{0.809378in}}%
\pgfpathlineto{\pgfqpoint{1.135600in}{0.810009in}}%
\pgfpathlineto{\pgfqpoint{1.145120in}{0.810665in}}%
\pgfpathlineto{\pgfqpoint{1.154640in}{0.811344in}}%
\pgfpathlineto{\pgfqpoint{1.164160in}{0.812022in}}%
\pgfpathlineto{\pgfqpoint{1.173680in}{0.812638in}}%
\pgfpathlineto{\pgfqpoint{1.183200in}{0.813346in}}%
\pgfpathlineto{\pgfqpoint{1.192720in}{0.814139in}}%
\pgfpathlineto{\pgfqpoint{1.202240in}{0.814814in}}%
\pgfpathlineto{\pgfqpoint{1.211760in}{0.815819in}}%
\pgfpathlineto{\pgfqpoint{1.221280in}{0.816711in}}%
\pgfpathlineto{\pgfqpoint{1.230800in}{0.817698in}}%
\pgfpathlineto{\pgfqpoint{1.240320in}{0.818738in}}%
\pgfpathlineto{\pgfqpoint{1.249840in}{0.820329in}}%
\pgfpathlineto{\pgfqpoint{1.259360in}{0.822237in}}%
\pgfpathlineto{\pgfqpoint{1.268880in}{0.825299in}}%
\pgfpathlineto{\pgfqpoint{1.278400in}{0.843986in}}%
\pgfusepath{stroke}%
\end{pgfscope}%
\begin{pgfscope}%
\pgfpathrectangle{\pgfqpoint{0.326400in}{0.336000in}}{\pgfqpoint{0.952000in}{1.232000in}}%
\pgfusepath{clip}%
\pgfsetrectcap%
\pgfsetroundjoin%
\pgfsetlinewidth{1.204500pt}%
\definecolor{currentstroke}{rgb}{0.300000,0.300000,0.300000}%
\pgfsetstrokecolor{currentstroke}%
\pgfsetdash{}{0pt}%
\pgfpathmoveto{\pgfqpoint{0.326400in}{0.719035in}}%
\pgfpathlineto{\pgfqpoint{0.335920in}{0.735472in}}%
\pgfpathlineto{\pgfqpoint{0.345440in}{0.738267in}}%
\pgfpathlineto{\pgfqpoint{0.354960in}{0.740065in}}%
\pgfpathlineto{\pgfqpoint{0.364480in}{0.741448in}}%
\pgfpathlineto{\pgfqpoint{0.374000in}{0.742558in}}%
\pgfpathlineto{\pgfqpoint{0.383520in}{0.743654in}}%
\pgfpathlineto{\pgfqpoint{0.393040in}{0.744551in}}%
\pgfpathlineto{\pgfqpoint{0.402560in}{0.745389in}}%
\pgfpathlineto{\pgfqpoint{0.412080in}{0.746093in}}%
\pgfpathlineto{\pgfqpoint{0.421600in}{0.746807in}}%
\pgfpathlineto{\pgfqpoint{0.431120in}{0.747396in}}%
\pgfpathlineto{\pgfqpoint{0.440640in}{0.747992in}}%
\pgfpathlineto{\pgfqpoint{0.450160in}{0.748546in}}%
\pgfpathlineto{\pgfqpoint{0.459680in}{0.749095in}}%
\pgfpathlineto{\pgfqpoint{0.469200in}{0.749638in}}%
\pgfpathlineto{\pgfqpoint{0.478720in}{0.750183in}}%
\pgfpathlineto{\pgfqpoint{0.488240in}{0.750668in}}%
\pgfpathlineto{\pgfqpoint{0.497760in}{0.751145in}}%
\pgfpathlineto{\pgfqpoint{0.507280in}{0.751581in}}%
\pgfpathlineto{\pgfqpoint{0.516800in}{0.752108in}}%
\pgfpathlineto{\pgfqpoint{0.526320in}{0.752564in}}%
\pgfpathlineto{\pgfqpoint{0.535840in}{0.753016in}}%
\pgfpathlineto{\pgfqpoint{0.545360in}{0.753451in}}%
\pgfpathlineto{\pgfqpoint{0.554880in}{0.753869in}}%
\pgfpathlineto{\pgfqpoint{0.564400in}{0.754255in}}%
\pgfpathlineto{\pgfqpoint{0.573920in}{0.754710in}}%
\pgfpathlineto{\pgfqpoint{0.583440in}{0.755136in}}%
\pgfpathlineto{\pgfqpoint{0.592960in}{0.755535in}}%
\pgfpathlineto{\pgfqpoint{0.602480in}{0.755892in}}%
\pgfpathlineto{\pgfqpoint{0.612000in}{0.756264in}}%
\pgfpathlineto{\pgfqpoint{0.621520in}{0.756631in}}%
\pgfpathlineto{\pgfqpoint{0.631040in}{0.756991in}}%
\pgfpathlineto{\pgfqpoint{0.640560in}{0.757334in}}%
\pgfpathlineto{\pgfqpoint{0.650080in}{0.757662in}}%
\pgfpathlineto{\pgfqpoint{0.659600in}{0.758074in}}%
\pgfpathlineto{\pgfqpoint{0.669120in}{0.758487in}}%
\pgfpathlineto{\pgfqpoint{0.678640in}{0.758870in}}%
\pgfpathlineto{\pgfqpoint{0.688160in}{0.759282in}}%
\pgfpathlineto{\pgfqpoint{0.697680in}{0.759651in}}%
\pgfpathlineto{\pgfqpoint{0.707200in}{0.760031in}}%
\pgfpathlineto{\pgfqpoint{0.716720in}{0.760436in}}%
\pgfpathlineto{\pgfqpoint{0.726240in}{0.760795in}}%
\pgfpathlineto{\pgfqpoint{0.735760in}{0.761185in}}%
\pgfpathlineto{\pgfqpoint{0.745280in}{0.761571in}}%
\pgfpathlineto{\pgfqpoint{0.754800in}{0.761938in}}%
\pgfpathlineto{\pgfqpoint{0.764320in}{0.762333in}}%
\pgfpathlineto{\pgfqpoint{0.773840in}{0.762675in}}%
\pgfpathlineto{\pgfqpoint{0.783360in}{0.763071in}}%
\pgfpathlineto{\pgfqpoint{0.792880in}{0.763487in}}%
\pgfpathlineto{\pgfqpoint{0.802400in}{0.763867in}}%
\pgfpathlineto{\pgfqpoint{0.811920in}{0.764227in}}%
\pgfpathlineto{\pgfqpoint{0.821440in}{0.764592in}}%
\pgfpathlineto{\pgfqpoint{0.830960in}{0.764970in}}%
\pgfpathlineto{\pgfqpoint{0.840480in}{0.765362in}}%
\pgfpathlineto{\pgfqpoint{0.850000in}{0.765810in}}%
\pgfpathlineto{\pgfqpoint{0.859520in}{0.766223in}}%
\pgfpathlineto{\pgfqpoint{0.869040in}{0.766609in}}%
\pgfpathlineto{\pgfqpoint{0.878560in}{0.767021in}}%
\pgfpathlineto{\pgfqpoint{0.888080in}{0.767408in}}%
\pgfpathlineto{\pgfqpoint{0.897600in}{0.767834in}}%
\pgfpathlineto{\pgfqpoint{0.907120in}{0.768214in}}%
\pgfpathlineto{\pgfqpoint{0.916640in}{0.768620in}}%
\pgfpathlineto{\pgfqpoint{0.926160in}{0.769071in}}%
\pgfpathlineto{\pgfqpoint{0.935680in}{0.769474in}}%
\pgfpathlineto{\pgfqpoint{0.945200in}{0.769901in}}%
\pgfpathlineto{\pgfqpoint{0.954720in}{0.770359in}}%
\pgfpathlineto{\pgfqpoint{0.964240in}{0.770808in}}%
\pgfpathlineto{\pgfqpoint{0.973760in}{0.771263in}}%
\pgfpathlineto{\pgfqpoint{0.983280in}{0.771701in}}%
\pgfpathlineto{\pgfqpoint{0.992800in}{0.772145in}}%
\pgfpathlineto{\pgfqpoint{1.002320in}{0.772592in}}%
\pgfpathlineto{\pgfqpoint{1.011840in}{0.773036in}}%
\pgfpathlineto{\pgfqpoint{1.021360in}{0.773488in}}%
\pgfpathlineto{\pgfqpoint{1.030880in}{0.773911in}}%
\pgfpathlineto{\pgfqpoint{1.040400in}{0.774376in}}%
\pgfpathlineto{\pgfqpoint{1.049920in}{0.774860in}}%
\pgfpathlineto{\pgfqpoint{1.059440in}{0.775404in}}%
\pgfpathlineto{\pgfqpoint{1.068960in}{0.775879in}}%
\pgfpathlineto{\pgfqpoint{1.078480in}{0.776405in}}%
\pgfpathlineto{\pgfqpoint{1.088000in}{0.776990in}}%
\pgfpathlineto{\pgfqpoint{1.097520in}{0.777548in}}%
\pgfpathlineto{\pgfqpoint{1.107040in}{0.778170in}}%
\pgfpathlineto{\pgfqpoint{1.116560in}{0.778728in}}%
\pgfpathlineto{\pgfqpoint{1.126080in}{0.779330in}}%
\pgfpathlineto{\pgfqpoint{1.135600in}{0.779905in}}%
\pgfpathlineto{\pgfqpoint{1.145120in}{0.780504in}}%
\pgfpathlineto{\pgfqpoint{1.154640in}{0.781225in}}%
\pgfpathlineto{\pgfqpoint{1.164160in}{0.781829in}}%
\pgfpathlineto{\pgfqpoint{1.173680in}{0.782588in}}%
\pgfpathlineto{\pgfqpoint{1.183200in}{0.783358in}}%
\pgfpathlineto{\pgfqpoint{1.192720in}{0.784071in}}%
\pgfpathlineto{\pgfqpoint{1.202240in}{0.784958in}}%
\pgfpathlineto{\pgfqpoint{1.211760in}{0.785801in}}%
\pgfpathlineto{\pgfqpoint{1.221280in}{0.786758in}}%
\pgfpathlineto{\pgfqpoint{1.230800in}{0.787872in}}%
\pgfpathlineto{\pgfqpoint{1.240320in}{0.789226in}}%
\pgfpathlineto{\pgfqpoint{1.249840in}{0.790960in}}%
\pgfpathlineto{\pgfqpoint{1.259360in}{0.793009in}}%
\pgfpathlineto{\pgfqpoint{1.268880in}{0.796201in}}%
\pgfpathlineto{\pgfqpoint{1.278400in}{0.818588in}}%
\pgfusepath{stroke}%
\end{pgfscope}%
\begin{pgfscope}%
\pgfpathrectangle{\pgfqpoint{0.326400in}{0.336000in}}{\pgfqpoint{0.952000in}{1.232000in}}%
\pgfusepath{clip}%
\pgfsetbuttcap%
\pgfsetroundjoin%
\pgfsetlinewidth{1.204500pt}%
\definecolor{currentstroke}{rgb}{0.100000,0.100000,0.100000}%
\pgfsetstrokecolor{currentstroke}%
\pgfsetdash{{4.440000pt}{1.920000pt}}{0.000000pt}%
\pgfpathmoveto{\pgfqpoint{0.326400in}{0.411199in}}%
\pgfpathlineto{\pgfqpoint{0.335920in}{0.427637in}}%
\pgfpathlineto{\pgfqpoint{0.345440in}{0.430432in}}%
\pgfpathlineto{\pgfqpoint{0.354960in}{0.432229in}}%
\pgfpathlineto{\pgfqpoint{0.364480in}{0.433613in}}%
\pgfpathlineto{\pgfqpoint{0.374000in}{0.434722in}}%
\pgfpathlineto{\pgfqpoint{0.383520in}{0.435819in}}%
\pgfpathlineto{\pgfqpoint{0.393040in}{0.436715in}}%
\pgfpathlineto{\pgfqpoint{0.402560in}{0.437553in}}%
\pgfpathlineto{\pgfqpoint{0.412080in}{0.438258in}}%
\pgfpathlineto{\pgfqpoint{0.421600in}{0.438971in}}%
\pgfpathlineto{\pgfqpoint{0.431120in}{0.439561in}}%
\pgfpathlineto{\pgfqpoint{0.440640in}{0.440156in}}%
\pgfpathlineto{\pgfqpoint{0.450160in}{0.440710in}}%
\pgfpathlineto{\pgfqpoint{0.459680in}{0.441260in}}%
\pgfpathlineto{\pgfqpoint{0.469200in}{0.441802in}}%
\pgfpathlineto{\pgfqpoint{0.478720in}{0.442347in}}%
\pgfpathlineto{\pgfqpoint{0.488240in}{0.442832in}}%
\pgfpathlineto{\pgfqpoint{0.497760in}{0.443309in}}%
\pgfpathlineto{\pgfqpoint{0.507280in}{0.443745in}}%
\pgfpathlineto{\pgfqpoint{0.516800in}{0.444272in}}%
\pgfpathlineto{\pgfqpoint{0.526320in}{0.444728in}}%
\pgfpathlineto{\pgfqpoint{0.535840in}{0.445180in}}%
\pgfpathlineto{\pgfqpoint{0.545360in}{0.445616in}}%
\pgfpathlineto{\pgfqpoint{0.554880in}{0.446033in}}%
\pgfpathlineto{\pgfqpoint{0.564400in}{0.446419in}}%
\pgfpathlineto{\pgfqpoint{0.573920in}{0.446875in}}%
\pgfpathlineto{\pgfqpoint{0.583440in}{0.447300in}}%
\pgfpathlineto{\pgfqpoint{0.592960in}{0.447699in}}%
\pgfpathlineto{\pgfqpoint{0.602480in}{0.448056in}}%
\pgfpathlineto{\pgfqpoint{0.612000in}{0.448428in}}%
\pgfpathlineto{\pgfqpoint{0.621520in}{0.448796in}}%
\pgfpathlineto{\pgfqpoint{0.631040in}{0.449156in}}%
\pgfpathlineto{\pgfqpoint{0.640560in}{0.449498in}}%
\pgfpathlineto{\pgfqpoint{0.650080in}{0.449826in}}%
\pgfpathlineto{\pgfqpoint{0.659600in}{0.450238in}}%
\pgfpathlineto{\pgfqpoint{0.669120in}{0.450651in}}%
\pgfpathlineto{\pgfqpoint{0.678640in}{0.451034in}}%
\pgfpathlineto{\pgfqpoint{0.688160in}{0.451446in}}%
\pgfpathlineto{\pgfqpoint{0.697680in}{0.451816in}}%
\pgfpathlineto{\pgfqpoint{0.707200in}{0.452195in}}%
\pgfpathlineto{\pgfqpoint{0.716720in}{0.452600in}}%
\pgfpathlineto{\pgfqpoint{0.726240in}{0.452960in}}%
\pgfpathlineto{\pgfqpoint{0.735760in}{0.453350in}}%
\pgfpathlineto{\pgfqpoint{0.745280in}{0.453735in}}%
\pgfpathlineto{\pgfqpoint{0.754800in}{0.454102in}}%
\pgfpathlineto{\pgfqpoint{0.764320in}{0.454497in}}%
\pgfpathlineto{\pgfqpoint{0.773840in}{0.454839in}}%
\pgfpathlineto{\pgfqpoint{0.783360in}{0.455235in}}%
\pgfpathlineto{\pgfqpoint{0.792880in}{0.455651in}}%
\pgfpathlineto{\pgfqpoint{0.802400in}{0.456031in}}%
\pgfpathlineto{\pgfqpoint{0.811920in}{0.456391in}}%
\pgfpathlineto{\pgfqpoint{0.821440in}{0.456756in}}%
\pgfpathlineto{\pgfqpoint{0.830960in}{0.457135in}}%
\pgfpathlineto{\pgfqpoint{0.840480in}{0.457527in}}%
\pgfpathlineto{\pgfqpoint{0.850000in}{0.457974in}}%
\pgfpathlineto{\pgfqpoint{0.859520in}{0.458387in}}%
\pgfpathlineto{\pgfqpoint{0.869040in}{0.458774in}}%
\pgfpathlineto{\pgfqpoint{0.878560in}{0.459186in}}%
\pgfpathlineto{\pgfqpoint{0.888080in}{0.459572in}}%
\pgfpathlineto{\pgfqpoint{0.897600in}{0.459998in}}%
\pgfpathlineto{\pgfqpoint{0.907120in}{0.460378in}}%
\pgfpathlineto{\pgfqpoint{0.916640in}{0.460784in}}%
\pgfpathlineto{\pgfqpoint{0.926160in}{0.461235in}}%
\pgfpathlineto{\pgfqpoint{0.935680in}{0.461639in}}%
\pgfpathlineto{\pgfqpoint{0.945200in}{0.462066in}}%
\pgfpathlineto{\pgfqpoint{0.954720in}{0.462524in}}%
\pgfpathlineto{\pgfqpoint{0.964240in}{0.462973in}}%
\pgfpathlineto{\pgfqpoint{0.973760in}{0.463428in}}%
\pgfpathlineto{\pgfqpoint{0.983280in}{0.463865in}}%
\pgfpathlineto{\pgfqpoint{0.992800in}{0.464309in}}%
\pgfpathlineto{\pgfqpoint{1.002320in}{0.464756in}}%
\pgfpathlineto{\pgfqpoint{1.011840in}{0.465201in}}%
\pgfpathlineto{\pgfqpoint{1.021360in}{0.465653in}}%
\pgfpathlineto{\pgfqpoint{1.030880in}{0.466075in}}%
\pgfpathlineto{\pgfqpoint{1.040400in}{0.466540in}}%
\pgfpathlineto{\pgfqpoint{1.049920in}{0.467024in}}%
\pgfpathlineto{\pgfqpoint{1.059440in}{0.467568in}}%
\pgfpathlineto{\pgfqpoint{1.068960in}{0.468044in}}%
\pgfpathlineto{\pgfqpoint{1.078480in}{0.468569in}}%
\pgfpathlineto{\pgfqpoint{1.088000in}{0.469154in}}%
\pgfpathlineto{\pgfqpoint{1.097520in}{0.469712in}}%
\pgfpathlineto{\pgfqpoint{1.107040in}{0.470334in}}%
\pgfpathlineto{\pgfqpoint{1.116560in}{0.470892in}}%
\pgfpathlineto{\pgfqpoint{1.126080in}{0.471495in}}%
\pgfpathlineto{\pgfqpoint{1.135600in}{0.472069in}}%
\pgfpathlineto{\pgfqpoint{1.145120in}{0.472668in}}%
\pgfpathlineto{\pgfqpoint{1.154640in}{0.473389in}}%
\pgfpathlineto{\pgfqpoint{1.164160in}{0.473993in}}%
\pgfpathlineto{\pgfqpoint{1.173680in}{0.474753in}}%
\pgfpathlineto{\pgfqpoint{1.183200in}{0.475522in}}%
\pgfpathlineto{\pgfqpoint{1.192720in}{0.476235in}}%
\pgfpathlineto{\pgfqpoint{1.202240in}{0.477122in}}%
\pgfpathlineto{\pgfqpoint{1.211760in}{0.477966in}}%
\pgfpathlineto{\pgfqpoint{1.221280in}{0.478922in}}%
\pgfpathlineto{\pgfqpoint{1.230800in}{0.480036in}}%
\pgfpathlineto{\pgfqpoint{1.240320in}{0.481390in}}%
\pgfpathlineto{\pgfqpoint{1.249840in}{0.483124in}}%
\pgfpathlineto{\pgfqpoint{1.259360in}{0.485173in}}%
\pgfpathlineto{\pgfqpoint{1.268880in}{0.488365in}}%
\pgfpathlineto{\pgfqpoint{1.278400in}{0.510752in}}%
\pgfusepath{stroke}%
\end{pgfscope}%
\begin{pgfscope}%
\pgfsetrectcap%
\pgfsetmiterjoin%
\pgfsetlinewidth{0.803000pt}%
\definecolor{currentstroke}{rgb}{0.000000,0.000000,0.000000}%
\pgfsetstrokecolor{currentstroke}%
\pgfsetdash{}{0pt}%
\pgfpathmoveto{\pgfqpoint{0.326400in}{0.336000in}}%
\pgfpathlineto{\pgfqpoint{0.326400in}{1.568000in}}%
\pgfusepath{stroke}%
\end{pgfscope}%
\begin{pgfscope}%
\pgfsetrectcap%
\pgfsetmiterjoin%
\pgfsetlinewidth{0.803000pt}%
\definecolor{currentstroke}{rgb}{0.000000,0.000000,0.000000}%
\pgfsetstrokecolor{currentstroke}%
\pgfsetdash{}{0pt}%
\pgfpathmoveto{\pgfqpoint{1.278400in}{0.336000in}}%
\pgfpathlineto{\pgfqpoint{1.278400in}{1.568000in}}%
\pgfusepath{stroke}%
\end{pgfscope}%
\begin{pgfscope}%
\pgfsetrectcap%
\pgfsetmiterjoin%
\pgfsetlinewidth{0.803000pt}%
\definecolor{currentstroke}{rgb}{0.000000,0.000000,0.000000}%
\pgfsetstrokecolor{currentstroke}%
\pgfsetdash{}{0pt}%
\pgfpathmoveto{\pgfqpoint{0.326400in}{0.336000in}}%
\pgfpathlineto{\pgfqpoint{1.278400in}{0.336000in}}%
\pgfusepath{stroke}%
\end{pgfscope}%
\begin{pgfscope}%
\pgfsetrectcap%
\pgfsetmiterjoin%
\pgfsetlinewidth{0.803000pt}%
\definecolor{currentstroke}{rgb}{0.000000,0.000000,0.000000}%
\pgfsetstrokecolor{currentstroke}%
\pgfsetdash{}{0pt}%
\pgfpathmoveto{\pgfqpoint{0.326400in}{1.568000in}}%
\pgfpathlineto{\pgfqpoint{1.278400in}{1.568000in}}%
\pgfusepath{stroke}%
\end{pgfscope}%
\begin{pgfscope}%
\pgfsetbuttcap%
\pgfsetmiterjoin%
\pgfsetlinewidth{0.000000pt}%
\definecolor{currentstroke}{rgb}{0.800000,0.800000,0.800000}%
\pgfsetstrokecolor{currentstroke}%
\pgfsetstrokeopacity{0.000000}%
\pgfsetdash{}{0pt}%
\pgfpathmoveto{\pgfqpoint{0.388622in}{0.877629in}}%
\pgfpathlineto{\pgfqpoint{1.273907in}{0.877629in}}%
\pgfpathquadraticcurveto{\pgfqpoint{1.291685in}{0.877629in}}{\pgfqpoint{1.291685in}{0.895407in}}%
\pgfpathlineto{\pgfqpoint{1.291685in}{1.505778in}}%
\pgfpathquadraticcurveto{\pgfqpoint{1.291685in}{1.523556in}}{\pgfqpoint{1.273907in}{1.523556in}}%
\pgfpathlineto{\pgfqpoint{0.388622in}{1.523556in}}%
\pgfpathquadraticcurveto{\pgfqpoint{0.370844in}{1.523556in}}{\pgfqpoint{0.370844in}{1.505778in}}%
\pgfpathlineto{\pgfqpoint{0.370844in}{0.895407in}}%
\pgfpathquadraticcurveto{\pgfqpoint{0.370844in}{0.877629in}}{\pgfqpoint{0.388622in}{0.877629in}}%
\pgfpathclose%
\pgfusepath{}%
\end{pgfscope}%
\begin{pgfscope}%
\pgfsetrectcap%
\pgfsetroundjoin%
\pgfsetlinewidth{1.204500pt}%
\definecolor{currentstroke}{rgb}{0.200000,0.200000,0.950000}%
\pgfsetstrokecolor{currentstroke}%
\pgfsetdash{}{0pt}%
\pgfpathmoveto{\pgfqpoint{0.406400in}{1.456889in}}%
\pgfpathlineto{\pgfqpoint{0.584178in}{1.456889in}}%
\pgfusepath{stroke}%
\end{pgfscope}%
\begin{pgfscope}%
\pgftext[x=0.655289in,y=1.425778in,left,base]{\rmfamily\fontsize{6.400000}{7.680000}\selectfont 30s}%
\end{pgfscope}%
\begin{pgfscope}%
\pgfsetrectcap%
\pgfsetroundjoin%
\pgfsetlinewidth{1.204500pt}%
\definecolor{currentstroke}{rgb}{0.900000,0.100000,0.100000}%
\pgfsetstrokecolor{currentstroke}%
\pgfsetdash{}{0pt}%
\pgfpathmoveto{\pgfqpoint{0.406400in}{1.334018in}}%
\pgfpathlineto{\pgfqpoint{0.584178in}{1.334018in}}%
\pgfusepath{stroke}%
\end{pgfscope}%
\begin{pgfscope}%
\pgftext[x=0.655289in,y=1.302907in,left,base]{\rmfamily\fontsize{6.400000}{7.680000}\selectfont 300s}%
\end{pgfscope}%
\begin{pgfscope}%
\pgfsetrectcap%
\pgfsetroundjoin%
\pgfsetlinewidth{1.204500pt}%
\definecolor{currentstroke}{rgb}{0.200000,0.800000,0.300000}%
\pgfsetstrokecolor{currentstroke}%
\pgfsetdash{}{0pt}%
\pgfpathmoveto{\pgfqpoint{0.406400in}{1.211148in}}%
\pgfpathlineto{\pgfqpoint{0.584178in}{1.211148in}}%
\pgfusepath{stroke}%
\end{pgfscope}%
\begin{pgfscope}%
\pgftext[x=0.655289in,y=1.180037in,left,base]{\rmfamily\fontsize{6.400000}{7.680000}\selectfont 3600s}%
\end{pgfscope}%
\begin{pgfscope}%
\pgfsetrectcap%
\pgfsetroundjoin%
\pgfsetlinewidth{1.204500pt}%
\definecolor{currentstroke}{rgb}{0.300000,0.300000,0.300000}%
\pgfsetstrokecolor{currentstroke}%
\pgfsetdash{}{0pt}%
\pgfpathmoveto{\pgfqpoint{0.406400in}{1.088278in}}%
\pgfpathlineto{\pgfqpoint{0.584178in}{1.088278in}}%
\pgfusepath{stroke}%
\end{pgfscope}%
\begin{pgfscope}%
\pgftext[x=0.655289in,y=1.057167in,left,base]{\rmfamily\fontsize{6.400000}{7.680000}\selectfont 24H}%
\end{pgfscope}%
\begin{pgfscope}%
\pgfsetbuttcap%
\pgfsetroundjoin%
\pgfsetlinewidth{1.204500pt}%
\definecolor{currentstroke}{rgb}{0.100000,0.100000,0.100000}%
\pgfsetstrokecolor{currentstroke}%
\pgfsetdash{{4.440000pt}{1.920000pt}}{0.000000pt}%
\pgfpathmoveto{\pgfqpoint{0.406400in}{0.965129in}}%
\pgfpathlineto{\pgfqpoint{0.584178in}{0.965129in}}%
\pgfusepath{stroke}%
\end{pgfscope}%
\begin{pgfscope}%
\pgftext[x=0.655289in,y=0.934018in,left,base]{\rmfamily\fontsize{6.400000}{7.680000}\selectfont 30s (w/o SO)}%
\end{pgfscope}%
\end{pgfpicture}%
\makeatother%
\endgroup%

%% file: plot_data/packet_level_simulation/packet_level_average_fct.pgf
\begingroup%
\makeatletter%
\begin{pgfpicture}%
\pgfpathrectangle{\pgfpointorigin}{\pgfqpoint{1.360000in}{1.600000in}}%
\pgfusepath{use as bounding box, clip}%
\begin{pgfscope}%
\pgfsetbuttcap%
\pgfsetmiterjoin%
\definecolor{currentfill}{rgb}{1.000000,1.000000,1.000000}%
\pgfsetfillcolor{currentfill}%
\pgfsetlinewidth{0.000000pt}%
\definecolor{currentstroke}{rgb}{1.000000,1.000000,1.000000}%
\pgfsetstrokecolor{currentstroke}%
\pgfsetdash{}{0pt}%
\pgfpathmoveto{\pgfqpoint{0.000000in}{0.000000in}}%
\pgfpathlineto{\pgfqpoint{1.360000in}{0.000000in}}%
\pgfpathlineto{\pgfqpoint{1.360000in}{1.600000in}}%
\pgfpathlineto{\pgfqpoint{0.000000in}{1.600000in}}%
\pgfpathclose%
\pgfusepath{fill}%
\end{pgfscope}%
\begin{pgfscope}%
\pgfsetbuttcap%
\pgfsetmiterjoin%
\definecolor{currentfill}{rgb}{0.917647,0.917647,0.949020}%
\pgfsetfillcolor{currentfill}%
\pgfsetlinewidth{0.000000pt}%
\definecolor{currentstroke}{rgb}{0.000000,0.000000,0.000000}%
\pgfsetstrokecolor{currentstroke}%
\pgfsetstrokeopacity{0.000000}%
\pgfsetdash{}{0pt}%
\pgfpathmoveto{\pgfqpoint{0.340000in}{0.400000in}}%
\pgfpathlineto{\pgfqpoint{1.346400in}{0.400000in}}%
\pgfpathlineto{\pgfqpoint{1.346400in}{1.568000in}}%
\pgfpathlineto{\pgfqpoint{0.340000in}{1.568000in}}%
\pgfpathclose%
\pgfusepath{fill}%
\end{pgfscope}%
\begin{pgfscope}%
\definecolor{textcolor}{rgb}{0.150000,0.150000,0.150000}%
\pgfsetstrokecolor{textcolor}%
\pgfsetfillcolor{textcolor}%
\pgftext[x=0.444700in,y=0.136885in,left,base,rotate=90.000000]{\color{textcolor}\rmfamily\fontsize{6.500000}{7.800000}\selectfont 1.1}%
\end{pgfscope}%
\begin{pgfscope}%
\definecolor{textcolor}{rgb}{0.150000,0.150000,0.150000}%
\pgfsetstrokecolor{textcolor}%
\pgfsetfillcolor{textcolor}%
\pgftext[x=0.612433in,y=0.136885in,left,base,rotate=90.000000]{\color{textcolor}\rmfamily\fontsize{6.500000}{7.800000}\selectfont 1.2}%
\end{pgfscope}%
\begin{pgfscope}%
\definecolor{textcolor}{rgb}{0.150000,0.150000,0.150000}%
\pgfsetstrokecolor{textcolor}%
\pgfsetfillcolor{textcolor}%
\pgftext[x=0.780167in,y=0.136885in,left,base,rotate=90.000000]{\color{textcolor}\rmfamily\fontsize{6.500000}{7.800000}\selectfont 1.4}%
\end{pgfscope}%
\begin{pgfscope}%
\definecolor{textcolor}{rgb}{0.150000,0.150000,0.150000}%
\pgfsetstrokecolor{textcolor}%
\pgfsetfillcolor{textcolor}%
\pgftext[x=0.947900in,y=0.136885in,left,base,rotate=90.000000]{\color{textcolor}\rmfamily\fontsize{6.500000}{7.800000}\selectfont 1.6}%
\end{pgfscope}%
\begin{pgfscope}%
\definecolor{textcolor}{rgb}{0.150000,0.150000,0.150000}%
\pgfsetstrokecolor{textcolor}%
\pgfsetfillcolor{textcolor}%
\pgftext[x=1.115633in,y=0.136885in,left,base,rotate=90.000000]{\color{textcolor}\rmfamily\fontsize{6.500000}{7.800000}\selectfont 1.8}%
\end{pgfscope}%
\begin{pgfscope}%
\definecolor{textcolor}{rgb}{0.150000,0.150000,0.150000}%
\pgfsetstrokecolor{textcolor}%
\pgfsetfillcolor{textcolor}%
\pgftext[x=1.283367in,y=0.136885in,left,base,rotate=90.000000]{\color{textcolor}\rmfamily\fontsize{6.500000}{7.800000}\selectfont 1.9}%
\end{pgfscope}%
\begin{pgfscope}%
\definecolor{textcolor}{rgb}{0.150000,0.150000,0.150000}%
\pgfsetstrokecolor{textcolor}%
\pgfsetfillcolor{textcolor}%
\pgftext[x=0.843200in,y=0.118829in,,top]{\color{textcolor}\rmfamily\fontsize{7.400000}{8.880000}\selectfont Avg. Hop Count}%
\end{pgfscope}%
\begin{pgfscope}%
\definecolor{textcolor}{rgb}{0.150000,0.150000,0.150000}%
\pgfsetstrokecolor{textcolor}%
\pgfsetfillcolor{textcolor}%
\pgftext[x=0.191852in,y=1.319764in,left,base,rotate=90.000000]{\color{textcolor}\rmfamily\fontsize{6.500000}{7.800000}\selectfont 0.8}%
\end{pgfscope}%
\begin{pgfscope}%
\definecolor{textcolor}{rgb}{0.150000,0.150000,0.150000}%
\pgfsetstrokecolor{textcolor}%
\pgfsetfillcolor{textcolor}%
\pgftext[x=0.191852in,y=1.027764in,left,base,rotate=90.000000]{\color{textcolor}\rmfamily\fontsize{6.500000}{7.800000}\selectfont 0.6}%
\end{pgfscope}%
\begin{pgfscope}%
\definecolor{textcolor}{rgb}{0.150000,0.150000,0.150000}%
\pgfsetstrokecolor{textcolor}%
\pgfsetfillcolor{textcolor}%
\pgftext[x=0.191852in,y=0.735764in,left,base,rotate=90.000000]{\color{textcolor}\rmfamily\fontsize{6.500000}{7.800000}\selectfont 0.4}%
\end{pgfscope}%
\begin{pgfscope}%
\definecolor{textcolor}{rgb}{0.150000,0.150000,0.150000}%
\pgfsetstrokecolor{textcolor}%
\pgfsetfillcolor{textcolor}%
\pgftext[x=0.191852in,y=0.443764in,left,base,rotate=90.000000]{\color{textcolor}\rmfamily\fontsize{6.500000}{7.800000}\selectfont 0.2}%
\end{pgfscope}%
\begin{pgfscope}%
\definecolor{textcolor}{rgb}{0.150000,0.150000,0.150000}%
\pgfsetstrokecolor{textcolor}%
\pgfsetfillcolor{textcolor}%
\pgftext[x=0.120093in,y=0.984000in,,bottom,rotate=90.000000]{\color{textcolor}\rmfamily\fontsize{7.400000}{8.880000}\selectfont MLU}%
\end{pgfscope}%
\begin{pgfscope}%
\pgfpathrectangle{\pgfqpoint{0.340000in}{0.400000in}}{\pgfqpoint{1.006400in}{1.168000in}}%
\pgfusepath{clip}%
\pgfsetbuttcap%
\pgfsetroundjoin%
\definecolor{currentfill}{rgb}{0.441123,0.576532,0.954545}%
\pgfsetfillcolor{currentfill}%
\pgfsetlinewidth{0.602250pt}%
\definecolor{currentstroke}{rgb}{0.000000,0.000000,0.000000}%
\pgfsetstrokecolor{currentstroke}%
\pgfsetdash{}{0pt}%
\pgfpathmoveto{\pgfqpoint{0.340000in}{1.568000in}}%
\pgfpathlineto{\pgfqpoint{0.507733in}{1.568000in}}%
\pgfpathlineto{\pgfqpoint{0.507733in}{1.276000in}}%
\pgfpathlineto{\pgfqpoint{0.340000in}{1.276000in}}%
\pgfpathlineto{\pgfqpoint{0.340000in}{1.568000in}}%
\pgfusepath{stroke,fill}%
\end{pgfscope}%
\begin{pgfscope}%
\pgfpathrectangle{\pgfqpoint{0.340000in}{0.400000in}}{\pgfqpoint{1.006400in}{1.168000in}}%
\pgfusepath{clip}%
\pgfsetbuttcap%
\pgfsetroundjoin%
\definecolor{currentfill}{rgb}{0.613933,0.739923,0.999142}%
\pgfsetfillcolor{currentfill}%
\pgfsetlinewidth{0.602250pt}%
\definecolor{currentstroke}{rgb}{0.000000,0.000000,0.000000}%
\pgfsetstrokecolor{currentstroke}%
\pgfsetdash{}{0pt}%
\pgfpathmoveto{\pgfqpoint{0.507733in}{1.568000in}}%
\pgfpathlineto{\pgfqpoint{0.675467in}{1.568000in}}%
\pgfpathlineto{\pgfqpoint{0.675467in}{1.276000in}}%
\pgfpathlineto{\pgfqpoint{0.507733in}{1.276000in}}%
\pgfpathlineto{\pgfqpoint{0.507733in}{1.568000in}}%
\pgfusepath{stroke,fill}%
\end{pgfscope}%
\begin{pgfscope}%
\pgfpathrectangle{\pgfqpoint{0.340000in}{0.400000in}}{\pgfqpoint{1.006400in}{1.168000in}}%
\pgfusepath{clip}%
\pgfsetbuttcap%
\pgfsetroundjoin%
\definecolor{currentfill}{rgb}{0.902849,0.844796,0.811970}%
\pgfsetfillcolor{currentfill}%
\pgfsetlinewidth{0.602250pt}%
\definecolor{currentstroke}{rgb}{0.000000,0.000000,0.000000}%
\pgfsetstrokecolor{currentstroke}%
\pgfsetdash{}{0pt}%
\pgfpathmoveto{\pgfqpoint{0.675467in}{1.568000in}}%
\pgfpathlineto{\pgfqpoint{0.843200in}{1.568000in}}%
\pgfpathlineto{\pgfqpoint{0.843200in}{1.276000in}}%
\pgfpathlineto{\pgfqpoint{0.675467in}{1.276000in}}%
\pgfpathlineto{\pgfqpoint{0.675467in}{1.568000in}}%
\pgfusepath{stroke,fill}%
\end{pgfscope}%
\begin{pgfscope}%
\pgfpathrectangle{\pgfqpoint{0.340000in}{0.400000in}}{\pgfqpoint{1.006400in}{1.168000in}}%
\pgfusepath{clip}%
\pgfsetbuttcap%
\pgfsetroundjoin%
\definecolor{currentfill}{rgb}{0.951254,0.578799,0.459408}%
\pgfsetfillcolor{currentfill}%
\pgfsetlinewidth{0.602250pt}%
\definecolor{currentstroke}{rgb}{0.000000,0.000000,0.000000}%
\pgfsetstrokecolor{currentstroke}%
\pgfsetdash{}{0pt}%
\pgfpathmoveto{\pgfqpoint{0.843200in}{1.568000in}}%
\pgfpathlineto{\pgfqpoint{1.010933in}{1.568000in}}%
\pgfpathlineto{\pgfqpoint{1.010933in}{1.276000in}}%
\pgfpathlineto{\pgfqpoint{0.843200in}{1.276000in}}%
\pgfpathlineto{\pgfqpoint{0.843200in}{1.568000in}}%
\pgfusepath{stroke,fill}%
\end{pgfscope}%
\begin{pgfscope}%
\pgfpathrectangle{\pgfqpoint{0.340000in}{0.400000in}}{\pgfqpoint{1.006400in}{1.168000in}}%
\pgfusepath{clip}%
\pgfsetbuttcap%
\pgfsetroundjoin%
\definecolor{currentfill}{rgb}{0.815508,0.277781,0.240294}%
\pgfsetfillcolor{currentfill}%
\pgfsetlinewidth{0.602250pt}%
\definecolor{currentstroke}{rgb}{0.000000,0.000000,0.000000}%
\pgfsetstrokecolor{currentstroke}%
\pgfsetdash{}{0pt}%
\pgfpathmoveto{\pgfqpoint{1.010933in}{1.568000in}}%
\pgfpathlineto{\pgfqpoint{1.178667in}{1.568000in}}%
\pgfpathlineto{\pgfqpoint{1.178667in}{1.276000in}}%
\pgfpathlineto{\pgfqpoint{1.010933in}{1.276000in}}%
\pgfpathlineto{\pgfqpoint{1.010933in}{1.568000in}}%
\pgfusepath{stroke,fill}%
\end{pgfscope}%
\begin{pgfscope}%
\pgfpathrectangle{\pgfqpoint{0.340000in}{0.400000in}}{\pgfqpoint{1.006400in}{1.168000in}}%
\pgfusepath{clip}%
\pgfsetbuttcap%
\pgfsetroundjoin%
\definecolor{currentfill}{rgb}{0.705673,0.015556,0.150233}%
\pgfsetfillcolor{currentfill}%
\pgfsetlinewidth{0.602250pt}%
\definecolor{currentstroke}{rgb}{0.000000,0.000000,0.000000}%
\pgfsetstrokecolor{currentstroke}%
\pgfsetdash{}{0pt}%
\pgfpathmoveto{\pgfqpoint{1.178667in}{1.568000in}}%
\pgfpathlineto{\pgfqpoint{1.346400in}{1.568000in}}%
\pgfpathlineto{\pgfqpoint{1.346400in}{1.276000in}}%
\pgfpathlineto{\pgfqpoint{1.178667in}{1.276000in}}%
\pgfpathlineto{\pgfqpoint{1.178667in}{1.568000in}}%
\pgfusepath{stroke,fill}%
\end{pgfscope}%
\begin{pgfscope}%
\pgfpathrectangle{\pgfqpoint{0.340000in}{0.400000in}}{\pgfqpoint{1.006400in}{1.168000in}}%
\pgfusepath{clip}%
\pgfsetbuttcap%
\pgfsetroundjoin%
\definecolor{currentfill}{rgb}{0.257234,0.339661,0.789661}%
\pgfsetfillcolor{currentfill}%
\pgfsetlinewidth{0.602250pt}%
\definecolor{currentstroke}{rgb}{0.000000,0.000000,0.000000}%
\pgfsetstrokecolor{currentstroke}%
\pgfsetdash{}{0pt}%
\pgfpathmoveto{\pgfqpoint{0.340000in}{1.276000in}}%
\pgfpathlineto{\pgfqpoint{0.507733in}{1.276000in}}%
\pgfpathlineto{\pgfqpoint{0.507733in}{0.984000in}}%
\pgfpathlineto{\pgfqpoint{0.340000in}{0.984000in}}%
\pgfpathlineto{\pgfqpoint{0.340000in}{1.276000in}}%
\pgfusepath{stroke,fill}%
\end{pgfscope}%
\begin{pgfscope}%
\pgfpathrectangle{\pgfqpoint{0.340000in}{0.400000in}}{\pgfqpoint{1.006400in}{1.168000in}}%
\pgfusepath{clip}%
\pgfsetbuttcap%
\pgfsetroundjoin%
\definecolor{currentfill}{rgb}{0.280550,0.373423,0.818011}%
\pgfsetfillcolor{currentfill}%
\pgfsetlinewidth{0.602250pt}%
\definecolor{currentstroke}{rgb}{0.000000,0.000000,0.000000}%
\pgfsetstrokecolor{currentstroke}%
\pgfsetdash{}{0pt}%
\pgfpathmoveto{\pgfqpoint{0.507733in}{1.276000in}}%
\pgfpathlineto{\pgfqpoint{0.675467in}{1.276000in}}%
\pgfpathlineto{\pgfqpoint{0.675467in}{0.984000in}}%
\pgfpathlineto{\pgfqpoint{0.507733in}{0.984000in}}%
\pgfpathlineto{\pgfqpoint{0.507733in}{1.276000in}}%
\pgfusepath{stroke,fill}%
\end{pgfscope}%
\begin{pgfscope}%
\pgfpathrectangle{\pgfqpoint{0.340000in}{0.400000in}}{\pgfqpoint{1.006400in}{1.168000in}}%
\pgfusepath{clip}%
\pgfsetbuttcap%
\pgfsetroundjoin%
\definecolor{currentfill}{rgb}{0.430507,0.564883,0.948889}%
\pgfsetfillcolor{currentfill}%
\pgfsetlinewidth{0.602250pt}%
\definecolor{currentstroke}{rgb}{0.000000,0.000000,0.000000}%
\pgfsetstrokecolor{currentstroke}%
\pgfsetdash{}{0pt}%
\pgfpathmoveto{\pgfqpoint{0.675467in}{1.276000in}}%
\pgfpathlineto{\pgfqpoint{0.843200in}{1.276000in}}%
\pgfpathlineto{\pgfqpoint{0.843200in}{0.984000in}}%
\pgfpathlineto{\pgfqpoint{0.675467in}{0.984000in}}%
\pgfpathlineto{\pgfqpoint{0.675467in}{1.276000in}}%
\pgfusepath{stroke,fill}%
\end{pgfscope}%
\begin{pgfscope}%
\pgfpathrectangle{\pgfqpoint{0.340000in}{0.400000in}}{\pgfqpoint{1.006400in}{1.168000in}}%
\pgfusepath{clip}%
\pgfsetbuttcap%
\pgfsetroundjoin%
\definecolor{currentfill}{rgb}{0.718985,0.811993,0.977656}%
\pgfsetfillcolor{currentfill}%
\pgfsetlinewidth{0.602250pt}%
\definecolor{currentstroke}{rgb}{0.000000,0.000000,0.000000}%
\pgfsetstrokecolor{currentstroke}%
\pgfsetdash{}{0pt}%
\pgfpathmoveto{\pgfqpoint{0.843200in}{1.276000in}}%
\pgfpathlineto{\pgfqpoint{1.010933in}{1.276000in}}%
\pgfpathlineto{\pgfqpoint{1.010933in}{0.984000in}}%
\pgfpathlineto{\pgfqpoint{0.843200in}{0.984000in}}%
\pgfpathlineto{\pgfqpoint{0.843200in}{1.276000in}}%
\pgfusepath{stroke,fill}%
\end{pgfscope}%
\begin{pgfscope}%
\pgfpathrectangle{\pgfqpoint{0.340000in}{0.400000in}}{\pgfqpoint{1.006400in}{1.168000in}}%
\pgfusepath{clip}%
\pgfsetbuttcap%
\pgfsetroundjoin%
\definecolor{currentfill}{rgb}{0.925563,0.825517,0.771136}%
\pgfsetfillcolor{currentfill}%
\pgfsetlinewidth{0.602250pt}%
\definecolor{currentstroke}{rgb}{0.000000,0.000000,0.000000}%
\pgfsetstrokecolor{currentstroke}%
\pgfsetdash{}{0pt}%
\pgfpathmoveto{\pgfqpoint{1.010933in}{1.276000in}}%
\pgfpathlineto{\pgfqpoint{1.178667in}{1.276000in}}%
\pgfpathlineto{\pgfqpoint{1.178667in}{0.984000in}}%
\pgfpathlineto{\pgfqpoint{1.010933in}{0.984000in}}%
\pgfpathlineto{\pgfqpoint{1.010933in}{1.276000in}}%
\pgfusepath{stroke,fill}%
\end{pgfscope}%
\begin{pgfscope}%
\pgfpathrectangle{\pgfqpoint{0.340000in}{0.400000in}}{\pgfqpoint{1.006400in}{1.168000in}}%
\pgfusepath{clip}%
\pgfsetbuttcap%
\pgfsetroundjoin%
\definecolor{currentfill}{rgb}{0.968203,0.720844,0.612293}%
\pgfsetfillcolor{currentfill}%
\pgfsetlinewidth{0.602250pt}%
\definecolor{currentstroke}{rgb}{0.000000,0.000000,0.000000}%
\pgfsetstrokecolor{currentstroke}%
\pgfsetdash{}{0pt}%
\pgfpathmoveto{\pgfqpoint{1.178667in}{1.276000in}}%
\pgfpathlineto{\pgfqpoint{1.346400in}{1.276000in}}%
\pgfpathlineto{\pgfqpoint{1.346400in}{0.984000in}}%
\pgfpathlineto{\pgfqpoint{1.178667in}{0.984000in}}%
\pgfpathlineto{\pgfqpoint{1.178667in}{1.276000in}}%
\pgfusepath{stroke,fill}%
\end{pgfscope}%
\begin{pgfscope}%
\pgfpathrectangle{\pgfqpoint{0.340000in}{0.400000in}}{\pgfqpoint{1.006400in}{1.168000in}}%
\pgfusepath{clip}%
\pgfsetbuttcap%
\pgfsetroundjoin%
\definecolor{currentfill}{rgb}{0.234377,0.305542,0.759680}%
\pgfsetfillcolor{currentfill}%
\pgfsetlinewidth{0.602250pt}%
\definecolor{currentstroke}{rgb}{0.000000,0.000000,0.000000}%
\pgfsetstrokecolor{currentstroke}%
\pgfsetdash{}{0pt}%
\pgfpathmoveto{\pgfqpoint{0.340000in}{0.984000in}}%
\pgfpathlineto{\pgfqpoint{0.507733in}{0.984000in}}%
\pgfpathlineto{\pgfqpoint{0.507733in}{0.692000in}}%
\pgfpathlineto{\pgfqpoint{0.340000in}{0.692000in}}%
\pgfpathlineto{\pgfqpoint{0.340000in}{0.984000in}}%
\pgfusepath{stroke,fill}%
\end{pgfscope}%
\begin{pgfscope}%
\pgfpathrectangle{\pgfqpoint{0.340000in}{0.400000in}}{\pgfqpoint{1.006400in}{1.168000in}}%
\pgfusepath{clip}%
\pgfsetbuttcap%
\pgfsetroundjoin%
\definecolor{currentfill}{rgb}{0.238948,0.312365,0.765676}%
\pgfsetfillcolor{currentfill}%
\pgfsetlinewidth{0.602250pt}%
\definecolor{currentstroke}{rgb}{0.000000,0.000000,0.000000}%
\pgfsetstrokecolor{currentstroke}%
\pgfsetdash{}{0pt}%
\pgfpathmoveto{\pgfqpoint{0.507733in}{0.984000in}}%
\pgfpathlineto{\pgfqpoint{0.675467in}{0.984000in}}%
\pgfpathlineto{\pgfqpoint{0.675467in}{0.692000in}}%
\pgfpathlineto{\pgfqpoint{0.507733in}{0.692000in}}%
\pgfpathlineto{\pgfqpoint{0.507733in}{0.984000in}}%
\pgfusepath{stroke,fill}%
\end{pgfscope}%
\begin{pgfscope}%
\pgfpathrectangle{\pgfqpoint{0.340000in}{0.400000in}}{\pgfqpoint{1.006400in}{1.168000in}}%
\pgfusepath{clip}%
\pgfsetbuttcap%
\pgfsetroundjoin%
\definecolor{currentfill}{rgb}{0.248091,0.326013,0.777669}%
\pgfsetfillcolor{currentfill}%
\pgfsetlinewidth{0.602250pt}%
\definecolor{currentstroke}{rgb}{0.000000,0.000000,0.000000}%
\pgfsetstrokecolor{currentstroke}%
\pgfsetdash{}{0pt}%
\pgfpathmoveto{\pgfqpoint{0.675467in}{0.984000in}}%
\pgfpathlineto{\pgfqpoint{0.843200in}{0.984000in}}%
\pgfpathlineto{\pgfqpoint{0.843200in}{0.692000in}}%
\pgfpathlineto{\pgfqpoint{0.675467in}{0.692000in}}%
\pgfpathlineto{\pgfqpoint{0.675467in}{0.984000in}}%
\pgfusepath{stroke,fill}%
\end{pgfscope}%
\begin{pgfscope}%
\pgfpathrectangle{\pgfqpoint{0.340000in}{0.400000in}}{\pgfqpoint{1.006400in}{1.168000in}}%
\pgfusepath{clip}%
\pgfsetbuttcap%
\pgfsetroundjoin%
\definecolor{currentfill}{rgb}{0.271104,0.360011,0.807095}%
\pgfsetfillcolor{currentfill}%
\pgfsetlinewidth{0.602250pt}%
\definecolor{currentstroke}{rgb}{0.000000,0.000000,0.000000}%
\pgfsetstrokecolor{currentstroke}%
\pgfsetdash{}{0pt}%
\pgfpathmoveto{\pgfqpoint{0.843200in}{0.984000in}}%
\pgfpathlineto{\pgfqpoint{1.010933in}{0.984000in}}%
\pgfpathlineto{\pgfqpoint{1.010933in}{0.692000in}}%
\pgfpathlineto{\pgfqpoint{0.843200in}{0.692000in}}%
\pgfpathlineto{\pgfqpoint{0.843200in}{0.984000in}}%
\pgfusepath{stroke,fill}%
\end{pgfscope}%
\begin{pgfscope}%
\pgfpathrectangle{\pgfqpoint{0.340000in}{0.400000in}}{\pgfqpoint{1.006400in}{1.168000in}}%
\pgfusepath{clip}%
\pgfsetbuttcap%
\pgfsetroundjoin%
\definecolor{currentfill}{rgb}{0.333490,0.446265,0.874452}%
\pgfsetfillcolor{currentfill}%
\pgfsetlinewidth{0.602250pt}%
\definecolor{currentstroke}{rgb}{0.000000,0.000000,0.000000}%
\pgfsetstrokecolor{currentstroke}%
\pgfsetdash{}{0pt}%
\pgfpathmoveto{\pgfqpoint{1.010933in}{0.984000in}}%
\pgfpathlineto{\pgfqpoint{1.178667in}{0.984000in}}%
\pgfpathlineto{\pgfqpoint{1.178667in}{0.692000in}}%
\pgfpathlineto{\pgfqpoint{1.010933in}{0.692000in}}%
\pgfpathlineto{\pgfqpoint{1.010933in}{0.984000in}}%
\pgfusepath{stroke,fill}%
\end{pgfscope}%
\begin{pgfscope}%
\pgfpathrectangle{\pgfqpoint{0.340000in}{0.400000in}}{\pgfqpoint{1.006400in}{1.168000in}}%
\pgfusepath{clip}%
\pgfsetbuttcap%
\pgfsetroundjoin%
\definecolor{currentfill}{rgb}{0.378598,0.503856,0.913692}%
\pgfsetfillcolor{currentfill}%
\pgfsetlinewidth{0.602250pt}%
\definecolor{currentstroke}{rgb}{0.000000,0.000000,0.000000}%
\pgfsetstrokecolor{currentstroke}%
\pgfsetdash{}{0pt}%
\pgfpathmoveto{\pgfqpoint{1.178667in}{0.984000in}}%
\pgfpathlineto{\pgfqpoint{1.346400in}{0.984000in}}%
\pgfpathlineto{\pgfqpoint{1.346400in}{0.692000in}}%
\pgfpathlineto{\pgfqpoint{1.178667in}{0.692000in}}%
\pgfpathlineto{\pgfqpoint{1.178667in}{0.984000in}}%
\pgfusepath{stroke,fill}%
\end{pgfscope}%
\begin{pgfscope}%
\pgfpathrectangle{\pgfqpoint{0.340000in}{0.400000in}}{\pgfqpoint{1.006400in}{1.168000in}}%
\pgfusepath{clip}%
\pgfsetbuttcap%
\pgfsetroundjoin%
\definecolor{currentfill}{rgb}{0.229806,0.298718,0.753683}%
\pgfsetfillcolor{currentfill}%
\pgfsetlinewidth{0.602250pt}%
\definecolor{currentstroke}{rgb}{0.000000,0.000000,0.000000}%
\pgfsetstrokecolor{currentstroke}%
\pgfsetdash{}{0pt}%
\pgfpathmoveto{\pgfqpoint{0.340000in}{0.692000in}}%
\pgfpathlineto{\pgfqpoint{0.507733in}{0.692000in}}%
\pgfpathlineto{\pgfqpoint{0.507733in}{0.400000in}}%
\pgfpathlineto{\pgfqpoint{0.340000in}{0.400000in}}%
\pgfpathlineto{\pgfqpoint{0.340000in}{0.692000in}}%
\pgfusepath{stroke,fill}%
\end{pgfscope}%
\begin{pgfscope}%
\pgfpathrectangle{\pgfqpoint{0.340000in}{0.400000in}}{\pgfqpoint{1.006400in}{1.168000in}}%
\pgfusepath{clip}%
\pgfsetbuttcap%
\pgfsetroundjoin%
\definecolor{currentfill}{rgb}{0.229806,0.298718,0.753683}%
\pgfsetfillcolor{currentfill}%
\pgfsetlinewidth{0.602250pt}%
\definecolor{currentstroke}{rgb}{0.000000,0.000000,0.000000}%
\pgfsetstrokecolor{currentstroke}%
\pgfsetdash{}{0pt}%
\pgfpathmoveto{\pgfqpoint{0.507733in}{0.692000in}}%
\pgfpathlineto{\pgfqpoint{0.675467in}{0.692000in}}%
\pgfpathlineto{\pgfqpoint{0.675467in}{0.400000in}}%
\pgfpathlineto{\pgfqpoint{0.507733in}{0.400000in}}%
\pgfpathlineto{\pgfqpoint{0.507733in}{0.692000in}}%
\pgfusepath{stroke,fill}%
\end{pgfscope}%
\begin{pgfscope}%
\pgfpathrectangle{\pgfqpoint{0.340000in}{0.400000in}}{\pgfqpoint{1.006400in}{1.168000in}}%
\pgfusepath{clip}%
\pgfsetbuttcap%
\pgfsetroundjoin%
\definecolor{currentfill}{rgb}{0.229806,0.298718,0.753683}%
\pgfsetfillcolor{currentfill}%
\pgfsetlinewidth{0.602250pt}%
\definecolor{currentstroke}{rgb}{0.000000,0.000000,0.000000}%
\pgfsetstrokecolor{currentstroke}%
\pgfsetdash{}{0pt}%
\pgfpathmoveto{\pgfqpoint{0.675467in}{0.692000in}}%
\pgfpathlineto{\pgfqpoint{0.843200in}{0.692000in}}%
\pgfpathlineto{\pgfqpoint{0.843200in}{0.400000in}}%
\pgfpathlineto{\pgfqpoint{0.675467in}{0.400000in}}%
\pgfpathlineto{\pgfqpoint{0.675467in}{0.692000in}}%
\pgfusepath{stroke,fill}%
\end{pgfscope}%
\begin{pgfscope}%
\pgfpathrectangle{\pgfqpoint{0.340000in}{0.400000in}}{\pgfqpoint{1.006400in}{1.168000in}}%
\pgfusepath{clip}%
\pgfsetbuttcap%
\pgfsetroundjoin%
\definecolor{currentfill}{rgb}{0.234377,0.305542,0.759680}%
\pgfsetfillcolor{currentfill}%
\pgfsetlinewidth{0.602250pt}%
\definecolor{currentstroke}{rgb}{0.000000,0.000000,0.000000}%
\pgfsetstrokecolor{currentstroke}%
\pgfsetdash{}{0pt}%
\pgfpathmoveto{\pgfqpoint{0.843200in}{0.692000in}}%
\pgfpathlineto{\pgfqpoint{1.010933in}{0.692000in}}%
\pgfpathlineto{\pgfqpoint{1.010933in}{0.400000in}}%
\pgfpathlineto{\pgfqpoint{0.843200in}{0.400000in}}%
\pgfpathlineto{\pgfqpoint{0.843200in}{0.692000in}}%
\pgfusepath{stroke,fill}%
\end{pgfscope}%
\begin{pgfscope}%
\pgfpathrectangle{\pgfqpoint{0.340000in}{0.400000in}}{\pgfqpoint{1.006400in}{1.168000in}}%
\pgfusepath{clip}%
\pgfsetbuttcap%
\pgfsetroundjoin%
\definecolor{currentfill}{rgb}{0.238948,0.312365,0.765676}%
\pgfsetfillcolor{currentfill}%
\pgfsetlinewidth{0.602250pt}%
\definecolor{currentstroke}{rgb}{0.000000,0.000000,0.000000}%
\pgfsetstrokecolor{currentstroke}%
\pgfsetdash{}{0pt}%
\pgfpathmoveto{\pgfqpoint{1.010933in}{0.692000in}}%
\pgfpathlineto{\pgfqpoint{1.178667in}{0.692000in}}%
\pgfpathlineto{\pgfqpoint{1.178667in}{0.400000in}}%
\pgfpathlineto{\pgfqpoint{1.010933in}{0.400000in}}%
\pgfpathlineto{\pgfqpoint{1.010933in}{0.692000in}}%
\pgfusepath{stroke,fill}%
\end{pgfscope}%
\begin{pgfscope}%
\pgfpathrectangle{\pgfqpoint{0.340000in}{0.400000in}}{\pgfqpoint{1.006400in}{1.168000in}}%
\pgfusepath{clip}%
\pgfsetbuttcap%
\pgfsetroundjoin%
\definecolor{currentfill}{rgb}{0.238948,0.312365,0.765676}%
\pgfsetfillcolor{currentfill}%
\pgfsetlinewidth{0.602250pt}%
\definecolor{currentstroke}{rgb}{0.000000,0.000000,0.000000}%
\pgfsetstrokecolor{currentstroke}%
\pgfsetdash{}{0pt}%
\pgfpathmoveto{\pgfqpoint{1.178667in}{0.692000in}}%
\pgfpathlineto{\pgfqpoint{1.346400in}{0.692000in}}%
\pgfpathlineto{\pgfqpoint{1.346400in}{0.400000in}}%
\pgfpathlineto{\pgfqpoint{1.178667in}{0.400000in}}%
\pgfpathlineto{\pgfqpoint{1.178667in}{0.692000in}}%
\pgfusepath{stroke,fill}%
\end{pgfscope}%
\begin{pgfscope}%
\definecolor{textcolor}{rgb}{1.000000,1.000000,1.000000}%
\pgfsetstrokecolor{textcolor}%
\pgfsetfillcolor{textcolor}%
\pgftext[x=0.423867in,y=1.422000in,,]{\color{textcolor}\rmfamily\fontsize{6.000000}{7.200000}\selectfont 95}%
\end{pgfscope}%
\begin{pgfscope}%
\definecolor{textcolor}{rgb}{0.150000,0.150000,0.150000}%
\pgfsetstrokecolor{textcolor}%
\pgfsetfillcolor{textcolor}%
\pgftext[x=0.591600in,y=1.422000in,,]{\color{textcolor}\rmfamily\fontsize{6.000000}{7.200000}\selectfont 162}%
\end{pgfscope}%
\begin{pgfscope}%
\definecolor{textcolor}{rgb}{0.150000,0.150000,0.150000}%
\pgfsetstrokecolor{textcolor}%
\pgfsetfillcolor{textcolor}%
\pgftext[x=0.759333in,y=1.422000in,,]{\color{textcolor}\rmfamily\fontsize{6.000000}{7.200000}\selectfont 291}%
\end{pgfscope}%
\begin{pgfscope}%
\definecolor{textcolor}{rgb}{0.150000,0.150000,0.150000}%
\pgfsetstrokecolor{textcolor}%
\pgfsetfillcolor{textcolor}%
\pgftext[x=0.927067in,y=1.422000in,,]{\color{textcolor}\rmfamily\fontsize{6.000000}{7.200000}\selectfont 411}%
\end{pgfscope}%
\begin{pgfscope}%
\definecolor{textcolor}{rgb}{1.000000,1.000000,1.000000}%
\pgfsetstrokecolor{textcolor}%
\pgfsetfillcolor{textcolor}%
\pgftext[x=1.094800in,y=1.422000in,,]{\color{textcolor}\rmfamily\fontsize{6.000000}{7.200000}\selectfont 495}%
\end{pgfscope}%
\begin{pgfscope}%
\definecolor{textcolor}{rgb}{1.000000,1.000000,1.000000}%
\pgfsetstrokecolor{textcolor}%
\pgfsetfillcolor{textcolor}%
\pgftext[x=1.262533in,y=1.422000in,,]{\color{textcolor}\rmfamily\fontsize{6.000000}{7.200000}\selectfont 572}%
\end{pgfscope}%
\begin{pgfscope}%
\definecolor{textcolor}{rgb}{1.000000,1.000000,1.000000}%
\pgfsetstrokecolor{textcolor}%
\pgfsetfillcolor{textcolor}%
\pgftext[x=0.423867in,y=1.130000in,,]{\color{textcolor}\rmfamily\fontsize{6.000000}{7.200000}\selectfont 19}%
\end{pgfscope}%
\begin{pgfscope}%
\definecolor{textcolor}{rgb}{1.000000,1.000000,1.000000}%
\pgfsetstrokecolor{textcolor}%
\pgfsetfillcolor{textcolor}%
\pgftext[x=0.591600in,y=1.130000in,,]{\color{textcolor}\rmfamily\fontsize{6.000000}{7.200000}\selectfont 29}%
\end{pgfscope}%
\begin{pgfscope}%
\definecolor{textcolor}{rgb}{1.000000,1.000000,1.000000}%
\pgfsetstrokecolor{textcolor}%
\pgfsetfillcolor{textcolor}%
\pgftext[x=0.759333in,y=1.130000in,,]{\color{textcolor}\rmfamily\fontsize{6.000000}{7.200000}\selectfont 90}%
\end{pgfscope}%
\begin{pgfscope}%
\definecolor{textcolor}{rgb}{0.150000,0.150000,0.150000}%
\pgfsetstrokecolor{textcolor}%
\pgfsetfillcolor{textcolor}%
\pgftext[x=0.927067in,y=1.130000in,,]{\color{textcolor}\rmfamily\fontsize{6.000000}{7.200000}\selectfont 204}%
\end{pgfscope}%
\begin{pgfscope}%
\definecolor{textcolor}{rgb}{0.150000,0.150000,0.150000}%
\pgfsetstrokecolor{textcolor}%
\pgfsetfillcolor{textcolor}%
\pgftext[x=1.094800in,y=1.130000in,,]{\color{textcolor}\rmfamily\fontsize{6.000000}{7.200000}\selectfont 305}%
\end{pgfscope}%
\begin{pgfscope}%
\definecolor{textcolor}{rgb}{0.150000,0.150000,0.150000}%
\pgfsetstrokecolor{textcolor}%
\pgfsetfillcolor{textcolor}%
\pgftext[x=1.262533in,y=1.130000in,,]{\color{textcolor}\rmfamily\fontsize{6.000000}{7.200000}\selectfont 359}%
\end{pgfscope}%
\begin{pgfscope}%
\definecolor{textcolor}{rgb}{1.000000,1.000000,1.000000}%
\pgfsetstrokecolor{textcolor}%
\pgfsetfillcolor{textcolor}%
\pgftext[x=0.423867in,y=0.838000in,,]{\color{textcolor}\rmfamily\fontsize{6.000000}{7.200000}\selectfont 7}%
\end{pgfscope}%
\begin{pgfscope}%
\definecolor{textcolor}{rgb}{1.000000,1.000000,1.000000}%
\pgfsetstrokecolor{textcolor}%
\pgfsetfillcolor{textcolor}%
\pgftext[x=0.591600in,y=0.838000in,,]{\color{textcolor}\rmfamily\fontsize{6.000000}{7.200000}\selectfont 9}%
\end{pgfscope}%
\begin{pgfscope}%
\definecolor{textcolor}{rgb}{1.000000,1.000000,1.000000}%
\pgfsetstrokecolor{textcolor}%
\pgfsetfillcolor{textcolor}%
\pgftext[x=0.759333in,y=0.838000in,,]{\color{textcolor}\rmfamily\fontsize{6.000000}{7.200000}\selectfont 14}%
\end{pgfscope}%
\begin{pgfscope}%
\definecolor{textcolor}{rgb}{1.000000,1.000000,1.000000}%
\pgfsetstrokecolor{textcolor}%
\pgfsetfillcolor{textcolor}%
\pgftext[x=0.927067in,y=0.838000in,,]{\color{textcolor}\rmfamily\fontsize{6.000000}{7.200000}\selectfont 24}%
\end{pgfscope}%
\begin{pgfscope}%
\definecolor{textcolor}{rgb}{1.000000,1.000000,1.000000}%
\pgfsetstrokecolor{textcolor}%
\pgfsetfillcolor{textcolor}%
\pgftext[x=1.094800in,y=0.838000in,,]{\color{textcolor}\rmfamily\fontsize{6.000000}{7.200000}\selectfont 52}%
\end{pgfscope}%
\begin{pgfscope}%
\definecolor{textcolor}{rgb}{1.000000,1.000000,1.000000}%
\pgfsetstrokecolor{textcolor}%
\pgfsetfillcolor{textcolor}%
\pgftext[x=1.262533in,y=0.838000in,,]{\color{textcolor}\rmfamily\fontsize{6.000000}{7.200000}\selectfont 69}%
\end{pgfscope}%
\begin{pgfscope}%
\definecolor{textcolor}{rgb}{1.000000,1.000000,1.000000}%
\pgfsetstrokecolor{textcolor}%
\pgfsetfillcolor{textcolor}%
\pgftext[x=0.423867in,y=0.546000in,,]{\color{textcolor}\rmfamily\fontsize{6.000000}{7.200000}\selectfont 4}%
\end{pgfscope}%
\begin{pgfscope}%
\definecolor{textcolor}{rgb}{1.000000,1.000000,1.000000}%
\pgfsetstrokecolor{textcolor}%
\pgfsetfillcolor{textcolor}%
\pgftext[x=0.591600in,y=0.546000in,,]{\color{textcolor}\rmfamily\fontsize{6.000000}{7.200000}\selectfont 5}%
\end{pgfscope}%
\begin{pgfscope}%
\definecolor{textcolor}{rgb}{1.000000,1.000000,1.000000}%
\pgfsetstrokecolor{textcolor}%
\pgfsetfillcolor{textcolor}%
\pgftext[x=0.759333in,y=0.546000in,,]{\color{textcolor}\rmfamily\fontsize{6.000000}{7.200000}\selectfont 5}%
\end{pgfscope}%
\begin{pgfscope}%
\definecolor{textcolor}{rgb}{1.000000,1.000000,1.000000}%
\pgfsetstrokecolor{textcolor}%
\pgfsetfillcolor{textcolor}%
\pgftext[x=0.927067in,y=0.546000in,,]{\color{textcolor}\rmfamily\fontsize{6.000000}{7.200000}\selectfont 7}%
\end{pgfscope}%
\begin{pgfscope}%
\definecolor{textcolor}{rgb}{1.000000,1.000000,1.000000}%
\pgfsetstrokecolor{textcolor}%
\pgfsetfillcolor{textcolor}%
\pgftext[x=1.094800in,y=0.546000in,,]{\color{textcolor}\rmfamily\fontsize{6.000000}{7.200000}\selectfont 9}%
\end{pgfscope}%
\begin{pgfscope}%
\definecolor{textcolor}{rgb}{1.000000,1.000000,1.000000}%
\pgfsetstrokecolor{textcolor}%
\pgfsetfillcolor{textcolor}%
\pgftext[x=1.262533in,y=0.546000in,,]{\color{textcolor}\rmfamily\fontsize{6.000000}{7.200000}\selectfont 9}%
\end{pgfscope}%
\end{pgfpicture}%
\makeatother%
\endgroup%

%% file: plot_data/packet_level_simulation/packet_level_average_rtt.pgf
\begingroup%
\makeatletter%
\begin{pgfpicture}%
\pgfpathrectangle{\pgfpointorigin}{\pgfqpoint{1.360000in}{1.600000in}}%
\pgfusepath{use as bounding box, clip}%
\begin{pgfscope}%
\pgfsetbuttcap%
\pgfsetmiterjoin%
\definecolor{currentfill}{rgb}{1.000000,1.000000,1.000000}%
\pgfsetfillcolor{currentfill}%
\pgfsetlinewidth{0.000000pt}%
\definecolor{currentstroke}{rgb}{1.000000,1.000000,1.000000}%
\pgfsetstrokecolor{currentstroke}%
\pgfsetdash{}{0pt}%
\pgfpathmoveto{\pgfqpoint{0.000000in}{0.000000in}}%
\pgfpathlineto{\pgfqpoint{1.360000in}{0.000000in}}%
\pgfpathlineto{\pgfqpoint{1.360000in}{1.600000in}}%
\pgfpathlineto{\pgfqpoint{0.000000in}{1.600000in}}%
\pgfpathclose%
\pgfusepath{fill}%
\end{pgfscope}%
\begin{pgfscope}%
\pgfsetbuttcap%
\pgfsetmiterjoin%
\definecolor{currentfill}{rgb}{0.917647,0.917647,0.949020}%
\pgfsetfillcolor{currentfill}%
\pgfsetlinewidth{0.000000pt}%
\definecolor{currentstroke}{rgb}{0.000000,0.000000,0.000000}%
\pgfsetstrokecolor{currentstroke}%
\pgfsetstrokeopacity{0.000000}%
\pgfsetdash{}{0pt}%
\pgfpathmoveto{\pgfqpoint{0.340000in}{0.400000in}}%
\pgfpathlineto{\pgfqpoint{1.346400in}{0.400000in}}%
\pgfpathlineto{\pgfqpoint{1.346400in}{1.568000in}}%
\pgfpathlineto{\pgfqpoint{0.340000in}{1.568000in}}%
\pgfpathclose%
\pgfusepath{fill}%
\end{pgfscope}%
\begin{pgfscope}%
\definecolor{textcolor}{rgb}{0.150000,0.150000,0.150000}%
\pgfsetstrokecolor{textcolor}%
\pgfsetfillcolor{textcolor}%
\pgftext[x=0.444700in,y=0.136885in,left,base,rotate=90.000000]{\color{textcolor}\rmfamily\fontsize{6.500000}{7.800000}\selectfont 1.1}%
\end{pgfscope}%
\begin{pgfscope}%
\definecolor{textcolor}{rgb}{0.150000,0.150000,0.150000}%
\pgfsetstrokecolor{textcolor}%
\pgfsetfillcolor{textcolor}%
\pgftext[x=0.612433in,y=0.136885in,left,base,rotate=90.000000]{\color{textcolor}\rmfamily\fontsize{6.500000}{7.800000}\selectfont 1.2}%
\end{pgfscope}%
\begin{pgfscope}%
\definecolor{textcolor}{rgb}{0.150000,0.150000,0.150000}%
\pgfsetstrokecolor{textcolor}%
\pgfsetfillcolor{textcolor}%
\pgftext[x=0.780167in,y=0.136885in,left,base,rotate=90.000000]{\color{textcolor}\rmfamily\fontsize{6.500000}{7.800000}\selectfont 1.4}%
\end{pgfscope}%
\begin{pgfscope}%
\definecolor{textcolor}{rgb}{0.150000,0.150000,0.150000}%
\pgfsetstrokecolor{textcolor}%
\pgfsetfillcolor{textcolor}%
\pgftext[x=0.947900in,y=0.136885in,left,base,rotate=90.000000]{\color{textcolor}\rmfamily\fontsize{6.500000}{7.800000}\selectfont 1.6}%
\end{pgfscope}%
\begin{pgfscope}%
\definecolor{textcolor}{rgb}{0.150000,0.150000,0.150000}%
\pgfsetstrokecolor{textcolor}%
\pgfsetfillcolor{textcolor}%
\pgftext[x=1.115633in,y=0.136885in,left,base,rotate=90.000000]{\color{textcolor}\rmfamily\fontsize{6.500000}{7.800000}\selectfont 1.8}%
\end{pgfscope}%
\begin{pgfscope}%
\definecolor{textcolor}{rgb}{0.150000,0.150000,0.150000}%
\pgfsetstrokecolor{textcolor}%
\pgfsetfillcolor{textcolor}%
\pgftext[x=1.283367in,y=0.136885in,left,base,rotate=90.000000]{\color{textcolor}\rmfamily\fontsize{6.500000}{7.800000}\selectfont 1.9}%
\end{pgfscope}%
\begin{pgfscope}%
\definecolor{textcolor}{rgb}{0.150000,0.150000,0.150000}%
\pgfsetstrokecolor{textcolor}%
\pgfsetfillcolor{textcolor}%
\pgftext[x=0.843200in,y=0.118829in,,top]{\color{textcolor}\rmfamily\fontsize{7.400000}{8.880000}\selectfont Avg. Hop Count}%
\end{pgfscope}%
\begin{pgfscope}%
\definecolor{textcolor}{rgb}{0.150000,0.150000,0.150000}%
\pgfsetstrokecolor{textcolor}%
\pgfsetfillcolor{textcolor}%
\pgftext[x=0.191852in,y=1.319764in,left,base,rotate=90.000000]{\color{textcolor}\rmfamily\fontsize{6.500000}{7.800000}\selectfont 0.8}%
\end{pgfscope}%
\begin{pgfscope}%
\definecolor{textcolor}{rgb}{0.150000,0.150000,0.150000}%
\pgfsetstrokecolor{textcolor}%
\pgfsetfillcolor{textcolor}%
\pgftext[x=0.191852in,y=1.027764in,left,base,rotate=90.000000]{\color{textcolor}\rmfamily\fontsize{6.500000}{7.800000}\selectfont 0.6}%
\end{pgfscope}%
\begin{pgfscope}%
\definecolor{textcolor}{rgb}{0.150000,0.150000,0.150000}%
\pgfsetstrokecolor{textcolor}%
\pgfsetfillcolor{textcolor}%
\pgftext[x=0.191852in,y=0.735764in,left,base,rotate=90.000000]{\color{textcolor}\rmfamily\fontsize{6.500000}{7.800000}\selectfont 0.4}%
\end{pgfscope}%
\begin{pgfscope}%
\definecolor{textcolor}{rgb}{0.150000,0.150000,0.150000}%
\pgfsetstrokecolor{textcolor}%
\pgfsetfillcolor{textcolor}%
\pgftext[x=0.191852in,y=0.443764in,left,base,rotate=90.000000]{\color{textcolor}\rmfamily\fontsize{6.500000}{7.800000}\selectfont 0.2}%
\end{pgfscope}%
\begin{pgfscope}%
\definecolor{textcolor}{rgb}{0.150000,0.150000,0.150000}%
\pgfsetstrokecolor{textcolor}%
\pgfsetfillcolor{textcolor}%
\pgftext[x=0.120093in,y=0.984000in,,bottom,rotate=90.000000]{\color{textcolor}\rmfamily\fontsize{7.400000}{8.880000}\selectfont MLU}%
\end{pgfscope}%
\begin{pgfscope}%
\pgfpathrectangle{\pgfqpoint{0.340000in}{0.400000in}}{\pgfqpoint{1.006400in}{1.168000in}}%
\pgfusepath{clip}%
\pgfsetbuttcap%
\pgfsetroundjoin%
\definecolor{currentfill}{rgb}{0.483854,0.622050,0.974808}%
\pgfsetfillcolor{currentfill}%
\pgfsetlinewidth{0.602250pt}%
\definecolor{currentstroke}{rgb}{0.000000,0.000000,0.000000}%
\pgfsetstrokecolor{currentstroke}%
\pgfsetdash{}{0pt}%
\pgfpathmoveto{\pgfqpoint{0.340000in}{1.568000in}}%
\pgfpathlineto{\pgfqpoint{0.507733in}{1.568000in}}%
\pgfpathlineto{\pgfqpoint{0.507733in}{1.276000in}}%
\pgfpathlineto{\pgfqpoint{0.340000in}{1.276000in}}%
\pgfpathlineto{\pgfqpoint{0.340000in}{1.568000in}}%
\pgfusepath{stroke,fill}%
\end{pgfscope}%
\begin{pgfscope}%
\pgfpathrectangle{\pgfqpoint{0.340000in}{0.400000in}}{\pgfqpoint{1.006400in}{1.168000in}}%
\pgfusepath{clip}%
\pgfsetbuttcap%
\pgfsetroundjoin%
\definecolor{currentfill}{rgb}{0.619318,0.744121,0.998931}%
\pgfsetfillcolor{currentfill}%
\pgfsetlinewidth{0.602250pt}%
\definecolor{currentstroke}{rgb}{0.000000,0.000000,0.000000}%
\pgfsetstrokecolor{currentstroke}%
\pgfsetdash{}{0pt}%
\pgfpathmoveto{\pgfqpoint{0.507733in}{1.568000in}}%
\pgfpathlineto{\pgfqpoint{0.675467in}{1.568000in}}%
\pgfpathlineto{\pgfqpoint{0.675467in}{1.276000in}}%
\pgfpathlineto{\pgfqpoint{0.507733in}{1.276000in}}%
\pgfpathlineto{\pgfqpoint{0.507733in}{1.568000in}}%
\pgfusepath{stroke,fill}%
\end{pgfscope}%
\begin{pgfscope}%
\pgfpathrectangle{\pgfqpoint{0.340000in}{0.400000in}}{\pgfqpoint{1.006400in}{1.168000in}}%
\pgfusepath{clip}%
\pgfsetbuttcap%
\pgfsetroundjoin%
\definecolor{currentfill}{rgb}{0.831148,0.859513,0.903110}%
\pgfsetfillcolor{currentfill}%
\pgfsetlinewidth{0.602250pt}%
\definecolor{currentstroke}{rgb}{0.000000,0.000000,0.000000}%
\pgfsetstrokecolor{currentstroke}%
\pgfsetdash{}{0pt}%
\pgfpathmoveto{\pgfqpoint{0.675467in}{1.568000in}}%
\pgfpathlineto{\pgfqpoint{0.843200in}{1.568000in}}%
\pgfpathlineto{\pgfqpoint{0.843200in}{1.276000in}}%
\pgfpathlineto{\pgfqpoint{0.675467in}{1.276000in}}%
\pgfpathlineto{\pgfqpoint{0.675467in}{1.568000in}}%
\pgfusepath{stroke,fill}%
\end{pgfscope}%
\begin{pgfscope}%
\pgfpathrectangle{\pgfqpoint{0.340000in}{0.400000in}}{\pgfqpoint{1.006400in}{1.168000in}}%
\pgfusepath{clip}%
\pgfsetbuttcap%
\pgfsetroundjoin%
\definecolor{currentfill}{rgb}{0.969851,0.695830,0.581312}%
\pgfsetfillcolor{currentfill}%
\pgfsetlinewidth{0.602250pt}%
\definecolor{currentstroke}{rgb}{0.000000,0.000000,0.000000}%
\pgfsetstrokecolor{currentstroke}%
\pgfsetdash{}{0pt}%
\pgfpathmoveto{\pgfqpoint{0.843200in}{1.568000in}}%
\pgfpathlineto{\pgfqpoint{1.010933in}{1.568000in}}%
\pgfpathlineto{\pgfqpoint{1.010933in}{1.276000in}}%
\pgfpathlineto{\pgfqpoint{0.843200in}{1.276000in}}%
\pgfpathlineto{\pgfqpoint{0.843200in}{1.568000in}}%
\pgfusepath{stroke,fill}%
\end{pgfscope}%
\begin{pgfscope}%
\pgfpathrectangle{\pgfqpoint{0.340000in}{0.400000in}}{\pgfqpoint{1.006400in}{1.168000in}}%
\pgfusepath{clip}%
\pgfsetbuttcap%
\pgfsetroundjoin%
\definecolor{currentfill}{rgb}{0.820401,0.286765,0.245160}%
\pgfsetfillcolor{currentfill}%
\pgfsetlinewidth{0.602250pt}%
\definecolor{currentstroke}{rgb}{0.000000,0.000000,0.000000}%
\pgfsetstrokecolor{currentstroke}%
\pgfsetdash{}{0pt}%
\pgfpathmoveto{\pgfqpoint{1.010933in}{1.568000in}}%
\pgfpathlineto{\pgfqpoint{1.178667in}{1.568000in}}%
\pgfpathlineto{\pgfqpoint{1.178667in}{1.276000in}}%
\pgfpathlineto{\pgfqpoint{1.010933in}{1.276000in}}%
\pgfpathlineto{\pgfqpoint{1.010933in}{1.568000in}}%
\pgfusepath{stroke,fill}%
\end{pgfscope}%
\begin{pgfscope}%
\pgfpathrectangle{\pgfqpoint{0.340000in}{0.400000in}}{\pgfqpoint{1.006400in}{1.168000in}}%
\pgfusepath{clip}%
\pgfsetbuttcap%
\pgfsetroundjoin%
\definecolor{currentfill}{rgb}{0.705673,0.015556,0.150233}%
\pgfsetfillcolor{currentfill}%
\pgfsetlinewidth{0.602250pt}%
\definecolor{currentstroke}{rgb}{0.000000,0.000000,0.000000}%
\pgfsetstrokecolor{currentstroke}%
\pgfsetdash{}{0pt}%
\pgfpathmoveto{\pgfqpoint{1.178667in}{1.568000in}}%
\pgfpathlineto{\pgfqpoint{1.346400in}{1.568000in}}%
\pgfpathlineto{\pgfqpoint{1.346400in}{1.276000in}}%
\pgfpathlineto{\pgfqpoint{1.178667in}{1.276000in}}%
\pgfpathlineto{\pgfqpoint{1.178667in}{1.568000in}}%
\pgfusepath{stroke,fill}%
\end{pgfscope}%
\begin{pgfscope}%
\pgfpathrectangle{\pgfqpoint{0.340000in}{0.400000in}}{\pgfqpoint{1.006400in}{1.168000in}}%
\pgfusepath{clip}%
\pgfsetbuttcap%
\pgfsetroundjoin%
\definecolor{currentfill}{rgb}{0.294718,0.393542,0.834384}%
\pgfsetfillcolor{currentfill}%
\pgfsetlinewidth{0.602250pt}%
\definecolor{currentstroke}{rgb}{0.000000,0.000000,0.000000}%
\pgfsetstrokecolor{currentstroke}%
\pgfsetdash{}{0pt}%
\pgfpathmoveto{\pgfqpoint{0.340000in}{1.276000in}}%
\pgfpathlineto{\pgfqpoint{0.507733in}{1.276000in}}%
\pgfpathlineto{\pgfqpoint{0.507733in}{0.984000in}}%
\pgfpathlineto{\pgfqpoint{0.340000in}{0.984000in}}%
\pgfpathlineto{\pgfqpoint{0.340000in}{1.276000in}}%
\pgfusepath{stroke,fill}%
\end{pgfscope}%
\begin{pgfscope}%
\pgfpathrectangle{\pgfqpoint{0.340000in}{0.400000in}}{\pgfqpoint{1.006400in}{1.168000in}}%
\pgfusepath{clip}%
\pgfsetbuttcap%
\pgfsetroundjoin%
\definecolor{currentfill}{rgb}{0.318832,0.426605,0.859857}%
\pgfsetfillcolor{currentfill}%
\pgfsetlinewidth{0.602250pt}%
\definecolor{currentstroke}{rgb}{0.000000,0.000000,0.000000}%
\pgfsetstrokecolor{currentstroke}%
\pgfsetdash{}{0pt}%
\pgfpathmoveto{\pgfqpoint{0.507733in}{1.276000in}}%
\pgfpathlineto{\pgfqpoint{0.675467in}{1.276000in}}%
\pgfpathlineto{\pgfqpoint{0.675467in}{0.984000in}}%
\pgfpathlineto{\pgfqpoint{0.507733in}{0.984000in}}%
\pgfpathlineto{\pgfqpoint{0.507733in}{1.276000in}}%
\pgfusepath{stroke,fill}%
\end{pgfscope}%
\begin{pgfscope}%
\pgfpathrectangle{\pgfqpoint{0.340000in}{0.400000in}}{\pgfqpoint{1.006400in}{1.168000in}}%
\pgfusepath{clip}%
\pgfsetbuttcap%
\pgfsetroundjoin%
\definecolor{currentfill}{rgb}{0.430507,0.564883,0.948889}%
\pgfsetfillcolor{currentfill}%
\pgfsetlinewidth{0.602250pt}%
\definecolor{currentstroke}{rgb}{0.000000,0.000000,0.000000}%
\pgfsetstrokecolor{currentstroke}%
\pgfsetdash{}{0pt}%
\pgfpathmoveto{\pgfqpoint{0.675467in}{1.276000in}}%
\pgfpathlineto{\pgfqpoint{0.843200in}{1.276000in}}%
\pgfpathlineto{\pgfqpoint{0.843200in}{0.984000in}}%
\pgfpathlineto{\pgfqpoint{0.675467in}{0.984000in}}%
\pgfpathlineto{\pgfqpoint{0.675467in}{1.276000in}}%
\pgfusepath{stroke,fill}%
\end{pgfscope}%
\begin{pgfscope}%
\pgfpathrectangle{\pgfqpoint{0.340000in}{0.400000in}}{\pgfqpoint{1.006400in}{1.168000in}}%
\pgfusepath{clip}%
\pgfsetbuttcap%
\pgfsetroundjoin%
\definecolor{currentfill}{rgb}{0.608547,0.735725,0.999354}%
\pgfsetfillcolor{currentfill}%
\pgfsetlinewidth{0.602250pt}%
\definecolor{currentstroke}{rgb}{0.000000,0.000000,0.000000}%
\pgfsetstrokecolor{currentstroke}%
\pgfsetdash{}{0pt}%
\pgfpathmoveto{\pgfqpoint{0.843200in}{1.276000in}}%
\pgfpathlineto{\pgfqpoint{1.010933in}{1.276000in}}%
\pgfpathlineto{\pgfqpoint{1.010933in}{0.984000in}}%
\pgfpathlineto{\pgfqpoint{0.843200in}{0.984000in}}%
\pgfpathlineto{\pgfqpoint{0.843200in}{1.276000in}}%
\pgfusepath{stroke,fill}%
\end{pgfscope}%
\begin{pgfscope}%
\pgfpathrectangle{\pgfqpoint{0.340000in}{0.400000in}}{\pgfqpoint{1.006400in}{1.168000in}}%
\pgfusepath{clip}%
\pgfsetbuttcap%
\pgfsetroundjoin%
\definecolor{currentfill}{rgb}{0.782049,0.842864,0.942980}%
\pgfsetfillcolor{currentfill}%
\pgfsetlinewidth{0.602250pt}%
\definecolor{currentstroke}{rgb}{0.000000,0.000000,0.000000}%
\pgfsetstrokecolor{currentstroke}%
\pgfsetdash{}{0pt}%
\pgfpathmoveto{\pgfqpoint{1.010933in}{1.276000in}}%
\pgfpathlineto{\pgfqpoint{1.178667in}{1.276000in}}%
\pgfpathlineto{\pgfqpoint{1.178667in}{0.984000in}}%
\pgfpathlineto{\pgfqpoint{1.010933in}{0.984000in}}%
\pgfpathlineto{\pgfqpoint{1.010933in}{1.276000in}}%
\pgfusepath{stroke,fill}%
\end{pgfscope}%
\begin{pgfscope}%
\pgfpathrectangle{\pgfqpoint{0.340000in}{0.400000in}}{\pgfqpoint{1.006400in}{1.168000in}}%
\pgfusepath{clip}%
\pgfsetbuttcap%
\pgfsetroundjoin%
\definecolor{currentfill}{rgb}{0.847365,0.862472,0.885540}%
\pgfsetfillcolor{currentfill}%
\pgfsetlinewidth{0.602250pt}%
\definecolor{currentstroke}{rgb}{0.000000,0.000000,0.000000}%
\pgfsetstrokecolor{currentstroke}%
\pgfsetdash{}{0pt}%
\pgfpathmoveto{\pgfqpoint{1.178667in}{1.276000in}}%
\pgfpathlineto{\pgfqpoint{1.346400in}{1.276000in}}%
\pgfpathlineto{\pgfqpoint{1.346400in}{0.984000in}}%
\pgfpathlineto{\pgfqpoint{1.178667in}{0.984000in}}%
\pgfpathlineto{\pgfqpoint{1.178667in}{1.276000in}}%
\pgfusepath{stroke,fill}%
\end{pgfscope}%
\begin{pgfscope}%
\pgfpathrectangle{\pgfqpoint{0.340000in}{0.400000in}}{\pgfqpoint{1.006400in}{1.168000in}}%
\pgfusepath{clip}%
\pgfsetbuttcap%
\pgfsetroundjoin%
\definecolor{currentfill}{rgb}{0.252663,0.332837,0.783665}%
\pgfsetfillcolor{currentfill}%
\pgfsetlinewidth{0.602250pt}%
\definecolor{currentstroke}{rgb}{0.000000,0.000000,0.000000}%
\pgfsetstrokecolor{currentstroke}%
\pgfsetdash{}{0pt}%
\pgfpathmoveto{\pgfqpoint{0.340000in}{0.984000in}}%
\pgfpathlineto{\pgfqpoint{0.507733in}{0.984000in}}%
\pgfpathlineto{\pgfqpoint{0.507733in}{0.692000in}}%
\pgfpathlineto{\pgfqpoint{0.340000in}{0.692000in}}%
\pgfpathlineto{\pgfqpoint{0.340000in}{0.984000in}}%
\pgfusepath{stroke,fill}%
\end{pgfscope}%
\begin{pgfscope}%
\pgfpathrectangle{\pgfqpoint{0.340000in}{0.400000in}}{\pgfqpoint{1.006400in}{1.168000in}}%
\pgfusepath{clip}%
\pgfsetbuttcap%
\pgfsetroundjoin%
\definecolor{currentfill}{rgb}{0.266381,0.353304,0.801637}%
\pgfsetfillcolor{currentfill}%
\pgfsetlinewidth{0.602250pt}%
\definecolor{currentstroke}{rgb}{0.000000,0.000000,0.000000}%
\pgfsetstrokecolor{currentstroke}%
\pgfsetdash{}{0pt}%
\pgfpathmoveto{\pgfqpoint{0.507733in}{0.984000in}}%
\pgfpathlineto{\pgfqpoint{0.675467in}{0.984000in}}%
\pgfpathlineto{\pgfqpoint{0.675467in}{0.692000in}}%
\pgfpathlineto{\pgfqpoint{0.507733in}{0.692000in}}%
\pgfpathlineto{\pgfqpoint{0.507733in}{0.984000in}}%
\pgfusepath{stroke,fill}%
\end{pgfscope}%
\begin{pgfscope}%
\pgfpathrectangle{\pgfqpoint{0.340000in}{0.400000in}}{\pgfqpoint{1.006400in}{1.168000in}}%
\pgfusepath{clip}%
\pgfsetbuttcap%
\pgfsetroundjoin%
\definecolor{currentfill}{rgb}{0.304174,0.406945,0.845263}%
\pgfsetfillcolor{currentfill}%
\pgfsetlinewidth{0.602250pt}%
\definecolor{currentstroke}{rgb}{0.000000,0.000000,0.000000}%
\pgfsetstrokecolor{currentstroke}%
\pgfsetdash{}{0pt}%
\pgfpathmoveto{\pgfqpoint{0.675467in}{0.984000in}}%
\pgfpathlineto{\pgfqpoint{0.843200in}{0.984000in}}%
\pgfpathlineto{\pgfqpoint{0.843200in}{0.692000in}}%
\pgfpathlineto{\pgfqpoint{0.675467in}{0.692000in}}%
\pgfpathlineto{\pgfqpoint{0.675467in}{0.984000in}}%
\pgfusepath{stroke,fill}%
\end{pgfscope}%
\begin{pgfscope}%
\pgfpathrectangle{\pgfqpoint{0.340000in}{0.400000in}}{\pgfqpoint{1.006400in}{1.168000in}}%
\pgfusepath{clip}%
\pgfsetbuttcap%
\pgfsetroundjoin%
\definecolor{currentfill}{rgb}{0.328604,0.439712,0.869587}%
\pgfsetfillcolor{currentfill}%
\pgfsetlinewidth{0.602250pt}%
\definecolor{currentstroke}{rgb}{0.000000,0.000000,0.000000}%
\pgfsetstrokecolor{currentstroke}%
\pgfsetdash{}{0pt}%
\pgfpathmoveto{\pgfqpoint{0.843200in}{0.984000in}}%
\pgfpathlineto{\pgfqpoint{1.010933in}{0.984000in}}%
\pgfpathlineto{\pgfqpoint{1.010933in}{0.692000in}}%
\pgfpathlineto{\pgfqpoint{0.843200in}{0.692000in}}%
\pgfpathlineto{\pgfqpoint{0.843200in}{0.984000in}}%
\pgfusepath{stroke,fill}%
\end{pgfscope}%
\begin{pgfscope}%
\pgfpathrectangle{\pgfqpoint{0.340000in}{0.400000in}}{\pgfqpoint{1.006400in}{1.168000in}}%
\pgfusepath{clip}%
\pgfsetbuttcap%
\pgfsetroundjoin%
\definecolor{currentfill}{rgb}{0.383662,0.510183,0.917831}%
\pgfsetfillcolor{currentfill}%
\pgfsetlinewidth{0.602250pt}%
\definecolor{currentstroke}{rgb}{0.000000,0.000000,0.000000}%
\pgfsetstrokecolor{currentstroke}%
\pgfsetdash{}{0pt}%
\pgfpathmoveto{\pgfqpoint{1.010933in}{0.984000in}}%
\pgfpathlineto{\pgfqpoint{1.178667in}{0.984000in}}%
\pgfpathlineto{\pgfqpoint{1.178667in}{0.692000in}}%
\pgfpathlineto{\pgfqpoint{1.010933in}{0.692000in}}%
\pgfpathlineto{\pgfqpoint{1.010933in}{0.984000in}}%
\pgfusepath{stroke,fill}%
\end{pgfscope}%
\begin{pgfscope}%
\pgfpathrectangle{\pgfqpoint{0.340000in}{0.400000in}}{\pgfqpoint{1.006400in}{1.168000in}}%
\pgfusepath{clip}%
\pgfsetbuttcap%
\pgfsetroundjoin%
\definecolor{currentfill}{rgb}{0.404421,0.534643,0.932002}%
\pgfsetfillcolor{currentfill}%
\pgfsetlinewidth{0.602250pt}%
\definecolor{currentstroke}{rgb}{0.000000,0.000000,0.000000}%
\pgfsetstrokecolor{currentstroke}%
\pgfsetdash{}{0pt}%
\pgfpathmoveto{\pgfqpoint{1.178667in}{0.984000in}}%
\pgfpathlineto{\pgfqpoint{1.346400in}{0.984000in}}%
\pgfpathlineto{\pgfqpoint{1.346400in}{0.692000in}}%
\pgfpathlineto{\pgfqpoint{1.178667in}{0.692000in}}%
\pgfpathlineto{\pgfqpoint{1.178667in}{0.984000in}}%
\pgfusepath{stroke,fill}%
\end{pgfscope}%
\begin{pgfscope}%
\pgfpathrectangle{\pgfqpoint{0.340000in}{0.400000in}}{\pgfqpoint{1.006400in}{1.168000in}}%
\pgfusepath{clip}%
\pgfsetbuttcap%
\pgfsetroundjoin%
\definecolor{currentfill}{rgb}{0.229806,0.298718,0.753683}%
\pgfsetfillcolor{currentfill}%
\pgfsetlinewidth{0.602250pt}%
\definecolor{currentstroke}{rgb}{0.000000,0.000000,0.000000}%
\pgfsetstrokecolor{currentstroke}%
\pgfsetdash{}{0pt}%
\pgfpathmoveto{\pgfqpoint{0.340000in}{0.692000in}}%
\pgfpathlineto{\pgfqpoint{0.507733in}{0.692000in}}%
\pgfpathlineto{\pgfqpoint{0.507733in}{0.400000in}}%
\pgfpathlineto{\pgfqpoint{0.340000in}{0.400000in}}%
\pgfpathlineto{\pgfqpoint{0.340000in}{0.692000in}}%
\pgfusepath{stroke,fill}%
\end{pgfscope}%
\begin{pgfscope}%
\pgfpathrectangle{\pgfqpoint{0.340000in}{0.400000in}}{\pgfqpoint{1.006400in}{1.168000in}}%
\pgfusepath{clip}%
\pgfsetbuttcap%
\pgfsetroundjoin%
\definecolor{currentfill}{rgb}{0.234377,0.305542,0.759680}%
\pgfsetfillcolor{currentfill}%
\pgfsetlinewidth{0.602250pt}%
\definecolor{currentstroke}{rgb}{0.000000,0.000000,0.000000}%
\pgfsetstrokecolor{currentstroke}%
\pgfsetdash{}{0pt}%
\pgfpathmoveto{\pgfqpoint{0.507733in}{0.692000in}}%
\pgfpathlineto{\pgfqpoint{0.675467in}{0.692000in}}%
\pgfpathlineto{\pgfqpoint{0.675467in}{0.400000in}}%
\pgfpathlineto{\pgfqpoint{0.507733in}{0.400000in}}%
\pgfpathlineto{\pgfqpoint{0.507733in}{0.692000in}}%
\pgfusepath{stroke,fill}%
\end{pgfscope}%
\begin{pgfscope}%
\pgfpathrectangle{\pgfqpoint{0.340000in}{0.400000in}}{\pgfqpoint{1.006400in}{1.168000in}}%
\pgfusepath{clip}%
\pgfsetbuttcap%
\pgfsetroundjoin%
\definecolor{currentfill}{rgb}{0.252663,0.332837,0.783665}%
\pgfsetfillcolor{currentfill}%
\pgfsetlinewidth{0.602250pt}%
\definecolor{currentstroke}{rgb}{0.000000,0.000000,0.000000}%
\pgfsetstrokecolor{currentstroke}%
\pgfsetdash{}{0pt}%
\pgfpathmoveto{\pgfqpoint{0.675467in}{0.692000in}}%
\pgfpathlineto{\pgfqpoint{0.843200in}{0.692000in}}%
\pgfpathlineto{\pgfqpoint{0.843200in}{0.400000in}}%
\pgfpathlineto{\pgfqpoint{0.675467in}{0.400000in}}%
\pgfpathlineto{\pgfqpoint{0.675467in}{0.692000in}}%
\pgfusepath{stroke,fill}%
\end{pgfscope}%
\begin{pgfscope}%
\pgfpathrectangle{\pgfqpoint{0.340000in}{0.400000in}}{\pgfqpoint{1.006400in}{1.168000in}}%
\pgfusepath{clip}%
\pgfsetbuttcap%
\pgfsetroundjoin%
\definecolor{currentfill}{rgb}{0.261805,0.346484,0.795658}%
\pgfsetfillcolor{currentfill}%
\pgfsetlinewidth{0.602250pt}%
\definecolor{currentstroke}{rgb}{0.000000,0.000000,0.000000}%
\pgfsetstrokecolor{currentstroke}%
\pgfsetdash{}{0pt}%
\pgfpathmoveto{\pgfqpoint{0.843200in}{0.692000in}}%
\pgfpathlineto{\pgfqpoint{1.010933in}{0.692000in}}%
\pgfpathlineto{\pgfqpoint{1.010933in}{0.400000in}}%
\pgfpathlineto{\pgfqpoint{0.843200in}{0.400000in}}%
\pgfpathlineto{\pgfqpoint{0.843200in}{0.692000in}}%
\pgfusepath{stroke,fill}%
\end{pgfscope}%
\begin{pgfscope}%
\pgfpathrectangle{\pgfqpoint{0.340000in}{0.400000in}}{\pgfqpoint{1.006400in}{1.168000in}}%
\pgfusepath{clip}%
\pgfsetbuttcap%
\pgfsetroundjoin%
\definecolor{currentfill}{rgb}{0.280550,0.373423,0.818011}%
\pgfsetfillcolor{currentfill}%
\pgfsetlinewidth{0.602250pt}%
\definecolor{currentstroke}{rgb}{0.000000,0.000000,0.000000}%
\pgfsetstrokecolor{currentstroke}%
\pgfsetdash{}{0pt}%
\pgfpathmoveto{\pgfqpoint{1.010933in}{0.692000in}}%
\pgfpathlineto{\pgfqpoint{1.178667in}{0.692000in}}%
\pgfpathlineto{\pgfqpoint{1.178667in}{0.400000in}}%
\pgfpathlineto{\pgfqpoint{1.010933in}{0.400000in}}%
\pgfpathlineto{\pgfqpoint{1.010933in}{0.692000in}}%
\pgfusepath{stroke,fill}%
\end{pgfscope}%
\begin{pgfscope}%
\pgfpathrectangle{\pgfqpoint{0.340000in}{0.400000in}}{\pgfqpoint{1.006400in}{1.168000in}}%
\pgfusepath{clip}%
\pgfsetbuttcap%
\pgfsetroundjoin%
\definecolor{currentfill}{rgb}{0.285273,0.380129,0.823469}%
\pgfsetfillcolor{currentfill}%
\pgfsetlinewidth{0.602250pt}%
\definecolor{currentstroke}{rgb}{0.000000,0.000000,0.000000}%
\pgfsetstrokecolor{currentstroke}%
\pgfsetdash{}{0pt}%
\pgfpathmoveto{\pgfqpoint{1.178667in}{0.692000in}}%
\pgfpathlineto{\pgfqpoint{1.346400in}{0.692000in}}%
\pgfpathlineto{\pgfqpoint{1.346400in}{0.400000in}}%
\pgfpathlineto{\pgfqpoint{1.178667in}{0.400000in}}%
\pgfpathlineto{\pgfqpoint{1.178667in}{0.692000in}}%
\pgfusepath{stroke,fill}%
\end{pgfscope}%
\begin{pgfscope}%
\definecolor{textcolor}{rgb}{1.000000,1.000000,1.000000}%
\pgfsetstrokecolor{textcolor}%
\pgfsetfillcolor{textcolor}%
\pgftext[x=0.423867in,y=1.422000in,,]{\color{textcolor}\rmfamily\fontsize{6.000000}{7.200000}\selectfont 22}%
\end{pgfscope}%
\begin{pgfscope}%
\definecolor{textcolor}{rgb}{0.150000,0.150000,0.150000}%
\pgfsetstrokecolor{textcolor}%
\pgfsetfillcolor{textcolor}%
\pgftext[x=0.591600in,y=1.422000in,,]{\color{textcolor}\rmfamily\fontsize{6.000000}{7.200000}\selectfont 28}%
\end{pgfscope}%
\begin{pgfscope}%
\definecolor{textcolor}{rgb}{0.150000,0.150000,0.150000}%
\pgfsetstrokecolor{textcolor}%
\pgfsetfillcolor{textcolor}%
\pgftext[x=0.759333in,y=1.422000in,,]{\color{textcolor}\rmfamily\fontsize{6.000000}{7.200000}\selectfont 39}%
\end{pgfscope}%
\begin{pgfscope}%
\definecolor{textcolor}{rgb}{0.150000,0.150000,0.150000}%
\pgfsetstrokecolor{textcolor}%
\pgfsetfillcolor{textcolor}%
\pgftext[x=0.927067in,y=1.422000in,,]{\color{textcolor}\rmfamily\fontsize{6.000000}{7.200000}\selectfont 53}%
\end{pgfscope}%
\begin{pgfscope}%
\definecolor{textcolor}{rgb}{1.000000,1.000000,1.000000}%
\pgfsetstrokecolor{textcolor}%
\pgfsetfillcolor{textcolor}%
\pgftext[x=1.094800in,y=1.422000in,,]{\color{textcolor}\rmfamily\fontsize{6.000000}{7.200000}\selectfont 68}%
\end{pgfscope}%
\begin{pgfscope}%
\definecolor{textcolor}{rgb}{1.000000,1.000000,1.000000}%
\pgfsetstrokecolor{textcolor}%
\pgfsetfillcolor{textcolor}%
\pgftext[x=1.262533in,y=1.422000in,,]{\color{textcolor}\rmfamily\fontsize{6.000000}{7.200000}\selectfont 78}%
\end{pgfscope}%
\begin{pgfscope}%
\definecolor{textcolor}{rgb}{1.000000,1.000000,1.000000}%
\pgfsetstrokecolor{textcolor}%
\pgfsetfillcolor{textcolor}%
\pgftext[x=0.423867in,y=1.130000in,,]{\color{textcolor}\rmfamily\fontsize{6.000000}{7.200000}\selectfont 13}%
\end{pgfscope}%
\begin{pgfscope}%
\definecolor{textcolor}{rgb}{1.000000,1.000000,1.000000}%
\pgfsetstrokecolor{textcolor}%
\pgfsetfillcolor{textcolor}%
\pgftext[x=0.591600in,y=1.130000in,,]{\color{textcolor}\rmfamily\fontsize{6.000000}{7.200000}\selectfont 14}%
\end{pgfscope}%
\begin{pgfscope}%
\definecolor{textcolor}{rgb}{1.000000,1.000000,1.000000}%
\pgfsetstrokecolor{textcolor}%
\pgfsetfillcolor{textcolor}%
\pgftext[x=0.759333in,y=1.130000in,,]{\color{textcolor}\rmfamily\fontsize{6.000000}{7.200000}\selectfont 19}%
\end{pgfscope}%
\begin{pgfscope}%
\definecolor{textcolor}{rgb}{0.150000,0.150000,0.150000}%
\pgfsetstrokecolor{textcolor}%
\pgfsetfillcolor{textcolor}%
\pgftext[x=0.927067in,y=1.130000in,,]{\color{textcolor}\rmfamily\fontsize{6.000000}{7.200000}\selectfont 28}%
\end{pgfscope}%
\begin{pgfscope}%
\definecolor{textcolor}{rgb}{0.150000,0.150000,0.150000}%
\pgfsetstrokecolor{textcolor}%
\pgfsetfillcolor{textcolor}%
\pgftext[x=1.094800in,y=1.130000in,,]{\color{textcolor}\rmfamily\fontsize{6.000000}{7.200000}\selectfont 36}%
\end{pgfscope}%
\begin{pgfscope}%
\definecolor{textcolor}{rgb}{0.150000,0.150000,0.150000}%
\pgfsetstrokecolor{textcolor}%
\pgfsetfillcolor{textcolor}%
\pgftext[x=1.262533in,y=1.130000in,,]{\color{textcolor}\rmfamily\fontsize{6.000000}{7.200000}\selectfont 40}%
\end{pgfscope}%
\begin{pgfscope}%
\definecolor{textcolor}{rgb}{1.000000,1.000000,1.000000}%
\pgfsetstrokecolor{textcolor}%
\pgfsetfillcolor{textcolor}%
\pgftext[x=0.423867in,y=0.838000in,,]{\color{textcolor}\rmfamily\fontsize{6.000000}{7.200000}\selectfont 10}%
\end{pgfscope}%
\begin{pgfscope}%
\definecolor{textcolor}{rgb}{1.000000,1.000000,1.000000}%
\pgfsetstrokecolor{textcolor}%
\pgfsetfillcolor{textcolor}%
\pgftext[x=0.591600in,y=0.838000in,,]{\color{textcolor}\rmfamily\fontsize{6.000000}{7.200000}\selectfont 11}%
\end{pgfscope}%
\begin{pgfscope}%
\definecolor{textcolor}{rgb}{1.000000,1.000000,1.000000}%
\pgfsetstrokecolor{textcolor}%
\pgfsetfillcolor{textcolor}%
\pgftext[x=0.759333in,y=0.838000in,,]{\color{textcolor}\rmfamily\fontsize{6.000000}{7.200000}\selectfont 13}%
\end{pgfscope}%
\begin{pgfscope}%
\definecolor{textcolor}{rgb}{1.000000,1.000000,1.000000}%
\pgfsetstrokecolor{textcolor}%
\pgfsetfillcolor{textcolor}%
\pgftext[x=0.927067in,y=0.838000in,,]{\color{textcolor}\rmfamily\fontsize{6.000000}{7.200000}\selectfont 14}%
\end{pgfscope}%
\begin{pgfscope}%
\definecolor{textcolor}{rgb}{1.000000,1.000000,1.000000}%
\pgfsetstrokecolor{textcolor}%
\pgfsetfillcolor{textcolor}%
\pgftext[x=1.094800in,y=0.838000in,,]{\color{textcolor}\rmfamily\fontsize{6.000000}{7.200000}\selectfont 17}%
\end{pgfscope}%
\begin{pgfscope}%
\definecolor{textcolor}{rgb}{1.000000,1.000000,1.000000}%
\pgfsetstrokecolor{textcolor}%
\pgfsetfillcolor{textcolor}%
\pgftext[x=1.262533in,y=0.838000in,,]{\color{textcolor}\rmfamily\fontsize{6.000000}{7.200000}\selectfont 18}%
\end{pgfscope}%
\begin{pgfscope}%
\definecolor{textcolor}{rgb}{1.000000,1.000000,1.000000}%
\pgfsetstrokecolor{textcolor}%
\pgfsetfillcolor{textcolor}%
\pgftext[x=0.423867in,y=0.546000in,,]{\color{textcolor}\rmfamily\fontsize{6.000000}{7.200000}\selectfont 8.6}%
\end{pgfscope}%
\begin{pgfscope}%
\definecolor{textcolor}{rgb}{1.000000,1.000000,1.000000}%
\pgfsetstrokecolor{textcolor}%
\pgfsetfillcolor{textcolor}%
\pgftext[x=0.591600in,y=0.546000in,,]{\color{textcolor}\rmfamily\fontsize{6.000000}{7.200000}\selectfont 9}%
\end{pgfscope}%
\begin{pgfscope}%
\definecolor{textcolor}{rgb}{1.000000,1.000000,1.000000}%
\pgfsetstrokecolor{textcolor}%
\pgfsetfillcolor{textcolor}%
\pgftext[x=0.759333in,y=0.546000in,,]{\color{textcolor}\rmfamily\fontsize{6.000000}{7.200000}\selectfont 10}%
\end{pgfscope}%
\begin{pgfscope}%
\definecolor{textcolor}{rgb}{1.000000,1.000000,1.000000}%
\pgfsetstrokecolor{textcolor}%
\pgfsetfillcolor{textcolor}%
\pgftext[x=0.927067in,y=0.546000in,,]{\color{textcolor}\rmfamily\fontsize{6.000000}{7.200000}\selectfont 11}%
\end{pgfscope}%
\begin{pgfscope}%
\definecolor{textcolor}{rgb}{1.000000,1.000000,1.000000}%
\pgfsetstrokecolor{textcolor}%
\pgfsetfillcolor{textcolor}%
\pgftext[x=1.094800in,y=0.546000in,,]{\color{textcolor}\rmfamily\fontsize{6.000000}{7.200000}\selectfont 12}%
\end{pgfscope}%
\begin{pgfscope}%
\definecolor{textcolor}{rgb}{1.000000,1.000000,1.000000}%
\pgfsetstrokecolor{textcolor}%
\pgfsetfillcolor{textcolor}%
\pgftext[x=1.262533in,y=0.546000in,,]{\color{textcolor}\rmfamily\fontsize{6.000000}{7.200000}\selectfont 12}%
\end{pgfscope}%
\end{pgfpicture}%
\makeatother%
\endgroup%

%% file: plot_data/packet_level_simulation/packet_level_average_packet_loss.pgf
\begingroup%
\makeatletter%
\begin{pgfpicture}%
\pgfpathrectangle{\pgfpointorigin}{\pgfqpoint{1.360000in}{1.600000in}}%
\pgfusepath{use as bounding box, clip}%
\begin{pgfscope}%
\pgfsetbuttcap%
\pgfsetmiterjoin%
\definecolor{currentfill}{rgb}{1.000000,1.000000,1.000000}%
\pgfsetfillcolor{currentfill}%
\pgfsetlinewidth{0.000000pt}%
\definecolor{currentstroke}{rgb}{1.000000,1.000000,1.000000}%
\pgfsetstrokecolor{currentstroke}%
\pgfsetdash{}{0pt}%
\pgfpathmoveto{\pgfqpoint{0.000000in}{0.000000in}}%
\pgfpathlineto{\pgfqpoint{1.360000in}{0.000000in}}%
\pgfpathlineto{\pgfqpoint{1.360000in}{1.600000in}}%
\pgfpathlineto{\pgfqpoint{0.000000in}{1.600000in}}%
\pgfpathclose%
\pgfusepath{fill}%
\end{pgfscope}%
\begin{pgfscope}%
\pgfsetbuttcap%
\pgfsetmiterjoin%
\definecolor{currentfill}{rgb}{1.000000,1.000000,1.000000}%
\pgfsetfillcolor{currentfill}%
\pgfsetlinewidth{0.000000pt}%
\definecolor{currentstroke}{rgb}{0.000000,0.000000,0.000000}%
\pgfsetstrokecolor{currentstroke}%
\pgfsetstrokeopacity{0.000000}%
\pgfsetdash{}{0pt}%
\pgfpathmoveto{\pgfqpoint{0.306000in}{0.304000in}}%
\pgfpathlineto{\pgfqpoint{1.332800in}{0.304000in}}%
\pgfpathlineto{\pgfqpoint{1.332800in}{1.536000in}}%
\pgfpathlineto{\pgfqpoint{0.306000in}{1.536000in}}%
\pgfpathclose%
\pgfusepath{fill}%
\end{pgfscope}%
\begin{pgfscope}%
\pgfsetbuttcap%
\pgfsetroundjoin%
\definecolor{currentfill}{rgb}{0.000000,0.000000,0.000000}%
\pgfsetfillcolor{currentfill}%
\pgfsetlinewidth{0.803000pt}%
\definecolor{currentstroke}{rgb}{0.000000,0.000000,0.000000}%
\pgfsetstrokecolor{currentstroke}%
\pgfsetdash{}{0pt}%
\pgfsys@defobject{currentmarker}{\pgfqpoint{0.000000in}{-0.048611in}}{\pgfqpoint{0.000000in}{0.000000in}}{%
\pgfpathmoveto{\pgfqpoint{0.000000in}{0.000000in}}%
\pgfpathlineto{\pgfqpoint{0.000000in}{-0.048611in}}%
\pgfusepath{stroke,fill}%
}%
\begin{pgfscope}%
\pgfsys@transformshift{0.487200in}{0.304000in}%
\pgfsys@useobject{currentmarker}{}%
\end{pgfscope}%
\end{pgfscope}%
\begin{pgfscope}%
\pgftext[x=0.487200in,y=0.206778in,,top]{\rmfamily\fontsize{6.500000}{7.800000}\selectfont \(\displaystyle 1.25\)}%
\end{pgfscope}%
\begin{pgfscope}%
\pgfsetbuttcap%
\pgfsetroundjoin%
\definecolor{currentfill}{rgb}{0.000000,0.000000,0.000000}%
\pgfsetfillcolor{currentfill}%
\pgfsetlinewidth{0.803000pt}%
\definecolor{currentstroke}{rgb}{0.000000,0.000000,0.000000}%
\pgfsetstrokecolor{currentstroke}%
\pgfsetdash{}{0pt}%
\pgfsys@defobject{currentmarker}{\pgfqpoint{0.000000in}{-0.048611in}}{\pgfqpoint{0.000000in}{0.000000in}}{%
\pgfpathmoveto{\pgfqpoint{0.000000in}{0.000000in}}%
\pgfpathlineto{\pgfqpoint{0.000000in}{-0.048611in}}%
\pgfusepath{stroke,fill}%
}%
\begin{pgfscope}%
\pgfsys@transformshift{0.789200in}{0.304000in}%
\pgfsys@useobject{currentmarker}{}%
\end{pgfscope}%
\end{pgfscope}%
\begin{pgfscope}%
\pgftext[x=0.789200in,y=0.206778in,,top]{\rmfamily\fontsize{6.500000}{7.800000}\selectfont \(\displaystyle 1.50\)}%
\end{pgfscope}%
\begin{pgfscope}%
\pgfsetbuttcap%
\pgfsetroundjoin%
\definecolor{currentfill}{rgb}{0.000000,0.000000,0.000000}%
\pgfsetfillcolor{currentfill}%
\pgfsetlinewidth{0.803000pt}%
\definecolor{currentstroke}{rgb}{0.000000,0.000000,0.000000}%
\pgfsetstrokecolor{currentstroke}%
\pgfsetdash{}{0pt}%
\pgfsys@defobject{currentmarker}{\pgfqpoint{0.000000in}{-0.048611in}}{\pgfqpoint{0.000000in}{0.000000in}}{%
\pgfpathmoveto{\pgfqpoint{0.000000in}{0.000000in}}%
\pgfpathlineto{\pgfqpoint{0.000000in}{-0.048611in}}%
\pgfusepath{stroke,fill}%
}%
\begin{pgfscope}%
\pgfsys@transformshift{1.091200in}{0.304000in}%
\pgfsys@useobject{currentmarker}{}%
\end{pgfscope}%
\end{pgfscope}%
\begin{pgfscope}%
\pgftext[x=1.091200in,y=0.206778in,,top]{\rmfamily\fontsize{6.500000}{7.800000}\selectfont \(\displaystyle 1.75\)}%
\end{pgfscope}%
\begin{pgfscope}%
\pgftext[x=0.819400in,y=0.118815in,,top]{\rmfamily\fontsize{7.400000}{8.880000}\selectfont Avg. Hop Count}%
\end{pgfscope}%
\begin{pgfscope}%
\pgfpathrectangle{\pgfqpoint{0.306000in}{0.304000in}}{\pgfqpoint{1.026800in}{1.232000in}}%
\pgfusepath{clip}%
\pgfsetbuttcap%
\pgfsetroundjoin%
\pgfsetlinewidth{0.702625pt}%
\definecolor{currentstroke}{rgb}{0.690196,0.690196,0.690196}%
\pgfsetstrokecolor{currentstroke}%
\pgfsetdash{{4.480000pt}{1.120000pt}{0.700000pt}{1.120000pt}}{0.000000pt}%
\pgfpathmoveto{\pgfqpoint{0.306000in}{0.360000in}}%
\pgfpathlineto{\pgfqpoint{1.332800in}{0.360000in}}%
\pgfusepath{stroke}%
\end{pgfscope}%
\begin{pgfscope}%
\pgfsetbuttcap%
\pgfsetroundjoin%
\definecolor{currentfill}{rgb}{0.000000,0.000000,0.000000}%
\pgfsetfillcolor{currentfill}%
\pgfsetlinewidth{0.803000pt}%
\definecolor{currentstroke}{rgb}{0.000000,0.000000,0.000000}%
\pgfsetstrokecolor{currentstroke}%
\pgfsetdash{}{0pt}%
\pgfsys@defobject{currentmarker}{\pgfqpoint{0.000000in}{0.000000in}}{\pgfqpoint{0.048611in}{0.000000in}}{%
\pgfpathmoveto{\pgfqpoint{0.000000in}{0.000000in}}%
\pgfpathlineto{\pgfqpoint{0.048611in}{0.000000in}}%
\pgfusepath{stroke,fill}%
}%
\begin{pgfscope}%
\pgfsys@transformshift{0.306000in}{0.360000in}%
\pgfsys@useobject{currentmarker}{}%
\end{pgfscope}%
\end{pgfscope}%
\begin{pgfscope}%
\pgftext[x=0.206464in,y=0.331065in,left,base]{\rmfamily\fontsize{6.500000}{7.800000}\selectfont \(\displaystyle 0\)}%
\end{pgfscope}%
\begin{pgfscope}%
\pgfpathrectangle{\pgfqpoint{0.306000in}{0.304000in}}{\pgfqpoint{1.026800in}{1.232000in}}%
\pgfusepath{clip}%
\pgfsetbuttcap%
\pgfsetroundjoin%
\pgfsetlinewidth{0.702625pt}%
\definecolor{currentstroke}{rgb}{0.690196,0.690196,0.690196}%
\pgfsetstrokecolor{currentstroke}%
\pgfsetdash{{4.480000pt}{1.120000pt}{0.700000pt}{1.120000pt}}{0.000000pt}%
\pgfpathmoveto{\pgfqpoint{0.306000in}{0.664789in}}%
\pgfpathlineto{\pgfqpoint{1.332800in}{0.664789in}}%
\pgfusepath{stroke}%
\end{pgfscope}%
\begin{pgfscope}%
\pgfsetbuttcap%
\pgfsetroundjoin%
\definecolor{currentfill}{rgb}{0.000000,0.000000,0.000000}%
\pgfsetfillcolor{currentfill}%
\pgfsetlinewidth{0.803000pt}%
\definecolor{currentstroke}{rgb}{0.000000,0.000000,0.000000}%
\pgfsetstrokecolor{currentstroke}%
\pgfsetdash{}{0pt}%
\pgfsys@defobject{currentmarker}{\pgfqpoint{0.000000in}{0.000000in}}{\pgfqpoint{0.048611in}{0.000000in}}{%
\pgfpathmoveto{\pgfqpoint{0.000000in}{0.000000in}}%
\pgfpathlineto{\pgfqpoint{0.048611in}{0.000000in}}%
\pgfusepath{stroke,fill}%
}%
\begin{pgfscope}%
\pgfsys@transformshift{0.306000in}{0.664789in}%
\pgfsys@useobject{currentmarker}{}%
\end{pgfscope}%
\end{pgfscope}%
\begin{pgfscope}%
\pgftext[x=0.155539in,y=0.635854in,left,base]{\rmfamily\fontsize{6.500000}{7.800000}\selectfont \(\displaystyle 10\)}%
\end{pgfscope}%
\begin{pgfscope}%
\pgfpathrectangle{\pgfqpoint{0.306000in}{0.304000in}}{\pgfqpoint{1.026800in}{1.232000in}}%
\pgfusepath{clip}%
\pgfsetbuttcap%
\pgfsetroundjoin%
\pgfsetlinewidth{0.702625pt}%
\definecolor{currentstroke}{rgb}{0.690196,0.690196,0.690196}%
\pgfsetstrokecolor{currentstroke}%
\pgfsetdash{{4.480000pt}{1.120000pt}{0.700000pt}{1.120000pt}}{0.000000pt}%
\pgfpathmoveto{\pgfqpoint{0.306000in}{0.969578in}}%
\pgfpathlineto{\pgfqpoint{1.332800in}{0.969578in}}%
\pgfusepath{stroke}%
\end{pgfscope}%
\begin{pgfscope}%
\pgfsetbuttcap%
\pgfsetroundjoin%
\definecolor{currentfill}{rgb}{0.000000,0.000000,0.000000}%
\pgfsetfillcolor{currentfill}%
\pgfsetlinewidth{0.803000pt}%
\definecolor{currentstroke}{rgb}{0.000000,0.000000,0.000000}%
\pgfsetstrokecolor{currentstroke}%
\pgfsetdash{}{0pt}%
\pgfsys@defobject{currentmarker}{\pgfqpoint{0.000000in}{0.000000in}}{\pgfqpoint{0.048611in}{0.000000in}}{%
\pgfpathmoveto{\pgfqpoint{0.000000in}{0.000000in}}%
\pgfpathlineto{\pgfqpoint{0.048611in}{0.000000in}}%
\pgfusepath{stroke,fill}%
}%
\begin{pgfscope}%
\pgfsys@transformshift{0.306000in}{0.969578in}%
\pgfsys@useobject{currentmarker}{}%
\end{pgfscope}%
\end{pgfscope}%
\begin{pgfscope}%
\pgftext[x=0.155539in,y=0.940643in,left,base]{\rmfamily\fontsize{6.500000}{7.800000}\selectfont \(\displaystyle 20\)}%
\end{pgfscope}%
\begin{pgfscope}%
\pgfpathrectangle{\pgfqpoint{0.306000in}{0.304000in}}{\pgfqpoint{1.026800in}{1.232000in}}%
\pgfusepath{clip}%
\pgfsetbuttcap%
\pgfsetroundjoin%
\pgfsetlinewidth{0.702625pt}%
\definecolor{currentstroke}{rgb}{0.690196,0.690196,0.690196}%
\pgfsetstrokecolor{currentstroke}%
\pgfsetdash{{4.480000pt}{1.120000pt}{0.700000pt}{1.120000pt}}{0.000000pt}%
\pgfpathmoveto{\pgfqpoint{0.306000in}{1.274367in}}%
\pgfpathlineto{\pgfqpoint{1.332800in}{1.274367in}}%
\pgfusepath{stroke}%
\end{pgfscope}%
\begin{pgfscope}%
\pgfsetbuttcap%
\pgfsetroundjoin%
\definecolor{currentfill}{rgb}{0.000000,0.000000,0.000000}%
\pgfsetfillcolor{currentfill}%
\pgfsetlinewidth{0.803000pt}%
\definecolor{currentstroke}{rgb}{0.000000,0.000000,0.000000}%
\pgfsetstrokecolor{currentstroke}%
\pgfsetdash{}{0pt}%
\pgfsys@defobject{currentmarker}{\pgfqpoint{0.000000in}{0.000000in}}{\pgfqpoint{0.048611in}{0.000000in}}{%
\pgfpathmoveto{\pgfqpoint{0.000000in}{0.000000in}}%
\pgfpathlineto{\pgfqpoint{0.048611in}{0.000000in}}%
\pgfusepath{stroke,fill}%
}%
\begin{pgfscope}%
\pgfsys@transformshift{0.306000in}{1.274367in}%
\pgfsys@useobject{currentmarker}{}%
\end{pgfscope}%
\end{pgfscope}%
\begin{pgfscope}%
\pgftext[x=0.155539in,y=1.245432in,left,base]{\rmfamily\fontsize{6.500000}{7.800000}\selectfont \(\displaystyle 30\)}%
\end{pgfscope}%
\begin{pgfscope}%
\pgfsetbuttcap%
\pgfsetroundjoin%
\definecolor{currentfill}{rgb}{0.000000,0.000000,0.000000}%
\pgfsetfillcolor{currentfill}%
\pgfsetlinewidth{0.602250pt}%
\definecolor{currentstroke}{rgb}{0.000000,0.000000,0.000000}%
\pgfsetstrokecolor{currentstroke}%
\pgfsetdash{}{0pt}%
\pgfsys@defobject{currentmarker}{\pgfqpoint{0.000000in}{0.000000in}}{\pgfqpoint{0.027778in}{0.000000in}}{%
\pgfpathmoveto{\pgfqpoint{0.000000in}{0.000000in}}%
\pgfpathlineto{\pgfqpoint{0.027778in}{0.000000in}}%
\pgfusepath{stroke,fill}%
}%
\begin{pgfscope}%
\pgfsys@transformshift{0.306000in}{0.420958in}%
\pgfsys@useobject{currentmarker}{}%
\end{pgfscope}%
\end{pgfscope}%
\begin{pgfscope}%
\pgfsetbuttcap%
\pgfsetroundjoin%
\definecolor{currentfill}{rgb}{0.000000,0.000000,0.000000}%
\pgfsetfillcolor{currentfill}%
\pgfsetlinewidth{0.602250pt}%
\definecolor{currentstroke}{rgb}{0.000000,0.000000,0.000000}%
\pgfsetstrokecolor{currentstroke}%
\pgfsetdash{}{0pt}%
\pgfsys@defobject{currentmarker}{\pgfqpoint{0.000000in}{0.000000in}}{\pgfqpoint{0.027778in}{0.000000in}}{%
\pgfpathmoveto{\pgfqpoint{0.000000in}{0.000000in}}%
\pgfpathlineto{\pgfqpoint{0.027778in}{0.000000in}}%
\pgfusepath{stroke,fill}%
}%
\begin{pgfscope}%
\pgfsys@transformshift{0.306000in}{0.481916in}%
\pgfsys@useobject{currentmarker}{}%
\end{pgfscope}%
\end{pgfscope}%
\begin{pgfscope}%
\pgfsetbuttcap%
\pgfsetroundjoin%
\definecolor{currentfill}{rgb}{0.000000,0.000000,0.000000}%
\pgfsetfillcolor{currentfill}%
\pgfsetlinewidth{0.602250pt}%
\definecolor{currentstroke}{rgb}{0.000000,0.000000,0.000000}%
\pgfsetstrokecolor{currentstroke}%
\pgfsetdash{}{0pt}%
\pgfsys@defobject{currentmarker}{\pgfqpoint{0.000000in}{0.000000in}}{\pgfqpoint{0.027778in}{0.000000in}}{%
\pgfpathmoveto{\pgfqpoint{0.000000in}{0.000000in}}%
\pgfpathlineto{\pgfqpoint{0.027778in}{0.000000in}}%
\pgfusepath{stroke,fill}%
}%
\begin{pgfscope}%
\pgfsys@transformshift{0.306000in}{0.542873in}%
\pgfsys@useobject{currentmarker}{}%
\end{pgfscope}%
\end{pgfscope}%
\begin{pgfscope}%
\pgfsetbuttcap%
\pgfsetroundjoin%
\definecolor{currentfill}{rgb}{0.000000,0.000000,0.000000}%
\pgfsetfillcolor{currentfill}%
\pgfsetlinewidth{0.602250pt}%
\definecolor{currentstroke}{rgb}{0.000000,0.000000,0.000000}%
\pgfsetstrokecolor{currentstroke}%
\pgfsetdash{}{0pt}%
\pgfsys@defobject{currentmarker}{\pgfqpoint{0.000000in}{0.000000in}}{\pgfqpoint{0.027778in}{0.000000in}}{%
\pgfpathmoveto{\pgfqpoint{0.000000in}{0.000000in}}%
\pgfpathlineto{\pgfqpoint{0.027778in}{0.000000in}}%
\pgfusepath{stroke,fill}%
}%
\begin{pgfscope}%
\pgfsys@transformshift{0.306000in}{0.603831in}%
\pgfsys@useobject{currentmarker}{}%
\end{pgfscope}%
\end{pgfscope}%
\begin{pgfscope}%
\pgfsetbuttcap%
\pgfsetroundjoin%
\definecolor{currentfill}{rgb}{0.000000,0.000000,0.000000}%
\pgfsetfillcolor{currentfill}%
\pgfsetlinewidth{0.602250pt}%
\definecolor{currentstroke}{rgb}{0.000000,0.000000,0.000000}%
\pgfsetstrokecolor{currentstroke}%
\pgfsetdash{}{0pt}%
\pgfsys@defobject{currentmarker}{\pgfqpoint{0.000000in}{0.000000in}}{\pgfqpoint{0.027778in}{0.000000in}}{%
\pgfpathmoveto{\pgfqpoint{0.000000in}{0.000000in}}%
\pgfpathlineto{\pgfqpoint{0.027778in}{0.000000in}}%
\pgfusepath{stroke,fill}%
}%
\begin{pgfscope}%
\pgfsys@transformshift{0.306000in}{0.725747in}%
\pgfsys@useobject{currentmarker}{}%
\end{pgfscope}%
\end{pgfscope}%
\begin{pgfscope}%
\pgfsetbuttcap%
\pgfsetroundjoin%
\definecolor{currentfill}{rgb}{0.000000,0.000000,0.000000}%
\pgfsetfillcolor{currentfill}%
\pgfsetlinewidth{0.602250pt}%
\definecolor{currentstroke}{rgb}{0.000000,0.000000,0.000000}%
\pgfsetstrokecolor{currentstroke}%
\pgfsetdash{}{0pt}%
\pgfsys@defobject{currentmarker}{\pgfqpoint{0.000000in}{0.000000in}}{\pgfqpoint{0.027778in}{0.000000in}}{%
\pgfpathmoveto{\pgfqpoint{0.000000in}{0.000000in}}%
\pgfpathlineto{\pgfqpoint{0.027778in}{0.000000in}}%
\pgfusepath{stroke,fill}%
}%
\begin{pgfscope}%
\pgfsys@transformshift{0.306000in}{0.786705in}%
\pgfsys@useobject{currentmarker}{}%
\end{pgfscope}%
\end{pgfscope}%
\begin{pgfscope}%
\pgfsetbuttcap%
\pgfsetroundjoin%
\definecolor{currentfill}{rgb}{0.000000,0.000000,0.000000}%
\pgfsetfillcolor{currentfill}%
\pgfsetlinewidth{0.602250pt}%
\definecolor{currentstroke}{rgb}{0.000000,0.000000,0.000000}%
\pgfsetstrokecolor{currentstroke}%
\pgfsetdash{}{0pt}%
\pgfsys@defobject{currentmarker}{\pgfqpoint{0.000000in}{0.000000in}}{\pgfqpoint{0.027778in}{0.000000in}}{%
\pgfpathmoveto{\pgfqpoint{0.000000in}{0.000000in}}%
\pgfpathlineto{\pgfqpoint{0.027778in}{0.000000in}}%
\pgfusepath{stroke,fill}%
}%
\begin{pgfscope}%
\pgfsys@transformshift{0.306000in}{0.847663in}%
\pgfsys@useobject{currentmarker}{}%
\end{pgfscope}%
\end{pgfscope}%
\begin{pgfscope}%
\pgfsetbuttcap%
\pgfsetroundjoin%
\definecolor{currentfill}{rgb}{0.000000,0.000000,0.000000}%
\pgfsetfillcolor{currentfill}%
\pgfsetlinewidth{0.602250pt}%
\definecolor{currentstroke}{rgb}{0.000000,0.000000,0.000000}%
\pgfsetstrokecolor{currentstroke}%
\pgfsetdash{}{0pt}%
\pgfsys@defobject{currentmarker}{\pgfqpoint{0.000000in}{0.000000in}}{\pgfqpoint{0.027778in}{0.000000in}}{%
\pgfpathmoveto{\pgfqpoint{0.000000in}{0.000000in}}%
\pgfpathlineto{\pgfqpoint{0.027778in}{0.000000in}}%
\pgfusepath{stroke,fill}%
}%
\begin{pgfscope}%
\pgfsys@transformshift{0.306000in}{0.908620in}%
\pgfsys@useobject{currentmarker}{}%
\end{pgfscope}%
\end{pgfscope}%
\begin{pgfscope}%
\pgfsetbuttcap%
\pgfsetroundjoin%
\definecolor{currentfill}{rgb}{0.000000,0.000000,0.000000}%
\pgfsetfillcolor{currentfill}%
\pgfsetlinewidth{0.602250pt}%
\definecolor{currentstroke}{rgb}{0.000000,0.000000,0.000000}%
\pgfsetstrokecolor{currentstroke}%
\pgfsetdash{}{0pt}%
\pgfsys@defobject{currentmarker}{\pgfqpoint{0.000000in}{0.000000in}}{\pgfqpoint{0.027778in}{0.000000in}}{%
\pgfpathmoveto{\pgfqpoint{0.000000in}{0.000000in}}%
\pgfpathlineto{\pgfqpoint{0.027778in}{0.000000in}}%
\pgfusepath{stroke,fill}%
}%
\begin{pgfscope}%
\pgfsys@transformshift{0.306000in}{1.030536in}%
\pgfsys@useobject{currentmarker}{}%
\end{pgfscope}%
\end{pgfscope}%
\begin{pgfscope}%
\pgfsetbuttcap%
\pgfsetroundjoin%
\definecolor{currentfill}{rgb}{0.000000,0.000000,0.000000}%
\pgfsetfillcolor{currentfill}%
\pgfsetlinewidth{0.602250pt}%
\definecolor{currentstroke}{rgb}{0.000000,0.000000,0.000000}%
\pgfsetstrokecolor{currentstroke}%
\pgfsetdash{}{0pt}%
\pgfsys@defobject{currentmarker}{\pgfqpoint{0.000000in}{0.000000in}}{\pgfqpoint{0.027778in}{0.000000in}}{%
\pgfpathmoveto{\pgfqpoint{0.000000in}{0.000000in}}%
\pgfpathlineto{\pgfqpoint{0.027778in}{0.000000in}}%
\pgfusepath{stroke,fill}%
}%
\begin{pgfscope}%
\pgfsys@transformshift{0.306000in}{1.091494in}%
\pgfsys@useobject{currentmarker}{}%
\end{pgfscope}%
\end{pgfscope}%
\begin{pgfscope}%
\pgfsetbuttcap%
\pgfsetroundjoin%
\definecolor{currentfill}{rgb}{0.000000,0.000000,0.000000}%
\pgfsetfillcolor{currentfill}%
\pgfsetlinewidth{0.602250pt}%
\definecolor{currentstroke}{rgb}{0.000000,0.000000,0.000000}%
\pgfsetstrokecolor{currentstroke}%
\pgfsetdash{}{0pt}%
\pgfsys@defobject{currentmarker}{\pgfqpoint{0.000000in}{0.000000in}}{\pgfqpoint{0.027778in}{0.000000in}}{%
\pgfpathmoveto{\pgfqpoint{0.000000in}{0.000000in}}%
\pgfpathlineto{\pgfqpoint{0.027778in}{0.000000in}}%
\pgfusepath{stroke,fill}%
}%
\begin{pgfscope}%
\pgfsys@transformshift{0.306000in}{1.152452in}%
\pgfsys@useobject{currentmarker}{}%
\end{pgfscope}%
\end{pgfscope}%
\begin{pgfscope}%
\pgfsetbuttcap%
\pgfsetroundjoin%
\definecolor{currentfill}{rgb}{0.000000,0.000000,0.000000}%
\pgfsetfillcolor{currentfill}%
\pgfsetlinewidth{0.602250pt}%
\definecolor{currentstroke}{rgb}{0.000000,0.000000,0.000000}%
\pgfsetstrokecolor{currentstroke}%
\pgfsetdash{}{0pt}%
\pgfsys@defobject{currentmarker}{\pgfqpoint{0.000000in}{0.000000in}}{\pgfqpoint{0.027778in}{0.000000in}}{%
\pgfpathmoveto{\pgfqpoint{0.000000in}{0.000000in}}%
\pgfpathlineto{\pgfqpoint{0.027778in}{0.000000in}}%
\pgfusepath{stroke,fill}%
}%
\begin{pgfscope}%
\pgfsys@transformshift{0.306000in}{1.213409in}%
\pgfsys@useobject{currentmarker}{}%
\end{pgfscope}%
\end{pgfscope}%
\begin{pgfscope}%
\pgfsetbuttcap%
\pgfsetroundjoin%
\definecolor{currentfill}{rgb}{0.000000,0.000000,0.000000}%
\pgfsetfillcolor{currentfill}%
\pgfsetlinewidth{0.602250pt}%
\definecolor{currentstroke}{rgb}{0.000000,0.000000,0.000000}%
\pgfsetstrokecolor{currentstroke}%
\pgfsetdash{}{0pt}%
\pgfsys@defobject{currentmarker}{\pgfqpoint{0.000000in}{0.000000in}}{\pgfqpoint{0.027778in}{0.000000in}}{%
\pgfpathmoveto{\pgfqpoint{0.000000in}{0.000000in}}%
\pgfpathlineto{\pgfqpoint{0.027778in}{0.000000in}}%
\pgfusepath{stroke,fill}%
}%
\begin{pgfscope}%
\pgfsys@transformshift{0.306000in}{1.335325in}%
\pgfsys@useobject{currentmarker}{}%
\end{pgfscope}%
\end{pgfscope}%
\begin{pgfscope}%
\pgfsetbuttcap%
\pgfsetroundjoin%
\definecolor{currentfill}{rgb}{0.000000,0.000000,0.000000}%
\pgfsetfillcolor{currentfill}%
\pgfsetlinewidth{0.602250pt}%
\definecolor{currentstroke}{rgb}{0.000000,0.000000,0.000000}%
\pgfsetstrokecolor{currentstroke}%
\pgfsetdash{}{0pt}%
\pgfsys@defobject{currentmarker}{\pgfqpoint{0.000000in}{0.000000in}}{\pgfqpoint{0.027778in}{0.000000in}}{%
\pgfpathmoveto{\pgfqpoint{0.000000in}{0.000000in}}%
\pgfpathlineto{\pgfqpoint{0.027778in}{0.000000in}}%
\pgfusepath{stroke,fill}%
}%
\begin{pgfscope}%
\pgfsys@transformshift{0.306000in}{1.396283in}%
\pgfsys@useobject{currentmarker}{}%
\end{pgfscope}%
\end{pgfscope}%
\begin{pgfscope}%
\pgfsetbuttcap%
\pgfsetroundjoin%
\definecolor{currentfill}{rgb}{0.000000,0.000000,0.000000}%
\pgfsetfillcolor{currentfill}%
\pgfsetlinewidth{0.602250pt}%
\definecolor{currentstroke}{rgb}{0.000000,0.000000,0.000000}%
\pgfsetstrokecolor{currentstroke}%
\pgfsetdash{}{0pt}%
\pgfsys@defobject{currentmarker}{\pgfqpoint{0.000000in}{0.000000in}}{\pgfqpoint{0.027778in}{0.000000in}}{%
\pgfpathmoveto{\pgfqpoint{0.000000in}{0.000000in}}%
\pgfpathlineto{\pgfqpoint{0.027778in}{0.000000in}}%
\pgfusepath{stroke,fill}%
}%
\begin{pgfscope}%
\pgfsys@transformshift{0.306000in}{1.457241in}%
\pgfsys@useobject{currentmarker}{}%
\end{pgfscope}%
\end{pgfscope}%
\begin{pgfscope}%
\pgfsetbuttcap%
\pgfsetroundjoin%
\definecolor{currentfill}{rgb}{0.000000,0.000000,0.000000}%
\pgfsetfillcolor{currentfill}%
\pgfsetlinewidth{0.602250pt}%
\definecolor{currentstroke}{rgb}{0.000000,0.000000,0.000000}%
\pgfsetstrokecolor{currentstroke}%
\pgfsetdash{}{0pt}%
\pgfsys@defobject{currentmarker}{\pgfqpoint{0.000000in}{0.000000in}}{\pgfqpoint{0.027778in}{0.000000in}}{%
\pgfpathmoveto{\pgfqpoint{0.000000in}{0.000000in}}%
\pgfpathlineto{\pgfqpoint{0.027778in}{0.000000in}}%
\pgfusepath{stroke,fill}%
}%
\begin{pgfscope}%
\pgfsys@transformshift{0.306000in}{1.518198in}%
\pgfsys@useobject{currentmarker}{}%
\end{pgfscope}%
\end{pgfscope}%
\begin{pgfscope}%
\pgftext[x=0.141650in,y=0.920000in,,bottom,rotate=90.000000]{\rmfamily\fontsize{7.400000}{8.880000}\selectfont Avg. Packet Drops / Flow}%
\end{pgfscope}%
\begin{pgfscope}%
\pgfpathrectangle{\pgfqpoint{0.306000in}{0.304000in}}{\pgfqpoint{1.026800in}{1.232000in}}%
\pgfusepath{clip}%
\pgfsetrectcap%
\pgfsetroundjoin%
\pgfsetlinewidth{0.903375pt}%
\definecolor{currentstroke}{rgb}{0.121569,0.466667,0.705882}%
\pgfsetstrokecolor{currentstroke}%
\pgfsetdash{}{0pt}%
\pgfpathmoveto{\pgfqpoint{0.306000in}{0.360000in}}%
\pgfpathlineto{\pgfqpoint{0.426800in}{0.360000in}}%
\pgfpathlineto{\pgfqpoint{0.668400in}{0.360000in}}%
\pgfpathlineto{\pgfqpoint{0.910000in}{0.360000in}}%
\pgfpathlineto{\pgfqpoint{1.151600in}{0.362438in}}%
\pgfpathlineto{\pgfqpoint{1.272400in}{0.367620in}}%
\pgfusepath{stroke}%
\end{pgfscope}%
\begin{pgfscope}%
\pgfpathrectangle{\pgfqpoint{0.306000in}{0.304000in}}{\pgfqpoint{1.026800in}{1.232000in}}%
\pgfusepath{clip}%
\pgfsetbuttcap%
\pgfsetroundjoin%
\definecolor{currentfill}{rgb}{0.121569,0.466667,0.705882}%
\pgfsetfillcolor{currentfill}%
\pgfsetlinewidth{1.003750pt}%
\definecolor{currentstroke}{rgb}{0.121569,0.466667,0.705882}%
\pgfsetstrokecolor{currentstroke}%
\pgfsetdash{}{0pt}%
\pgfsys@defobject{currentmarker}{\pgfqpoint{-0.033333in}{-0.041667in}}{\pgfqpoint{0.033333in}{0.020833in}}{%
\pgfpathmoveto{\pgfqpoint{0.000000in}{0.000000in}}%
\pgfpathlineto{\pgfqpoint{0.000000in}{-0.041667in}}%
\pgfpathmoveto{\pgfqpoint{0.000000in}{0.000000in}}%
\pgfpathlineto{\pgfqpoint{0.033333in}{0.020833in}}%
\pgfpathmoveto{\pgfqpoint{0.000000in}{0.000000in}}%
\pgfpathlineto{\pgfqpoint{-0.033333in}{0.020833in}}%
\pgfusepath{stroke,fill}%
}%
\begin{pgfscope}%
\pgfsys@transformshift{0.306000in}{0.360000in}%
\pgfsys@useobject{currentmarker}{}%
\end{pgfscope}%
\begin{pgfscope}%
\pgfsys@transformshift{0.426800in}{0.360000in}%
\pgfsys@useobject{currentmarker}{}%
\end{pgfscope}%
\begin{pgfscope}%
\pgfsys@transformshift{0.668400in}{0.360000in}%
\pgfsys@useobject{currentmarker}{}%
\end{pgfscope}%
\begin{pgfscope}%
\pgfsys@transformshift{0.910000in}{0.360000in}%
\pgfsys@useobject{currentmarker}{}%
\end{pgfscope}%
\begin{pgfscope}%
\pgfsys@transformshift{1.151600in}{0.362438in}%
\pgfsys@useobject{currentmarker}{}%
\end{pgfscope}%
\begin{pgfscope}%
\pgfsys@transformshift{1.272400in}{0.367620in}%
\pgfsys@useobject{currentmarker}{}%
\end{pgfscope}%
\end{pgfscope}%
\begin{pgfscope}%
\pgfpathrectangle{\pgfqpoint{0.306000in}{0.304000in}}{\pgfqpoint{1.026800in}{1.232000in}}%
\pgfusepath{clip}%
\pgfsetbuttcap%
\pgfsetroundjoin%
\pgfsetlinewidth{0.903375pt}%
\definecolor{currentstroke}{rgb}{1.000000,0.498039,0.054902}%
\pgfsetstrokecolor{currentstroke}%
\pgfsetdash{{0.900000pt}{1.485000pt}}{0.000000pt}%
\pgfpathmoveto{\pgfqpoint{0.306000in}{0.360000in}}%
\pgfpathlineto{\pgfqpoint{0.426800in}{0.360000in}}%
\pgfpathlineto{\pgfqpoint{0.668400in}{0.360000in}}%
\pgfpathlineto{\pgfqpoint{0.910000in}{0.367315in}}%
\pgfpathlineto{\pgfqpoint{1.151600in}{0.389260in}}%
\pgfpathlineto{\pgfqpoint{1.272400in}{0.414252in}}%
\pgfusepath{stroke}%
\end{pgfscope}%
\begin{pgfscope}%
\pgfpathrectangle{\pgfqpoint{0.306000in}{0.304000in}}{\pgfqpoint{1.026800in}{1.232000in}}%
\pgfusepath{clip}%
\pgfsetbuttcap%
\pgfsetroundjoin%
\definecolor{currentfill}{rgb}{1.000000,0.498039,0.054902}%
\pgfsetfillcolor{currentfill}%
\pgfsetlinewidth{1.003750pt}%
\definecolor{currentstroke}{rgb}{1.000000,0.498039,0.054902}%
\pgfsetstrokecolor{currentstroke}%
\pgfsetdash{}{0pt}%
\pgfsys@defobject{currentmarker}{\pgfqpoint{-0.041667in}{-0.041667in}}{\pgfqpoint{0.041667in}{0.041667in}}{%
\pgfpathmoveto{\pgfqpoint{-0.041667in}{0.000000in}}%
\pgfpathlineto{\pgfqpoint{0.041667in}{0.000000in}}%
\pgfpathmoveto{\pgfqpoint{0.000000in}{-0.041667in}}%
\pgfpathlineto{\pgfqpoint{0.000000in}{0.041667in}}%
\pgfusepath{stroke,fill}%
}%
\begin{pgfscope}%
\pgfsys@transformshift{0.306000in}{0.360000in}%
\pgfsys@useobject{currentmarker}{}%
\end{pgfscope}%
\begin{pgfscope}%
\pgfsys@transformshift{0.426800in}{0.360000in}%
\pgfsys@useobject{currentmarker}{}%
\end{pgfscope}%
\begin{pgfscope}%
\pgfsys@transformshift{0.668400in}{0.360000in}%
\pgfsys@useobject{currentmarker}{}%
\end{pgfscope}%
\begin{pgfscope}%
\pgfsys@transformshift{0.910000in}{0.367315in}%
\pgfsys@useobject{currentmarker}{}%
\end{pgfscope}%
\begin{pgfscope}%
\pgfsys@transformshift{1.151600in}{0.389260in}%
\pgfsys@useobject{currentmarker}{}%
\end{pgfscope}%
\begin{pgfscope}%
\pgfsys@transformshift{1.272400in}{0.414252in}%
\pgfsys@useobject{currentmarker}{}%
\end{pgfscope}%
\end{pgfscope}%
\begin{pgfscope}%
\pgfpathrectangle{\pgfqpoint{0.306000in}{0.304000in}}{\pgfqpoint{1.026800in}{1.232000in}}%
\pgfusepath{clip}%
\pgfsetbuttcap%
\pgfsetroundjoin%
\pgfsetlinewidth{0.903375pt}%
\definecolor{currentstroke}{rgb}{0.172549,0.627451,0.172549}%
\pgfsetstrokecolor{currentstroke}%
\pgfsetdash{{5.760000pt}{1.440000pt}{0.900000pt}{1.440000pt}}{0.000000pt}%
\pgfpathmoveto{\pgfqpoint{0.306000in}{0.385360in}}%
\pgfpathlineto{\pgfqpoint{0.426800in}{0.392551in}}%
\pgfpathlineto{\pgfqpoint{0.668400in}{0.402199in}}%
\pgfpathlineto{\pgfqpoint{0.910000in}{0.424654in}}%
\pgfpathlineto{\pgfqpoint{1.151600in}{0.475435in}}%
\pgfpathlineto{\pgfqpoint{1.272400in}{0.621950in}}%
\pgfusepath{stroke}%
\end{pgfscope}%
\begin{pgfscope}%
\pgfpathrectangle{\pgfqpoint{0.306000in}{0.304000in}}{\pgfqpoint{1.026800in}{1.232000in}}%
\pgfusepath{clip}%
\pgfsetbuttcap%
\pgfsetbeveljoin%
\definecolor{currentfill}{rgb}{0.172549,0.627451,0.172549}%
\pgfsetfillcolor{currentfill}%
\pgfsetlinewidth{1.003750pt}%
\definecolor{currentstroke}{rgb}{0.172549,0.627451,0.172549}%
\pgfsetstrokecolor{currentstroke}%
\pgfsetdash{}{0pt}%
\pgfsys@defobject{currentmarker}{\pgfqpoint{-0.039627in}{-0.033709in}}{\pgfqpoint{0.039627in}{0.041667in}}{%
\pgfpathmoveto{\pgfqpoint{0.000000in}{0.041667in}}%
\pgfpathlineto{\pgfqpoint{-0.009355in}{0.012876in}}%
\pgfpathlineto{\pgfqpoint{-0.039627in}{0.012876in}}%
\pgfpathlineto{\pgfqpoint{-0.015136in}{-0.004918in}}%
\pgfpathlineto{\pgfqpoint{-0.024491in}{-0.033709in}}%
\pgfpathlineto{\pgfqpoint{-0.000000in}{-0.015915in}}%
\pgfpathlineto{\pgfqpoint{0.024491in}{-0.033709in}}%
\pgfpathlineto{\pgfqpoint{0.015136in}{-0.004918in}}%
\pgfpathlineto{\pgfqpoint{0.039627in}{0.012876in}}%
\pgfpathlineto{\pgfqpoint{0.009355in}{0.012876in}}%
\pgfpathclose%
\pgfusepath{stroke,fill}%
}%
\begin{pgfscope}%
\pgfsys@transformshift{0.306000in}{0.385360in}%
\pgfsys@useobject{currentmarker}{}%
\end{pgfscope}%
\begin{pgfscope}%
\pgfsys@transformshift{0.426800in}{0.392551in}%
\pgfsys@useobject{currentmarker}{}%
\end{pgfscope}%
\begin{pgfscope}%
\pgfsys@transformshift{0.668400in}{0.402199in}%
\pgfsys@useobject{currentmarker}{}%
\end{pgfscope}%
\begin{pgfscope}%
\pgfsys@transformshift{0.910000in}{0.424654in}%
\pgfsys@useobject{currentmarker}{}%
\end{pgfscope}%
\begin{pgfscope}%
\pgfsys@transformshift{1.151600in}{0.475435in}%
\pgfsys@useobject{currentmarker}{}%
\end{pgfscope}%
\begin{pgfscope}%
\pgfsys@transformshift{1.272400in}{0.621950in}%
\pgfsys@useobject{currentmarker}{}%
\end{pgfscope}%
\end{pgfscope}%
\begin{pgfscope}%
\pgfpathrectangle{\pgfqpoint{0.306000in}{0.304000in}}{\pgfqpoint{1.026800in}{1.232000in}}%
\pgfusepath{clip}%
\pgfsetbuttcap%
\pgfsetroundjoin%
\pgfsetlinewidth{0.903375pt}%
\definecolor{currentstroke}{rgb}{0.839216,0.152941,0.156863}%
\pgfsetstrokecolor{currentstroke}%
\pgfsetdash{{3.330000pt}{1.440000pt}}{0.000000pt}%
\pgfpathmoveto{\pgfqpoint{0.306000in}{0.447440in}}%
\pgfpathlineto{\pgfqpoint{0.426800in}{0.455672in}}%
\pgfpathlineto{\pgfqpoint{0.668400in}{0.506247in}}%
\pgfpathlineto{\pgfqpoint{0.910000in}{0.735949in}}%
\pgfpathlineto{\pgfqpoint{1.151600in}{1.039759in}}%
\pgfpathlineto{\pgfqpoint{1.272400in}{1.480000in}}%
\pgfusepath{stroke}%
\end{pgfscope}%
\begin{pgfscope}%
\pgfpathrectangle{\pgfqpoint{0.306000in}{0.304000in}}{\pgfqpoint{1.026800in}{1.232000in}}%
\pgfusepath{clip}%
\pgfsetbuttcap%
\pgfsetroundjoin%
\definecolor{currentfill}{rgb}{0.839216,0.152941,0.156863}%
\pgfsetfillcolor{currentfill}%
\pgfsetlinewidth{1.003750pt}%
\definecolor{currentstroke}{rgb}{0.839216,0.152941,0.156863}%
\pgfsetstrokecolor{currentstroke}%
\pgfsetdash{}{0pt}%
\pgfsys@defobject{currentmarker}{\pgfqpoint{-0.041667in}{-0.041667in}}{\pgfqpoint{0.041667in}{0.041667in}}{%
\pgfpathmoveto{\pgfqpoint{-0.041667in}{-0.041667in}}%
\pgfpathlineto{\pgfqpoint{0.041667in}{0.041667in}}%
\pgfpathmoveto{\pgfqpoint{-0.041667in}{0.041667in}}%
\pgfpathlineto{\pgfqpoint{0.041667in}{-0.041667in}}%
\pgfusepath{stroke,fill}%
}%
\begin{pgfscope}%
\pgfsys@transformshift{0.306000in}{0.447440in}%
\pgfsys@useobject{currentmarker}{}%
\end{pgfscope}%
\begin{pgfscope}%
\pgfsys@transformshift{0.426800in}{0.455672in}%
\pgfsys@useobject{currentmarker}{}%
\end{pgfscope}%
\begin{pgfscope}%
\pgfsys@transformshift{0.668400in}{0.506247in}%
\pgfsys@useobject{currentmarker}{}%
\end{pgfscope}%
\begin{pgfscope}%
\pgfsys@transformshift{0.910000in}{0.735949in}%
\pgfsys@useobject{currentmarker}{}%
\end{pgfscope}%
\begin{pgfscope}%
\pgfsys@transformshift{1.151600in}{1.039759in}%
\pgfsys@useobject{currentmarker}{}%
\end{pgfscope}%
\begin{pgfscope}%
\pgfsys@transformshift{1.272400in}{1.480000in}%
\pgfsys@useobject{currentmarker}{}%
\end{pgfscope}%
\end{pgfscope}%
\begin{pgfscope}%
\pgfsetrectcap%
\pgfsetmiterjoin%
\pgfsetlinewidth{0.803000pt}%
\definecolor{currentstroke}{rgb}{0.000000,0.000000,0.000000}%
\pgfsetstrokecolor{currentstroke}%
\pgfsetdash{}{0pt}%
\pgfpathmoveto{\pgfqpoint{0.306000in}{0.304000in}}%
\pgfpathlineto{\pgfqpoint{0.306000in}{1.536000in}}%
\pgfusepath{stroke}%
\end{pgfscope}%
\begin{pgfscope}%
\pgfsetrectcap%
\pgfsetmiterjoin%
\pgfsetlinewidth{0.803000pt}%
\definecolor{currentstroke}{rgb}{0.000000,0.000000,0.000000}%
\pgfsetstrokecolor{currentstroke}%
\pgfsetdash{}{0pt}%
\pgfpathmoveto{\pgfqpoint{1.332800in}{0.304000in}}%
\pgfpathlineto{\pgfqpoint{1.332800in}{1.536000in}}%
\pgfusepath{stroke}%
\end{pgfscope}%
\begin{pgfscope}%
\pgfsetrectcap%
\pgfsetmiterjoin%
\pgfsetlinewidth{0.803000pt}%
\definecolor{currentstroke}{rgb}{0.000000,0.000000,0.000000}%
\pgfsetstrokecolor{currentstroke}%
\pgfsetdash{}{0pt}%
\pgfpathmoveto{\pgfqpoint{0.306000in}{0.304000in}}%
\pgfpathlineto{\pgfqpoint{1.332800in}{0.304000in}}%
\pgfusepath{stroke}%
\end{pgfscope}%
\begin{pgfscope}%
\pgfsetrectcap%
\pgfsetmiterjoin%
\pgfsetlinewidth{0.803000pt}%
\definecolor{currentstroke}{rgb}{0.000000,0.000000,0.000000}%
\pgfsetstrokecolor{currentstroke}%
\pgfsetdash{}{0pt}%
\pgfpathmoveto{\pgfqpoint{0.306000in}{1.536000in}}%
\pgfpathlineto{\pgfqpoint{1.332800in}{1.536000in}}%
\pgfusepath{stroke}%
\end{pgfscope}%
\begin{pgfscope}%
\pgfsetbuttcap%
\pgfsetmiterjoin%
\definecolor{currentfill}{rgb}{1.000000,1.000000,1.000000}%
\pgfsetfillcolor{currentfill}%
\pgfsetfillopacity{0.800000}%
\pgfsetlinewidth{1.003750pt}%
\definecolor{currentstroke}{rgb}{0.800000,0.800000,0.800000}%
\pgfsetstrokecolor{currentstroke}%
\pgfsetstrokeopacity{0.800000}%
\pgfsetdash{}{0pt}%
\pgfpathmoveto{\pgfqpoint{0.364333in}{0.888777in}}%
\pgfpathlineto{\pgfqpoint{0.762171in}{0.888777in}}%
\pgfpathquadraticcurveto{\pgfqpoint{0.778838in}{0.888777in}}{\pgfqpoint{0.778838in}{0.905444in}}%
\pgfpathlineto{\pgfqpoint{0.778838in}{1.477667in}}%
\pgfpathquadraticcurveto{\pgfqpoint{0.778838in}{1.494333in}}{\pgfqpoint{0.762171in}{1.494333in}}%
\pgfpathlineto{\pgfqpoint{0.364333in}{1.494333in}}%
\pgfpathquadraticcurveto{\pgfqpoint{0.347667in}{1.494333in}}{\pgfqpoint{0.347667in}{1.477667in}}%
\pgfpathlineto{\pgfqpoint{0.347667in}{0.905444in}}%
\pgfpathquadraticcurveto{\pgfqpoint{0.347667in}{0.888777in}}{\pgfqpoint{0.364333in}{0.888777in}}%
\pgfpathclose%
\pgfusepath{stroke,fill}%
\end{pgfscope}%
\begin{pgfscope}%
\pgftext[x=0.449056in,y=1.403130in,left,base]{\rmfamily\fontsize{6.000000}{7.200000}\selectfont MLU}%
\end{pgfscope}%
\begin{pgfscope}%
\pgfsetrectcap%
\pgfsetroundjoin%
\pgfsetlinewidth{0.903375pt}%
\definecolor{currentstroke}{rgb}{0.121569,0.466667,0.705882}%
\pgfsetstrokecolor{currentstroke}%
\pgfsetdash{}{0pt}%
\pgfpathmoveto{\pgfqpoint{0.381000in}{1.316092in}}%
\pgfpathlineto{\pgfqpoint{0.547667in}{1.316092in}}%
\pgfusepath{stroke}%
\end{pgfscope}%
\begin{pgfscope}%
\pgfsetbuttcap%
\pgfsetroundjoin%
\definecolor{currentfill}{rgb}{0.121569,0.466667,0.705882}%
\pgfsetfillcolor{currentfill}%
\pgfsetlinewidth{1.003750pt}%
\definecolor{currentstroke}{rgb}{0.121569,0.466667,0.705882}%
\pgfsetstrokecolor{currentstroke}%
\pgfsetdash{}{0pt}%
\pgfsys@defobject{currentmarker}{\pgfqpoint{-0.033333in}{-0.041667in}}{\pgfqpoint{0.033333in}{0.020833in}}{%
\pgfpathmoveto{\pgfqpoint{0.000000in}{0.000000in}}%
\pgfpathlineto{\pgfqpoint{0.000000in}{-0.041667in}}%
\pgfpathmoveto{\pgfqpoint{0.000000in}{0.000000in}}%
\pgfpathlineto{\pgfqpoint{0.033333in}{0.020833in}}%
\pgfpathmoveto{\pgfqpoint{0.000000in}{0.000000in}}%
\pgfpathlineto{\pgfqpoint{-0.033333in}{0.020833in}}%
\pgfusepath{stroke,fill}%
}%
\begin{pgfscope}%
\pgfsys@transformshift{0.464333in}{1.316092in}%
\pgfsys@useobject{currentmarker}{}%
\end{pgfscope}%
\end{pgfscope}%
\begin{pgfscope}%
\pgftext[x=0.614333in,y=1.286926in,left,base]{\rmfamily\fontsize{6.000000}{7.200000}\selectfont 0.2}%
\end{pgfscope}%
\begin{pgfscope}%
\pgfsetbuttcap%
\pgfsetroundjoin%
\pgfsetlinewidth{0.903375pt}%
\definecolor{currentstroke}{rgb}{1.000000,0.498039,0.054902}%
\pgfsetstrokecolor{currentstroke}%
\pgfsetdash{{0.900000pt}{1.485000pt}}{0.000000pt}%
\pgfpathmoveto{\pgfqpoint{0.381000in}{1.199889in}}%
\pgfpathlineto{\pgfqpoint{0.547667in}{1.199889in}}%
\pgfusepath{stroke}%
\end{pgfscope}%
\begin{pgfscope}%
\pgfsetbuttcap%
\pgfsetroundjoin%
\definecolor{currentfill}{rgb}{1.000000,0.498039,0.054902}%
\pgfsetfillcolor{currentfill}%
\pgfsetlinewidth{1.003750pt}%
\definecolor{currentstroke}{rgb}{1.000000,0.498039,0.054902}%
\pgfsetstrokecolor{currentstroke}%
\pgfsetdash{}{0pt}%
\pgfsys@defobject{currentmarker}{\pgfqpoint{-0.041667in}{-0.041667in}}{\pgfqpoint{0.041667in}{0.041667in}}{%
\pgfpathmoveto{\pgfqpoint{-0.041667in}{0.000000in}}%
\pgfpathlineto{\pgfqpoint{0.041667in}{0.000000in}}%
\pgfpathmoveto{\pgfqpoint{0.000000in}{-0.041667in}}%
\pgfpathlineto{\pgfqpoint{0.000000in}{0.041667in}}%
\pgfusepath{stroke,fill}%
}%
\begin{pgfscope}%
\pgfsys@transformshift{0.464333in}{1.199889in}%
\pgfsys@useobject{currentmarker}{}%
\end{pgfscope}%
\end{pgfscope}%
\begin{pgfscope}%
\pgftext[x=0.614333in,y=1.170722in,left,base]{\rmfamily\fontsize{6.000000}{7.200000}\selectfont 0.4}%
\end{pgfscope}%
\begin{pgfscope}%
\pgfsetbuttcap%
\pgfsetroundjoin%
\pgfsetlinewidth{0.903375pt}%
\definecolor{currentstroke}{rgb}{0.172549,0.627451,0.172549}%
\pgfsetstrokecolor{currentstroke}%
\pgfsetdash{{5.760000pt}{1.440000pt}{0.900000pt}{1.440000pt}}{0.000000pt}%
\pgfpathmoveto{\pgfqpoint{0.381000in}{1.083685in}}%
\pgfpathlineto{\pgfqpoint{0.547667in}{1.083685in}}%
\pgfusepath{stroke}%
\end{pgfscope}%
\begin{pgfscope}%
\pgfsetbuttcap%
\pgfsetbeveljoin%
\definecolor{currentfill}{rgb}{0.172549,0.627451,0.172549}%
\pgfsetfillcolor{currentfill}%
\pgfsetlinewidth{1.003750pt}%
\definecolor{currentstroke}{rgb}{0.172549,0.627451,0.172549}%
\pgfsetstrokecolor{currentstroke}%
\pgfsetdash{}{0pt}%
\pgfsys@defobject{currentmarker}{\pgfqpoint{-0.039627in}{-0.033709in}}{\pgfqpoint{0.039627in}{0.041667in}}{%
\pgfpathmoveto{\pgfqpoint{0.000000in}{0.041667in}}%
\pgfpathlineto{\pgfqpoint{-0.009355in}{0.012876in}}%
\pgfpathlineto{\pgfqpoint{-0.039627in}{0.012876in}}%
\pgfpathlineto{\pgfqpoint{-0.015136in}{-0.004918in}}%
\pgfpathlineto{\pgfqpoint{-0.024491in}{-0.033709in}}%
\pgfpathlineto{\pgfqpoint{-0.000000in}{-0.015915in}}%
\pgfpathlineto{\pgfqpoint{0.024491in}{-0.033709in}}%
\pgfpathlineto{\pgfqpoint{0.015136in}{-0.004918in}}%
\pgfpathlineto{\pgfqpoint{0.039627in}{0.012876in}}%
\pgfpathlineto{\pgfqpoint{0.009355in}{0.012876in}}%
\pgfpathclose%
\pgfusepath{stroke,fill}%
}%
\begin{pgfscope}%
\pgfsys@transformshift{0.464333in}{1.083685in}%
\pgfsys@useobject{currentmarker}{}%
\end{pgfscope}%
\end{pgfscope}%
\begin{pgfscope}%
\pgftext[x=0.614333in,y=1.054518in,left,base]{\rmfamily\fontsize{6.000000}{7.200000}\selectfont 0.6}%
\end{pgfscope}%
\begin{pgfscope}%
\pgfsetbuttcap%
\pgfsetroundjoin%
\pgfsetlinewidth{0.903375pt}%
\definecolor{currentstroke}{rgb}{0.839216,0.152941,0.156863}%
\pgfsetstrokecolor{currentstroke}%
\pgfsetdash{{3.330000pt}{1.440000pt}}{0.000000pt}%
\pgfpathmoveto{\pgfqpoint{0.381000in}{0.967481in}}%
\pgfpathlineto{\pgfqpoint{0.547667in}{0.967481in}}%
\pgfusepath{stroke}%
\end{pgfscope}%
\begin{pgfscope}%
\pgfsetbuttcap%
\pgfsetroundjoin%
\definecolor{currentfill}{rgb}{0.839216,0.152941,0.156863}%
\pgfsetfillcolor{currentfill}%
\pgfsetlinewidth{1.003750pt}%
\definecolor{currentstroke}{rgb}{0.839216,0.152941,0.156863}%
\pgfsetstrokecolor{currentstroke}%
\pgfsetdash{}{0pt}%
\pgfsys@defobject{currentmarker}{\pgfqpoint{-0.041667in}{-0.041667in}}{\pgfqpoint{0.041667in}{0.041667in}}{%
\pgfpathmoveto{\pgfqpoint{-0.041667in}{-0.041667in}}%
\pgfpathlineto{\pgfqpoint{0.041667in}{0.041667in}}%
\pgfpathmoveto{\pgfqpoint{-0.041667in}{0.041667in}}%
\pgfpathlineto{\pgfqpoint{0.041667in}{-0.041667in}}%
\pgfusepath{stroke,fill}%
}%
\begin{pgfscope}%
\pgfsys@transformshift{0.464333in}{0.967481in}%
\pgfsys@useobject{currentmarker}{}%
\end{pgfscope}%
\end{pgfscope}%
\begin{pgfscope}%
\pgftext[x=0.614333in,y=0.938315in,left,base]{\rmfamily\fontsize{6.000000}{7.200000}\selectfont 0.8}%
\end{pgfscope}%
\end{pgfpicture}%
\makeatother%
\endgroup%

%% file: plot_data/packet_level_simulation/packet_level_frac_flows_completed.pgf
\begingroup%
\makeatletter%
\begin{pgfpicture}%
\pgfpathrectangle{\pgfpointorigin}{\pgfqpoint{1.360000in}{1.600000in}}%
\pgfusepath{use as bounding box, clip}%
\begin{pgfscope}%
\pgfsetbuttcap%
\pgfsetmiterjoin%
\definecolor{currentfill}{rgb}{1.000000,1.000000,1.000000}%
\pgfsetfillcolor{currentfill}%
\pgfsetlinewidth{0.000000pt}%
\definecolor{currentstroke}{rgb}{1.000000,1.000000,1.000000}%
\pgfsetstrokecolor{currentstroke}%
\pgfsetdash{}{0pt}%
\pgfpathmoveto{\pgfqpoint{0.000000in}{0.000000in}}%
\pgfpathlineto{\pgfqpoint{1.360000in}{0.000000in}}%
\pgfpathlineto{\pgfqpoint{1.360000in}{1.600000in}}%
\pgfpathlineto{\pgfqpoint{0.000000in}{1.600000in}}%
\pgfpathclose%
\pgfusepath{fill}%
\end{pgfscope}%
\begin{pgfscope}%
\pgfsetbuttcap%
\pgfsetmiterjoin%
\definecolor{currentfill}{rgb}{1.000000,1.000000,1.000000}%
\pgfsetfillcolor{currentfill}%
\pgfsetlinewidth{0.000000pt}%
\definecolor{currentstroke}{rgb}{0.000000,0.000000,0.000000}%
\pgfsetstrokecolor{currentstroke}%
\pgfsetstrokeopacity{0.000000}%
\pgfsetdash{}{0pt}%
\pgfpathmoveto{\pgfqpoint{0.306000in}{0.304000in}}%
\pgfpathlineto{\pgfqpoint{1.332800in}{0.304000in}}%
\pgfpathlineto{\pgfqpoint{1.332800in}{1.536000in}}%
\pgfpathlineto{\pgfqpoint{0.306000in}{1.536000in}}%
\pgfpathclose%
\pgfusepath{fill}%
\end{pgfscope}%
\begin{pgfscope}%
\pgfsetbuttcap%
\pgfsetroundjoin%
\definecolor{currentfill}{rgb}{0.000000,0.000000,0.000000}%
\pgfsetfillcolor{currentfill}%
\pgfsetlinewidth{0.803000pt}%
\definecolor{currentstroke}{rgb}{0.000000,0.000000,0.000000}%
\pgfsetstrokecolor{currentstroke}%
\pgfsetdash{}{0pt}%
\pgfsys@defobject{currentmarker}{\pgfqpoint{0.000000in}{-0.048611in}}{\pgfqpoint{0.000000in}{0.000000in}}{%
\pgfpathmoveto{\pgfqpoint{0.000000in}{0.000000in}}%
\pgfpathlineto{\pgfqpoint{0.000000in}{-0.048611in}}%
\pgfusepath{stroke,fill}%
}%
\begin{pgfscope}%
\pgfsys@transformshift{0.487200in}{0.304000in}%
\pgfsys@useobject{currentmarker}{}%
\end{pgfscope}%
\end{pgfscope}%
\begin{pgfscope}%
\pgftext[x=0.487200in,y=0.206778in,,top]{\rmfamily\fontsize{6.500000}{7.800000}\selectfont \(\displaystyle 1.25\)}%
\end{pgfscope}%
\begin{pgfscope}%
\pgfsetbuttcap%
\pgfsetroundjoin%
\definecolor{currentfill}{rgb}{0.000000,0.000000,0.000000}%
\pgfsetfillcolor{currentfill}%
\pgfsetlinewidth{0.803000pt}%
\definecolor{currentstroke}{rgb}{0.000000,0.000000,0.000000}%
\pgfsetstrokecolor{currentstroke}%
\pgfsetdash{}{0pt}%
\pgfsys@defobject{currentmarker}{\pgfqpoint{0.000000in}{-0.048611in}}{\pgfqpoint{0.000000in}{0.000000in}}{%
\pgfpathmoveto{\pgfqpoint{0.000000in}{0.000000in}}%
\pgfpathlineto{\pgfqpoint{0.000000in}{-0.048611in}}%
\pgfusepath{stroke,fill}%
}%
\begin{pgfscope}%
\pgfsys@transformshift{0.789200in}{0.304000in}%
\pgfsys@useobject{currentmarker}{}%
\end{pgfscope}%
\end{pgfscope}%
\begin{pgfscope}%
\pgftext[x=0.789200in,y=0.206778in,,top]{\rmfamily\fontsize{6.500000}{7.800000}\selectfont \(\displaystyle 1.50\)}%
\end{pgfscope}%
\begin{pgfscope}%
\pgfsetbuttcap%
\pgfsetroundjoin%
\definecolor{currentfill}{rgb}{0.000000,0.000000,0.000000}%
\pgfsetfillcolor{currentfill}%
\pgfsetlinewidth{0.803000pt}%
\definecolor{currentstroke}{rgb}{0.000000,0.000000,0.000000}%
\pgfsetstrokecolor{currentstroke}%
\pgfsetdash{}{0pt}%
\pgfsys@defobject{currentmarker}{\pgfqpoint{0.000000in}{-0.048611in}}{\pgfqpoint{0.000000in}{0.000000in}}{%
\pgfpathmoveto{\pgfqpoint{0.000000in}{0.000000in}}%
\pgfpathlineto{\pgfqpoint{0.000000in}{-0.048611in}}%
\pgfusepath{stroke,fill}%
}%
\begin{pgfscope}%
\pgfsys@transformshift{1.091200in}{0.304000in}%
\pgfsys@useobject{currentmarker}{}%
\end{pgfscope}%
\end{pgfscope}%
\begin{pgfscope}%
\pgftext[x=1.091200in,y=0.206778in,,top]{\rmfamily\fontsize{6.500000}{7.800000}\selectfont \(\displaystyle 1.75\)}%
\end{pgfscope}%
\begin{pgfscope}%
\pgftext[x=0.819400in,y=0.118815in,,top]{\rmfamily\fontsize{7.400000}{8.880000}\selectfont Avg. Hop Count}%
\end{pgfscope}%
\begin{pgfscope}%
\pgfpathrectangle{\pgfqpoint{0.306000in}{0.304000in}}{\pgfqpoint{1.026800in}{1.232000in}}%
\pgfusepath{clip}%
\pgfsetbuttcap%
\pgfsetroundjoin%
\pgfsetlinewidth{0.702625pt}%
\definecolor{currentstroke}{rgb}{0.690196,0.690196,0.690196}%
\pgfsetstrokecolor{currentstroke}%
\pgfsetdash{{4.480000pt}{1.120000pt}{0.700000pt}{1.120000pt}}{0.000000pt}%
\pgfpathmoveto{\pgfqpoint{0.306000in}{0.304000in}}%
\pgfpathlineto{\pgfqpoint{1.332800in}{0.304000in}}%
\pgfusepath{stroke}%
\end{pgfscope}%
\begin{pgfscope}%
\pgfsetbuttcap%
\pgfsetroundjoin%
\definecolor{currentfill}{rgb}{0.000000,0.000000,0.000000}%
\pgfsetfillcolor{currentfill}%
\pgfsetlinewidth{0.803000pt}%
\definecolor{currentstroke}{rgb}{0.000000,0.000000,0.000000}%
\pgfsetstrokecolor{currentstroke}%
\pgfsetdash{}{0pt}%
\pgfsys@defobject{currentmarker}{\pgfqpoint{0.000000in}{0.000000in}}{\pgfqpoint{0.048611in}{0.000000in}}{%
\pgfpathmoveto{\pgfqpoint{0.000000in}{0.000000in}}%
\pgfpathlineto{\pgfqpoint{0.048611in}{0.000000in}}%
\pgfusepath{stroke,fill}%
}%
\begin{pgfscope}%
\pgfsys@transformshift{0.306000in}{0.304000in}%
\pgfsys@useobject{currentmarker}{}%
\end{pgfscope}%
\end{pgfscope}%
\begin{pgfscope}%
\pgftext[x=0.123903in,y=0.275065in,left,base]{\rmfamily\fontsize{6.500000}{7.800000}\selectfont \(\displaystyle 0.0\)}%
\end{pgfscope}%
\begin{pgfscope}%
\pgfpathrectangle{\pgfqpoint{0.306000in}{0.304000in}}{\pgfqpoint{1.026800in}{1.232000in}}%
\pgfusepath{clip}%
\pgfsetbuttcap%
\pgfsetroundjoin%
\pgfsetlinewidth{0.702625pt}%
\definecolor{currentstroke}{rgb}{0.690196,0.690196,0.690196}%
\pgfsetstrokecolor{currentstroke}%
\pgfsetdash{{4.480000pt}{1.120000pt}{0.700000pt}{1.120000pt}}{0.000000pt}%
\pgfpathmoveto{\pgfqpoint{0.306000in}{0.545569in}}%
\pgfpathlineto{\pgfqpoint{1.332800in}{0.545569in}}%
\pgfusepath{stroke}%
\end{pgfscope}%
\begin{pgfscope}%
\pgfsetbuttcap%
\pgfsetroundjoin%
\definecolor{currentfill}{rgb}{0.000000,0.000000,0.000000}%
\pgfsetfillcolor{currentfill}%
\pgfsetlinewidth{0.803000pt}%
\definecolor{currentstroke}{rgb}{0.000000,0.000000,0.000000}%
\pgfsetstrokecolor{currentstroke}%
\pgfsetdash{}{0pt}%
\pgfsys@defobject{currentmarker}{\pgfqpoint{0.000000in}{0.000000in}}{\pgfqpoint{0.048611in}{0.000000in}}{%
\pgfpathmoveto{\pgfqpoint{0.000000in}{0.000000in}}%
\pgfpathlineto{\pgfqpoint{0.048611in}{0.000000in}}%
\pgfusepath{stroke,fill}%
}%
\begin{pgfscope}%
\pgfsys@transformshift{0.306000in}{0.545569in}%
\pgfsys@useobject{currentmarker}{}%
\end{pgfscope}%
\end{pgfscope}%
\begin{pgfscope}%
\pgftext[x=0.123903in,y=0.516633in,left,base]{\rmfamily\fontsize{6.500000}{7.800000}\selectfont \(\displaystyle 0.2\)}%
\end{pgfscope}%
\begin{pgfscope}%
\pgfpathrectangle{\pgfqpoint{0.306000in}{0.304000in}}{\pgfqpoint{1.026800in}{1.232000in}}%
\pgfusepath{clip}%
\pgfsetbuttcap%
\pgfsetroundjoin%
\pgfsetlinewidth{0.702625pt}%
\definecolor{currentstroke}{rgb}{0.690196,0.690196,0.690196}%
\pgfsetstrokecolor{currentstroke}%
\pgfsetdash{{4.480000pt}{1.120000pt}{0.700000pt}{1.120000pt}}{0.000000pt}%
\pgfpathmoveto{\pgfqpoint{0.306000in}{0.787137in}}%
\pgfpathlineto{\pgfqpoint{1.332800in}{0.787137in}}%
\pgfusepath{stroke}%
\end{pgfscope}%
\begin{pgfscope}%
\pgfsetbuttcap%
\pgfsetroundjoin%
\definecolor{currentfill}{rgb}{0.000000,0.000000,0.000000}%
\pgfsetfillcolor{currentfill}%
\pgfsetlinewidth{0.803000pt}%
\definecolor{currentstroke}{rgb}{0.000000,0.000000,0.000000}%
\pgfsetstrokecolor{currentstroke}%
\pgfsetdash{}{0pt}%
\pgfsys@defobject{currentmarker}{\pgfqpoint{0.000000in}{0.000000in}}{\pgfqpoint{0.048611in}{0.000000in}}{%
\pgfpathmoveto{\pgfqpoint{0.000000in}{0.000000in}}%
\pgfpathlineto{\pgfqpoint{0.048611in}{0.000000in}}%
\pgfusepath{stroke,fill}%
}%
\begin{pgfscope}%
\pgfsys@transformshift{0.306000in}{0.787137in}%
\pgfsys@useobject{currentmarker}{}%
\end{pgfscope}%
\end{pgfscope}%
\begin{pgfscope}%
\pgftext[x=0.123903in,y=0.758202in,left,base]{\rmfamily\fontsize{6.500000}{7.800000}\selectfont \(\displaystyle 0.4\)}%
\end{pgfscope}%
\begin{pgfscope}%
\pgfpathrectangle{\pgfqpoint{0.306000in}{0.304000in}}{\pgfqpoint{1.026800in}{1.232000in}}%
\pgfusepath{clip}%
\pgfsetbuttcap%
\pgfsetroundjoin%
\pgfsetlinewidth{0.702625pt}%
\definecolor{currentstroke}{rgb}{0.690196,0.690196,0.690196}%
\pgfsetstrokecolor{currentstroke}%
\pgfsetdash{{4.480000pt}{1.120000pt}{0.700000pt}{1.120000pt}}{0.000000pt}%
\pgfpathmoveto{\pgfqpoint{0.306000in}{1.028706in}}%
\pgfpathlineto{\pgfqpoint{1.332800in}{1.028706in}}%
\pgfusepath{stroke}%
\end{pgfscope}%
\begin{pgfscope}%
\pgfsetbuttcap%
\pgfsetroundjoin%
\definecolor{currentfill}{rgb}{0.000000,0.000000,0.000000}%
\pgfsetfillcolor{currentfill}%
\pgfsetlinewidth{0.803000pt}%
\definecolor{currentstroke}{rgb}{0.000000,0.000000,0.000000}%
\pgfsetstrokecolor{currentstroke}%
\pgfsetdash{}{0pt}%
\pgfsys@defobject{currentmarker}{\pgfqpoint{0.000000in}{0.000000in}}{\pgfqpoint{0.048611in}{0.000000in}}{%
\pgfpathmoveto{\pgfqpoint{0.000000in}{0.000000in}}%
\pgfpathlineto{\pgfqpoint{0.048611in}{0.000000in}}%
\pgfusepath{stroke,fill}%
}%
\begin{pgfscope}%
\pgfsys@transformshift{0.306000in}{1.028706in}%
\pgfsys@useobject{currentmarker}{}%
\end{pgfscope}%
\end{pgfscope}%
\begin{pgfscope}%
\pgftext[x=0.123903in,y=0.999771in,left,base]{\rmfamily\fontsize{6.500000}{7.800000}\selectfont \(\displaystyle 0.6\)}%
\end{pgfscope}%
\begin{pgfscope}%
\pgfpathrectangle{\pgfqpoint{0.306000in}{0.304000in}}{\pgfqpoint{1.026800in}{1.232000in}}%
\pgfusepath{clip}%
\pgfsetbuttcap%
\pgfsetroundjoin%
\pgfsetlinewidth{0.702625pt}%
\definecolor{currentstroke}{rgb}{0.690196,0.690196,0.690196}%
\pgfsetstrokecolor{currentstroke}%
\pgfsetdash{{4.480000pt}{1.120000pt}{0.700000pt}{1.120000pt}}{0.000000pt}%
\pgfpathmoveto{\pgfqpoint{0.306000in}{1.270275in}}%
\pgfpathlineto{\pgfqpoint{1.332800in}{1.270275in}}%
\pgfusepath{stroke}%
\end{pgfscope}%
\begin{pgfscope}%
\pgfsetbuttcap%
\pgfsetroundjoin%
\definecolor{currentfill}{rgb}{0.000000,0.000000,0.000000}%
\pgfsetfillcolor{currentfill}%
\pgfsetlinewidth{0.803000pt}%
\definecolor{currentstroke}{rgb}{0.000000,0.000000,0.000000}%
\pgfsetstrokecolor{currentstroke}%
\pgfsetdash{}{0pt}%
\pgfsys@defobject{currentmarker}{\pgfqpoint{0.000000in}{0.000000in}}{\pgfqpoint{0.048611in}{0.000000in}}{%
\pgfpathmoveto{\pgfqpoint{0.000000in}{0.000000in}}%
\pgfpathlineto{\pgfqpoint{0.048611in}{0.000000in}}%
\pgfusepath{stroke,fill}%
}%
\begin{pgfscope}%
\pgfsys@transformshift{0.306000in}{1.270275in}%
\pgfsys@useobject{currentmarker}{}%
\end{pgfscope}%
\end{pgfscope}%
\begin{pgfscope}%
\pgftext[x=0.123903in,y=1.241339in,left,base]{\rmfamily\fontsize{6.500000}{7.800000}\selectfont \(\displaystyle 0.8\)}%
\end{pgfscope}%
\begin{pgfscope}%
\pgfpathrectangle{\pgfqpoint{0.306000in}{0.304000in}}{\pgfqpoint{1.026800in}{1.232000in}}%
\pgfusepath{clip}%
\pgfsetbuttcap%
\pgfsetroundjoin%
\pgfsetlinewidth{0.702625pt}%
\definecolor{currentstroke}{rgb}{0.690196,0.690196,0.690196}%
\pgfsetstrokecolor{currentstroke}%
\pgfsetdash{{4.480000pt}{1.120000pt}{0.700000pt}{1.120000pt}}{0.000000pt}%
\pgfpathmoveto{\pgfqpoint{0.306000in}{1.511843in}}%
\pgfpathlineto{\pgfqpoint{1.332800in}{1.511843in}}%
\pgfusepath{stroke}%
\end{pgfscope}%
\begin{pgfscope}%
\pgfsetbuttcap%
\pgfsetroundjoin%
\definecolor{currentfill}{rgb}{0.000000,0.000000,0.000000}%
\pgfsetfillcolor{currentfill}%
\pgfsetlinewidth{0.803000pt}%
\definecolor{currentstroke}{rgb}{0.000000,0.000000,0.000000}%
\pgfsetstrokecolor{currentstroke}%
\pgfsetdash{}{0pt}%
\pgfsys@defobject{currentmarker}{\pgfqpoint{0.000000in}{0.000000in}}{\pgfqpoint{0.048611in}{0.000000in}}{%
\pgfpathmoveto{\pgfqpoint{0.000000in}{0.000000in}}%
\pgfpathlineto{\pgfqpoint{0.048611in}{0.000000in}}%
\pgfusepath{stroke,fill}%
}%
\begin{pgfscope}%
\pgfsys@transformshift{0.306000in}{1.511843in}%
\pgfsys@useobject{currentmarker}{}%
\end{pgfscope}%
\end{pgfscope}%
\begin{pgfscope}%
\pgftext[x=0.123903in,y=1.482908in,left,base]{\rmfamily\fontsize{6.500000}{7.800000}\selectfont \(\displaystyle 1.0\)}%
\end{pgfscope}%
\begin{pgfscope}%
\pgfsetbuttcap%
\pgfsetroundjoin%
\definecolor{currentfill}{rgb}{0.000000,0.000000,0.000000}%
\pgfsetfillcolor{currentfill}%
\pgfsetlinewidth{0.602250pt}%
\definecolor{currentstroke}{rgb}{0.000000,0.000000,0.000000}%
\pgfsetstrokecolor{currentstroke}%
\pgfsetdash{}{0pt}%
\pgfsys@defobject{currentmarker}{\pgfqpoint{0.000000in}{0.000000in}}{\pgfqpoint{0.027778in}{0.000000in}}{%
\pgfpathmoveto{\pgfqpoint{0.000000in}{0.000000in}}%
\pgfpathlineto{\pgfqpoint{0.027778in}{0.000000in}}%
\pgfusepath{stroke,fill}%
}%
\begin{pgfscope}%
\pgfsys@transformshift{0.306000in}{0.364392in}%
\pgfsys@useobject{currentmarker}{}%
\end{pgfscope}%
\end{pgfscope}%
\begin{pgfscope}%
\pgfsetbuttcap%
\pgfsetroundjoin%
\definecolor{currentfill}{rgb}{0.000000,0.000000,0.000000}%
\pgfsetfillcolor{currentfill}%
\pgfsetlinewidth{0.602250pt}%
\definecolor{currentstroke}{rgb}{0.000000,0.000000,0.000000}%
\pgfsetstrokecolor{currentstroke}%
\pgfsetdash{}{0pt}%
\pgfsys@defobject{currentmarker}{\pgfqpoint{0.000000in}{0.000000in}}{\pgfqpoint{0.027778in}{0.000000in}}{%
\pgfpathmoveto{\pgfqpoint{0.000000in}{0.000000in}}%
\pgfpathlineto{\pgfqpoint{0.027778in}{0.000000in}}%
\pgfusepath{stroke,fill}%
}%
\begin{pgfscope}%
\pgfsys@transformshift{0.306000in}{0.424784in}%
\pgfsys@useobject{currentmarker}{}%
\end{pgfscope}%
\end{pgfscope}%
\begin{pgfscope}%
\pgfsetbuttcap%
\pgfsetroundjoin%
\definecolor{currentfill}{rgb}{0.000000,0.000000,0.000000}%
\pgfsetfillcolor{currentfill}%
\pgfsetlinewidth{0.602250pt}%
\definecolor{currentstroke}{rgb}{0.000000,0.000000,0.000000}%
\pgfsetstrokecolor{currentstroke}%
\pgfsetdash{}{0pt}%
\pgfsys@defobject{currentmarker}{\pgfqpoint{0.000000in}{0.000000in}}{\pgfqpoint{0.027778in}{0.000000in}}{%
\pgfpathmoveto{\pgfqpoint{0.000000in}{0.000000in}}%
\pgfpathlineto{\pgfqpoint{0.027778in}{0.000000in}}%
\pgfusepath{stroke,fill}%
}%
\begin{pgfscope}%
\pgfsys@transformshift{0.306000in}{0.485176in}%
\pgfsys@useobject{currentmarker}{}%
\end{pgfscope}%
\end{pgfscope}%
\begin{pgfscope}%
\pgfsetbuttcap%
\pgfsetroundjoin%
\definecolor{currentfill}{rgb}{0.000000,0.000000,0.000000}%
\pgfsetfillcolor{currentfill}%
\pgfsetlinewidth{0.602250pt}%
\definecolor{currentstroke}{rgb}{0.000000,0.000000,0.000000}%
\pgfsetstrokecolor{currentstroke}%
\pgfsetdash{}{0pt}%
\pgfsys@defobject{currentmarker}{\pgfqpoint{0.000000in}{0.000000in}}{\pgfqpoint{0.027778in}{0.000000in}}{%
\pgfpathmoveto{\pgfqpoint{0.000000in}{0.000000in}}%
\pgfpathlineto{\pgfqpoint{0.027778in}{0.000000in}}%
\pgfusepath{stroke,fill}%
}%
\begin{pgfscope}%
\pgfsys@transformshift{0.306000in}{0.605961in}%
\pgfsys@useobject{currentmarker}{}%
\end{pgfscope}%
\end{pgfscope}%
\begin{pgfscope}%
\pgfsetbuttcap%
\pgfsetroundjoin%
\definecolor{currentfill}{rgb}{0.000000,0.000000,0.000000}%
\pgfsetfillcolor{currentfill}%
\pgfsetlinewidth{0.602250pt}%
\definecolor{currentstroke}{rgb}{0.000000,0.000000,0.000000}%
\pgfsetstrokecolor{currentstroke}%
\pgfsetdash{}{0pt}%
\pgfsys@defobject{currentmarker}{\pgfqpoint{0.000000in}{0.000000in}}{\pgfqpoint{0.027778in}{0.000000in}}{%
\pgfpathmoveto{\pgfqpoint{0.000000in}{0.000000in}}%
\pgfpathlineto{\pgfqpoint{0.027778in}{0.000000in}}%
\pgfusepath{stroke,fill}%
}%
\begin{pgfscope}%
\pgfsys@transformshift{0.306000in}{0.666353in}%
\pgfsys@useobject{currentmarker}{}%
\end{pgfscope}%
\end{pgfscope}%
\begin{pgfscope}%
\pgfsetbuttcap%
\pgfsetroundjoin%
\definecolor{currentfill}{rgb}{0.000000,0.000000,0.000000}%
\pgfsetfillcolor{currentfill}%
\pgfsetlinewidth{0.602250pt}%
\definecolor{currentstroke}{rgb}{0.000000,0.000000,0.000000}%
\pgfsetstrokecolor{currentstroke}%
\pgfsetdash{}{0pt}%
\pgfsys@defobject{currentmarker}{\pgfqpoint{0.000000in}{0.000000in}}{\pgfqpoint{0.027778in}{0.000000in}}{%
\pgfpathmoveto{\pgfqpoint{0.000000in}{0.000000in}}%
\pgfpathlineto{\pgfqpoint{0.027778in}{0.000000in}}%
\pgfusepath{stroke,fill}%
}%
\begin{pgfscope}%
\pgfsys@transformshift{0.306000in}{0.726745in}%
\pgfsys@useobject{currentmarker}{}%
\end{pgfscope}%
\end{pgfscope}%
\begin{pgfscope}%
\pgfsetbuttcap%
\pgfsetroundjoin%
\definecolor{currentfill}{rgb}{0.000000,0.000000,0.000000}%
\pgfsetfillcolor{currentfill}%
\pgfsetlinewidth{0.602250pt}%
\definecolor{currentstroke}{rgb}{0.000000,0.000000,0.000000}%
\pgfsetstrokecolor{currentstroke}%
\pgfsetdash{}{0pt}%
\pgfsys@defobject{currentmarker}{\pgfqpoint{0.000000in}{0.000000in}}{\pgfqpoint{0.027778in}{0.000000in}}{%
\pgfpathmoveto{\pgfqpoint{0.000000in}{0.000000in}}%
\pgfpathlineto{\pgfqpoint{0.027778in}{0.000000in}}%
\pgfusepath{stroke,fill}%
}%
\begin{pgfscope}%
\pgfsys@transformshift{0.306000in}{0.847529in}%
\pgfsys@useobject{currentmarker}{}%
\end{pgfscope}%
\end{pgfscope}%
\begin{pgfscope}%
\pgfsetbuttcap%
\pgfsetroundjoin%
\definecolor{currentfill}{rgb}{0.000000,0.000000,0.000000}%
\pgfsetfillcolor{currentfill}%
\pgfsetlinewidth{0.602250pt}%
\definecolor{currentstroke}{rgb}{0.000000,0.000000,0.000000}%
\pgfsetstrokecolor{currentstroke}%
\pgfsetdash{}{0pt}%
\pgfsys@defobject{currentmarker}{\pgfqpoint{0.000000in}{0.000000in}}{\pgfqpoint{0.027778in}{0.000000in}}{%
\pgfpathmoveto{\pgfqpoint{0.000000in}{0.000000in}}%
\pgfpathlineto{\pgfqpoint{0.027778in}{0.000000in}}%
\pgfusepath{stroke,fill}%
}%
\begin{pgfscope}%
\pgfsys@transformshift{0.306000in}{0.907922in}%
\pgfsys@useobject{currentmarker}{}%
\end{pgfscope}%
\end{pgfscope}%
\begin{pgfscope}%
\pgfsetbuttcap%
\pgfsetroundjoin%
\definecolor{currentfill}{rgb}{0.000000,0.000000,0.000000}%
\pgfsetfillcolor{currentfill}%
\pgfsetlinewidth{0.602250pt}%
\definecolor{currentstroke}{rgb}{0.000000,0.000000,0.000000}%
\pgfsetstrokecolor{currentstroke}%
\pgfsetdash{}{0pt}%
\pgfsys@defobject{currentmarker}{\pgfqpoint{0.000000in}{0.000000in}}{\pgfqpoint{0.027778in}{0.000000in}}{%
\pgfpathmoveto{\pgfqpoint{0.000000in}{0.000000in}}%
\pgfpathlineto{\pgfqpoint{0.027778in}{0.000000in}}%
\pgfusepath{stroke,fill}%
}%
\begin{pgfscope}%
\pgfsys@transformshift{0.306000in}{0.968314in}%
\pgfsys@useobject{currentmarker}{}%
\end{pgfscope}%
\end{pgfscope}%
\begin{pgfscope}%
\pgfsetbuttcap%
\pgfsetroundjoin%
\definecolor{currentfill}{rgb}{0.000000,0.000000,0.000000}%
\pgfsetfillcolor{currentfill}%
\pgfsetlinewidth{0.602250pt}%
\definecolor{currentstroke}{rgb}{0.000000,0.000000,0.000000}%
\pgfsetstrokecolor{currentstroke}%
\pgfsetdash{}{0pt}%
\pgfsys@defobject{currentmarker}{\pgfqpoint{0.000000in}{0.000000in}}{\pgfqpoint{0.027778in}{0.000000in}}{%
\pgfpathmoveto{\pgfqpoint{0.000000in}{0.000000in}}%
\pgfpathlineto{\pgfqpoint{0.027778in}{0.000000in}}%
\pgfusepath{stroke,fill}%
}%
\begin{pgfscope}%
\pgfsys@transformshift{0.306000in}{1.089098in}%
\pgfsys@useobject{currentmarker}{}%
\end{pgfscope}%
\end{pgfscope}%
\begin{pgfscope}%
\pgfsetbuttcap%
\pgfsetroundjoin%
\definecolor{currentfill}{rgb}{0.000000,0.000000,0.000000}%
\pgfsetfillcolor{currentfill}%
\pgfsetlinewidth{0.602250pt}%
\definecolor{currentstroke}{rgb}{0.000000,0.000000,0.000000}%
\pgfsetstrokecolor{currentstroke}%
\pgfsetdash{}{0pt}%
\pgfsys@defobject{currentmarker}{\pgfqpoint{0.000000in}{0.000000in}}{\pgfqpoint{0.027778in}{0.000000in}}{%
\pgfpathmoveto{\pgfqpoint{0.000000in}{0.000000in}}%
\pgfpathlineto{\pgfqpoint{0.027778in}{0.000000in}}%
\pgfusepath{stroke,fill}%
}%
\begin{pgfscope}%
\pgfsys@transformshift{0.306000in}{1.149490in}%
\pgfsys@useobject{currentmarker}{}%
\end{pgfscope}%
\end{pgfscope}%
\begin{pgfscope}%
\pgfsetbuttcap%
\pgfsetroundjoin%
\definecolor{currentfill}{rgb}{0.000000,0.000000,0.000000}%
\pgfsetfillcolor{currentfill}%
\pgfsetlinewidth{0.602250pt}%
\definecolor{currentstroke}{rgb}{0.000000,0.000000,0.000000}%
\pgfsetstrokecolor{currentstroke}%
\pgfsetdash{}{0pt}%
\pgfsys@defobject{currentmarker}{\pgfqpoint{0.000000in}{0.000000in}}{\pgfqpoint{0.027778in}{0.000000in}}{%
\pgfpathmoveto{\pgfqpoint{0.000000in}{0.000000in}}%
\pgfpathlineto{\pgfqpoint{0.027778in}{0.000000in}}%
\pgfusepath{stroke,fill}%
}%
\begin{pgfscope}%
\pgfsys@transformshift{0.306000in}{1.209882in}%
\pgfsys@useobject{currentmarker}{}%
\end{pgfscope}%
\end{pgfscope}%
\begin{pgfscope}%
\pgfsetbuttcap%
\pgfsetroundjoin%
\definecolor{currentfill}{rgb}{0.000000,0.000000,0.000000}%
\pgfsetfillcolor{currentfill}%
\pgfsetlinewidth{0.602250pt}%
\definecolor{currentstroke}{rgb}{0.000000,0.000000,0.000000}%
\pgfsetstrokecolor{currentstroke}%
\pgfsetdash{}{0pt}%
\pgfsys@defobject{currentmarker}{\pgfqpoint{0.000000in}{0.000000in}}{\pgfqpoint{0.027778in}{0.000000in}}{%
\pgfpathmoveto{\pgfqpoint{0.000000in}{0.000000in}}%
\pgfpathlineto{\pgfqpoint{0.027778in}{0.000000in}}%
\pgfusepath{stroke,fill}%
}%
\begin{pgfscope}%
\pgfsys@transformshift{0.306000in}{1.330667in}%
\pgfsys@useobject{currentmarker}{}%
\end{pgfscope}%
\end{pgfscope}%
\begin{pgfscope}%
\pgfsetbuttcap%
\pgfsetroundjoin%
\definecolor{currentfill}{rgb}{0.000000,0.000000,0.000000}%
\pgfsetfillcolor{currentfill}%
\pgfsetlinewidth{0.602250pt}%
\definecolor{currentstroke}{rgb}{0.000000,0.000000,0.000000}%
\pgfsetstrokecolor{currentstroke}%
\pgfsetdash{}{0pt}%
\pgfsys@defobject{currentmarker}{\pgfqpoint{0.000000in}{0.000000in}}{\pgfqpoint{0.027778in}{0.000000in}}{%
\pgfpathmoveto{\pgfqpoint{0.000000in}{0.000000in}}%
\pgfpathlineto{\pgfqpoint{0.027778in}{0.000000in}}%
\pgfusepath{stroke,fill}%
}%
\begin{pgfscope}%
\pgfsys@transformshift{0.306000in}{1.391059in}%
\pgfsys@useobject{currentmarker}{}%
\end{pgfscope}%
\end{pgfscope}%
\begin{pgfscope}%
\pgfsetbuttcap%
\pgfsetroundjoin%
\definecolor{currentfill}{rgb}{0.000000,0.000000,0.000000}%
\pgfsetfillcolor{currentfill}%
\pgfsetlinewidth{0.602250pt}%
\definecolor{currentstroke}{rgb}{0.000000,0.000000,0.000000}%
\pgfsetstrokecolor{currentstroke}%
\pgfsetdash{}{0pt}%
\pgfsys@defobject{currentmarker}{\pgfqpoint{0.000000in}{0.000000in}}{\pgfqpoint{0.027778in}{0.000000in}}{%
\pgfpathmoveto{\pgfqpoint{0.000000in}{0.000000in}}%
\pgfpathlineto{\pgfqpoint{0.027778in}{0.000000in}}%
\pgfusepath{stroke,fill}%
}%
\begin{pgfscope}%
\pgfsys@transformshift{0.306000in}{1.451451in}%
\pgfsys@useobject{currentmarker}{}%
\end{pgfscope}%
\end{pgfscope}%
\begin{pgfscope}%
\pgftext[x=0.110014in,y=0.920000in,,bottom,rotate=90.000000]{\rmfamily\fontsize{7.400000}{8.880000}\selectfont Frac. of Flows Completed}%
\end{pgfscope}%
\begin{pgfscope}%
\pgfpathrectangle{\pgfqpoint{0.306000in}{0.304000in}}{\pgfqpoint{1.026800in}{1.232000in}}%
\pgfusepath{clip}%
\pgfsetrectcap%
\pgfsetroundjoin%
\pgfsetlinewidth{0.903375pt}%
\definecolor{currentstroke}{rgb}{0.121569,0.466667,0.705882}%
\pgfsetstrokecolor{currentstroke}%
\pgfsetdash{}{0pt}%
\pgfpathmoveto{\pgfqpoint{0.306000in}{1.511843in}}%
\pgfpathlineto{\pgfqpoint{0.426800in}{1.511843in}}%
\pgfpathlineto{\pgfqpoint{0.668400in}{1.511843in}}%
\pgfpathlineto{\pgfqpoint{0.910000in}{1.511843in}}%
\pgfpathlineto{\pgfqpoint{1.151600in}{1.511843in}}%
\pgfpathlineto{\pgfqpoint{1.272400in}{1.511843in}}%
\pgfusepath{stroke}%
\end{pgfscope}%
\begin{pgfscope}%
\pgfpathrectangle{\pgfqpoint{0.306000in}{0.304000in}}{\pgfqpoint{1.026800in}{1.232000in}}%
\pgfusepath{clip}%
\pgfsetbuttcap%
\pgfsetroundjoin%
\definecolor{currentfill}{rgb}{0.121569,0.466667,0.705882}%
\pgfsetfillcolor{currentfill}%
\pgfsetlinewidth{1.003750pt}%
\definecolor{currentstroke}{rgb}{0.121569,0.466667,0.705882}%
\pgfsetstrokecolor{currentstroke}%
\pgfsetdash{}{0pt}%
\pgfsys@defobject{currentmarker}{\pgfqpoint{-0.033333in}{-0.041667in}}{\pgfqpoint{0.033333in}{0.020833in}}{%
\pgfpathmoveto{\pgfqpoint{0.000000in}{0.000000in}}%
\pgfpathlineto{\pgfqpoint{0.000000in}{-0.041667in}}%
\pgfpathmoveto{\pgfqpoint{0.000000in}{0.000000in}}%
\pgfpathlineto{\pgfqpoint{0.033333in}{0.020833in}}%
\pgfpathmoveto{\pgfqpoint{0.000000in}{0.000000in}}%
\pgfpathlineto{\pgfqpoint{-0.033333in}{0.020833in}}%
\pgfusepath{stroke,fill}%
}%
\begin{pgfscope}%
\pgfsys@transformshift{0.306000in}{1.511843in}%
\pgfsys@useobject{currentmarker}{}%
\end{pgfscope}%
\begin{pgfscope}%
\pgfsys@transformshift{0.426800in}{1.511843in}%
\pgfsys@useobject{currentmarker}{}%
\end{pgfscope}%
\begin{pgfscope}%
\pgfsys@transformshift{0.668400in}{1.511843in}%
\pgfsys@useobject{currentmarker}{}%
\end{pgfscope}%
\begin{pgfscope}%
\pgfsys@transformshift{0.910000in}{1.511843in}%
\pgfsys@useobject{currentmarker}{}%
\end{pgfscope}%
\begin{pgfscope}%
\pgfsys@transformshift{1.151600in}{1.511843in}%
\pgfsys@useobject{currentmarker}{}%
\end{pgfscope}%
\begin{pgfscope}%
\pgfsys@transformshift{1.272400in}{1.511843in}%
\pgfsys@useobject{currentmarker}{}%
\end{pgfscope}%
\end{pgfscope}%
\begin{pgfscope}%
\pgfpathrectangle{\pgfqpoint{0.306000in}{0.304000in}}{\pgfqpoint{1.026800in}{1.232000in}}%
\pgfusepath{clip}%
\pgfsetbuttcap%
\pgfsetroundjoin%
\pgfsetlinewidth{0.903375pt}%
\definecolor{currentstroke}{rgb}{1.000000,0.498039,0.054902}%
\pgfsetstrokecolor{currentstroke}%
\pgfsetdash{{0.900000pt}{1.485000pt}}{0.000000pt}%
\pgfpathmoveto{\pgfqpoint{0.306000in}{1.510818in}}%
\pgfpathlineto{\pgfqpoint{0.426800in}{1.510614in}}%
\pgfpathlineto{\pgfqpoint{0.668400in}{1.506925in}}%
\pgfpathlineto{\pgfqpoint{0.910000in}{1.499138in}}%
\pgfpathlineto{\pgfqpoint{1.151600in}{1.469423in}}%
\pgfpathlineto{\pgfqpoint{1.272400in}{1.438069in}}%
\pgfusepath{stroke}%
\end{pgfscope}%
\begin{pgfscope}%
\pgfpathrectangle{\pgfqpoint{0.306000in}{0.304000in}}{\pgfqpoint{1.026800in}{1.232000in}}%
\pgfusepath{clip}%
\pgfsetbuttcap%
\pgfsetroundjoin%
\definecolor{currentfill}{rgb}{1.000000,0.498039,0.054902}%
\pgfsetfillcolor{currentfill}%
\pgfsetlinewidth{1.003750pt}%
\definecolor{currentstroke}{rgb}{1.000000,0.498039,0.054902}%
\pgfsetstrokecolor{currentstroke}%
\pgfsetdash{}{0pt}%
\pgfsys@defobject{currentmarker}{\pgfqpoint{-0.041667in}{-0.041667in}}{\pgfqpoint{0.041667in}{0.041667in}}{%
\pgfpathmoveto{\pgfqpoint{-0.041667in}{0.000000in}}%
\pgfpathlineto{\pgfqpoint{0.041667in}{0.000000in}}%
\pgfpathmoveto{\pgfqpoint{0.000000in}{-0.041667in}}%
\pgfpathlineto{\pgfqpoint{0.000000in}{0.041667in}}%
\pgfusepath{stroke,fill}%
}%
\begin{pgfscope}%
\pgfsys@transformshift{0.306000in}{1.510818in}%
\pgfsys@useobject{currentmarker}{}%
\end{pgfscope}%
\begin{pgfscope}%
\pgfsys@transformshift{0.426800in}{1.510614in}%
\pgfsys@useobject{currentmarker}{}%
\end{pgfscope}%
\begin{pgfscope}%
\pgfsys@transformshift{0.668400in}{1.506925in}%
\pgfsys@useobject{currentmarker}{}%
\end{pgfscope}%
\begin{pgfscope}%
\pgfsys@transformshift{0.910000in}{1.499138in}%
\pgfsys@useobject{currentmarker}{}%
\end{pgfscope}%
\begin{pgfscope}%
\pgfsys@transformshift{1.151600in}{1.469423in}%
\pgfsys@useobject{currentmarker}{}%
\end{pgfscope}%
\begin{pgfscope}%
\pgfsys@transformshift{1.272400in}{1.438069in}%
\pgfsys@useobject{currentmarker}{}%
\end{pgfscope}%
\end{pgfscope}%
\begin{pgfscope}%
\pgfpathrectangle{\pgfqpoint{0.306000in}{0.304000in}}{\pgfqpoint{1.026800in}{1.232000in}}%
\pgfusepath{clip}%
\pgfsetbuttcap%
\pgfsetroundjoin%
\pgfsetlinewidth{0.903375pt}%
\definecolor{currentstroke}{rgb}{0.172549,0.627451,0.172549}%
\pgfsetstrokecolor{currentstroke}%
\pgfsetdash{{5.760000pt}{1.440000pt}{0.900000pt}{1.440000pt}}{0.000000pt}%
\pgfpathmoveto{\pgfqpoint{0.306000in}{1.498548in}}%
\pgfpathlineto{\pgfqpoint{0.426800in}{1.479495in}}%
\pgfpathlineto{\pgfqpoint{0.668400in}{1.355723in}}%
\pgfpathlineto{\pgfqpoint{0.910000in}{1.060067in}}%
\pgfpathlineto{\pgfqpoint{1.151600in}{0.836509in}}%
\pgfpathlineto{\pgfqpoint{1.272400in}{0.734942in}}%
\pgfusepath{stroke}%
\end{pgfscope}%
\begin{pgfscope}%
\pgfpathrectangle{\pgfqpoint{0.306000in}{0.304000in}}{\pgfqpoint{1.026800in}{1.232000in}}%
\pgfusepath{clip}%
\pgfsetbuttcap%
\pgfsetbeveljoin%
\definecolor{currentfill}{rgb}{0.172549,0.627451,0.172549}%
\pgfsetfillcolor{currentfill}%
\pgfsetlinewidth{1.003750pt}%
\definecolor{currentstroke}{rgb}{0.172549,0.627451,0.172549}%
\pgfsetstrokecolor{currentstroke}%
\pgfsetdash{}{0pt}%
\pgfsys@defobject{currentmarker}{\pgfqpoint{-0.039627in}{-0.033709in}}{\pgfqpoint{0.039627in}{0.041667in}}{%
\pgfpathmoveto{\pgfqpoint{0.000000in}{0.041667in}}%
\pgfpathlineto{\pgfqpoint{-0.009355in}{0.012876in}}%
\pgfpathlineto{\pgfqpoint{-0.039627in}{0.012876in}}%
\pgfpathlineto{\pgfqpoint{-0.015136in}{-0.004918in}}%
\pgfpathlineto{\pgfqpoint{-0.024491in}{-0.033709in}}%
\pgfpathlineto{\pgfqpoint{-0.000000in}{-0.015915in}}%
\pgfpathlineto{\pgfqpoint{0.024491in}{-0.033709in}}%
\pgfpathlineto{\pgfqpoint{0.015136in}{-0.004918in}}%
\pgfpathlineto{\pgfqpoint{0.039627in}{0.012876in}}%
\pgfpathlineto{\pgfqpoint{0.009355in}{0.012876in}}%
\pgfpathclose%
\pgfusepath{stroke,fill}%
}%
\begin{pgfscope}%
\pgfsys@transformshift{0.306000in}{1.498548in}%
\pgfsys@useobject{currentmarker}{}%
\end{pgfscope}%
\begin{pgfscope}%
\pgfsys@transformshift{0.426800in}{1.479495in}%
\pgfsys@useobject{currentmarker}{}%
\end{pgfscope}%
\begin{pgfscope}%
\pgfsys@transformshift{0.668400in}{1.355723in}%
\pgfsys@useobject{currentmarker}{}%
\end{pgfscope}%
\begin{pgfscope}%
\pgfsys@transformshift{0.910000in}{1.060067in}%
\pgfsys@useobject{currentmarker}{}%
\end{pgfscope}%
\begin{pgfscope}%
\pgfsys@transformshift{1.151600in}{0.836509in}%
\pgfsys@useobject{currentmarker}{}%
\end{pgfscope}%
\begin{pgfscope}%
\pgfsys@transformshift{1.272400in}{0.734942in}%
\pgfsys@useobject{currentmarker}{}%
\end{pgfscope}%
\end{pgfscope}%
\begin{pgfscope}%
\pgfpathrectangle{\pgfqpoint{0.306000in}{0.304000in}}{\pgfqpoint{1.026800in}{1.232000in}}%
\pgfusepath{clip}%
\pgfsetbuttcap%
\pgfsetroundjoin%
\pgfsetlinewidth{0.903375pt}%
\definecolor{currentstroke}{rgb}{0.839216,0.152941,0.156863}%
\pgfsetstrokecolor{currentstroke}%
\pgfsetdash{{3.330000pt}{1.440000pt}}{0.000000pt}%
\pgfpathmoveto{\pgfqpoint{0.306000in}{1.331417in}}%
\pgfpathlineto{\pgfqpoint{0.426800in}{1.149956in}}%
\pgfpathlineto{\pgfqpoint{0.668400in}{0.838451in}}%
\pgfpathlineto{\pgfqpoint{0.910000in}{0.613539in}}%
\pgfpathlineto{\pgfqpoint{1.151600in}{0.477185in}}%
\pgfpathlineto{\pgfqpoint{1.272400in}{0.407662in}}%
\pgfusepath{stroke}%
\end{pgfscope}%
\begin{pgfscope}%
\pgfpathrectangle{\pgfqpoint{0.306000in}{0.304000in}}{\pgfqpoint{1.026800in}{1.232000in}}%
\pgfusepath{clip}%
\pgfsetbuttcap%
\pgfsetroundjoin%
\definecolor{currentfill}{rgb}{0.839216,0.152941,0.156863}%
\pgfsetfillcolor{currentfill}%
\pgfsetlinewidth{1.003750pt}%
\definecolor{currentstroke}{rgb}{0.839216,0.152941,0.156863}%
\pgfsetstrokecolor{currentstroke}%
\pgfsetdash{}{0pt}%
\pgfsys@defobject{currentmarker}{\pgfqpoint{-0.041667in}{-0.041667in}}{\pgfqpoint{0.041667in}{0.041667in}}{%
\pgfpathmoveto{\pgfqpoint{-0.041667in}{-0.041667in}}%
\pgfpathlineto{\pgfqpoint{0.041667in}{0.041667in}}%
\pgfpathmoveto{\pgfqpoint{-0.041667in}{0.041667in}}%
\pgfpathlineto{\pgfqpoint{0.041667in}{-0.041667in}}%
\pgfusepath{stroke,fill}%
}%
\begin{pgfscope}%
\pgfsys@transformshift{0.306000in}{1.331417in}%
\pgfsys@useobject{currentmarker}{}%
\end{pgfscope}%
\begin{pgfscope}%
\pgfsys@transformshift{0.426800in}{1.149956in}%
\pgfsys@useobject{currentmarker}{}%
\end{pgfscope}%
\begin{pgfscope}%
\pgfsys@transformshift{0.668400in}{0.838451in}%
\pgfsys@useobject{currentmarker}{}%
\end{pgfscope}%
\begin{pgfscope}%
\pgfsys@transformshift{0.910000in}{0.613539in}%
\pgfsys@useobject{currentmarker}{}%
\end{pgfscope}%
\begin{pgfscope}%
\pgfsys@transformshift{1.151600in}{0.477185in}%
\pgfsys@useobject{currentmarker}{}%
\end{pgfscope}%
\begin{pgfscope}%
\pgfsys@transformshift{1.272400in}{0.407662in}%
\pgfsys@useobject{currentmarker}{}%
\end{pgfscope}%
\end{pgfscope}%
\begin{pgfscope}%
\pgfsetrectcap%
\pgfsetmiterjoin%
\pgfsetlinewidth{0.803000pt}%
\definecolor{currentstroke}{rgb}{0.000000,0.000000,0.000000}%
\pgfsetstrokecolor{currentstroke}%
\pgfsetdash{}{0pt}%
\pgfpathmoveto{\pgfqpoint{0.306000in}{0.304000in}}%
\pgfpathlineto{\pgfqpoint{0.306000in}{1.536000in}}%
\pgfusepath{stroke}%
\end{pgfscope}%
\begin{pgfscope}%
\pgfsetrectcap%
\pgfsetmiterjoin%
\pgfsetlinewidth{0.803000pt}%
\definecolor{currentstroke}{rgb}{0.000000,0.000000,0.000000}%
\pgfsetstrokecolor{currentstroke}%
\pgfsetdash{}{0pt}%
\pgfpathmoveto{\pgfqpoint{1.332800in}{0.304000in}}%
\pgfpathlineto{\pgfqpoint{1.332800in}{1.536000in}}%
\pgfusepath{stroke}%
\end{pgfscope}%
\begin{pgfscope}%
\pgfsetrectcap%
\pgfsetmiterjoin%
\pgfsetlinewidth{0.803000pt}%
\definecolor{currentstroke}{rgb}{0.000000,0.000000,0.000000}%
\pgfsetstrokecolor{currentstroke}%
\pgfsetdash{}{0pt}%
\pgfpathmoveto{\pgfqpoint{0.306000in}{0.304000in}}%
\pgfpathlineto{\pgfqpoint{1.332800in}{0.304000in}}%
\pgfusepath{stroke}%
\end{pgfscope}%
\begin{pgfscope}%
\pgfsetrectcap%
\pgfsetmiterjoin%
\pgfsetlinewidth{0.803000pt}%
\definecolor{currentstroke}{rgb}{0.000000,0.000000,0.000000}%
\pgfsetstrokecolor{currentstroke}%
\pgfsetdash{}{0pt}%
\pgfpathmoveto{\pgfqpoint{0.306000in}{1.536000in}}%
\pgfpathlineto{\pgfqpoint{1.332800in}{1.536000in}}%
\pgfusepath{stroke}%
\end{pgfscope}%
\begin{pgfscope}%
\pgfsetbuttcap%
\pgfsetmiterjoin%
\definecolor{currentfill}{rgb}{1.000000,1.000000,1.000000}%
\pgfsetfillcolor{currentfill}%
\pgfsetfillopacity{0.800000}%
\pgfsetlinewidth{1.003750pt}%
\definecolor{currentstroke}{rgb}{0.800000,0.800000,0.800000}%
\pgfsetstrokecolor{currentstroke}%
\pgfsetstrokeopacity{0.800000}%
\pgfsetdash{}{0pt}%
\pgfpathmoveto{\pgfqpoint{0.364333in}{0.345667in}}%
\pgfpathlineto{\pgfqpoint{0.762171in}{0.345667in}}%
\pgfpathquadraticcurveto{\pgfqpoint{0.778838in}{0.345667in}}{\pgfqpoint{0.778838in}{0.362333in}}%
\pgfpathlineto{\pgfqpoint{0.778838in}{0.934556in}}%
\pgfpathquadraticcurveto{\pgfqpoint{0.778838in}{0.951222in}}{\pgfqpoint{0.762171in}{0.951222in}}%
\pgfpathlineto{\pgfqpoint{0.364333in}{0.951222in}}%
\pgfpathquadraticcurveto{\pgfqpoint{0.347667in}{0.951222in}}{\pgfqpoint{0.347667in}{0.934556in}}%
\pgfpathlineto{\pgfqpoint{0.347667in}{0.362333in}}%
\pgfpathquadraticcurveto{\pgfqpoint{0.347667in}{0.345667in}}{\pgfqpoint{0.364333in}{0.345667in}}%
\pgfpathclose%
\pgfusepath{stroke,fill}%
\end{pgfscope}%
\begin{pgfscope}%
\pgftext[x=0.449056in,y=0.860019in,left,base]{\rmfamily\fontsize{6.000000}{7.200000}\selectfont MLU}%
\end{pgfscope}%
\begin{pgfscope}%
\pgfsetrectcap%
\pgfsetroundjoin%
\pgfsetlinewidth{0.903375pt}%
\definecolor{currentstroke}{rgb}{0.121569,0.466667,0.705882}%
\pgfsetstrokecolor{currentstroke}%
\pgfsetdash{}{0pt}%
\pgfpathmoveto{\pgfqpoint{0.381000in}{0.772982in}}%
\pgfpathlineto{\pgfqpoint{0.547667in}{0.772982in}}%
\pgfusepath{stroke}%
\end{pgfscope}%
\begin{pgfscope}%
\pgfsetbuttcap%
\pgfsetroundjoin%
\definecolor{currentfill}{rgb}{0.121569,0.466667,0.705882}%
\pgfsetfillcolor{currentfill}%
\pgfsetlinewidth{1.003750pt}%
\definecolor{currentstroke}{rgb}{0.121569,0.466667,0.705882}%
\pgfsetstrokecolor{currentstroke}%
\pgfsetdash{}{0pt}%
\pgfsys@defobject{currentmarker}{\pgfqpoint{-0.033333in}{-0.041667in}}{\pgfqpoint{0.033333in}{0.020833in}}{%
\pgfpathmoveto{\pgfqpoint{0.000000in}{0.000000in}}%
\pgfpathlineto{\pgfqpoint{0.000000in}{-0.041667in}}%
\pgfpathmoveto{\pgfqpoint{0.000000in}{0.000000in}}%
\pgfpathlineto{\pgfqpoint{0.033333in}{0.020833in}}%
\pgfpathmoveto{\pgfqpoint{0.000000in}{0.000000in}}%
\pgfpathlineto{\pgfqpoint{-0.033333in}{0.020833in}}%
\pgfusepath{stroke,fill}%
}%
\begin{pgfscope}%
\pgfsys@transformshift{0.464333in}{0.772982in}%
\pgfsys@useobject{currentmarker}{}%
\end{pgfscope}%
\end{pgfscope}%
\begin{pgfscope}%
\pgftext[x=0.614333in,y=0.743815in,left,base]{\rmfamily\fontsize{6.000000}{7.200000}\selectfont 0.2}%
\end{pgfscope}%
\begin{pgfscope}%
\pgfsetbuttcap%
\pgfsetroundjoin%
\pgfsetlinewidth{0.903375pt}%
\definecolor{currentstroke}{rgb}{1.000000,0.498039,0.054902}%
\pgfsetstrokecolor{currentstroke}%
\pgfsetdash{{0.900000pt}{1.485000pt}}{0.000000pt}%
\pgfpathmoveto{\pgfqpoint{0.381000in}{0.656778in}}%
\pgfpathlineto{\pgfqpoint{0.547667in}{0.656778in}}%
\pgfusepath{stroke}%
\end{pgfscope}%
\begin{pgfscope}%
\pgfsetbuttcap%
\pgfsetroundjoin%
\definecolor{currentfill}{rgb}{1.000000,0.498039,0.054902}%
\pgfsetfillcolor{currentfill}%
\pgfsetlinewidth{1.003750pt}%
\definecolor{currentstroke}{rgb}{1.000000,0.498039,0.054902}%
\pgfsetstrokecolor{currentstroke}%
\pgfsetdash{}{0pt}%
\pgfsys@defobject{currentmarker}{\pgfqpoint{-0.041667in}{-0.041667in}}{\pgfqpoint{0.041667in}{0.041667in}}{%
\pgfpathmoveto{\pgfqpoint{-0.041667in}{0.000000in}}%
\pgfpathlineto{\pgfqpoint{0.041667in}{0.000000in}}%
\pgfpathmoveto{\pgfqpoint{0.000000in}{-0.041667in}}%
\pgfpathlineto{\pgfqpoint{0.000000in}{0.041667in}}%
\pgfusepath{stroke,fill}%
}%
\begin{pgfscope}%
\pgfsys@transformshift{0.464333in}{0.656778in}%
\pgfsys@useobject{currentmarker}{}%
\end{pgfscope}%
\end{pgfscope}%
\begin{pgfscope}%
\pgftext[x=0.614333in,y=0.627611in,left,base]{\rmfamily\fontsize{6.000000}{7.200000}\selectfont 0.4}%
\end{pgfscope}%
\begin{pgfscope}%
\pgfsetbuttcap%
\pgfsetroundjoin%
\pgfsetlinewidth{0.903375pt}%
\definecolor{currentstroke}{rgb}{0.172549,0.627451,0.172549}%
\pgfsetstrokecolor{currentstroke}%
\pgfsetdash{{5.760000pt}{1.440000pt}{0.900000pt}{1.440000pt}}{0.000000pt}%
\pgfpathmoveto{\pgfqpoint{0.381000in}{0.540574in}}%
\pgfpathlineto{\pgfqpoint{0.547667in}{0.540574in}}%
\pgfusepath{stroke}%
\end{pgfscope}%
\begin{pgfscope}%
\pgfsetbuttcap%
\pgfsetbeveljoin%
\definecolor{currentfill}{rgb}{0.172549,0.627451,0.172549}%
\pgfsetfillcolor{currentfill}%
\pgfsetlinewidth{1.003750pt}%
\definecolor{currentstroke}{rgb}{0.172549,0.627451,0.172549}%
\pgfsetstrokecolor{currentstroke}%
\pgfsetdash{}{0pt}%
\pgfsys@defobject{currentmarker}{\pgfqpoint{-0.039627in}{-0.033709in}}{\pgfqpoint{0.039627in}{0.041667in}}{%
\pgfpathmoveto{\pgfqpoint{0.000000in}{0.041667in}}%
\pgfpathlineto{\pgfqpoint{-0.009355in}{0.012876in}}%
\pgfpathlineto{\pgfqpoint{-0.039627in}{0.012876in}}%
\pgfpathlineto{\pgfqpoint{-0.015136in}{-0.004918in}}%
\pgfpathlineto{\pgfqpoint{-0.024491in}{-0.033709in}}%
\pgfpathlineto{\pgfqpoint{-0.000000in}{-0.015915in}}%
\pgfpathlineto{\pgfqpoint{0.024491in}{-0.033709in}}%
\pgfpathlineto{\pgfqpoint{0.015136in}{-0.004918in}}%
\pgfpathlineto{\pgfqpoint{0.039627in}{0.012876in}}%
\pgfpathlineto{\pgfqpoint{0.009355in}{0.012876in}}%
\pgfpathclose%
\pgfusepath{stroke,fill}%
}%
\begin{pgfscope}%
\pgfsys@transformshift{0.464333in}{0.540574in}%
\pgfsys@useobject{currentmarker}{}%
\end{pgfscope}%
\end{pgfscope}%
\begin{pgfscope}%
\pgftext[x=0.614333in,y=0.511408in,left,base]{\rmfamily\fontsize{6.000000}{7.200000}\selectfont 0.6}%
\end{pgfscope}%
\begin{pgfscope}%
\pgfsetbuttcap%
\pgfsetroundjoin%
\pgfsetlinewidth{0.903375pt}%
\definecolor{currentstroke}{rgb}{0.839216,0.152941,0.156863}%
\pgfsetstrokecolor{currentstroke}%
\pgfsetdash{{3.330000pt}{1.440000pt}}{0.000000pt}%
\pgfpathmoveto{\pgfqpoint{0.381000in}{0.424370in}}%
\pgfpathlineto{\pgfqpoint{0.547667in}{0.424370in}}%
\pgfusepath{stroke}%
\end{pgfscope}%
\begin{pgfscope}%
\pgfsetbuttcap%
\pgfsetroundjoin%
\definecolor{currentfill}{rgb}{0.839216,0.152941,0.156863}%
\pgfsetfillcolor{currentfill}%
\pgfsetlinewidth{1.003750pt}%
\definecolor{currentstroke}{rgb}{0.839216,0.152941,0.156863}%
\pgfsetstrokecolor{currentstroke}%
\pgfsetdash{}{0pt}%
\pgfsys@defobject{currentmarker}{\pgfqpoint{-0.041667in}{-0.041667in}}{\pgfqpoint{0.041667in}{0.041667in}}{%
\pgfpathmoveto{\pgfqpoint{-0.041667in}{-0.041667in}}%
\pgfpathlineto{\pgfqpoint{0.041667in}{0.041667in}}%
\pgfpathmoveto{\pgfqpoint{-0.041667in}{0.041667in}}%
\pgfpathlineto{\pgfqpoint{0.041667in}{-0.041667in}}%
\pgfusepath{stroke,fill}%
}%
\begin{pgfscope}%
\pgfsys@transformshift{0.464333in}{0.424370in}%
\pgfsys@useobject{currentmarker}{}%
\end{pgfscope}%
\end{pgfscope}%
\begin{pgfscope}%
\pgftext[x=0.614333in,y=0.395204in,left,base]{\rmfamily\fontsize{6.000000}{7.200000}\selectfont 0.8}%
\end{pgfscope}%
\end{pgfpicture}%
\makeatother%
\endgroup%